\documentclass[12pt]{article}
\usepackage{amsfonts,amsthm,amsmath,amssymb,upgreek,bm}

\usepackage{float}
\usepackage{cite}
\usepackage{color}
\usepackage{xcolor}
\definecolor{darkblue}{rgb}{0.1,0.1,.7}
\usepackage[colorlinks=true, linkcolor=darkblue, citecolor=darkblue, urlcolor=darkblue]{hyperref}
\usepackage[paper=letterpaper,margin=.85in]{geometry}
\usepackage{graphicx}
\usepackage{units}
\usepackage[export]{adjustbox}
\usepackage[shortlabels]{enumitem}
\usepackage[mathscr]{euscript}
\usepackage[normalem]{ulem}
\usepackage{tikz}
\usepackage{tikzsymbols}
\usepackage{pgfplots}
\pgfplotsset{compat=1.18}
\usepackage{ytableau}
\usepackage{slashed}
\usepackage{mathtools}
\usepackage{bbold}
\usepackage{comment}
\usepackage{physics}
\usepackage{bbm}
\usepackage{subcaption}
\numberwithin{equation}{section}

\def\ii{\mathrm{i}}
\def\O{\mathcal{O}}
\def\cV{\mathcal V}
\def\cP{\mathcal P}

\def\cH{\mathcal H}

\def\cO{\mathcal O}

\def\cC{\mathcal C}
\def\cF{\mathcal F}
\def\cG{\mathcal G}
\def\cN{\mathcal N}
\def\cU{\mathcal U}

\def\R{\mathbb R}
\def\Z{\mathbb Z}
\def\C{\mathbb C}
\def\T{\mathbb T}
\def\S{\mathbb{S}}
\def\K{\mathbb{K}}
\def\H{\mathbb H}
\def\bE{\mathbb{E}}

\def\p{\partial}
\def\nn{\nonumber}
\DeclareMathOperator{\Cov}{Cov}
\DeclareMathOperator{\diag}{diag}

\newcommand{\inlinefig}[2][4]{%
    \ensuremath{\vcenter{\hbox{\includegraphics[height=#1\fontcharht\font`\B]{#2}}}}%
}

\NewDocumentCommand{\bker}{s m m m}{
    \IfBooleanTF{#1}
        {
            \mathbb{B}^*^{#2}_{#3 #4}
        }
        {
            \mathbb{B}^{#2}_{#3 #4}
        }
}
\NewDocumentCommand{\fker}{s m m g g g g}{
    \IfBooleanTF{#1}
        {
            \IfNoValueTF{#4}
                {\mathbb{F}^*_{#2 #3}}
                {\mathbb{F}^*_{#2 #3}
                {\begin{bmatrix}
                    #4 & #5 \\ 
                    #6 & #7
                \end{bmatrix}}}
        }
        {
            \IfNoValueTF{#4}
                {\mathbb{F}_{#2 #3}}
                {\mathbb{F}_{#2 #3}{\begin{bmatrix}
                    #4 & #5 \\ 
                    #6 & #7
                \end{bmatrix}}}
        }    
}
\NewDocumentCommand{\sker}{s m m g}{
    \IfBooleanTF{#1}
        {
            \IfNoValueTF{#4}
                {\mathbb{S}^*_{#2 #3}}
                {\mathbb{S}^*_{#2 #3}[#4]}
        }
        {
            \IfNoValueTF{#4}
                {\mathbb{S}_{#2 #3}}
                {\mathbb{S}_{#2 #3}[#4]}
        }
}

\newcommand{\appropto}{\mathrel{\vcenter{
  \offinterlineskip\halign{\hfil$##$\cr
    \propto\cr\noalign{\kern2pt}\sim\cr\noalign{\kern-2pt}}}}}

\newcommand{\es}[2] {\begin{equation} \label{#1} \begin{split} #2 \end{split} \end{equation}}

\newcommand{\indentpar}[1]{%
    \hfill\begin{minipage}{\dimexpr\textwidth-3cm}
    #1\par
    \xdef\tpd{\the\prevdepth}
    \end{minipage}\par
}

\definecolor{caribbeangreen}{rgb}{0.0, 0.8, 0.6}

\begin{document}
\thispagestyle{empty}
\vspace*{2.5cm}
\begin{center}
{\bfseries\Large The Multi-trace Eigenstate Thermalization Hypothesis}
\begin{center}
\vspace{1cm}
Ahmed I. Abdalla${}^{1}$, Gabriele Di Ubaldo${}^{1,2}$, Elisa Tabor${}^{3}$\\
\bigskip \rm
\bigskip \small
{${}^1$  Leinweber Institute for Theoretical Physics and Department of Physics,\\ University of California, Berkeley, CA 94720, USA,}
\,\vspace{0.5cm}\\
{${}^2$ RIKEN iTHEMS Center for Interdisciplinary Theoretical and Mathematical Sciences, 2-1 Hirosawa, Wako, Saitama 351-0198 Japan,}
\,\vspace{0.5cm}\\       
{${}^3$ Leinweber Institute for Theoretical Physics, Stanford University, Stanford, CA 94305, USA}
\end{center}
\vspace{1.5cm}
{\bfseries Abstract}
\end{center}
\begin{quotation}
\noindent
We introduce a multi-trace eigenstate thermalization hypothesis, an extension of ETH that governs connected correlations of operator matrix elements with arbitrary index structures, including several disjoint index cycles. We derive multi-trace ETH from local rotational invariance, relate its independent data to higher-order free cumulants, and construct a cyclic matrix model that encodes the resulting multi-cycle data. We provide evidence for multi-trace ETH in chaotic quantum systems and compute multi-trace correlations in low-dimensional gravity. In few-body systems, multi-trace contributions arise from correlated periodic orbits, including a linear ramp contribution
to the one-point spectral form factor. In two dimensions, we compute interaction-induced multi-trace correlations from a quartically coupled bulk scalar in JT gravity, relating winding matter diagrams to corrections of matter-channel weights and anomalous dimensions. In three dimensions, we show that multi-trace non-Gaussianities arise from both topology-changing saddles and light matter. Using Virasoro TQFT and conformal mean-field theory, we compute examples involving knot and link complements and torus wormholes with matter loops. Across these settings, multi-trace ETH provides a common framework for both connected correlators in chaotic quantum systems and multi-boundary wormhole amplitudes in gravity.
\end{quotation}
\setcounter{page}{0}
\setcounter{tocdepth}{2}
\setcounter{footnote}{0}
\newpage
\parskip 0.1in
\setcounter{page}{2}
\begingroup
\setlength{\parskip}{2pt}
\tableofcontents
\endgroup

\section{Introduction}
The interplay between holography and quantum chaos has taught us a great deal about both quantum gravity and chaotic many-body systems. Familiar examples include the behavior of out-of-time-order correlators \cite{Shenker:2013pqa,Hayden:2007cs,Sekino:2008he}, the bound on chaos ~\cite{Maldacena:2015waa}, and the random-matrix statistics of black hole microstates~\cite{Cotler:2016fpe,Saad:2018bqo, Saad:2019lba}. These examples reveal that semiclassical bulk calculations can analytically describe observables which are difficult to address directly in a strongly interacting many-body system and correctly capture universal features present in generic chaotic systems.

The connection between many-body quantum chaos and gravity provides useful insights in both directions. On one hand, the phenomena we observe in many-body physics provide a way to interpret and understand results obtained using the gravitational path integral. For example, a priori puzzling wormhole contributions \cite{Maldacena:2004rf} can be understood through random matrix theory (RMT) \cite{Cotler:2016fpe,Saad:2019pqd,Cotler:2020ugk, Cotler:2020hgz,DiUbaldo:2023qli} and the eigenstate thermalization hypothesis (ETH) for the dual chaotic conformal field theory (CFT). We can then interpret wormholes as providing a precise coarse-grained or statistical description of the microscopic data of the dual CFT~\cite{Saad:2018bqo,Belin:2020hea,Belin:2020jxr}. On the other hand, gravity is a clean laboratory to study quantum chaos since results are often simple analytic functions with manifest structures and symmetries. For example, the semiclassical expansion clearly distinguishes effects of different orders in Newton's constant $G_N$ and contributions from different geometries allow us to isolate and study the physical consequences of corrections that are non-perturbative in $G_N$.

To be more concrete, in the simple case of Jackiw-Teitelboim (JT) gravity, the topological coupling $S_0$ assigns a weight $e^{S_0\chi}$ to a connected surface of Euler characteristic $\chi$~\cite{Saad:2019lba}. This models the entropic expansion in quantum systems, which is weighted by the microcanonical entropy $S(E)$. Gravity can provide predictions for universal phenomena in many-body systems and act as a guide in searching for novel effects. In this broader sense, we are not restricting to a precise holographic duality between specific theories but are instead using simple theories of gravity as a tractable model which captures universal phenomena in many-body systems and CFTs. ETH is one such phenomenon that describes thermalization through the statistical properties of simple operators in the energy eigenbasis~\cite{Srednicki:1994mfb,Rigol:2007juv}; see~\cite{DAlessio:2015qtq, Deutsch:2018ulr, Mori:2017qhg} for reviews. In this context, gravity can give explicit analytic predictions for correlations between matrix elements, including index contractions, entropy scaling, and energy dependence.

In this paper we propose new contributions to the statistics of matrix elements that we name multi-trace ETH. Such contributions are defined by two properties:
\begin{enumerate}
\item They are not captured by standard ETH or by the single-cycle correlations specified by full ETH~\cite{Foini:2018sdb,Pappalardi:2022aaz,Pappalardi:2023nsj}.
\item They contribute at the same leading order in $e^{-S(E)}$ as standard ETH to connected 
multi-trace observables, despite their additional suppression at fixed matrix-element indices.
\end{enumerate}
For certain classes of observables, these corrections appear at leading order. In particular, they describe statistics of operator matrix elements that are relevant to reproducing multi-boundary gravitational amplitudes~\cite{Belin:2021ryy,deBoer:2024mqg}.

We denote such contributions as multi-trace ETH since they appear when considering observables which are products of traces over the Hilbert space, such as
\begin{equation}
\Tr(\O e^{-\beta_1 H})\Tr(\O e^{-\beta_2 H}),
\qquad
\Tr(\O e^{-\beta_1 H}\O e^{-\beta_2 H})\Tr(\O e^{-\beta_3 H}\O e^{-\beta_4 H}),\quad\ldots\,
\end{equation}
Whereas single-trace $n$-point functions probe cyclic products $\O_{i_1i_2}\O_{i_2i_3}\cdots\O_{i_ni_1}$, products of traces instead probe several disjoint cycles and require statistical data beyond these cyclic amplitudes. The simplest observable, which will be our main paradigmatic example, is the product of thermal one-point functions:
\begin{equation}
\Tr(\O e^{-\beta_1 H})\Tr(\O e^{-\beta_2 H})
=\sum_{i,j}\O_{ii}\O_{jj}e^{-\beta_1E_i-\beta_2E_j}.
\end{equation}
For this observable, multi-trace ETH predicts nontrivial statistics between different diagonal elements,
\begin{equation}\label{eq:intro-F11}
    \overline{\O_{ii}\O_{jj}} = e^{-S(E_i)} |f_\O(E_i,0)|^2\delta_{ij} + e^{-S(E_i)-S(E_j)} F_\O^{(1|1)}(E_i|E_j)+\cdots,
\end{equation}
where $f_\O$ is the standard Gaussian ETH function and $F_\O^{(1|1)}$ correlates two separate one-cycles. We summarize in Table~\ref{tab:intro-examples} the physical origins of the explicit multi-trace function $F_\O^{(1|1)}(E_i|E_j)$ in various systems, ranging from matrix models and semiclassical chaotic systems to low-dimensional gravity.

\begin{table}[H]
\centering
\small
\setlength{\tabcolsep}{5pt}
\renewcommand{\arraystretch}{1.14}
\begin{tabular}{@{}p{.235\linewidth}p{.605\linewidth}p{.09\linewidth}@{}}
\hline
\textbf{Example system} & \textbf{What produces the multi-trace correlations?} & \textbf{$F_\O^{(1|1)}$} \\
\hline
\raggedright Randomly rotated operators
& \raggedright A common random change of basis, with fixed operator traces
& \eqref{eq:F11-gri}, \eqref{eq:F11-globalhaar} \\[3pt]
\raggedright Correlated Gaussian energy blocks
& \raggedright Joint fluctuations of the energies in different windows, with Gaussian matrix statistics in each individual energy window
& \eqref{eq:F11-gue} \\[3pt]
\raggedright Non-Gaussian cyclic matrix model
& \raggedright Self-energy loop diagrams 
& \eqref{eq:F11-matrix} \\[3pt]
\raggedright Chaotic few-body systems
& \raggedright Correlated semiclassical periodic orbits in the Gutzwiller trace formula
& \eqref{eq:sc-multitrace-eth} \\[3pt]
\raggedright JT gravity with free matter
& \raggedright Matter loops winding around the throat of a two-boundary wormhole
& \eqref{eq:F11-jkms} \\[3pt]
\raggedright JT gravity with interacting matter
& \raggedright Bulk interactions: general matter, weak interactions, and an explicit $\phi^4$ example
& \eqref{eq:PW-wormhole-paper}, \eqref{eq:weak-F11-paper}, \eqref{eq:F11-interactingJT} \\[3pt]
\raggedright 3d gravity: nontrivial bulk topology
& \raggedright Topologically different ways of joining the boundaries through the bulk
& \eqref{eq:ip_poincare}, \eqref{eq:hopf_contraction} \\[3pt]
\raggedright Holographic CFTs: crossing invariance
& \raggedright Combining vacuum dominance in a particular channel with crossing symmetry
& \eqref{eq:F11-3dgen}, \eqref{eq:sphere_hopf} \\[3pt]
\raggedright 3d gravity: dynamical matter
& \raggedright Dynamical matter in wormhole geometries; near extremality, these multi-trace correlations reduce to the corresponding JT contributions
& \eqref{eq:F11-3dinteracting} \\
\hline
\end{tabular}
\caption{Example systems studied in this paper, the physical origin of their multi-trace correlations and the equation reference to the explicit form of the $F_\O^{(1|1)}$.}
\label{tab:intro-examples}
\end{table}

Considering the one-point spectral form factor obtained by analytically continuing the product of thermal one point functions reveals two key properties of multi-trace ETH correlations: 
\begin{equation}\label{eq:intro-one-point-time}
\begin{split}
K_\O(\beta,T)\simeq{}&
\sum_i e^{-2\beta E_i-S(E_i)}|f_\O(E_i,0)|^2\\
&+\sum_{i,j}e^{-\beta(E_i+E_j)-S(E_i)-S(E_j)}
 e^{-\ii T(E_i-E_j)}F_\O^{(1|1)}(E_i|E_j).
\end{split}
\end{equation}
The first property is that both contributions are of the same order in $e^{S}$: the first Gaussian ETH term has one independent index sum, compensating its $e^{-S}$ suppression, while the multi-trace ETH term has two independent sums, compensating its $e^{-2S}$ suppression. The second property is that the time-dependence of the two terms is qualitatively different: since Gaussian ETH gives a constant in time contribution, the time dependence of the one-point SFF directly probes correlations between different diagonal matrix elements. Moreover, the late-time behavior also depends on correlations between energy levels, which, together with multi-trace ETH, can produce a linear ramp in the SFF. A similar interplay between operator and spectral statistics underlies the late-time behavior of single-trace correlation functions~\cite{Saad:2019pqd,Bouverot-Dupuis:2024nxk}.

These observations raise the more general question of which multi-trace correlations are allowed in a chaotic system. Motivated by the statistical typicality of eigenstates within narrow energy windows~\cite{Pappalardi:2022aaz}, we derive the index structures and entropy scaling of multi-trace ETH correlations from local rotational invariance. We identify the additional multi-trace data with higher-order free cumulants, which isolate correlations not captured by lower-order contributions. This extends the relation between ETH and free probability~\cite{Pappalardi:2022aaz} to fluctuations involving multiple traces~\cite{collins2006secondorderfreenessfluctuations}, and connects to recent work on disjoint-cycle cumulants and microcanonical operator statistics~\cite{Bernard_2024, bernard2025additionstructuredrandommatrices,Belin:2026ssc}. While rotational invariance fixes the index structures, it does not determine the energy dependent coefficients (such as $F_\O^{(1|1)}$), as we will illustrate by computing examples of these functions in various ensembles consistent with rotational invariance.

\subsubsection*{Outline}
The paper is structured as follows. Section~\ref{sec:general} presents the general theoretical basis of multi-trace ETH. Section~\ref{sec:typicality} derives its relation to typicality and higher-order free probability, studies Haar and Gaussian ensembles, and constructs an operator matrix model. We then study few-body chaotic systems via periodic orbit theory in section~\ref{sec:periodic-orbits}, JT gravity with matter and related models in section~\ref{sec:JT}, and 3d gravity and holographic 2d CFTs in section~\ref{sec:3d}. These applications are fairly independent and provide complementary realizations of the general framework.

\section{Multi-trace ETH: general theory} \label{sec:general}
In this section, we develop the general statistical framework underlying multi-trace ETH. We begin with Gaussian ETH, review the full ETH extension to higher-point single-trace correlations, and finally introduce multi-trace ETH to describe connected products of traces.

\subsection{Review of full ETH} \label{subsec:full}
We consider a simple Hermitian operator $\O$ in the energy eigenbasis and subtract its smooth diagonal expectation value, so that the remaining matrix elements have vanishing local mean. The notation $\bE$ will denote an ordinary average and the overline will denote a connected statistical average, both of which can be understood either as an average over a probability distribution or ensemble, or as an empirical average in a narrow microcanonical window. The usual Gaussian ETH ansatz is
\begin{equation} \label{eq:gaussian-eth}
    \O_{ij} = 
    e^{-S(E_{ij})/2}f_\O(E_{ij}\,,\omega_{ij})R_{ij}\,,\qquad E_{ij}=\frac12(E_i+E_j)\,,\qquad \omega_{ij}=E_i-E_j\,,
\end{equation}
where $R_{ij}$ is a unit-variance random matrix satisfying
\begin{equation}
    \overline{R_{ij}} =0\,,\qquad \overline{R_{ij}R_{kl}}=\delta_{il}\delta_{jk}\,.
\end{equation}
Equivalently, the leading connected two-point function is
\begin{equation} \label{eq:gaussian-eth-2pt}
    \overline{\O_{ij} \O_{kl}} = e^{-S(E_{ij})}\abs{f_{\O}(E_{ij},\omega_{ij})}^{2}\delta_{il}\delta_{jk}+\text{subleading}\,,
\end{equation}
where the Kronecker deltas identify the two matrix elements into a single oriented two-cycle, $i\to j\to i$. Gaussian ETH is a sufficient ansatz to describe  thermalization\footnote{For a state narrowly peaked in energy, the diagonal ETH term (subtracted here) gives the thermal value, while off-diagonal contributions dephase and produce only small late-time fluctuations. This basic mechanism of thermalization relies only on the first two matrix-element moments.} 
and two-point correlators, but it does not fix the matrix elements that determine higher-point functions.

In particular, thermal $n$-point functions and OTOCs involve longer closed sequences of transitions between energy eigenstates, such as $i_1\to i_2\to\cdots\to i_n\to i_1$. For example, consider the thermal trace
\begin{equation} \label{eq:single-trace-correlator}
    G_n(\boldsymbol\beta) \equiv \frac{1}{Z}\bE\qty[\Tr\!\left(e^{-\beta_1H}\O e^{-\beta_2H}\O\cdots e^{-\beta_nH}\O\right)] = \frac{1}{Z}\sum_{i_1,\ldots,i_n}e^{-\sum_{a=1}^{n}\beta_aE_{i_a}}\bE[\O_{i_1i_2}\cdots\O_{i_ni_1}]\,,
\end{equation}
where $Z=\Tr\qty(e^{-\sum_{a=1}^n\beta_aH})$, and the sums run over every assignment of the indices, including configurations in which some of them coincide. Full ETH \cite{Foini:2018sdb} extends the Gaussian ansatz by assigning a connected statistical amplitude to each closed path with mutually distinct indices:
\begin{equation}
    \overline{\O_{i_1i_2} \O_{i_2i_3}\cdots \O_{i_ni_1}} = e^{-(n-1)S\qty(E_{\rm avg})}F_\O^{(n)}(E_{i_1},\ldots,E_{i_n}) + \text{subleading}
    \label{eq:full-eth}
\end{equation}
where $E_{\rm avg}=\frac{1}{n}\sum_{a=1}^nE_{i_a}$. For $n=2$, this recovers the Gaussian ETH function:
\begin{equation}
    F_\O^{(2)}(E_{i},E_{j})=|f_\O(E_{ij},\omega_{ij})|^2\,.
\end{equation}
The entropy factor in \eqref{eq:full-eth} is fixed by the requirement that the corresponding thermal trace \eqref{eq:single-trace-correlator} be of order one: the $n$ independent energy sums contribute $e^{nS}$, and dividing by $Z\sim e^S$ leaves $e^{(n-1)S}$.

When two indices coincide, such as for example if $i_k=i_1$ while all other indices remain mutually distinct, then the original $n$-cycle splits into two closed cycles that share the state $i_1$. At leading order in the entropy expansion, the ordinary statistical average of these two cycles factorizes,
\begin{equation} \label{eq:full-factorization}
    \bE[\O_{i_1i_2} \cdots \O_{i_{k-1}i_1}\O_{i_1i_{k+1}}\cdots O_{i_ni_1}] = \overline{\O_{i_1i_2}\cdots \O_{i_{k-1}i_1}}~\overline{\O_{i_1i_{k+1}}\cdots \O_{i_ni_1}} + \text{subleading} \,.
\end{equation}
At leading order, the surviving coincidence patterns are the noncrossing, or cactus, contractions that maximize the number of independent energy sums \cite{Foini:2018sdb}. This is the origin of the first-order free-probability organization \cite{Pappalardi:2022aaz, Vallini:2025vvq, Pappalardi:2023nsj} reviewed in subsection \ref{subsec:free-prob}.

\subsection{Multi-trace ETH} \label{subsec:MT}
Full ETH describes fluctuations accumulated along a single closed cycle of energy eigenstates. Products of traces instead involve several such cycles, which can be correlated even without sharing an eigenstate index. For example, $\O_{ii}$ and $\O_{jj}$ may fluctuate together even for $i\neq j$. Multi-trace ETH captures the additional data needed to describe these correlations. Let $C_a$ be $b$ disjoint cycles of lengths $\ell_a$, and define
\begin{equation}
    \boldsymbol{\O}_{\ell} \equiv \O_{i_{1}i_{2}}\O_{i_{2}i_{3}}\cdots \O_{i_{\ell}i_{1}} \,,\qquad N \equiv \sum_{a=1}^b \ell_a\,,
\end{equation}
where $N$ denotes the total number of operator insertions and the number of cycles $b$ is equal to the number of traces.\footnote{We denote the number of traces by $b$ because it corresponds holographically to the number of boundaries of spacetime.} The multi-trace ETH conjectures that matrix elements of a connected $b$-trace observable with all indices distinct are given by
\begin{equation}\label{eq:MT}
    \overline{\boldsymbol{\O}_{\ell_1}\cdots \boldsymbol{\O}_{\ell_b}} = e^{-(N+b-2) S_{b}(\mathbf{E})} F_\O^{(\ell_1|\cdots|\ell_b)}\qty(\mathbf{E}_{\ell_1}|\ldots|\mathbf{E}_{\ell_b}) + \text{subleading}\,,
\end{equation}
where
\begin{equation}
    S_{b}(\mathbf{E}) \equiv\sum_{a=1}^b \frac{\ell_a}{N}S\qty(\frac{E_{i_{a,1}}+\cdots+E_{i_{a,\ell_a}}}{\ell_a}) \,.
\end{equation}
Excluding additional identifications between indices in each $\boldsymbol{\O}_\ell$ removes contact contributions inherited from fewer than $b$ cycles. The $b=1$ case of~\eqref{eq:MT} reduces to full ETH~\eqref{eq:full-eth}, with the single-trace cycle denoted by $\overline{\boldsymbol{\O}_n}$ and the connected amplitude denoted by $F_\O^{(\ell_1)}\qty(\mathbf{E}_{\ell_1})$.

The scaling in \eqref{eq:MT} has the same state-counting origin as for full ETH. Summing the $N$ independent indices gives $e^{NS_b}$, so a connected planar correlator of $b$ unnormalized traces scales as
\begin{equation}
    e^{NS_b}\,e^{-(N+b-2)S_b}=e^{(2-b)S_b}\,.
\end{equation}
The exponent $2-b$ is the Euler characteristic of a genus-zero surface with $b$ boundaries, matching the planar counting in the gravitational examples below.

We now consider the simplest example of multi-trace ETH, which will reappear throughout the paper: the connected correlation of two matrix elements. For arbitrary external indices, it is given by
\begin{equation}\label{eq:two-point-MT}
    \overline{\O_{ij}\O_{kl}} = e^{-S(E_{ij})}\abs{f_{\O}(E_{ij},\omega_{ij})}^{2}\delta_{il}\delta_{jk} + e^{-S(E_i)-S(E_k)}F_{\O}^{(1|1)}(E_i|E_k)\delta_{ij}\delta_{kl}+\text{subleading}\,,
\end{equation}
where the first term joins the two insertions in a single cycle $i\to j\to i$ and is fixed by Gaussian/full ETH, while the second term instead correlates two separate closed paths and is the simplest multi-trace ETH function. In particular, it can be nonzero for $i\neq l$, for which the full-ETH contact term vanishes.

Applying \eqref{eq:two-point-MT} to the connected square of a thermal one-point function gives
\begin{equation}\label{eq:mtt_eth_oo}
\begin{split}
    \overline{\Tr(e^{-\beta_1 H}\O) \Tr(e^{-\beta_2 H}\O)}
    &= \sum_i e^{-(\beta_1+\beta_2)E_i}e^{-S(E_i)}\abs{f_{\O}(E_i,0)}^{2}\,\\
    &\qquad + \sum_{i,j}e^{-\beta_1E_i-\beta_2E_j}e^{-S(E_i)-S(E_j)}F_{\O}^{(1|1)}(E_i|E_j)+\text{subleading}\,.
\end{split}
\end{equation}
The first term identifies $i=j$, leaving one independent sum and contributing $e^S e^{-S}\sim 1$. The multi-trace term leaves both indices independent, giving $e^{2S}e^{-2S}\sim 1$. Thus, despite its additional suppression at fixed indices, the multi-trace term contributes at the same leading order after summing over energies.

\subsection{RMT and ETH together: correlator spectral form factors} \label{subsec:rmt-eth}

Random matrix theory and ETH describe complementary aspects of quantum
systems. RMT describes the universal correlations of nearby energy levels in chaotic systems, whereas
full-ETH supplies the operator weights with which those levels enter a correlation
function. Their combination gives a useful and precise description of the late-time dynamics of thermal two-point functions,
as developed in JT gravity \cite{Saad:2019pqd,Iliesiu:2021ari}
and in many-body systems \cite{Bouverot-Dupuis:2024nxk}.

For a product of one-point traces, the operator data are different: the relevant
quantity is the covariance of two diagonal matrix elements, rather than the
variance of an off-diagonal element. This makes the one-point spectral form
factor a direct probe of the multi-trace ETH correlations for two separate one-cycles encoded by
$F_\O^{(1|1)}$.
We take $\O$ to be a Hermitian operator and subtract its smooth diagonal
one-point function $\bE_{\O|H}[\O_{ii}]=0$ at fixed spectrum. This centering removes the
contribution of the smooth operator mean, which would
otherwise also enter the connected observable. 
The one-point SFF $K_\O(\beta_1,\beta_2)$ is defined as:
\begin{equation}\label{eq:opsff-def}
\begin{split}
 K_\O(\beta_1,\beta_2)
 &\equiv\overline{Z_\O(\beta_1)Z_\O(\beta_2)}
 =\bE\!\left[\sum_{i,j}e^{-\beta_1E_i-\beta_2E_j}\O_{ii}\O_{jj}\right]\\
 Z_\O(\beta)&=\Tr(\O e^{-\beta H})=\sum_i e^{-\beta E_i}\O_{ii}, \qquad \bE[Z_\O]=0.
\end{split}
\end{equation}
The last equality follows from the vanishing conditional mean. On continuing
to $\beta_{1,2}=\beta\pm\mathrm iT$, we write
$K_\O(\beta,T)=\bE[|Z_\O(\beta+\mathrm iT)|^2]$.

Here and throughout the paper it will be useful to introduce continuum notation by first writing observables in terms of the exact density of states $\rho(E)$ and the density $d_\O(E)$ with an operator insertion,
\begin{equation}\label{eq:opsff-density}
 \rho(E)=\sum_i\delta(E-E_i),\qquad
 \bar\rho(E)=\bE[\rho(E)],\qquad
 d_\O(E)=\sum_i\O_{ii}\delta(E-E_i),
\end{equation}
where $\bar\rho$ denotes the smooth, mean density. We will be agnostic about the smoothing procedure here, it can be  either coarse-graining over a microcanonical window for an individual system or averaging over disorder realizations or a large $N$ approximation for example.
The discrete sum is then exactly rewritten as:
\begin{equation}\label{eq:opsff-exact-continuum}
 K_\O(\beta_1,\beta_2)
 =\int dE_1dE_2\,e^{-\beta_1E_1-\beta_2E_2}
       \bE[\rho(E_1)\rho(E_2)\O_{E_1E_1}\O_{E_2E_2}].
\end{equation}
The notation $\O_{EE}$ denotes the diagonal matrix element at an energy $E$.
We can factorize the spectral and operator statistics by a conditional average
\begin{equation}\label{eq:opsff-conditional-factorization}
 \bE[\rho(E_1)\rho(E_2)\O_{E_1E_1}\O_{E_2E_2}]
 =\bE[\rho(E_1)\rho(E_2)]\,
       \overline{\O_{E_1E_1}\O_{E_2E_2}}
\end{equation}
where $\overline{\O_{E_1E_1}\O_{E_2E_2}}$ denotes the operator statistics conditioned on the energy levels $E_1,E_2$. 

The continuum notation is useful as long as we correctly account for the contact terms coming from coincident energies $E_1=E_2$. We can first select them in the discrete case, then pass to the continuum:
\es{}{
\sum_{i} e^{-(\beta_1+\beta_2)E_i} O_{ii}^2= \int dE e^{-(\beta_1+\beta_2)E}\rho(E)O_{EE}^2\,.
}
The average can be taken in either continuum or discrete yielding identical results: 
\begin{align}
\sum_{i} e^{-(\beta_1+\beta_2)E_i}\,  \bE[O_{ii}^2]&= \sum_i e^{-(\beta_1+\beta_2)E_i} \, e^{-S(E_i)}|f_{\O}(E_i,0)|^2 \label{eq:rmt-eth-discrete}\\
\int dE e^{-(\beta_1+\beta_2)E}\rho(E)\bE[O_{EE}^2]&=\int dE e^{-(\beta_1+\beta_2)E}|f_{\O}(E,0)|^2. \label{eq:rmt-eth-cont}
\end{align}
Matching discrete and continuum is equivalent to the following Kronecker delta to Dirac delta substitution:
\begin{equation}\label{eq:opsff-delta-dictionary}
 \delta_{ij}\longrightarrow\frac{\delta(E_1-E_2)}{\bar\rho(E_1)},\qquad \sum_{j}h(E_j)\delta_{ij}=
 \int dE_2\,\bar\rho(E_2)
 \frac{\delta(E_i-E_2)}{\bar\rho(E_2)}h(E_2)=h(E_i).
\end{equation}
The matrix element correlations in~\eqref{eq:rmt-eth-discrete} and~\eqref{eq:rmt-eth-cont} have contributions from both Gaussian and multi-trace ETH,
\begin{equation}\label{eq:opsff-conditional-eth}
 \overline{\O_{ii}\O_{jj}}
 =e^{-S(E_i)}|f_\O(E_i,0)|^2\delta_{ij}
 +e^{-S(E_i)-S(E_j)}F_\O^{(1|1)}(E_i|E_j)+\cdots.
\end{equation}
We thus isolate the contact terms and define the two-point correlation of distinct energy levels:
\begin{equation}\label{eq:opsff-distinct-level-density}
\begin{split}
 \rho_2^{\ne}(E_1,E_2)
 &\equiv\bE\!\left[\sum_{i\ne j}
       \delta(E_1-E_i)\delta(E_2-E_j)\right]\\
 &=\bE[\rho(E_1)\rho(E_2)]
       -\bar\rho(E_1)\delta(E_1-E_2).
\end{split}
\end{equation}
The joint RMT and operator statistics are then given by:
\begin{equation}\label{eq:opsff-weighted-decomposition}
\begin{split}
\bE[\rho(E_1)\rho(E_2)\O_{E_1E_1}\O_{E_2E_2}]\simeq{}&
 \left[|f_\O(E_1,0)|^2
       +e^{-S(E_1)}F_\O^{(1|1)}(E_1|E_1)\right]\delta(E_1-E_2)\\
 &+\frac{\rho_2^{\ne}(E_1,E_2)}{\bar\rho(E_1)\bar\rho(E_2)}
       F_\O^{(1|1)}(E_1|E_2).
\end{split}
\end{equation}
where the first line is the contact term, including the Gaussian ETH and the diagonal part of the multi-trace ETH term, and the second line gives the combined RMT and multi-trace ETH statistics. Note that the multi-trace ETH contribution is suppressed only in the contact term but is leading for the case of distinct energies $E_1\neq E_2$ where the Gaussian ETH vanishes. 

Combining the two terms of \eqref{eq:opsff-weighted-decomposition} we obtain
the compact expression
\begin{equation}\label{eq:opsff-combined-master}
\begin{split}
 K_\O(\beta_1,\beta_2)\simeq{}&
 \int dE\,e^{-(\beta_1+\beta_2)E}|f_\O(E,0)|^2\\
 &+\int dE_1dE_2\,e^{-\beta_1E_1-\beta_2E_2}
       F_\O^{(1|1)}(E_1|E_2)\frac{\bE[\rho(E_1)\rho(E_2)]}{\bar{\rho}(E_1)\bar{\rho}(E_2)}.
\end{split}
\end{equation}
The second line contains the \emph{full} spectral two-point function:
\begin{equation}\label{eq:opsff-sine-kernel}
\begin{split}
 \bE[\rho(E_1)\rho(E_2)]
 &=\bar\rho(E_1)\bar\rho(E_2)+\overline{\rho(E_1)\rho(E_2)}\\
 &\simeq\bar\rho(E_1)\bar\rho(E_2)+\bar\rho(E)\delta(\omega)
 -\left[\frac{\sin(\pi\bar\rho(E)\omega)}{\pi\omega}\right]^2\,,
\end{split}
\end{equation}
where in the second line we substituted the sine-kernel for the connected spectral correlator. This term is universal to the statistics of random matrices \cite{Saad:2019lba}.

Substituting~\eqref{eq:opsff-sine-kernel} into the one-point SFF \eqref{eq:opsff-combined-master} gives
\begin{equation}\label{eq:opsff-explicit-rmt-eth}
\begin{split}
 K_\O(\beta,T)&\simeq{}
 \int dE\,e^{-2\beta E}|f_\O(E,0)|^2+\int dE\,e^{-2\beta E-S(E)}F_\O^{(1|1)}(E|E)\\
 &+\int dE_1dE_2\,e^{-\beta(E_1+E_2)-\mathrm iT(E_1-E_2)}
       F_\O^{(1|1)}(E_1|E_2)\\
 &-\int dE\,d\omega\,
 e^{-2\beta E-\mathrm iT\omega-S(E+\frac{\omega}{2})-S(E-\frac{\omega}{2})}F_\O^{(1|1)}\qty(E+\frac{\omega}{2}|E-\frac{\omega}{2})\left(\frac{\sin(\pi\bar\rho(E)\omega)}{\pi\omega}\right)^2.
\end{split}
\end{equation}
The first line includes the contact terms in the first line of \eqref{eq:opsff-weighted-decomposition}. The second line comes from the disconnected $\bar{\rho}(E_1)\bar{\rho}(E_2)$ piece and the last term includes both multi-trace ETH and RMT statistics. We have kept explicitly all the factors of $e^{S(E)}=\bar{\rho}(E)$ so it is manifest which terms are suppressed by simply counting the factors. 
The leading $\mathcal{O}(1)$ terms are the contact Gaussian ETH term and the multi-trace ETH contribution in the second line. Even though the multi-trace matrix element fluctuations are suppressed, in physical observables they are enhanced and become $\mathcal{O}(1)$ since we must sum over different energies. Thus in the one-point SFF, multi-trace ETH is not a neglectable correction to ETH, but rather it contributes at the same order. In other words, Gaussian ETH is generally an incorrect approximation for the one-point SFF, even at leading order in $e^{S}$.\footnote{We will see that in some specific models, the multi-trace ETH contribution can be suppressed in a different parameter than $e^S$, such as $\hbar$ in a loop expansion.}

The spectral RMT statistics contribute at first subleading order, just as for the standard SFF, and are only probed in the presence of multi-trace ETH. Conversely, the presence of RMT statistics in the one-point SFF is then a clear signal of multi-trace ETH and we will use it in subsection \ref{subsec:sc-extracting-multitrace} to extract $F_\O^{(1|1)}$. We can obtain the time dependence of the one-point SFF by evaluating the sine kernel integral with the smooth weight $F_\O^{(1|1)}$:
\begin{equation}\label{eq:opsff-final}
\begin{aligned}
 K_\O(\beta,T)\simeq{}&
 \int dE\,e^{-2\beta E}|f_\O(E,0)|^2\\
 &+\int dE_1dE_2\,e^{-\beta(E_1+E_2)-\mathrm iT(E_1-E_2)}
       F_\O^{(1|1)}(E_1|E_2)\\
 &+\int dE\,e^{-2\beta E-2S(E)}F_\O^{(1|1)}(E|E)
       \min\!\left(\frac{|T|}{2\pi},\bar\rho(E)\right).
 \end{aligned}
\end{equation}
At fixed energy we have:
\begin{equation}
\begin{aligned}
 K_\O(E,T)\simeq{}&
 |f_\O(E,0)|^2\\
 &+\int d\omega \,e^{-\mathrm iT\omega}
       F_\O^{(1|1)}(E-\omega/2|E+\omega/2)\\
 &+e^{-2S(E)}F_\O^{(1|1)}(E|E)
       \min\!\left(\frac{|T|}{2\pi},\bar\rho(E)\right).
 \end{aligned}
\end{equation}
The different contributions can be understood in terms of their time dependence:
\begin{itemize}
    \item \textbf{Gaussian ETH contact term.} Produces a completely time-independent $\mathcal{O}(1)$ contribution since time dependence arises from different energies. This makes time-dependence of the one-point SFF a smoking gun signature of multi-trace ETH. 
    \item \textbf{Multi-trace ETH term.} Since $F_\O^{(1|1)}$ is a smooth function of the energy difference $\omega$, its Fourier transform will produce a decaying function of time, analogously to the slope in the standard SFF.
    \item \textbf{RMT $\times$ multi-trace ETH.} The time dependence is the same as the standard SFF, now weighted by $F_\O^{(1|1)}$. This term is analogous to late time behavior of the thermal two-point function, in which case it is weighted by the ETH function $|f_\O(E,0)|^2$\cite{Saad:2019pqd,Bouverot-Dupuis:2024nxk}. Similar mixed ETH-RMT contributions in the OTOC have been studied in ~\cite{Garcia-Garcia:2026nlo}. 
\end{itemize}

The unitary class provides the simplest illustration, but the spectral
part of the argument applies in other symmetry classes. The multi-trace ETH $\times$ RMT contribution for a general symmetry class is 
\begin{equation}\label{eq:opsff-general-symmetry}
 \int dE\,e^{-2\beta E-S(E)}F_\O^{(1|1)}(E|E)K(T;E).
\end{equation}
For GUE, $K_{\beta}(T;E)=\min(\tau,1)$, reproducing the last
line of~\eqref{eq:opsff-final}. For  time-reversal invariant systems
in the orthogonal GOE class,
\begin{equation}\label{eq:opsff-goe-form-factor}
 K_{\rm GOE}(\tau)=2\tau-\tau\log(1+2\tau)
 =2\tau-2\tau^2+O(\tau^3),\qquad 0<\tau<1.
\end{equation}

\section{From typicality to multi-trace fluctuations}
\label{sec:typicality}

In chaotic many-body systems, simple observables are expected to treat nearby eigenstates as statistically interchangeable.\footnote{This is referred to as typicality in \cite{Foini:2018sdb, deBoer:2024mqg}, local typicality in \cite{Blommaert:2020seb}, and rotational invariance in \cite{Foini:2019tii, Pappalardi:2022aaz, Vallini:2025vvq}.} In this section, we use this notion of typicality to study the correlations between disjoint index cycles that we introduced in subsection \ref{subsec:MT}. It will be useful to start from the stronger assumption of global rotational invariance, in which we Haar randomize all basis directions, which erases any energy dependence. We will then weaken this assumption to an independent rotational invariance in each energy window, in which only nearby energy levels are treated as statistically interchangeable and where we do not impose the same statistics across different energy windows \cite{Foini:2019tii}.

We will consider what this rotational typicality implies for trace observables and their joint fluctuations, quantities which at large entropy are naturally organized by free probability.\footnote{Free probability organizes the large-$N$ statistics of noncommuting observables and the planar diagrammatics of matrix models \cite{Brezin:1977sv, Gopakumar:1994iq}. Similar ideas have appeared in studies of replica wormholes, double-scaled SYK and gravitational scrambling \cite{Penington:2019kki, Wang:2022ots, Penington:2025hrc, Berkooz:2018jqr, Wu:2023kin, Espindola:2026xuc, Espindola:2026qfo}.} Whereas first-order free probability organizes the planar contractions of single-trace observables by noncrossing partitions (see e.g. \cite{Voiculescu_1992, Speicher1994, Nica_Speicher_2006}), higher-order free probability describes the planar contractions for the fluctuations of single-trace observables using a multi-boundary generalization of noncrossing partitions \cite{mingo2003annularnoncrossingpermutationspartitions, collins2006secondorderfreenessfluctuations, Mingo_2006, Mingo_2007, Mingo_2017}. We will see that higher-order cumulants naturally encode connected products of traces, and, in the present setting, the independent functions of multi-trace ETH.

In subsection \ref{subsec:free-prob}, we briefly review first- and higher-order free probability, and we identify their free cumulants with the single- and multi-cycle data of ETH. In subsection \ref{subsec:global}, we review global rotational invariance and explicitly compute the second-order moments and free cumulants in fixed-spectrum and Gaussian ensembles. In subsection \ref{subsec:local}, we review local rotational invariance and show how the global free cumulants become energy-dependent ETH functions in several examples of ensembles. Finally, in subsection \ref{subsec:matrix-model}, we describe these local statistics with a matrix model in which the same elementary cyclic interactions generate both single-trace correlations and connected fluctuations of several traces.

\subsection{Higher-order free probability theory} \label{subsec:free-prob}

The matrix-element amplitudes introduced in subsections \ref{subsec:full} and \ref{subsec:MT} have a natural interpretation in first-order \cite{Voiculescu_1992, Speicher1994, Nica_Speicher_2006} and higher-order \cite{mingo2003annularnoncrossingpermutationspartitions, collins2006secondorderfreenessfluctuations, Mingo_2006, Mingo_2007, Mingo_2017} free probability theory, respectively. We will see that first-order free cumulants isolate the single-cycle contributions to a single trace, while higher-order free cumulants isolate the additional data in connected products of traces, after contributions assembled from lower-order cumulants have been removed. We first explain this distinction in an energy-independent microcanonical shell, where all entropy factors are powers of $L=e^S$. For a given operator $\O$, the single-trace and $b$-trace moments are
\begin{equation}
    m_n \equiv \lim_{L\to\infty} \frac{1}{L}\,\bE[\Tr\O^n]\,,\qquad M_{\ell_1,\ldots,\ell_b}^{(L)} \equiv L^{b-2}~ \bE[\Tr\O^{\ell_1}\cdots \Tr\O^{\ell_b}] \,,
\label{eq:moments}
\end{equation}
where $L$ denotes the dimension of the Hilbert space and $\bE$ denotes the average with respect to some probability measure.\footnote{For $b>1$, the $b$-trace moment $M^{(L)}_{\ell_1,\ldots,\ell_b}$ includes disconnected products of lower-order traces that grow with larger positive powers of $L$. We could normalize the full $b$-trace moment by enough powers of $L$ to obtain a finite factorized limit, but this would suppress the connected fluctuations that are relevant for higher-order freeness.} For $b=1$, this simplifies to $m_{\ell_1}=\lim_{L\to\infty} M_{\ell_1}^{(L)}$. Whereas moments as defined by~\eqref{eq:moments} contain both connected data and products of lower-point data, we will be mostly working with cumulants, which are built only from connected data.

More precisely, in the large $L$ limit, the single-trace moments $m_n$ can be reconstructed either by summing classical cumulants over all partitions $\mathcal{P}(n)$, or by summing free cumulants over only the noncrossing partitions $NC(n)$,
\begin{equation}
    m_n = \sum_{\pi\in \mathcal{P}(n)} \prod_{B\in\pi} c_{|B|} = \sum_{\pi\in NC(n)} \prod_{B\in\pi} \kappa_{|B|}^{(1)} \,,
\label{eq:first-order-free-moment-cumulant}
\end{equation}
where $B$ denotes a block of the partition $\pi$ and $|B|$ counts the number of elements in a block $B$. This formula can be inverted to reproduce either the first-order classical cumulants $c_{n}$, which correspond to the connected part of the moments $m_n$, or the first-order free cumulants $\kappa_{n}^{(1)}$, which capture the planar noncrossing contributions of $m_n$.

In higher-order free probability, it ends up being convenient to define the connected higher-order moments\footnote{The higher-order ``moments'' $\alpha_{\ell_1,\ldots,\ell_b}$ are already connected (i.e. act as cumulants) in the ordinary probabilistic sense, but can still contain planar contributions assembled from lower-order cyclic data, so they act as moments in the higher-order sense. For $b=1$, they reduce to ordinary normalized single-trace moments $\alpha_{\ell_1} = m_{\ell_1}$.} 
\begin{equation}\label{eq:alpha-def}
    \alpha_{\ell_1,\ldots,\ell_b}^{(L)} \equiv L^{b-2}\,  \frac{\p^b}{\p t_1\ldots\p t_b}\log \left.\bE\qty[\exp\qty(\sum_{a=1}^bt_a\Tr\O^{\ell_a})] \right|_{\mathbf{t}=0}\,,\quad \alpha_{\ell_1,\ldots,\ell_b} = \lim_{L\to\infty}\alpha_{\ell_1,\ldots,\ell_b}^{(L)}\,,
\end{equation}
where the scaling is meant to cancel the $L^{2-b}$ scaling of the connected planar $b$-trace contribution in matrix ensembles.\footnote{We assume that the scaled trace cumulants defined in this section have finite large-$L$ limits. This is an assumption about the fluctuation scaling of the ensemble, rather than a consequence of rotational invariance alone.} For example for $b=1$, this simplifies to $\alpha_{\ell_1}=m_{\ell_1}$, and for $b=2$,
\begin{equation}\label{eq:alpha-b2}
    \alpha_{\ell_1,\ell_2}^{(L)} = \bE[\Tr\O^{\ell_1}\Tr\O^{\ell_2}] - \bE[\Tr\O^{\ell_1}]~\bE[\Tr\O^{\ell_2}] \equiv \Cov\qty(\Tr\O^{\ell_1}, \Tr\O^{\ell_2})\,.
\end{equation}
Then the unconnected $b$-trace moments can be reconstructed by summing connected higher-order moments over all partitions $\mathcal{P}(b)$,
\begin{equation}
    M_{\ell_1,\ldots,\ell_b}^{(L)} = \sum_{\pi\in\mathcal P(b)} L^{2|\pi|-2} \prod_{B\in\pi} \alpha_{\{\ell_a\}_{a\in B}}^{(L)} \,,
\end{equation}
where $|\pi|$ denotes the number of blocks in $\pi$. For $b=1$, this simplifies to $M_{\ell_1}^{(L)}=\alpha_{\ell_1}^{(L)}$. For $b=2$,
\begin{equation} \label{eq:M-alpha}
    M_{\ell_1,\ell_2}^{(L)} = L^2\alpha_{\ell_1}^{(L)}\alpha_{\ell_2}^{(L)} + \alpha_{\ell_1,\ell_2}^{(L)} 
\end{equation}
where the left-hand side is simply $\bE[\Tr\O^{\ell_1}\Tr\O^{\ell_2}]$ and the right-hand side is $\bE[\Tr\O^{\ell_1}]~\bE[\Tr\O^{\ell_2}] + \Cov\qty(\Tr\O^{\ell_1}, \Tr\O^{\ell_2})$, so we see~\eqref{eq:M-alpha} is equivalent to~\eqref{eq:alpha-b2}.

The higher-order free cumulants are defined by subtracting planar sewings of lower-order cumulants from \eqref{eq:alpha-def}, or equivalently by inverting
\begin{equation} \label{eq:PS-NS}
    \alpha_{\ell_1,\ldots,\ell_b} = \sum_{\Theta\in \mathcal{PS}_{NC}(\ell_1,\ldots,\ell_b)}\kappa_{\Theta}\,.
\end{equation}
Here $\mathcal{PS}_{NC}(\ell_1,\ldots,\ell_b)$ denotes the genus-zero noncrossing partitioned permutations compatible with $b$ boundary cycles of lengths $\ell_1,\ldots,\ell_b$ with $N=\sum_a \ell_a$ \cite{collins2006secondorderfreenessfluctuations}.\footnote{For $b>1$, a noncrossing partitioned permutation contains more information than the set partition formed by its cycles, since it also records their cyclic order relative to the different boundaries \cite{mingo2003annularnoncrossingpermutationspartitions}.} A partitioned permutation $\Theta\in \mathcal{PS}_{NC}(\ell_1,\ldots,\ell_b)$ is a pair $(\cV,\pi)$ consisting of two layers of information. The first is an ordinary permutation $\pi\in S_N$ that specifies how the operator insertions are wired into oriented cyclic components. The second is a set partition $\cV\in\cP(N)$ that can group cycles together but not split them apart, and that specifies which of the cycles in $\pi$ are grouped into the same connected cumulant. Noncrossing compatibility is imposed relative to the external boundary permutation $\gamma$.

\begin{figure}
    \centering
    \vspace{-1.5cm}
    \includegraphics[width=\linewidth]{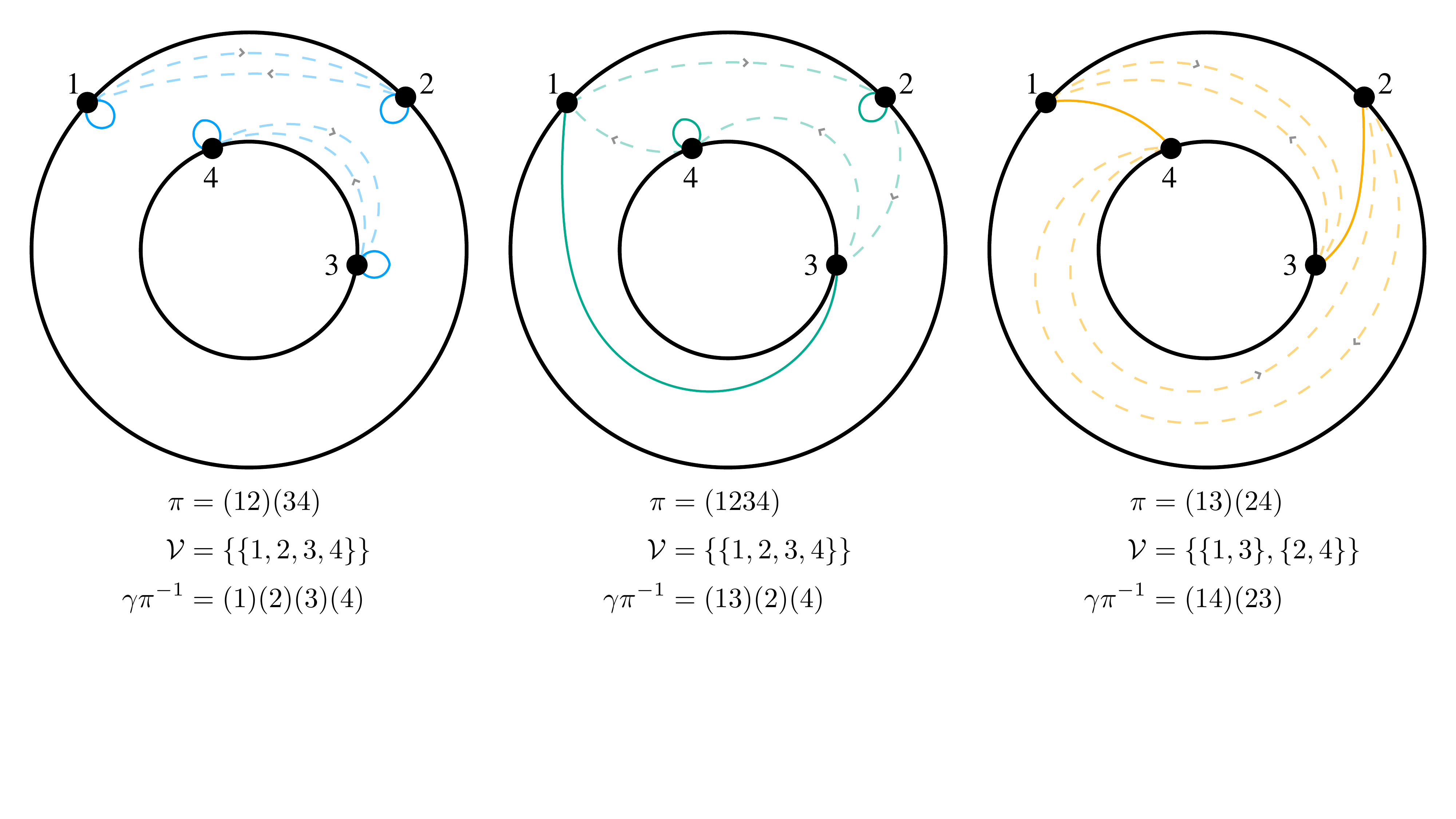}
    \vspace{-3.cm}
    \caption{Annular noncrossing decomposition $\alpha_{2,2} = \kappa^{(2)}_{2,2} + 4\kappa^{(1)}_4 + 2(\kappa^{(1)}_2)^2$ \cite{mingo2003annularnoncrossingpermutationspartitions}. The black circles correspond to two trace boundaries with permutation $\gamma=(12)(34)$, numbered points label the insertions $\O_{i_r i_{\gamma(r)}}$, dashed arrows label the permutation $\pi$, and solid colored curves show the identifications of energies encoded by $\gamma\pi^{-1}$. \textit{Left:} the two cycles of $\pi$ belong to the same $\cV$, giving $\kappa_{2,2}^{(2)}$ with four independent energy sums. \textit{Middle:} one of the four annular cyclic permutations giving $\kappa_4^{(1)}$, with $i_1=i_3$ leaving three independent sums. \textit{Right:} one of the two cross-boundary pairings giving $(\kappa_2^{(1)})^2$, with $i_1=i_4$ and $i_2=i_3$ leaving two independent sums.}
    \label{fig:annular}

    \vspace{.5cm}
    \includegraphics[width=\linewidth]{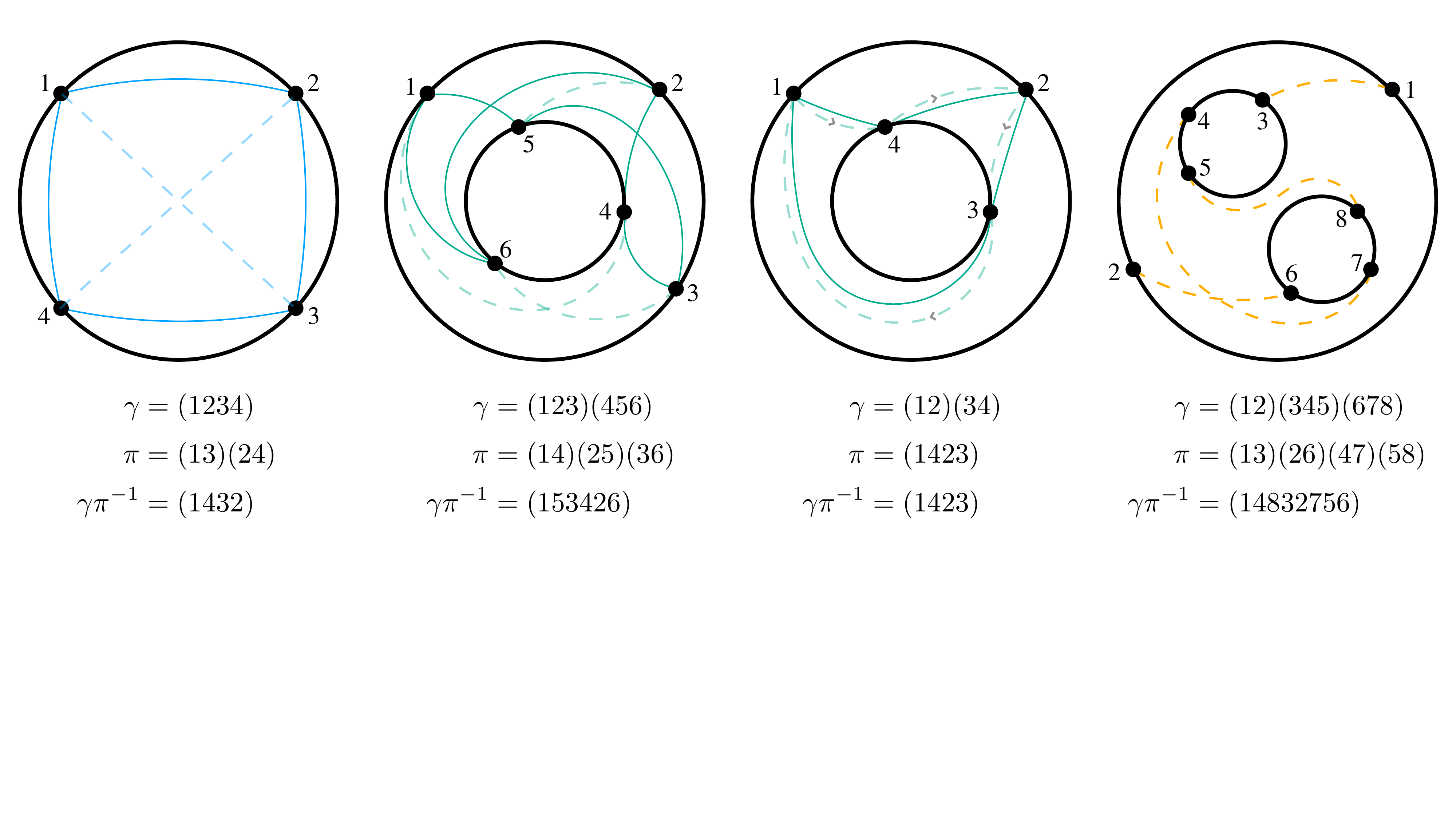}
    \vspace{-4cm}
    \caption{Examples of nonplanar contractions for one, two, two, and three external trace boundaries, using the conventions of figure~\ref{fig:annular}. Numbered points label operator insertions, black circles show the external cycles $\gamma$, dashed colored curves show the support wiring $\pi$, and solid colored curves show the energy index identifications encoded by $\gamma\pi^{-1}$. Each cycle of $\pi$ forms its own cumulant block in $\cV$, and in the fourth panel, $\gamma\pi^{-1}$ is listed rather than drawn for clarity. In all four examples, $\gamma\pi^{-1}$ has a single cycle, leaving one independent energy sum, whereas the corresponding planar contractions leave three independent sums, giving a relative suppression of $L^{-2}$. The third panel shows that a partitioned permutation can be nonplanar even with no visible crossings: for $\gamma=(12)(34)$, the ordering $\pi=(1423)$ leaves only one independent sum, while the planar four-cycle in figure \ref{fig:annular} leaves three. Note that the outer boundary is oriented clockwise while the inner boundaries are oriented counterclockwise.}
    \label{fig:crossing}
\end{figure}

For the single-trace case $b=1$, \eqref{eq:PS-NS} reduces to \eqref{eq:first-order-free-moment-cumulant}, and for the two-trace case $b=2$, it provides the annular noncrossing expansion with primitive second-order free cumulant $\kappa^{(2)}_{\ell_1,\ell_2}$. For example, for a centered, energy-independent operator we have
\begin{equation} \label{eq:annular-expansion}
    \alpha_{1,1} = \kappa^{(1)}_2 + \kappa^{(2)}_{1,1} \,,\qquad \alpha_{2,2} = \kappa^{(2)}_{2,2} + 4\kappa^{(1)}_4 + 2\left(\kappa^{(1)}_2\right)^2 \,,
\end{equation}
where the three contributions to $\alpha_{2,2}$ correspond to the following partitioned permutations,
\begin{equation} \label{eq:partitioned-example}
\begin{split}
    \Theta_A &= (\cV,\pi) = (\{\{1,2,3,4\}\},(12)(34))\,,\\
    \Theta_{B_i} &= (\{\{1,2,3,4\}\},\pi_i)\,,\qquad \pi_i\in\{(1243),(1342),(1234),(1432)\} \\
    \Theta_{C_1} &= (\{\{1,3\},\{2,4\}\},(13)(24))\quad\text{and}\quad \Theta_{C_2} = (\{\{1,4\},\{2,3\}\},(14)(23))\,,
\end{split}
\end{equation}
respectively, for fixed external boundary permutation $\gamma=(12)(34)$. An example of each type of contribution is shown in figure \ref{fig:annular}. For the first line of~\eqref{eq:partitioned-example}, the partition $\cV$ groups the separate two-cycles together, resulting in the primitive second-order cumulant $\kappa^{(2)}_{2,2}$. The permutations in the second line of~\eqref{eq:partitioned-example} already involve all four operators, so there is no nontrivial extra grouping for $\cV$ to perform for the contributions to $\kappa^{(1)}_4$. For the third line of~\eqref{eq:partitioned-example}, $\cV$ has two separate blocks, each of which corresponds to a single cycle, resulting in the product $(\kappa^{(1)}_2)^2$ of first-order cumulants. To summarize, each block $V\in\mathcal V$ must be a union of cycles of $\pi$; if it contains $r$ cycles of lengths $m_1,\ldots,m_r$, it contributes to the free cumulant $\kappa_{m_1,\ldots,m_r}^{(r)}$.

The left panel of figure \ref{fig:annular} is an example of a partitioned permutation that corresponds to the independent data of multi-trace ETH~\eqref{eq:MT}. More generally, this data arises from partitioned permutations $\Theta=(\cV,\pi)$ which correlate the $b$ trace cycles without identifying their individual energy indices, i.e. where the permutation $\pi$ leaves the $b$ boundary cycles intact, while the partition $\cV$ places all of them in a single cumulant block. The corresponding cumulant is given by $\kappa_{\ell_1,\ldots,\ell_b}^{(b)}$, with energy-resolved counterpart given by $F_\O^{(\ell_1|\cdots|\ell_b)}\qty(\mathbf{E}_{\ell_1}|\ldots|\mathbf{E}_{\ell_b})$. The remaining contributions to $\alpha_{\ell_1,\ldots,\ell_b}$ in the noncrossing expansion~\eqref{eq:PS-NS} are generated from lower-order cumulants.

Examples of one, two, and three-boundary nonplanar permutations, which are excluded from $\mathcal{PS}_{NC}(\ell_1,\ldots,\ell_b)$, are shown in figure \ref{fig:crossing}. The second and third panels show two different ways in which an annular contraction can be nonplanar: for $\pi=(14)(25)(36)$, the energy identifications cross, while for $\pi=(1423)$ (or similarly for $\pi=(1324)$), the four-cycle instead alternates between the two boundaries. Both cases increase the genus by one and lead to a factor of $L^{-2}$ relative to the planar permutations. 

\subsubsection*{Full ETH from first order free cumulants}

We now show that the first-order free cumulants $\kappa_n^{(1)}$ correspond to the independent data specified by full ETH in~\eqref{eq:full-eth}. Consider the thermal $n$-point correlator as in \eqref{eq:single-trace-correlator},
\begin{equation} \label{eq:single-trace-correlator2}
    G_n(\boldsymbol\beta) \equiv \frac{1}{Z}\bE\qty[\Tr\!\left(e^{-\beta_1H}\O e^{-\beta_2H}\O\cdots e^{-\beta_nH}\O\right)] = \frac{1}{Z}\sum_{i_1,\ldots,i_n}e^{-\sum_{a=1}^{n}\beta_aE_{i_a}}\bE[\O_{i_1i_2}\cdots\O_{i_ni_1}]\,,
\end{equation}
in which the trace fixes one oriented matrix-index boundary,
\begin{equation}
    i_1\longrightarrow i_2\longrightarrow\cdots\longrightarrow i_n\longrightarrow i_1 \,,
\end{equation}
corresponding to the permutation $\gamma=(1\,2\,\cdots\,n)$. The sums over $i_1,\ldots,i_n$ include configurations both with mutually distinct indices and with repeated indices, where at leading order the latter pinch the boundary into smaller cyclic components as in~\eqref{eq:full-factorization}. At large entropy, the dominant configurations are those that leave the maximal number of independent index loops, which we will now see correspond to the noncrossing partitions.

Consider the Laplace transform of the full ETH functions from \eqref{eq:full-eth}:
\begin{equation}
    \kappa_n^{(1)}(\O,\boldsymbol\beta) \equiv \frac{1}{Z} \sum_{i_1,\ldots,i_n}e^{-\sum\beta_aE_{i_a}} e^{-(n-1)S(E_{\rm avg})}F_{\O}^{(n)}(E_{i_1},\ldots,E_{i_n}) \,,
\label{eq:thermal-transform-F-to-kappa}
\end{equation}
and consider a fixed index contraction contributing to \eqref{eq:single-trace-correlator2}, which we label by a partition $\pi$ such that there are $N_{\rm loop}(\pi)=|\gamma^{-1}\pi|$ independent index loops. Then for fixed $\pi$, at leading order in $e^{S(E_{\rm avg})}$,
\begin{equation}
    \underbrace{\frac{1}{Z}}_{e^{-S}} \times \underbrace{\sum_{i_1,\ldots,i_n|_\pi} e^{-\sum\beta_aE_{i_a}}}_{e^{N_{\mathrm{loop}}(\pi)S}} \times \underbrace{\bE\qty[\O_{i_1i_2} \O_{i_2i_3}\cdots\O_{i_ni_1}]}_{e^{-(n-|\pi|)S}} \sim e^{[N_{\rm loop}(\pi)-n+|\pi|-1]S} \prod_{B\in\pi} \underbrace{\kappa_{|B|}^{(1)}(\O, \boldsymbol{\beta}_B)}_{O(1)}\,,
\end{equation}
where the sum is over independent indices $i_1,\ldots i_n$ that are compatible with the partition $\pi$. The scaling of the matrix elements follows from applying the leading scaling~\eqref{eq:full-eth} to each component in the factorization~\eqref{eq:full-factorization}:
$$\prod_{B\in\pi}e^{-(|B|-1)S} = e^{-\sum_{B\in\pi}(|B|-1)S} = e^{-(n-|\pi|)S}\,,$$
where $|B|$ denotes the number of operator insertions in each cycle of $\pi$. Planar noncrossing partitions are precisely the contractions that maximize the number of index loops:
\begin{equation}
    N_{\rm loop}(\pi)=n-|\pi|+1 \,.
\end{equation}
In contrast, crossing contractions impose additional identifications among the matrix indices, leaving fewer independent energy sums with each loop lost costing a factor of $L^{-1}=e^{-S}$.\footnote{In ribbon-graph language, unavoidable crossings correspond to nonplanar diagrams: a genus-$g$ contraction has two fewer index loops per handle, so is suppressed by $L^{-2g}=e^{-2gS}$.} Thus the leading contractions in \eqref{eq:single-trace-correlator2} are given by planar noncrossing partitions, and the leading single-trace moments can be written as
\begin{equation} \label{eq:single-trace-energy-dependent}
    \cG_n(\boldsymbol\beta) \approx \sum_{\pi\in NC(n)}\prod_{B\in\pi} \kappa_{|B|}^{(1)}(\O,\boldsymbol{\beta}_B) + \text{subleading} \,.
\end{equation}
In a fixed energy shell, \eqref{eq:single-trace-energy-dependent} simply reduces to \eqref{eq:first-order-free-moment-cumulant}, from which we see that $\kappa_n^{(1)}(\O,\boldsymbol\beta_B)$ are thermal first-order free cumulants.\footnote{For thermal correlators, the factorization into products of free cumulants holds at leading order in the thermodynamic limit, as in \cite{Pappalardi:2022aaz}. Since ETH kernels suppress large energy differences and vary smoothly with energy density, the sums can be evaluated at a common thermal saddle.} Hence the full ETH functions $F_{\O}^{(n)}(E_{i_1},\ldots,E_{i_n})$ contain the same primitive data as thermal free cumulants in different thermodynamic ensembles: they correspond to the genuinely new cyclic data left after subtracting all noncrossing products of lower-point functions \cite{Pappalardi:2022aaz, Vallini:2025vvq, Pappalardi:2023nsj}.

Finally, we briefly review the first subleading corrections to the factorization rule~\eqref{eq:full-factorization} given by the noncrossing expansion \cite{Vallini:2025vvq}, and we will see why these corrections differ from the data of multi-trace ETH. The first subleading correction is obtained by moving one step upward in the lattice $NC(n)$, namely by merging one pair of cycles. Each such ``step-up'' correction is smaller by one additional factor $L^{-1}=e^{-S}$ at fixed matrix elements,
\begin{equation}
\begin{split}
    \bE[\O_{i_1i_2} \cdots \O_{i_{k-1}i_1}\O_{i_1i_{k+1}}\cdots O_{i_ni_1}] &= \frac{F_\O^{(k-1)}(\mathbf{E}_{k-1})F_\O^{(n-k+1)}(\mathbf{E}_{n-k+1})}{L^{n-2}} \\
    &\quad+ \frac{F_\O^{(n)}(E_{i_1},\ldots,E_{i_{k-1}},E_{i_1},E_{i_{k+1}},\ldots,E_{i_n})}{L^{n-1}} + \cdots \,.
\end{split}
\end{equation}
For example, for pairwise distinct $i,j,k$, 
\begin{equation} \label{eq:step-up-example}
    \bE\qty[\O_{ij}\O_{ji}\O_{ik}\O_{ki}] = e^{-2S}F_{\O}^{(2)}(E_i,E_j)F_{\O}^{(2)}(E_i,E_k) + e^{-3S}F_{\O}^{(4)}(E_i,E_j,E_i,E_k)+\cdots\,.
\end{equation}
The leading cactus factorization is determined by a product of ordinary two-point cumulants. In the decomposition of $\alpha_{2,2}$ shown in \eqref{eq:annular-expansion}, the same data reappear in the $(\kappa_2^{(1)})^2$ terms, even though these are not exactly the same pairings. The subleading correction merges the two blocks into a single four-cycle, and when embedded in the annulus this contributes to the $\kappa_4^{(1)}$ term. Both terms in \eqref{eq:step-up-example} arise from single-trace cumulants, in contrast with $\kappa_{2,2}^{(2)}$ which describes a connected correlation between two independent cycles.

The correlation between two diagonal matrix elements also has a step-up correction given by
\begin{equation}
    \overline{\O_{ii}\O_{jj}}\big|_{\text{step-up}} \sim e^{-S}\delta_{ij}F^{(2)}(E_i,E_i)
\end{equation}
which corresponds to $\kappa_2^{(1)}$. This step-up term therefore vanishes for distinct energy eigenstates, and a smooth covariance for $i\neq j$ requires instead a nontrivial $\kappa^{(2)}_{1,1}$.

\subsubsection*{Multi-trace ETH from higher order free cumulants}
Now consider the connected correlator of $b$ traces,
\begin{equation} \label{eq:b-trace-correlator}
    \cG_N(\boldsymbol\beta_1|\cdots|\boldsymbol\beta_b) \equiv \frac{L^{2b-2}}{\boldsymbol{Z}}\overline{\prod_{a=1}^b\Tr\!\left(e^{-\beta_{a,1}H}\O \cdots e^{-\beta_{a,\ell_a}H}\O\right)}\,,
\end{equation}
where $\boldsymbol{Z}=\prod_{a=1}^b\Tr\qty(e^{-\sum_{j=1}^{\ell_a}\beta_{a,j}H})$ is the product of partition functions on each boundary. The normalization is fixed by the Hilbert space dimension $L$ such that in a fixed energy shell the correlator reduces to the higher-order moment $\alpha_{\ell_1,\ldots,\ell_b}$. Equivalently,
\begin{equation} \label{eq:b-trace-correlator2}
    \cG_N(\boldsymbol\beta_1|\cdots|\boldsymbol\beta_b) = \frac{L^{2b-2}}{\boldsymbol{Z}}\sum_{i_1,\ldots,i_N}e^{-\sum_{r=1}^{N}\beta_rE_{i_r}} \overline{\boldsymbol{\O}_{\ell_1}\cdots \boldsymbol{\O}_{\ell_b}}\,,
\end{equation}
in which each of the $b$ disjoint index cycles of length $\ell_a$ fixes an oriented matrix-index boundary,
\begin{equation}
    i_{1,1} \longrightarrow \cdots\longrightarrow i_{1,\ell_1}\longrightarrow i_{1,1} ~~, ~\ldots~,~~ i_{b,1} \longrightarrow \cdots\longrightarrow i_{b,\ell_b}\longrightarrow i_{b,1}
\end{equation}
with corresponding boundary permutation $\gamma$, and where $N=\sum_{a=1}^b\ell_a$ is the total number of operator insertions. Just as in the single-trace case, the indices $i_1,\ldots,i_N$ might have additional index identifications, but we will see that the dominant contributions to \eqref{eq:b-trace-correlator2} arise from noncrossing partitioned permutations introduced in \eqref{eq:PS-NS}.

Analogously to \eqref{eq:thermal-transform-F-to-kappa}, we can Laplace transform the multi-trace ETH functions from \eqref{eq:MT}:
\begin{equation} \label{eq:thermal-higher-order}
    \tilde\kappa_{\ell_1,\ldots,\ell_b}^{(b)} \qty(\O;\boldsymbol\beta_1|\cdots|\boldsymbol\beta_b) \equiv \frac{L^{2b-2}}{\boldsymbol{Z}} \sum_{i_1,\ldots,i_N} e^{-\sum\beta_{a}E_{i_a}} e^{-(N+b-2)S_{b}} F_{\O}^{(\ell_1|\cdots|\ell_b)} \qty(\mathbf E_{\ell_1}|\cdots|\mathbf E_{\ell_b}) \,,
\end{equation}
where the sum is over mutually distinct indices. In an energy-independent shell, \eqref{eq:thermal-higher-order} is order one and agrees at leading order with $\kappa_{\ell_1,\ldots,\ell_b}^{(b)}$.\footnote{For $b=1$, \eqref{eq:thermal-higher-order} agrees at leading order with the thermal free cumulant defined in \eqref{eq:thermal-transform-F-to-kappa}, but for $b>1$, the thermal transform need not coincide with the scalar higher-order free cumulant of the corresponding trace observables.} 

For a fixed partitioned permutation $\Theta=(\cV,\pi)$, at leading order in $L$,
\begin{equation}
    \underbrace{\frac{L^{2b-2}}{\boldsymbol{Z}}}_{L^{b-2}} \times \underbrace{\sum_{i_1,\ldots,i_N|_\Theta} e^{-\sum\beta_aE_{i_a}}}_{L^{N_{\mathrm{loop}}(\Theta)}} \times \underbrace{\overline{\boldsymbol{\O}_{\ell_1} \boldsymbol{\O}_{\ell_2}\cdots \boldsymbol{\O}_{\ell_b}}}_{L^{-(N+|\pi|-2|\cV|)}} \sim L^{b-2+N_{\rm loop}(\Theta)-N-|\pi|+2|\cV|} ~\underbrace{\tilde\kappa_{\Theta}(\O,\boldsymbol{\beta})}_{O(1)}\,,
\end{equation}
where the sum is over independent indices $i_1,\ldots i_N$ that are compatible with $\Theta$, $N_{\rm loop}(\Theta)=|\gamma^{-1}\pi|$ denotes the number of independent index loops, and $|\cV|$ denotes the number of partition blocks in $\cV$.\footnote{Here $\tilde\kappa_\Theta$ denotes the joint thermal transform of the contraction $\Theta$: the primitive ETH kernels are multiplied and their index identifications imposed before summing over energies. In general, $\tilde\kappa_\Theta\neq \prod_{V\in\mathcal V}\tilde\kappa_V$, since the energy sums need not factorize. In a common energy-independent shell, however, $\tilde\kappa_\Theta\to\kappa_\Theta$, recovering the higher-order free-cumulant relation \eqref{eq:PS-NS}.} The scaling of the matrix element correlations follows from \eqref{eq:MT},
$$\prod_{V\in\cV}L^{-(\ell_V+b_V-2)} = L^{-\sum_V(\ell_V+b_V-2)} = L^{-(N+|\pi|-2|\cV|)}\,.$$
For a fixed cumulant structure (fixed $b$, $N$, $|\pi|$, and $|\cV|$), the planar noncrossing partitioned permutations precisely maximize the number of independent sums:
\begin{equation}
    N_{\mathrm{loop}}(\Theta)
    =
    N+|\pi|-2|\cV|-b+2\,.
\end{equation}
The leading connected $b$-trace moment can then be written as
\begin{equation}\label{eq:b-trace-noncrossing}
    \cG_N(\boldsymbol\beta_1|\cdots|\boldsymbol\beta_b) \approx \sum_{\Theta\in \mathcal{PS}_{NC}(\ell_1,\ldots,\ell_b)} \tilde\kappa_{\Theta}(\O,\boldsymbol{\beta}) + \text{subleading} \,.
\end{equation}
In a fixed energy shell, this reduces to \eqref{eq:PS-NS} and the independent multi-trace data precisely corresponds to the higher-order free cumulants.

\begin{figure}
    \centering
    \vspace{-1cm}
    \includegraphics[width=.9\linewidth]{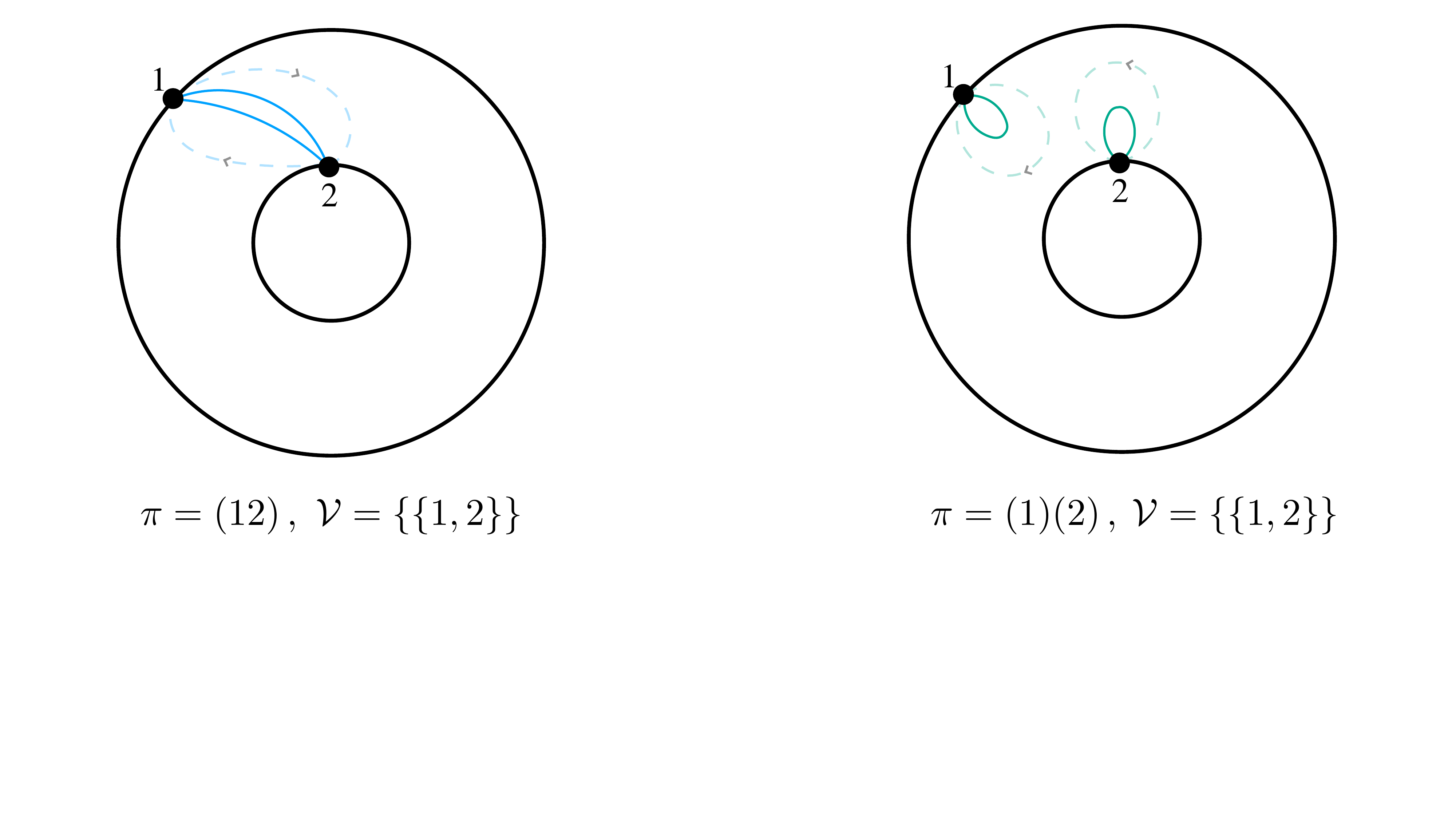}
    \vspace{-3cm}
    \caption{Annular noncrossing decomposition $\alpha_{1,1} = \kappa^{(1)}_{2} + \kappa^{(2)}_{1,1}$ \cite{mingo2003annularnoncrossingpermutationspartitions}. The black circles correspond to two trace boundaries with permutation $\gamma=(1)(2)$, numbered points label $\O_{ii}$ and $\O_{jj}$, dashed arrows label the permutation $\pi$, and solid curves show the identifications of energies encoded by $\gamma\pi^{-1}$. \textit{Left:} the cyclic permutation $\pi=(12)$ gives $\kappa_2^{(1)}$, with $i_1=i_2$ leaving one independent energy sum. \textit{Right:} the two singleton cycles are joined into one block $\mathcal V$, giving $\kappa_{1,1}^{(2)}$ with two independent sums.}
    \label{fig:F11}
\end{figure}

The simplest example is the square of a centered one-point trace, illustrated in Fig. \ref{fig:F11}. In this case,
\begin{align}
    \cG_2(\beta_1|\beta_2) &= \frac{L^2}{\Tr\qty(e^{-\beta_1 H})\Tr\qty(e^{-\beta_2 H})} \overline{\Tr\qty(e^{-\beta_1 H} \O)\Tr\qty(e^{-\beta_2 H} \O)} \nonumber\\
    &= \sum_{\Theta\in \mathcal{PS}_{NC}(1,1)} \tilde\kappa_{\Theta}(\O,\boldsymbol{\beta}) + \cdots \\
    &= \frac{L^2}{Z_{\beta_1}Z_{\beta_2}}\sum_{i,j} e^{-\beta_1E_i-\beta_2E_j} \qty[ e^{-S(E_i)}F_O^{(2)}(E_i,E_i)\delta_{ij} + e^{-S(E_i)-S(E_j)} F_O^{(1|1)}(E_i|E_j) ] + \cdots \,,\nonumber
\end{align}
where there are two partitioned permutations: the first involves the cyclic permutation $\pi=(12)$, which merges the two one-point insertions into an ordinary two-cycle, and the second involves
\begin{equation}
    \Theta_{1|1} = \qty(\{\{1,2\}\},(1)(2))\,,
\end{equation}
which keeps the two one-cycles distinct as permutation cycles but places them in a single connected block $\cV=\{1,2\}$.

Summing over $i$ and $j$ in a fixed energy shell gives
\begin{equation}
    \alpha_{1,1} = \lim_{L\to\infty}\mathbb{E}[\Tr\O\,\Tr\O] = \kappa_2^{(1)}+\kappa_{1,1}^{(2)}\,,\qquad F_{\O}^{(1|1)} \equiv \kappa_{1,1}^{(2)} = \alpha_{1,1}-\kappa_2^{(1)}\,,
\end{equation}
as expected.

\subsection{Global rotational invariance} \label{subsec:global}
In this subsection, we study globally rotationally invariant ensembles in order to illustrate more concretely the relationship between multi-trace ETH and higher-order free probability. In these ensembles, the operator statistics are invariant under arbitrary unitary rotations of the full $L$-dimensional Hilbert space, so the ETH functions have no energy dependence. This is too strong for a realistic many-body system, but the simplicity of global rotational invariance provides a useful model for illustrating the index contractions allowed in a unitary system. Since no basis is preferred, the ensemble must assign the same probability to $\O$ and to any rotated operator $V\O V^\dagger$,
\begin{equation}
    P(\O) = P(V\O V^\dagger)\,,\qquad V\in U(L)\,,
\label{eq:global-unitary-invariance}
\end{equation}
where $P(\O)$ denotes the probability distribution of the operator. Any invariant Hermitian ensemble may be written schematically as
\begin{equation}
    \O = U\Lambda U^\dag\,,
\end{equation}
where $U$ is Haar distributed conditional on the eigenvalue matrix $\Lambda$. We refer to $U$ and $\Lambda$ as the angular and energy variables, respectively: $U$ rotates the eigenvectors within the Hilbert space while $\Lambda$ determines the eigenvalue dependence of $\O$. The allowed index contractions are determined by the integral over $U$, which can be computed using Weingarten calculus \cite{collins2002momentscumulantspolynomialrandom, Collins:2006jgn, Collins_2022}. The remaining integral over the eigenvalue ensemble $P(\Lambda)$ fixes the moments $\Tr\Lambda^n$ and their fluctuations. Then correlations between matrix-elements can be obtained from
\begin{align} \label{eq:wein-exp}
    \bE(\boldsymbol{\O}_{\ell_1} \boldsymbol{\O}_{\ell_2}\cdots\boldsymbol{\O}_{\ell_b}) &= \int dP(\Lambda) \int_{U(L)} dU\,\prod_{r=1}^N\,(U\Lambda U^\dag)_{i_ri_{\gamma(r)}} \nonumber\\
    &= \int dP(\Lambda) \sum_{\pi,\sigma\in S_N} \delta_{\gamma\pi^{-1}} \operatorname{Wg}_L\left(\sigma^{-1}\pi\right)\Tr_{\sigma}(\Lambda) \,,
\end{align}
where $\operatorname{Wg}_L$ are Weingarten functions dependent on the dimension $L$ of the Hilbert space and
\begin{equation}
    \Tr_{\sigma}(\Lambda) = \prod_{c\in\operatorname{Cyc(\sigma)}}\operatorname{Tr}\left(\Lambda^{|c|}\right)\,,\qquad \delta_{\gamma\pi^{-1}} = \prod_{r=1}^N\delta_{i_r,i_{\gamma(\pi^{-1}(r))}}\,,
\end{equation}
for a permutation $\gamma\in S_N$ whose $b$ cycles encode $\ell_1,\ldots,\ell_b$. The Haar integral fixes the allowed index dependence, while the $dP(\Lambda)$ integral fixes the coefficients and can correlate otherwise separate traces. Let $\expval{\cdots}_\Lambda$ denote the average over $dP(\Lambda)$, while the overline denotes the connected ensemble average over both $U$ and $\Lambda$.

We now consider the application of~\eqref{eq:wein-exp} to the correlation between two matrix elements. It will be useful to first separate the trace singlet $t_1$ from the traceless matrix $\tilde\Lambda$,
\begin{equation} \label{eq:traceless-decomp}
    \O = t_1\boldsymbol{1} + U\tilde\Lambda U^\dag\,,\qquad \tilde\Lambda \equiv \Lambda-t_1\boldsymbol{1}\,,\qquad t_1 \equiv \frac1L\Tr\Lambda\,,\qquad t_2 \equiv \frac1L\Tr\qty(\tilde\Lambda^2)\,.
\end{equation}
Then we have
\begin{align} \label{eq:2-op-haar}
    \bE(\O_{ab}\O_{cd}) &= \expval{\frac{1}{L^2-1} \left[\left(\delta_{ab}\delta_{cd} - \frac1L \delta_{ad}\delta_{bc}\right) \left(\Tr[\Lambda]\right)^2 + \left(\delta_{ad}\delta_{bc} - \frac1L\delta_{ab}\delta_{cd}\right) \Tr[\Lambda^2] \right]}_\Lambda \,,\nn\\
    &= \left(\frac{L}{L^2-1} \expval{t_2}_\Lambda\right)\delta_{ad}\delta_{bc}+\left(\expval{t_1^2}_\Lambda - \frac{1}{L^2-1} \expval{t_2}_\Lambda\right)\delta_{ab}\delta_{cd}\,,
\end{align}
where the only allowed tensor structures $\delta_{ad}\delta_{bc}$ and $\delta_{ab}\delta_{cd}$ correspond to the single-trace two-point function and the two-trace one-point function squared, respectively. The connected two-point correlations are given by
\begin{align} \label{eq:global-AB}
    \overline{\O_{ab}\O_{cd}} &= A\,\delta_{ad}\delta_{bc}+B\,\delta_{ab}\delta_{cd}\\
    A &= \frac{L}{L^2-1} \expval{t_2}_\Lambda = \frac{\kappa_2^{(1)}}{L}+\cdots \nonumber\\
    B &= \expval{t_1^2}_\Lambda - \expval{t_1}^2_\Lambda - \frac{1}{L^2-1} \expval{t_2}_\Lambda = \frac{\kappa_{1,1}^{(2)}}{L^2}+\cdots \nonumber\,,
\end{align}
where $A$ and $B$ are independent of the energy eigenbasis and are given by free cumulants in the large $L$ limit. Whereas $\kappa_2^{(1)}$ measures the average spread of the eigenvalues within one realization, $\alpha_{1,1}=\kappa_2^{(1)}+\kappa_{1,1}^{(2)}$ measures how the trace singlet fluctuates between realizations: 
\begin{equation}
    \alpha_{1,1} = \lim_{L\to\infty} {\rm Var}(\Tr\O) = \lim_{L\to\infty} L^2{\rm Var}(t_1)\,.
\end{equation}
The (energy-independent) multi-trace ETH data $F_{\O}^{(1|1)}=\kappa_{1,1}^{(2)}$ is therefore determined by a competition between spectral trace fluctuations and the universal correlations imposed by Haar-distributed eigenvectors.

We next examine two illustrative cases: an operator with fixed spectrum, which has vanishing trace fluctuations $\alpha_{1,1}$, and an operator sampled from the Gaussian Unitary Ensemble (GUE), which has trace fluctuations that precisely cancel the universal Haar correlations. This is summarized in Table~\ref{tab:global}.

\begin{table}[h]
    \centering
    \begin{tabular}{rcc}
        & $\qquad\alpha_{1,1}\qquad$ & $F_\O^{(1|1)} = \kappa_{1,1}^{(2)}$ \vspace{.05cm}\\ \hline\hline\vspace{-.35cm} \\
        fixed spectrum: & $0$ & $-\kappa^{(1)}_2$ \\
        GUE: & $\kappa^{(1)}_2$ & $0$
    \end{tabular}
    \caption{Comparison of the second-order trace moment $\alpha_{1,1}$ and the primitive multi-trace data $F_{\O}^{(1|1)}=\kappa_{1,1}^{(2)}$ in globally rotationally invariant ensembles. For a fixed spectrum, trace fluctuations vanish and the nonzero $1|1$ cumulant is fixed entirely by the ordinary free cumulant $\kappa_2^{(1)}$. In the GUE, spectral trace fluctuations reproduce $\kappa_2^{(1)}$ and cancel the primitive two-boundary contribution. In contrast, a generic non-Gaussian probability distribution $P(\Lambda)$ can support both nonzero trace fluctuations and independent multi-trace data.}
    \label{tab:global}
\end{table}

\subsubsection*{Fixed-spectrum Haar} \label{subsec:fixed-haar}
A pure fixed-orbit Haar average is the special case
\begin{equation}
    \O=U\Lambda_0U^\dagger\,,\qquad dP_{\rm Haar}(\Lambda)=\delta(\Lambda-\Lambda_0)d\Lambda\,,
\label{eq:fixed-orbit-delta-measure}
\end{equation}
for which the energy integral is trivial, and Weingarten calculus supplies the complete randomness of the ensemble. In this case, 
\begin{equation} \label{eq:global-haar-t1}
    t_1 = \frac1L\Tr\Lambda_0\,,\qquad \tilde\Lambda_0 = \Lambda_0-t_1\mathbbm{1}\,,\qquad t_2 = \frac1L\Tr\qty(\tilde\Lambda_0^2) = \kappa_2^{(1)} + \cdots\,,
\end{equation}
where there are no fluctuations of $t_1$ or $t_2$, so the coefficients~\eqref{eq:global-AB} are simply given by
\begin{equation}\label{eq:global-haar-AB}
    A_{\rm Haar} = \frac{L}{L^2-1}t_2\,,\qquad B_{\rm Haar} = -\frac{1}{L^2-1}t_2 = -\frac{A_{\rm Haar}}{L}\,,
\end{equation}
such that
\begin{equation} \label{eq:global-haar-OO}
    \overline{\O_{ab}\O_{cd}} = \frac{L}{L^2-1}t_2 \qty(\delta_{ad}\delta_{bc}-\frac1L\delta_{ab}\delta_{cd})\,,
\end{equation}
and in the large $L$ limit we then have
\begin{equation}\label{eq:F11-gri}
    F_\O^{(1|1)} = \kappa_{1,1}^{(2)} = -\kappa_2^{(1)}\,,\quad \alpha_{1,1} = \kappa_2^{(1)} + \kappa_{1,1}^{(2)} = 0\,.
\end{equation}
In particular, we can confirm explicitly by summing over indices in~\eqref{eq:global-AB} that the variance of the trace vanishes:
\begin{equation}
    {\rm Var}(\Tr\O) = \overline{(\Tr\O)^2} - \qty(\overline{\Tr\O})^2 = \sum_{a,b} \qty(\overline{\O_{aa}\O_{bb}} - \overline{\O_{aa}}~\overline{\O_{bb}}) = LA+L^2B = 0\,.
\end{equation}
More generally, the Haar rotation changes only the eigenvectors of $\mathcal O$, while leaving
\begin{equation}
    \Tr\O^n = \Tr\Lambda_0^n
\end{equation}
fixed. Since each trace observable is constant over the Haar ensemble, classical cumulants of two or more traces vanish, so we conclude from~\eqref{eq:alpha-def} that
\begin{equation}
    \alpha_{\ell_1,\ldots,\ell_b} = 0 \qquad \text{for } b\geq2 \,.
\end{equation}
The fixed-spectrum Haar orbit is therefore the simplest example in which the multi-trace ETH functions~\eqref{eq:MT} are generically nonzero but contain no independent multi-trace data: they are completely determined by the ordinary single-trace cumulants $\kappa_n^{(1)}$.

\subsubsection*{GUE ensemble}
The Gaussian Unitary Ensemble (GUE) provides a complementary globally invariant example in which not just the angular but also the energy variables fluctuate. We consider a centered GUE operator with measure
\begin{equation}
    d\mu_{\rm GUE}(\O)\propto\exp\!\left[-\frac{L}{2\sigma^2}\Tr\O^2\right]d\O\,.
\label{eq:global-GUE-measure}
\end{equation}
Equivalently, after writing $\O=U\Lambda U^\dag$, the eigenvalues $\lambda_i$ in $\Lambda=\operatorname{diag}(\lambda_1,\ldots,\lambda_L)$ are no longer fixed, but rather fluctuate jointly between realizations, with probability measure
\begin{equation}
    dP_{\rm GUE}(\Lambda)\propto\prod_{a=1}^{L}d\lambda_a\,\Delta(\lambda)^2\exp\!\left[-\frac{L}{2\sigma^2}\sum_{a=1}^{L}\lambda_a^2\right] \,.
\label{eq:GUE-spectral-measure}
\end{equation}
where $\Delta(\lambda)$ is the Vandermonde determinant. Thus the GUE expansions are obtained by averaging the fixed-spectrum Haar integral over fluctuating eigenvalues. At large $L$, its eigenvalue density is semicircular, with the following free cumulants
\begin{equation} \label{eq:GUE-kappa}
    \kappa^{(b)}_{\ell_1,\ldots,\ell_b} = \begin{cases}
        \sigma^2\quad & b=1,\,\ell_1=2\\
        0 & \text{otherwise}\,.
    \end{cases}
\end{equation}
As expected from~\eqref{eq:PS-NS}, the connected higher-order moments are then simply given by functions of $\kappa_2^{(1)}=\sigma^2$, e.g.
\begin{equation} \label{eq:GUE-alpha}
    \alpha_{1,1} = \sigma^2 \,, \qquad \alpha_{2,2} = 2\sigma^4 \,, \qquad \alpha_{\ell_1,\ldots,\ell_b} = \left|NC_{2}^{(b)}(\ell_1,\ldots,\ell_b)\right| \sigma^N
\end{equation}
where $N=\ell_1+\cdots+\ell_b$ and $NC_{2}^{(b)}(\ell_1,\ldots,\ell_b)$ denotes the connected genus-zero pairings of the $N$ points distributed over $b$ oriented boundaries, which vanishes if $N$ is odd \cite{Mingo_2006}.

Unlike the fixed Haar orbit, both $t_1$ and $t_2$ defined by~\eqref{eq:traceless-decomp} fluctuate over the Gaussian ensemble, with fluctuations given by
\begin{equation}
    \expval{t_1}_\Lambda = 0\,,\qquad \expval{t_1^2}_\Lambda = \frac{\sigma^2}{L^2}\,,\qquad \expval{t_2}_\Lambda = \sigma^2\left(1-\frac{1}{L^2}\right)\,.
\end{equation}
Substituting these quantities into the general globally invariant two-point coefficients~\eqref{eq:global-AB} gives
\begin{equation}
    A_{\rm GUE} = \frac{L}{L^2-1}\expval{t_2}_\Lambda=\frac{\sigma^2}{L}\,,\qquad B_{\rm GUE} = \expval{t_1^2}_\Lambda - \frac{1}{L^2-1}\expval{t_2}_\Lambda=0\,,
\end{equation}
where
\begin{equation}
    \kappa_2^{(1)} = \lim_{L\to\infty}\expval{t_2} = \sigma^2\,,\qquad F_{\O}^{(1|1)} = \kappa_{1,1}^{(2)} = \lim_{L\to\infty} L^2B_{\rm GUE} = 0\,,\qquad \alpha_{1,1} = \kappa_2^{(1)} + \kappa_{1,1}^{(2)} = \sigma^2
\end{equation}
is fully consistent with~\eqref{eq:GUE-kappa} and~\eqref{eq:GUE-alpha}. Hence the full connected covariance~\eqref{eq:global-AB} is purely cyclic,
\begin{equation}
    \overline{\O_{ab}\O_{cd}} = \frac{\sigma^2}{L}\delta_{ad}\delta_{bc} \,,
\end{equation}
and in particular,
\begin{equation}
    \overline{(\Tr\O)^2}-\left(\overline{\Tr\O}\right)^2=LA_{\mathrm{GUE}}+L^2B_{\mathrm{GUE}}=\sigma^2\,.
\end{equation}
The one-point trace therefore fluctuates in the ordinary GUE, in contrast with the fixed-spectrum Haar orbit. However, this covariance is not independent multi-trace data: it is entirely reconstructed from $\kappa_2^{(1)}$, leaving a vanishing primitive two-cycle cumulant $F_{\O}^{(1|1)}=0$.

\subsection{Local rotational invariance} \label{subsec:local}
We have shown that in globally rotationally invariant ensembles, the independent data of multi-trace ETH corresponds to energy-independent higher-order free cumulants. However, in many-body systems, the ETH functions generally have nontrivial energy dependence, which is made possible by restricting eigenstate typicality to sufficiently narrow energy windows~\cite{Foini:2018sdb, deBoer:2024mqg, Blommaert:2020seb, Foini:2019tii, Pappalardi:2022aaz, Vallini:2025vvq, Pappalardi:2023nsj}. We therefore consider locally rotationally invariant ensembles, where Haar rotations act separately within each microcanonical window while operator correlations can vary smoothly with energy. We decompose the Hilbert space as
\begin{equation}
    \cH = \bigoplus_I\cH_I\,,\qquad L_I = \dim\cH_I\sim e^{S(E_I)}\,.
\end{equation}
Here the windows are disjoint energy intervals of width $\Delta$, chosen much larger than the local mean level spacing but much smaller than the scales on which thermodynamic quantities and ETH kernels vary, so that each $\cH_I$ contains $L_I\simeq \rho(E_I)\Delta\sim e^{S(E_I)}$ states.\footnote{In \cite{Vallini:2025vvq}, local rotational invariance is refined by replacing the sharp boxes with smoothed microcanonical projectors
\begin{equation}
    \widehat\sigma_E=\frac{1}{D_E}\sum_\alpha\delta_\Delta(E-E_\alpha)\ket{\alpha}\bra{\alpha}\,,\qquad D_E=\sum_\alpha\delta_\Delta(E-E_\alpha)\,,
\end{equation}
but for simplicity we will stick to the sharp windows here.} The local Haar unitary is
\begin{equation}
    U_{\mathrm{loc}} = \bigoplus_IU_I\,,\qquad U_I\in U(L_I)\,,
\end{equation}
where each $\cH_I$ carries an independent Haar rotation $U_I$ that does not mix different windows, so in particular, for the projector $P_I$ to the Hilbert space $\cH_I$, we have
\begin{equation}
    P_I = \sum_{i\in I}\ket{E_i}\bra{E_i}\,,\qquad [U_{\rm loc},P_I]=0
\end{equation}
with $\Tr P_I=L_I$. Local rotational invariance then imposes that the ensemble must assign the same probability to $\O$ and to any operator rotated from $\O$ by a block-diagonal unitary $U_{\rm loc}$,
\begin{equation}
    P(\O) = P(U_{\rm loc}\O U_{\rm loc}^\dag)\,.
\end{equation}
On a block $\O_{IJ}\equiv P_I\O P_J$ that connects $\mathcal H_J$ to $\mathcal H_I$, the action of $U_{\rm loc}$ reduces to local rotations acting independently on each index,
\begin{equation}
    \O_{IJ}' = P_I\O' P_J = P_I U_{\rm loc}\O U_{\rm loc}^\dag P_J = U_I\O_{IJ}U_J^\dag\,.
\end{equation}
The window labels $I,J$ are therefore preserved by the local rotations, while the microscopic indices $i\in I$, $j\in J$ inside each window are Haar randomized.

Local Haar rotations hence change only the choice of basis within each energy window, without affecting how the operator connects different energy windows. This information is encoded in single-trace observables such as $\Tr(\O_{I_1I_2}\O_{I_2I_3}\ldots \O_{I_nI_1})$, for which the unitary rotations cancel around the cycle. Hence whereas in the global rotational invariance case the data left unfixed was just the eigenvalue distribution, in the local case the analogous free data is both the spectral distribution and the closed sequences encoding angular correlations between blocks. We will use angle brackets $\expval{\cdot}$ to denote averages over this additional data, after carrying out the integral over each $U_I$.

We again focus on the simplest multi-trace observable, the product of two one-point functions, involving only diagonal blocks $\O_{II}$. Define the local trace singlet $\tau_I$ and trace-subtracted $X_I$ by
\begin{equation} \label{eq:local-t1-t2}
    \tau_I \equiv \Tr(\O_{II}) \,,\qquad 
    X_I \equiv \O_{II} - \frac{1}{L_I}\tau_IP_I\,,\qquad t_{2,I}\equiv\frac{1}{L_I}\Tr(X_I^2) \,.
\end{equation}
The generalization of~\eqref{eq:global-AB} is given by
\begin{align} \label{eq:local-AB}
    \overline{(\O_{II})_{ab} (\O_{JJ})_{cd}} &= \delta_{IJ}A_I\,\delta_{ad}\delta_{bc}+B_{IJ}\,\delta_{ab}\delta_{cd} \\
    A_I &= \frac{L_I}{L_I^2-1}\expval{t_{2,I}} = \frac{\kappa_2^{(1)}(I)}{L_I} + \cdots \,,\nn\\
    B_{IJ} &= \frac{\overline{\tau_I\tau_J}}{L_IL_J} - \delta_{IJ}\frac{1}{L_I^2-1}\expval{t_{2,I}} = \frac{F_\O^{(1|1)}(E_I|E_J)}{L_IL_J} + \cdots\nn
\end{align}
where we have also written the leading terms in the large $L_I,L_J$ limits, and we defined
\begin{equation} \label{eq:local-alpha11}
\begin{split}
    \kappa_2^{(1)}(I) &\equiv \lim_{L_I\to\infty}\expval{t_{2,I}} \,,\qquad \alpha_{1,1}(I,J) \equiv \lim_{L_I,L_J\to\infty}\overline{\tau_I\tau_J}\,, \\
    &F_{\O}^{(1|1)}(E_I|E_J) = \alpha_{1,1}(I,J) - \delta_{IJ}\kappa_2^{(1)}(I)\,,
\end{split}
\end{equation}
which provides an energy-resolved generalization of~\eqref{eq:annular-expansion}. We see from~\eqref{eq:local-AB} that between two different diagonal blocks $I\neq J$, only the singlet components $\tau_I,\tau_J$ can have nonzero correlations. Summing the exact matrix elements~\eqref{eq:local-AB} over the diagonal entries of the two blocks gives
\begin{equation}
\begin{split}
    \overline{\tau_I\tau_J} &= \sum_{i\in I,j\in J} \overline{(\O_{II})_{ii} (\O_{JJ})_{jj}} = \delta_{IJ}A_IL_I + B_{IJ} L_IL_J = \delta_{IJ}\kappa_2^{(1)}(I)+F_{\O}^{(1|1)}(E_I|E_J)+\cdots\,.
\end{split}
\end{equation}

\begin{figure}
    \centering
    \includegraphics[width=\linewidth]{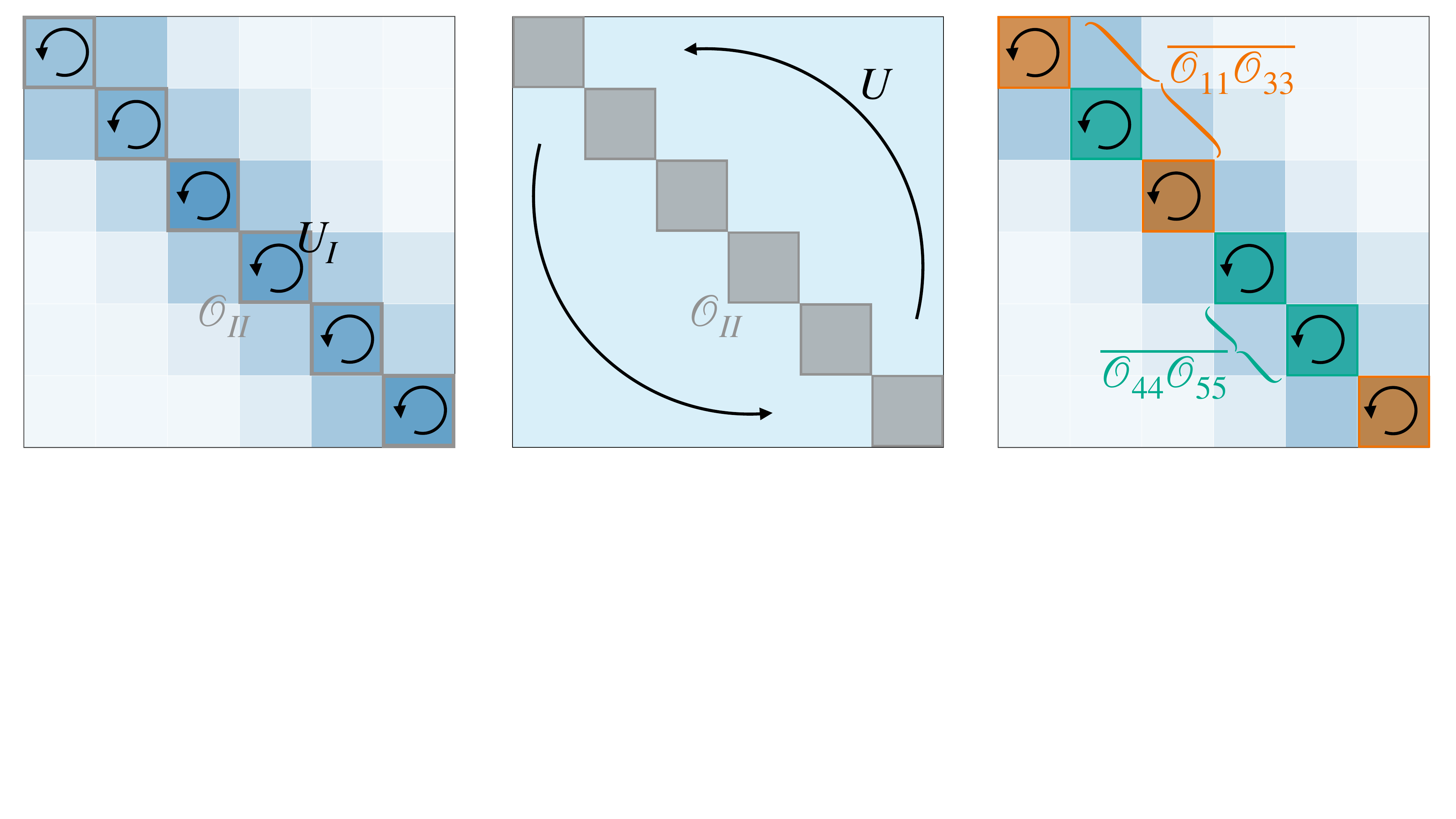}
    \vspace{-5cm}
    \caption{Sources of trace-singlet correlations in locally rotationally invariant ensembles. The blue shading schematically shows the ETH variance of matrix elements of $\O$ in the energy basis, with darker shading corresponding to larger $e^{-S(E_{ij})}|f_{\O}(E_{ij},\omega_{ij})|^2$. The diagonal squares represent microcanonical blocks. \textit{Left:} independent local Haar unitaries $U_I$ rotate the eigenvectors within each window. \textit{Middle:} a global Haar unitary $U$ rotates the entire basis. \textit{Right:} each window is invariant under local $U_I$, while the trace singlets of different blocks fluctuate jointly.} \label{fig:lri}
\end{figure}

The examples below illustrate two mechanisms for generating correlations between the trace singlets. We will see that a local Haar rotation leaves each $\tau_I$ fixed, while correlations can arise either from a shared eigenvector rotation that redistributes fixed spectral data among the windows or from correlated fluctuations of the eigenvalues assigned to different windows. Table~\ref{tab:local} summarizes the resulting covariances, while figure \ref{fig:lri} illustrates these mechanisms. 

\begin{table}[h]
    \centering
    \begin{tabular}{rcc}
        & $\overline{\tau_I\tau_J}$ & $F_{\O}^{(1|1)}(E_I|E_J)$ \vspace{.05cm}\\ \hline\hline\vspace{-.35cm} \\
        fixed spectrum Haar: & $0$ & $-\delta_{IJ}\kappa_2^{(1)}(I)$ \\
        global Haar embedding: & $\quad\frac{L}{L^2-1}\left(L_I\delta_{IJ}-\frac{L_IL_J}{L}\right)t_2\quad$ & $-\frac{L_IL_J}{L^2}\kappa_2^{(1)}$ \\
        correlated local GUE: & nonzero & nonzero
    \end{tabular}
    \caption{Trace-singlet fluctuations in the three locally invariant ensembles discussed in the text. Local fixed-spectrum Haar rotations leave all $\tau_I$ fixed; global Haar mixing correlates them by redistributing a fixed spectrum among the windows; correlated local GUE instead correlates their local spectral data without mixing their eigenbases.}
    \label{tab:local}
\end{table}

\subsubsection*{Fixed-spectrum Haar}
The local analogue of a fixed-spectrum Haar orbit is
\begin{equation}  \label{eq:fixed-local}
    \O = U_{\rm loc}\O_0U_{\rm loc}^\dag\,,\qquad U_{\mathrm{loc}} = \bigoplus_IU_I\,,\qquad U_I\in U(L_I)\,,
\end{equation}
in which we fix $\O_0$, thereby fixing both the spectral data and the correlation between windows. Since $U_{\rm loc}$ commutes with every $P_I$, the trace $\tau_I$ in each block is completely fixed by $P_I$ and $\O_0$,
\begin{equation}
    \tau_I = \Tr(P_I\O)=\Tr(P_I\O_0)\,,
\end{equation}
so both $\tau_I$ and $t_{2,I}$ defined as in~\eqref{eq:local-t1-t2} do not fluctuate. Then \eqref{eq:local-AB} has coefficients
\begin{equation}
    A_I^{\rm loc} = \frac{L_I}{L_I^2-1}t_{2,I}\,,\qquad B_{IJ}^{\rm loc} = -\delta_{IJ}\frac{1}{L_I^2-1}t_{2,I} = -\delta_{IJ}\frac{A_I^{\rm loc}}{L_I}\,.
\end{equation}
and using the definitions~\eqref{eq:local-alpha11},
\begin{equation}
    \alpha_{1,1}(I,J) = \lim_{L_I,L_J\to\infty}\overline{\tau_I\tau_J} = 0\,,\qquad F_{\O}^{(1|1)}(E_I|E_J) = -\delta_{IJ}\kappa_2^{(1)}(I)\,.
\end{equation}
Hence in this ensemble, matrix elements associated to blocks $I$ and $J$ can only be correlated when $I=J$,
\begin{equation} \label{eq:2-op-loc-loc}
    \overline{(\O_{II})_{ab}(\O_{JJ})_{cd}} = \delta_{IJ}\frac{L_I}{L_I^2-1}t_{2,I} \qty(\delta_{ad}\delta_{bc}-\frac{1}{L_I}\delta_{ab}\delta_{cd})\,.
\end{equation}
This is in contrast with the single-cycle two-point correlator, for which with $i\in I$ and $j\in J$,
\begin{equation}
    \overline{(\O_{IJ})_{ij} (\O_{JI})_{ji}} = \frac{1}{L_IL_J}\sum_{\bar i\in I,\bar j\in J}|\O_{\bar i\bar j}|^2\,,\qquad \text{for } I\neq J
\end{equation}
is generically nonvanishing due to closed matrix-element cycles connecting different windows \cite{Pappalardi:2022aaz, Vallini:2025vvq}.

This conclusion cannot be changed just by correlating the block diagonal internal rotations $U_I$. Indeed, for an arbitrary joint probability law $d\mu(U_1,U_2,\ldots)$, we still have $\tau_I=\Tr(P_I\O_0)$ in every realization, because these rotations preserve each window projector.

\subsubsection*{Global Haar embedding}
We next allow a global Haar rotation to mix the energy windows,
\begin{equation} \label{eq:global-Haar-embedding}
    \O = U\O_0U^\dag\,,\qquad U\sim {\rm Haar}\bigl(U(L)\bigr)\,,
\end{equation}
and keep $\O_0$ fixed.\footnote{We can equivalently understand this as starting from the local Haar construction, allowing an additional global rotation $V$ that changes the orientation of the fixed operator $\O_0$ relative to the energy windows, and then independently Haar randomizing within each window, $\O = U_{\rm loc}V\O_0V^\dag U_{\rm loc}^\dag$ where $V$ is a random unitary sampled from the Haar measure on $U(L)$.} Note that this satisfies local rotational invariance since
\begin{equation}
    P(U_{\rm loc}\O U_{\rm loc}^\dag) = P\qty((U_{\rm loc}U)\O_0(U_{\rm loc}U)^\dag) = P(\O)\,,
\end{equation}
where the global Haar measure is invariant under multiplication by any $U_{\rm loc}\subset U(L)$.

Then the global spectrum is fixed, but the spectra of the projected blocks $P_I\O P_I$ are not. Unlike $U_{\rm loc}$, $U$ does not commute with the projectors $P_I$, rather it moves eigenvectors of $\O_0$ between energy windows and leads to nontrivial fluctuations of the block traces
\begin{equation}
    \tau_I = \Tr(P_IU\O_0U^\dag) \,.
\end{equation}
Whereas the local fixed-spectrum Haar rotation $U_{\rm loc}$ preserved every $\tau_I$ separately, the global fixed-Haar rotation $U$ preserves only their sum, $\sum_I\tau_I=\Tr\O_0$ (since $\sum_IP_I=\mathbf1$).

To evaluate these fluctuations, it is useful to separate the global trace from the centered spectrum, just as we did when we considered the globally rotational invariant fixed spectrum in~\eqref{eq:global-haar-t1},
\begin{equation}
    t_1 = \frac1L\Tr\O_0\,,\qquad \widetilde\O_0 \equiv \O_0-t_1\mathbbm 1_L\,,\qquad t_2 = \frac{1}{L} \Tr(\widetilde\O_0^2) = \kappa_{2}^{(1)}+\cdots\,.
\end{equation}
When the external indices are restricted to windows $I$ and $J$, the two-point correlator~\eqref{eq:local-AB} simplifies to
\begin{equation} \label{eq:2-op-loc-glob}
    \overline{(\O_{II})_{ab}(\O_{JJ})_{cd}} = \frac{L}{L^2-1} t_2\qty(\delta_{IJ}\,\delta_{ad}\delta_{bc} - \frac{1}{L}\,\delta_{ab}\delta_{cd})\,,
\end{equation}
where $t_2$ is energy independent and the coefficients~\eqref{eq:local-AB} are the same as in the global rotational invariance case~\eqref{eq:global-AB},
\begin{equation}
    A_I^{\rm glob} = \frac{L}{L^2-1}t_2 \,,\qquad B_{IJ}^{\rm glob}=-\frac{A_I^{\rm glob}}{L}\,,
\end{equation}
which in particular are independent of the window labels $I$ and $J$. The window dependence enters only through the number of states included in each singlet trace,
\begin{equation} \label{eq:F11-globalhaar}
    \overline{\tau_I\tau_J} = \frac{L}{L^2-1} L_I\qty(\delta_{IJ}-\frac{L_J}{L})t_2\,,\qquad F_{\O}^{(1|1)}(E_I|E_J) = -\frac{L_IL_J}{L^2}\kappa_{2}^{(1)}\,.
\end{equation}
Hence in contrast to the independent local-Haar result~\eqref{eq:2-op-loc-loc}, where distinct windows were uncorrelated, the global Haar embedding result~\eqref{eq:2-op-loc-glob} generates a nonvanishing but energy-independent correlation between windows.

\subsubsection*{Correlated local GUE blocks}
Consider instead keeping the rotations local in $\O=U_{\rm loc}\O_0U_{\rm loc}^\dag$, but allowing the spectral data of $\O_0$ in different energy windows to fluctuate jointly. We choose these fluctuations so that each diagonal block $\O_{II}$ is individually an $L_I\times L_I$ GUE matrix with variance $\sigma_I^2/L_I$. This fixes
\begin{equation}
    A_I^{\rm GUE} = \frac{\sigma_I^2}{L_I}\,,\qquad B_{II}^{\rm GUE} = 0\,,\qquad \overline{\tau_I^2}=\sigma_I^2
\end{equation}
in each block, while the trace singlets of different windows can remain correlated,\footnote{We note that we do not fix the correlations between trace singlets, but that the matrix defined by $\Gamma_{IJ}\equiv\overline{\tau_I\tau_J}$ is a covariance matrix and so must be positive semidefinite. In particular, for any two by two principal submatrix $\Gamma^{(I,J)}$, this imposes $\det\Gamma^{(I,J)}=\sigma_I^2\sigma_J^2-\Gamma_{IJ}^2\geq0$, such that $|\Gamma_{IJ}|\leq\sigma_I\sigma_J$. Hence the fluctuations of two trace singlets cannot be more strongly correlated than their individual variances allow.}
\begin{equation}
    \overline{\tau_I\tau_J} = \delta_{IJ}\sigma_I^2 + B_{IJ}^{\rm GUE}L_IL_J\,,\qquad B_{IJ}^{\rm GUE} = \frac{\overline{\tau_I\tau_J}}{L_IL_J}-\delta_{IJ}\frac{\sigma_I^2}{L_I^2}\,.
\end{equation}
We choose the remaining Gaussian fluctuations so that the traceless components $X_I$ defined in~\eqref{eq:local-t1-t2} are independent between windows and independent of the singlet variables $\tau_J$
\begin{equation}
\begin{split}
    \overline{X_{I;ab}X_{J;cd}} &= \delta_{IJ}\frac{\sigma_I^2}{L_I}\left(\delta_{ad}\delta_{bc}-\frac{1}{L_I}\delta_{ab}\delta_{cd}\right)\,,\qquad \overline{X_{I;ab}\tau_J}=0\,,\\
    &\expval{t_{2,I}} = \frac{1}{L_I}\expval{\Tr(X_I^2)} = \sigma_I^2\qty(1-\frac{1}{L_I^2})\,.
\end{split}
\end{equation}
The correlation between two matrix elements sampled from two diagonal blocks is given by
\begin{equation}
\begin{split}
    \overline{(\O_{II})_{ab}(\O_{JJ})_{cd}} &= \delta_{IJ}\frac{\sigma_I^2}{L_I}\left(\delta_{ad}\delta_{bc} - \frac{1}{L_I}\delta_{ab}\delta_{cd}\right) + \frac{\overline{\tau_I\tau_J}}{L_IL_J} \delta_{ab}\delta_{cd}\,,
\end{split}
\end{equation}
where the nontrivial off-diagonal contribution arises from the nonvanishing correlation between trace singlets of different windows. The corresponding large $L_I,L_J$ quantities as defined in~\eqref{eq:local-alpha11} are
\begin{equation} \label{eq:F11-gue}
\begin{split}
    \kappa_2^{(1)}(I) &\equiv \lim_{L_I\to\infty}\expval{t_{2,I}} = \sigma_I^2 \,,\qquad \alpha_{1,1}(I,J) \equiv \lim_{L_I,L_J\to\infty}\overline{\tau_I\tau_J}\,, \\
    &F_{\O}^{(1|1)}(E_I|E_J) = \alpha_{1,1}(I,J) - \delta_{IJ}\sigma_I^2\,,
\end{split}
\end{equation}
where for a single block we recover $\alpha_{1,1}(I,I)=\sigma_I^2$ and $F_{\O}^{(1|1)}(E_I|E_I) = 0$. Thus, each diagonal block is GUE, but the traces $\tau_I$ can still fluctuate jointly across different energy windows. These correlations arise from the trace singlets rather than a common basis rotation, as in the global Haar construction. We now turn to a more general matrix model for $\O$ at fixed Hamiltonian.

\subsection{A matrix model for multi-trace ETH} \label{subsec:matrix-model}
We now introduce a matrix ensemble for $\O$ at fixed $H=\diag(E_1,\ldots,E_L)$ that generates both single- and multi-cycle correlations. We will see that a cyclic potential is sufficient to generate connected observables with multiple trace cycles (see e.g. \cite{Jafferis:2022uhu}). After subtracting the smooth one-point function, we define
\begin{equation}
    Z_{\O}[H]\equiv\int d\O\,e^{-V[\O;H]}\,,\qquad V[\O;H] = V_2[\O;H]+V_{\rm cyc}[\O;H]\,,
\end{equation}
with averages taken with respect to the normalized measure $d\O\,e^{-V}/Z_{\O}[H]$. The quadratic potential fixes the two-point ETH function,
\begin{equation} \label{eq:matrix-propagator}
    V_2[\O;H] = \frac{1}{2}\sum_{i,j}\frac{e^{S(E_{ij})}}{|f_{\O}(E_i,E_j)|^2}\O_{ij}\O_{ji}\,,\qquad \overline{\O_{ij}\O_{kl}} = e^{-S(E_{ij})}|f_{\O}(E_i,E_j)|^2 \,\delta_{il}\delta_{jk}\,.
\end{equation}
The higher ETH functions are generated by cyclic interactions,
\begin{equation} \label{eq:V-cyc}
    V_{\rm cyc}[\O;H] = -\sum_{n\geq3}\frac{1}{n}\sum_{i_1,\ldots,i_n} e^{\sum_{r=1}^n S(E_{i_ri_{r+1}})-(n-1)S(E_{\rm avg})} A_{\O}^{(n)}(E_{i_1},\ldots,E_{i_n})\,\O_{i_2i_1}\O_{i_3i_2}\cdots\O_{i_1i_n}
\end{equation}
where $i_{n+1}\equiv i_1$.\footnote{In a narrow energy window, the entropy prefactor reduces to $e^S=L$, giving the usual scaling of single-trace matrix models, as in e.g. \cite{Jafferis:2022uhu}.} 
The kernels $A_{\O}^{(n)}$ are fixed recursively by matching the connected $n$-cycle correlators to the ETH functions $F_{\O}^{(n)}$, subtracting contributions already generated by lower-order vertices.\footnote{Interactions can generate tadpoles and corrections to the two-point function. These are canceled by linear and quadratic counterterms, keeping the one-point function zero and the Gaussian ETH propagator fixed.} The reversed ordering in \eqref{eq:V-cyc} ensures that the external matrix elements contract with the corresponding oppositely oriented vertex legs: for an external cycle with $\gamma=(1,2,\cdots,n)$, the vertex has the opposite orientation $\gamma^{-1}$.

\subsubsection*{Cyclic vertices and full ETH}
Consider a simple cycle with mutually distinct indices. At first order in $A_{\O}^{(n)}$, each external matrix element contracts with one leg of the
$n$-point vertex. There are $n$ cyclic attachments, which cancel the factor $1/n$ in the potential. Using the propagator in \eqref{eq:matrix-propagator}, this gives
\begin{equation}
    \overline{\O_{i_1i_2} \O_{i_2i_3}\cdots\O_{i_ni_1}}\big|_{\rm tree} = e^{-(n-1)S(E_{\rm avg})} A_{\O}^{(n)}(E_{i_1},\ldots,E_{i_n}) \prod_{r=1}^{n}|f_{\O}(E_{i_r},E_{i_{r+1}})|^2\,.
\end{equation}
This has precisely the entropy scaling of full ETH~\cite{Foini:2018sdb}. At this order, the corresponding single-cycle ETH function is therefore
\begin{equation}
    F_{\O}^{(n)}(E_{i_1},\ldots,E_{i_n})|_{\rm tree} = A_{\O}^{(n)}(E_{i_1},\ldots,E_{i_n})\prod_{r=1}^{n}|f_{\O}(E_{i_r},E_{i_{r+1}})|^2\,.
\end{equation}
The couplings $A_{\O}^{(n)}$ should therefore not be identified directly with the ETH functions $F_{\O}^{(n)}$ or free cumulants, since the latter also receive planar contributions from diagrams with multiple interaction vertices.

\subsubsection*{Emergent multi-trace fluctuations}
The same cyclic interaction can contribute to different external index structures. The quartic term is the simplest example,
\begin{equation} \label{eq:V4}
    V_4[\O] = -\frac{1}{4}\sum_{a,b,c,d} e^{S(E_{ab})+S(E_{bc})+S(E_{cd})+S(E_{da})-3S(E_{\rm avg})} A_{\O}^{(4)}(E_a,E_b,E_c,E_d)\,\O_{ba}\O_{cb}\O_{dc}\O_{ad}\,.
\end{equation}
Let $a=b=i$ and $c=d=j$, where for $i\neq j$, we attach $\O_{ii}$ and $\O_{jj}$ to opposite legs and contract the remaining two legs. The three propagators contribute the entropy factors $e^{-S(E_{i})}$, $e^{-S(E_{j})}$, and $e^{-S(E_{ij})}$, while the specialized quartic vertex contributes $e^{S(E_{i})+S(E_{j})-S(E_{ij})}$. Then we have
\begin{equation}\label{eq:matrix-OO}
    \overline{\O_{ii}\O_{jj}}\big|_{V_4} = e^{-2S(E_{ij})} |f_{\O}(E_i,E_i)|^2|f_{\O}(E_j,E_j)|^2|f_{\O}(E_i,E_j)|^2\,A_{\O}^{(4)}(E_i,E_i,E_j,E_j)+\cdots\,,
\end{equation}
and comparing with the multi-trace ETH ansatz gives
\begin{equation}
    F_{\O}^{(1|1)}(E_i|E_j)\big|_{V_4} = e^{S(E_{i})+S(E_{j})-2S(E_{ij})} |f_{\O}(E_i,E_i)|^2|f_{\O}(E_j,E_j)|^2|f_{\O}(E_i,E_j)|^2\,A_{\O}^{(4)}(E_i,E_i,E_j,E_j)+\cdots\,.
\label{eq:F11-matrix}
\end{equation}
The ellipsis includes other vertices and diagrams with several interactions.

The same quartic vertex also generates planar corrections to the cyclic two-point function. By our prescription above, these lower-point contributions are removed by the corresponding quadratic subtraction, so that $V_2$ continues to fix the prescribed two-point function $f_{\O}$. This subtraction has the cyclic contact structure and therefore does not remove the smooth $i\neq j$ contribution in \eqref{eq:matrix-OO}. The diagonal covariance consequently takes the ETH form in~\eqref{eq:opsff-conditional-eth}.

Planar contractions of cyclic vertices can also connect several external index cycles, so matching the single-cycle ETH functions determines part of the multi-trace data. These relations are not universal and more general models may require additional multi-cycle interaction terms.

\section{Multi-trace ETH in semiclassical chaotic systems}
\label{sec:periodic-orbits}
The first class of systems we consider are quantum systems with a semiclassical limit. The stereotypical examples are quantum particles in a billiard, in a $d$ dimensional potential or on a curved manifold, as well as few body atomic systems.
In such systems, we will identify the multi-trace ETH correlations analytically, showing they precisely satisfy the general structure we presented in subsection \ref{subsec:MT}. We will then compute the one-point SFF and show that it exhibits a linear ramp at late times, in agreement with the prediction obtained by the combined RMT statistics of the spectrum plus multi-trace ETH in subsection \ref{subsec:rmt-eth}.
To achieve this we use the Gutzwiller trace formula\footnote{A great review of the Gutzwiller trace formula and applications can be found in \cite{bogomolny2003quantumarithmeticalchaos}. Other reviews include \cite{Gutzwiller_1990,Haake:2010fgh,Muller:2026ynp,Richter:2022sik}} and its generalization to matrix elements of operators, which approximates the path integral for the density of states $\rho(E)$ as: 
\es{eq:tracefor}{
\rho(E)&=\bar{\rho}(E)+\rho_{osc}(E).
}
The first term $\bar{\rho}(E)$ is the leading smooth contribution to the density of states and captures the average number of states in a microcanonical window. The second term $\rho_{osc}(E)$ contains an infinite sum over highly oscillatory contributions which encode the fine-grained features of the spectrum, including level statistics and discreteness. Each periodic orbit is non-perturbative in $\hbar$ and can be thought of as an oscillatory analogue of an instanton contribution. As a consequence, they are $\mathcal{O}(1)$ effects, unsuppressed in the $\hbar\rightarrow 0$  limit and should not be neglected.  The power of the Gutzwiller trace formula lies precisely in retaining this infinite set of non-perturbative yet large contributions which reconstruct the fine-grained features of the discrete spectrum, otherwise invisible in $\bar{\rho}(E)$.

Trace formulas have proven remarkably effective to obtain precise predictions for concrete systems, ranging from highly excited energy levels in chaotic systems such as atoms in external fields, electromagnetic cavities, and graphene sheets, to transport and response properties of mesoscopic systems \cite{epub1483,brouwer2008andersonlocalizationclassicaltrajectories,Dietz:2020vip}. For a particle on a hyperbolic manifold, the analogous Selberg trace formula is exact rather than asymptotic, making it a cornerstone of modern hyperbolic geometry and analytic number theory \cite{1571417124075027200,1570009749218165120,marklof2008selbergstraceformulaintroduction}.  

Beyond their application to specific systems, trace formulas can be used to derive universal results for chaotic systems, including for example the spectral statistics. The most celebrated example is Berry's semiclassical derivation of the linear ramp in the SFF of chaotic quantum systems \cite{Berry_1985}. By substituting Gutzwiller's trace formula into the product $\rho_{osc}(E_1)\rho_{osc}(E_2)$ one obtains a double sum over periodic orbits involving phase differences $e^{\frac{i}{\hbar}(S_{\gamma_1}-S_{\gamma_2})}$. Berry showed that the leading $\hbar\rightarrow 0$ constructive interference between the large phases is captured by the diagonal part of the double sum and this diagonal approximation produces the linear ramp. 
This provided a derivation of the leading order of the long-standing BGS conjecture \cite{Bohigas:1983er}, which states that the spectrum of a quantum system with a chaotic classical limit will exhibit RMT statistics.

Conceptually, this mechanism shows how a manifestly factorized microscopic quantity can develop a smooth connected component through interference between rapidly oscillating contributions.
In holography this mechanism provides a solution to the factorization problem, as it demonstrates that smooth wormhole contributions can arise from the interference of highly oscillatory terms in the CFT density of states, without the need for ensemble averaging. This can be realized explicitly in AdS$_3$/CFT$_2$ where a CFT trace formula and the analogue of Berry's diagonal approximation were shown to exactly reproduce the partition function of wormholes in three-dimensional gravity \cite{DiUbaldo:2023qli, Boruch:2025ilr}.

We will similarly use the trace formula to derive the multi-trace ETH correlations between diagonal matrix elements $O_{ii}O_{jj}$ in this class of systems. Gaussian ETH will follow from Berry's diagonal approximation and multi-trace ETH  will appear as the leading off-diagonal contribution. 
The derivation is rather universal and will essentially apply to any quantum system or field theory with a semiclassical limit $\hbar_{\rm eff}\rightarrow 0 $, in which the path integral admits a trace formula as an infinite sum over classical contributions of the form $e^{iS_{\rm cl}/{\hbar_{\rm eff}}}$.

Beyond its original application to few-particle systems, the semiclassical trace formula has been developed and applied to many other systems of which we mention a few: $N$-particle systems with local interactions \cite{gutkin2015classicalfoundationsmanyparticlequantum}, many-body systems \cite{Richter:2022sik, dubertrand2016spectralstatisticschaoticmanybody,Engl:2015hul, Rammensee:2018pyk,Liao:2021uwx}, spin chains \cite{Ermakov:2024pta}, Floquet systems \cite{Chan:2017kzq},  Dirac fermions and graphene\cite{Dietz:2020vip}, fermionic many-body systems \cite{gaspard1996semiclassicaltheorymanybodyfermionic}, Anderson localization \cite{ brouwer2008andersonlocalizationclassicaltrajectories} and the zeros of the Riemann zeta function \cite{Berry:1999qhj}. Trace formulas in continuum quantum field theories have been less developed, despite the early works of Goldstone, Jackiw and of Dashen, Hasslacher, Neveu \cite{Goldstone:1974gf,Dashen:1974ci,Dashen:1975hd}. Recently there have been several interesting and promising applications to quantum scars \cite{Cotler:2022syr}, to the spectrum of large charge \cite{Sannino:2026ena}, to resurgence and exact quantization \cite{Sueishi:2020rug,Ture:2024nbi} to 2d integrable QFTs like Sine-Gordon QFT and non-linear Schrodinger QFT \cite{Mussardo:2004zn, DelVecchio:2020lxf}. However the formulation of a trace formula for generic large $N$ quantum field theories or many-body quantum systems has yet to be understood.  In the context of 2d conformal field theories for central charge $c>1$, a CFT trace formula was presented in \cite{DiUbaldo:2023qli,Haehl:2023tkr} and applied to describe spectral statistics of chaotic CFTs.

\subsection{A brief introduction to semiclassical trace formulas}
\label{subsec:trace-formula-review}
The derivation of a trace formula starts from computing $\Tr(e^{-iHT/\hbar})$ in two different ways: in the energy basis and in the complete basis of an operator such as position $\hat{q}$,
\es{}{\label{eq:per-orb-tr}
\Tr(e^{-iHT/\hbar})= \int dE e^{-iET/\hbar}\rho(E)= \int dq \mel{q}{e^{-iHT/\hbar}}{q}.
}
The trace in the position basis can be computed using the path integral propagator from an initial position $q_i$ to a final position $q_f$. In the semiclassical limit the propagator is given by an infinite sum over classical paths connecting the endpoints \cite{Gutzwiller_1990,Haake:2010fgh, Muller:2026ynp}:
\begin{equation}
  \mel{q_f}{e^{-iHT/\hbar}}{q_i}
  \simeq
  \sum_{\text{classical paths }\alpha:q_i\to q_f}
  D_{\alpha}^{1/2}
  \exp\!\left[\frac{i}{\hbar}S_{\alpha}(q_f,q_i;T)
  -\frac{i\pi}{2}\nu_{\alpha}\right] .
  \label{eq:van-vleck-schematic}
\end{equation}
Here $S_{\alpha}$ is the classical action, $D_{\alpha}$ is the Van Vleck stability one-loop determinant, and
$\nu_{\alpha}$ is the Maslov index of the open trajectory.  We can then set $q_f=q_i=q$ and integrate over the common endpoint $q$ to obtain the trace.
The stationary configurations are then classical periodic trajectories that close after a time
$T$.  Fourier transforming from time $T$ to energy $E$ produces the Gutzwiller trace formula for a $D$-dimensional system \eqref{eq:tracefor}:
\begin{align}
  \overline\rho(E)&=\frac{1}{(2\pi\hbar)^D}\int d^{2D}x\,\delta(E-H(x)),
  \label{eq:weyl-density}\\
  \rho_{\rm osc}(E)&=
  \frac{1}{\pi\hbar}\operatorname{Re}
  \sum_{\gamma}w_{\gamma}
  e^{\frac{i}{\hbar}S_{\gamma}(E)} 
  \label{eq:gutzwiller-trace}
\end{align}
which defines the state densities in \eqref{eq:tracefor}. The smooth term \eqref{eq:weyl-density} is equal to the volume of phase-space at energy $E$ and can be thought of as the contribution of the trivial periodic orbit of vanishing length.
The oscillatory term sum runs over isolated classical periodic orbits of energy $E$, including primitive orbits and repetitions.  Their action and
period are
\begin{equation}
  S_{\gamma}(E)=\oint_{\gamma}p\,dq,
  \qquad
  T_{\gamma}(E)=\frac{\partial S_{\gamma}}{\partial E},
\end{equation}
and the stability amplitude is equal to
\begin{equation}
  w_{\gamma}=\frac{T_{\gamma}/r_{\gamma}}
  {\sqrt{\left|\det(M_{\gamma}-\mathbbm 1)\right|}}
  e^{-i\pi\mu_{\gamma}/2} .
  \label{eq:gutzwiller-amplitude}
\end{equation}
Here $M_\gamma$ is the monodromy matrix that describes small perturbations around the periodic orbit, $r_\gamma$ counts how many times the orbit is traversed, and $\mu_\gamma$ is the closed periodic-orbit Maslov index \cite{bogomolny2003quantumarithmeticalchaos, Haake:2010fgh}. The stability determinant $M_\gamma$ and Maslov index $\mu_\gamma$ are related to $D_\alpha$ and $\nu_\alpha$ in \eqref{eq:van-vleck-schematic} through a saddlepoint approximation of the trace \eqref{eq:per-orb-tr}.

Periodic orbit contributions  are not individually small, instead the semiclassical limit makes their
phases rapidly oscillatory. We shall now see how  spectral correlations and RMT statistics arise from coherent interference between infinitely many oscillatory terms. 
In what follows we keep $\hbar$ explicit to distinguish the semiclassical limit and the relevant time scales, recovering the conventions of the preceding section by setting $\hbar=1$. We assume ergodic and hyperbolic classical dynamics with isolated periodic orbits. Individual contributions generally disappear upon coarse graining, but pairs of orbits with correlated actions can survive. It is these surviving pairs that give rise to the linear ramp and, with operator insertions, to the multi-trace correlations of interest here.

To see this mechanism, consider the local connected spectral form factor
\begin{equation}
  K(T;E)=\frac{1}{\bar\rho(E)}\int d\omega\,
  e^{-i\omega T/\hbar}
  \overline{\rho(E+\omega/2)\rho(E-\omega/2)},
  \qquad
  T_H(E)=2\pi\hbar\bar\rho(E),\quad \tau=\frac{T}{T_H}.
  \label{eq:sc-spectral-form-factor}
\end{equation}
The Heisenberg or plateau time $T_H$ is the scale  to resolve the mean level spacing. The overline denotes a connected average, where only the product of oscillatory terms $\rho_{\rm osc}$ remains. For a single Hamiltonian these averages can be implemented by smoothing over an energy window containing many levels, but sufficiently narrow that the classical dynamics and $\bar\rho$ vary little across it.  We work within one symmetry sector and assume no systematic degeneracies of orbit actions. 

Substituting the trace formula into~\eqref{eq:sc-spectral-form-factor} turns the product of spectral densities into a sum over pairs of periodic orbits. Expanding in small energy difference, we have
\begin{equation}
 S_\gamma(E+\omega/2)-S_{\gamma'}(E-\omega/2)
 \simeq S_\gamma(E)-S_{\gamma'}(E)
       +\frac{\omega}{2}(T_\gamma+T_{\gamma'}),
\end{equation}
where we used $\partial_E S_\gamma=T_\gamma$. Fourier transformation therefore selects pairs with mean period $T$:
\begin{equation}
  K(T;E)\simeq\frac{1}{T_H}
  \bE\!\left[\sum_{\gamma,\gamma'}w_\gamma w_{\gamma'}^*
  e^{i(S_\gamma-S_{\gamma'})/\hbar}
  \delta_{\Delta T}\!\left(T-\frac{T_\gamma+T_{\gamma'}}{2}\right)\right].
  \label{eq:sc-spectral-orbit-sum}
\end{equation}
Here $\delta_{\Delta T}$ is a delta function broadened by the time smoothing. In equation~\eqref{eq:sc-spectral-orbit-sum}, two orbits with action difference of order $\hbar$ contribute coherently, while generic pairs have rapidly fluctuating phases and average to zero.

The simplest coherent pairs are an orbit with itself and, in a time-reversal invariant system, with its time reverse. Keeping these pairs is Berry's diagonal approximation~\cite{Berry_1985}, illustrated in figure \ref{fig:sc-orbit-pairs}(a). The word \emph{diagonal} here refers to the orbit labels $\gamma=\gamma'$; it is distinct  from equality of the energy indices $i=j$ which gives the plateau and, we shall see, Gaussian ETH. The sum over the surviving orbits is evaluated with the Hannay-Ozorio de Almeida sum rule~\cite{1362544419804635136} 
\begin{equation}
  \bE\!\left[\sum_\gamma |w_\gamma|^2
  \delta_{\Delta T}(T-T_\gamma)\right]\simeq T.
  \label{eq:sc-orbit-sum-rule}
\end{equation}
Physically, this sum rule expresses the equidistribution of long periodic orbits with their stability weights. Chaotic dynamics produces an increasing number of long orbits, while instability reduces the weight of each individual orbit. Their combined contribution has the simple asymptotic form~\eqref{eq:sc-orbit-sum-rule}. Consequently,
\begin{equation}
  K_{\rm diag}(\tau)=\kappa\tau,
  \qquad
  \kappa=\begin{cases}
    1,&\text{without time-reversal symmetry},\\
    2,&\text{with spinless time-reversal symmetry}.
  \end{cases}
  \label{eq:sc-berry-ramp}
\end{equation}
The factor of two in the second case counts the identical and time-reversed pairings. This is the leading GUE or GOE ramp, respectively, and gives the first semiclassical realization of the spectral universality anticipated by the BGS conjecture~\cite{Bohigas:1983er}. 

The diagonal approximation is not the end of the story. In a time-reversal invariant system, a long orbit can contain two stretches that pass close to one another with almost opposite velocities. Within this encounter, the stretches can be reconnected to form a distinct periodic orbit. The partner follows almost the same long loops as the original orbit, reversing the direction of one of them, but differs in the encounter region. Such a Sieber-Richter pair has an action difference of order $\hbar$ and thus survives energy averaging~\cite{epub1464}. A schematic example appears in figure \ref{fig:sc-orbit-pairs}(b). Counting these encounters and integrating their interference phases gives the first correction to the GOE ramp,
\begin{equation}
  K_{\rm partner}(\tau)=-2\tau^2,
  \qquad K(\tau)=2\tau-2\tau^2+O(\tau^3).
  \label{eq:sc-spectral-partner}
\end{equation}
More complicated encounters generate higher orders~\cite{Muller:2004nb, Muller:2026ynp, Muller:2005fgw} which have been systematically studied and shown to reproduce the infinite series predicted by RMT. In JT gravity, the same series has been reproduced by taking the $\uptau$-scaling limit of wormhole amplitudes \cite{Blommaert:2022lbh,Saad:2022kfe} 
and the relation between geometries and periodic orbits has been discussed in \cite{Saad:2022kfe,DiUbaldo:2025cmg}. The richer and more complex  GOE class dual to unorientable JT gravity has been studied in \cite{DiUbaldo:2025cmg,Tall:2024fij,Haneder:2024ccc}. For our purposes, the main point is already visible at this order: pairs of distinct classical orbits can contribute coherently and correct the orbit-diagonal result. We next explain how the same mechanism appears in the one-point SFF and diagonal matrix elements.

\subsection{The one-point SFF }
\label{subsec:sc-one-point-sff}
The one-point SFF differs from the ordinary spectral form factor by the insertion of an observable in each trace. Its semiclassical description is obtained by replacing the density of states with the operator-weighted density
\begin{equation}
  d_\O(E)=\Tr\!\left(\O\delta(E-H)\right)
         =\sum_i\O_{ii}\delta(E-E_i).
  \label{eq:sc-weighted-density}
\end{equation}
We take $\O$ to be Hermitian and subtract its smooth diagonal expectation value. This removes the contribution already determined by the ordinary spectral form factor, leaving the fluctuations of the matrix elements. The local form factor is 
\begin{equation}
  \frac{K_\O(T;E)}{\bar\rho(E)}=\frac{1}{\bar\rho(E)}\int d\omega\,
  e^{-i\omega T/\hbar}
  \overline{d_\O(E+\omega/2)d_\O(E-\omega/2)}.
  \label{eq:sc-weighted-form-factor}
\end{equation}
It is related to the thermal one-point SFF of the preceding section by
\begin{align}
  &\overline{\Tr(e^{-(\beta+iT/\hbar)H}\O)
              \Tr(e^{-(\beta-iT/\hbar)H}\O)}
  \nonumber\\
  &\hspace{1cm}=\int dE\,d\omega\,e^{-2\beta E-i\omega T/\hbar}
  \overline{d_\O(E+\omega/2)d_\O(E-\omega/2)}
  =\int dE\,e^{-2\beta E}\bar\rho(E)K_\O(T;E).
  \label{eq:sc-canonical-one-point}
\end{align}
The microcanonical calculation therefore supplies the local building block of the canonical observable.

The weighted trace formula~\cite{PhysRevA.45.3531, turek2005semiclassicalformfactorspectral,Feingold:1986fu,Wilkinson1987} has a particularly simple interpretation. Inserting $\O$ in the trace weighs a periodic orbit by the average of its classical observable along that orbit:
\begin{equation}
  d_{\O,{\rm osc}}(E)\simeq\frac{1}{\pi\hbar}\operatorname{Re}
  \sum_\gamma w_\gamma\O_\gamma e^{iS_\gamma/\hbar},
  \qquad
  \O_\gamma=\frac{1}{T_\gamma}
  \int_0^{T_\gamma}dt\,\O_{\rm cl}(x_\gamma(t)).
  \label{eq:sc-weighted-trace-formula}
\end{equation}
Here $\O_{\rm cl}$ is the Weyl symbol of $\O$, approaching the corresponding phase-space observable in the semiclassical limit. The actions and stability amplitudes are the same as in the spectral trace formula, with the operator dependence entering through $\O_\gamma$. The orbit-pair expression then becomes
\begin{equation}
 \frac{K_\O(T;E)}{\bar\rho(E)}\simeq\frac{1}{T_H}
  \bE\!\left[\sum_{\gamma,\gamma'}w_\gamma w_{\gamma'}^*
  \O_\gamma\O_{\gamma'}
  e^{i(S_\gamma-S_{\gamma'})/\hbar}
  \delta_{\Delta T}\!\left(T-\frac{T_\gamma+T_{\gamma'}}{2}\right)\right].
  \label{eq:sc-weighted-orbit-sum}
\end{equation}
Thus the same coherent pairs govern both form factors, but their contributions now carry correlated orbit averages of the observable.

The derivation of the one-point SFF from periodic orbits was performed in \cite{ turek2005semiclassicalformfactorspectral}, to which we refer for details. We consider time-reversal invariant observables $\cO$, since this is the case in which the GOE orbit pairings contribute to nonzero diagonal matrix elements.\footnote{For a general observable, the result involves the time-reversal projection $\O_{\rm cl}^{S}(x)=[\O_{\rm cl}(x)+\O_{\rm cl}(\Theta x)]/2$, with $\Theta(q,p)=(q,-p)$. A time-reversal odd Hermitian observable has vanishing diagonal elements in a nondegenerate time-reversal invariant eigenbasis.} The relevant classical input is the microcanonical autocorrelation,
\begin{equation}
  C_\O(t;E)=\langle\O_{\rm cl}(x)\O_{\rm cl}(x_t)\rangle_E,
  \qquad
  I_\O(E)=\int_0^\infty dt\,C_\O(t;E).
  \label{eq:sc-classical-correlator}
\end{equation}
The average $\langle\cdot\rangle_E$ is the classical phase space average on the energy shell, normalized by its phase-space volume. We assume that $\O$ has zero classical mean and that $C_\O$ decays sufficiently rapidly for the integral to converge. Conserved quantities and operators with non-decaying correlations require a separate treatment.

The role of $I_\O$ can be understood without the full encounter calculation. 
For a long periodic orbit, averaging $\O$ over its trajectory suppresses its fluctuations by a factor of $1/T$. In the weighted periodic-orbit sum, this suppression cancels the factor of $T$ responsible for the ordinary linear ramp, leaving a time-independent contribution to $K_\O$. Evaluating the identical and time-reversed pairings gives~\cite{eckhardt1995semiclassicalformfactormatrix,eckhardt1995approachergodicityquantumwave, turek2005semiclassicalformfactorspectral}
\begin{equation}
 \frac{K_{\O,{\rm diag}}(\tau;E)}{\bar\rho(E)} \simeq\frac{2I_\O(E)}{T_H(E)}.
  \label{eq:sc-turek-diagonal}
\end{equation}
Here $I_\O$ arises from the stability-weighted periodic-orbit average, in the normalization of~\eqref{eq:sc-weighted-orbit-sum}.

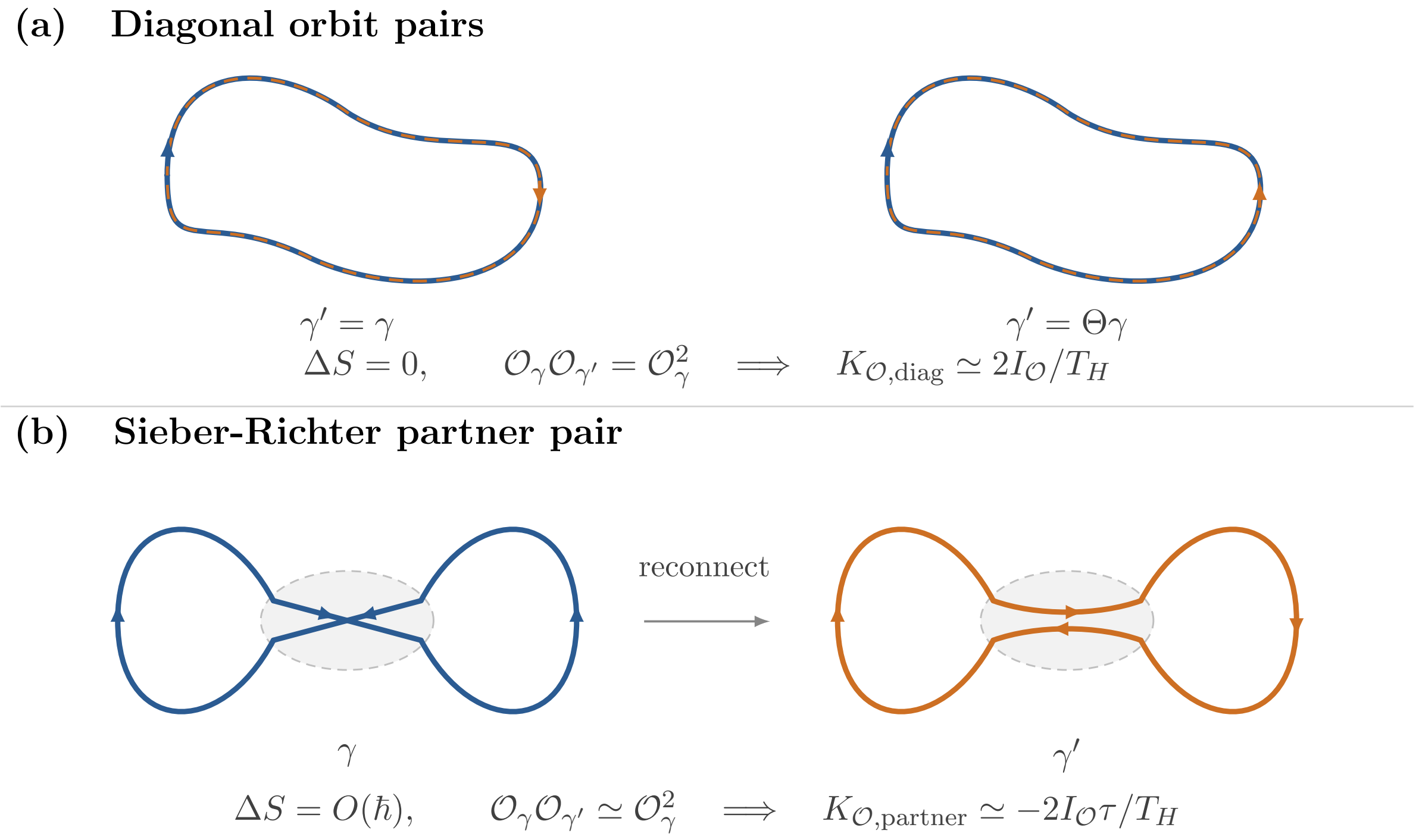
\begin{figure}[t]
\centering
\begin{tikzpicture}[x=1cm,y=1cm,font=\small,
  orbit/.style={line width=1.7pt,line cap=round,line join=round},
  flow/.style={-latex,line width=1.1pt},
  encounter/.style={fill=black!5,draw=black!25,densely dashed,line width=.55pt},
  labeltext/.style={text=black!75}]
\definecolor{scOrbitBlue}{RGB}{43,91,146}
\definecolor{scOrbitOrange}{RGB}{205,111,35}
\path[use as bounding box] (0,-.65) rectangle (16.0,8.5);
\node[anchor=west,font=\bfseries] at (.15,8.25) {(a)\quad Diagonal orbit pairs};
\begin{scope}[shift={(4,6.6)}]
  \def\scLoop{(-2,0) .. controls (-2,1.35) and (-.75,1.25) .. (0,.7)
   .. controls (1.0,.05) and (2.15,.75) .. (2.15,-.15)
   .. controls (2.15,-1.35) and (.45,-1.35) .. (-.45,-.9)
   .. controls (-1.6,-.35) and (-2,-1.0) .. (-2,0)}
  \draw[orbit,scOrbitBlue] \scLoop;
  \draw[line width=.9pt,scOrbitOrange,dash pattern=on 4.5pt off 3pt] \scLoop;
  \draw[flow,scOrbitBlue] (-2,.10) -- (-1.99,.42);
  \draw[flow,scOrbitOrange] (2.14,-.06) -- (2.14,-.36);
  \node[labeltext] at (0,-1.65) {$\gamma'=\gamma$};
\end{scope}
\begin{scope}[shift={(12,6.6)}]
  \def\scLoop{(-2,0) .. controls (-2,1.35) and (-.75,1.25) .. (0,.7)
   .. controls (1.0,.05) and (2.15,.75) .. (2.15,-.15)
   .. controls (2.15,-1.35) and (.45,-1.35) .. (-.45,-.9)
   .. controls (-1.6,-.35) and (-2,-1.0) .. (-2,0)}
  \draw[orbit,scOrbitBlue] \scLoop;
  \draw[line width=.9pt,scOrbitOrange,dash pattern=on 4.5pt off 3pt] \scLoop;
  \draw[flow,scOrbitBlue] (-2,.10) -- (-1.99,.42);
  \draw[flow,scOrbitOrange] (2.14,-.36) -- (2.14,-.06);
  \node[labeltext] at (0,-1.65) {$\gamma'=\Theta\gamma$};
\end{scope}
\node[labeltext] at (8,4.47)
 {$\Delta S=0,\qquad \O_\gamma\O_{\gamma'}=\O_\gamma^2
   \quad\Longrightarrow\quad K_{\O,\mathrm{diag}}\simeq {2I_\O}/{T_H}$};
\draw[black!18,line width=.5pt] (.15,4.03) -- (15.85,4.03);
\node[anchor=west,font=\bfseries] at (.15,3.72) {(b)\quad Sieber-Richter partner pair};
\begin{scope}[shift={(4,1.65)}]
  \draw[encounter] (0,0) ellipse (.96 and .55);
  \draw[orbit,scOrbitBlue]
   (-.82,.22) .. controls (-1.42,1.35) and (-2.55,1.27) .. (-2.55,0)
   .. controls (-2.55,-1.27) and (-1.42,-1.35) .. (-.82,-.22);
  \draw[orbit,scOrbitBlue]
   (.82,-.22) .. controls (1.42,-1.35) and (2.55,-1.27) .. (2.55,0)
   .. controls (2.55,1.27) and (1.42,1.35) .. (.82,.22);
  \draw[orbit,scOrbitBlue] (-.82,.22) -- (.82,-.22);
  \draw[orbit,scOrbitBlue] (.82,.22) -- (-.82,-.22);
  \draw[flow,scOrbitBlue] (-2.55,-.19) -- (-2.55,.20);
  \draw[flow,scOrbitBlue] (2.55,-.19) -- (2.55,.20);
  \draw[flow,scOrbitBlue] (-.45,.121) -- (-.10,.027);
  \draw[flow,scOrbitBlue] (.45,.121) -- (.10,.027);
  \node[labeltext] at (0,-1.50) {$\gamma$};
\end{scope}
\begin{scope}[shift={(12,1.65)}]
  \draw[encounter] (0,0) ellipse (.96 and .55);
  \draw[orbit,scOrbitOrange]
   (-.82,.22) .. controls (-1.42,1.35) and (-2.55,1.27) .. (-2.55,0)
   .. controls (-2.55,-1.27) and (-1.42,-1.35) .. (-.82,-.22);
  \draw[orbit,scOrbitOrange]
   (.82,-.22) .. controls (1.42,-1.35) and (2.55,-1.27) .. (2.55,0)
   .. controls (2.55,1.27) and (1.42,1.35) .. (.82,.22);
  \draw[orbit,scOrbitOrange]
   (-.82,.22) .. controls (-.35,.05) and (.35,.05) .. (.82,.22);
  \draw[orbit,scOrbitOrange]
   (.82,-.22) .. controls (.35,-.05) and (-.35,-.05) .. (-.82,-.22);
  \draw[flow,scOrbitOrange] (-2.55,-.19) -- (-2.55,.20);
  \draw[flow,scOrbitOrange] (2.55,.20) -- (2.55,-.19);
  \draw[flow,scOrbitOrange] (-.19,.0925) -- (.20,.0925);
  \draw[flow,scOrbitOrange] (.19,-.0925) -- (-.20,-.0925);
  \node[labeltext] at (0,-1.50) {$\gamma'$};
\end{scope}
\draw[-latex,black!48,line width=.7pt] (7.3,1.64) -- (8.7,1.64);
\node[labeltext,font=\footnotesize,align=center] at (8,2.25) {reconnect\,};
\node[labeltext] at (8,-.48)
 {$\Delta S=O(\hbar),\qquad \O_\gamma\O_{\gamma'}\simeq\O_\gamma^2
   \quad\Longrightarrow\quad K_{\O,\mathrm{partner}}\simeq -{2I_\O\tau}/{T_H}$};
\end{tikzpicture}

\caption{Periodic orbit pairs contributing to the one-point SFF of a mean-subtracted, time-reversal even observable $\cO$. Blue denotes the orbit $\gamma$ and orange denotes $\gamma'$. (a) Diagonal pair: identical and time-reversed orbits have equal actions and overlapping configuration-space projections; they give the  constant in time contribution. (b) Sieber-Richter pair: the two orbits follow the same two large loops, with one loop traversed in the opposite direction, and differ only within the shaded encounter. Reconnection replaces the crossing by an avoided crossing. The two orbits are equally weighted by the operator $\cO$ to leading order, giving the same factor $I_{\cO}$, while the encounter derivation gives the negative linear correction $-\tau/T_{H}$.}
\label{fig:sc-orbit-pairs}
\end{figure}

We now consider pairs of distinct but closely related partner orbits. A Sieber-Richter partner follows almost the same long stretches as the original orbit, sometimes in the opposite direction. Since $\O_{\rm cl}$ is time-reversal even, reversing a stretch does not change its contribution to the orbit average. To the order retained in the semiclassical calculation, the two orbit weights are therefore correlated as illustrated in figure \ref{fig:sc-orbit-pairs}(b). In evaluating the weighted encounter sum, we find that the observable enters through precisely the same integral $I_\O$ as in~\eqref{eq:sc-turek-diagonal}, while the interference and encounter counting supply the additional factor $-\tau$:
\begin{equation}
\frac{K_{\O,{\rm diag}}(\tau;E)}{\bar\rho(E)}
  \simeq-\frac{2I_\O(E)}{T_H(E)}\,\tau.
  \label{eq:sc-turek-contributions}
\end{equation}
Combining the two contributions gives the result for the one-point SFF:
\begin{equation}
  \frac{K_{\O}(\tau;E)}{\bar\rho(E)}\simeq
  \frac{I_\O(E)}{\tau T_H(E)}
  \left[K_{\rm GOE}(\tau)+O(\tau^3)\right]
  =\frac{2I_\O(E)}{T_H(E)}
    \left[1-\tau+O(\tau^2)\right].
  \label{eq:sc-turek-result}
\end{equation}
We immediately recognize the two signatures of Gaussian and multi-trace ETH, respectively a constant term and a linear ramp term as predicted in subsection \ref{subsec:rmt-eth}.

Equation~\eqref{eq:sc-turek-result} separates the classical information about the observable from the universal interference of orbit pairs. The dependence on the observable in \eqref{eq:sc-turek-result} enters only through $I_\O(E)$. The constant and linear terms arise from the same orbit-pair contributions as the first two orders of the GOE spectral form factor, but the orbit weights change their time dependence. As a result, the one-point SFF starts with the orbit-diagonal constant and receives a negative linear ramp correction.

It is interesting to note that in the one-point SFF, the diagonal approximation is associated with the Gaussian ETH term rather than with the spectral statistics $\overline{\rho(E_1)\rho(E_2)}$ term, i.e. with the first line in \eqref{eq:opsff-final} rather than the third line. The spectral statistics are instead captured by the leading off-diagonal Sieber-Richter pair of partner orbits. 

 The result holds at fixed small $\tau$ in the semiclassical limit, with $T$ large compared with the classical correlation and encounter times. Although the exact GOE spectral form factor is $K_{\rm GOE}(\tau)=2\tau-\tau\log(1+2\tau)$ for $0<\tau<1$, our calculation establishes only the orders displayed in~\eqref{eq:sc-turek-result}. Extending $K_{\rm GOE}(\tau)/\tau$ to arbitrary times would require additional orbit correlations. We will keep this restriction in mind when extracting the matrix-element correlations below.

\subsection{Multi-trace ETH from the one-point SFF}
\label{subsec:sc-extracting-multitrace}
The implication of this semiclassical calculation for ETH is that the matrix element correlations must go beyond Gaussian ETH. The partner-orbit term in~\eqref{eq:sc-turek-contributions} requires a correlation between $O_{ii}$ and $O_{jj}$ for $i\neq j$ as described by $F_\O^{(1|1)}$ in multi-trace ETH.

We now show this more explicitly. The diagonal covariance takes the form
\begin{equation}
  \overline{\O_{ii}\O_{jj}}
  =2e^{-S(E_i)}|f_\O(E_i,0)|^2\delta_{ij}
   +e^{-S(E_i)-S(E_j)}F_\O^{(1|1)}(E_i|E_j)
   +\cdots.
  \label{eq:sc-conditional-eth-dictionary}
\end{equation}
where the factor of two is due to GOE time-reversal symmetry. 
The pair correlation of distinct energy levels is
\begin{equation}
 \rho^{\neq}_2(E_1,E_2)
  =\bE\!\left[\sum_{i\ne j}\delta(E_1-E_i)\delta(E_2-E_j)\right]
  =\bE[\rho(E_1)\rho(E_2)]
    -\bar\rho(E_1)\delta(E_1-E_2).
  \label{eq:sc-distinct-level-density}
\end{equation}
Combining the spectral pair density with~\eqref{eq:sc-conditional-eth-dictionary} gives
\begin{align}
  \overline{d_\O(E_1)d_\O(E_2)}
  &\simeq 2|f_\O(E_1,0)|^2\delta(E_1-E_2)
   +\frac{\rho_2^{\ne}(E_1,E_2)}
          {\bar\rho(E_1)\bar\rho(E_2)}
       F_\O^{(1|1)}(E_1|E_2).
  \label{eq:sc-exact-weighted-decomposition}
\end{align}
as shown in \eqref{eq:opsff-weighted-decomposition}.
Equation~\eqref{eq:sc-exact-weighted-decomposition} displays the quantity that the one-point SFF actually probes: the diagonal matrix-element covariance multiplied by the probability of finding the corresponding pair of levels.
Late times correspond to small energy differences $E_1-E_2\ll1$, where we can approximate the mean densities $\bar\rho(E_1)\bar\rho(E_2)\sim \bar\rho(E)^2$ and obtain the prediction for the one-point SFF:
\begin{equation}
  K_\O(T;E)\simeq 2e^{-S(E)}|f_\O(E,0)|^2
  +\frac{1}{\bar\rho(E)}
    \int d\omega\,e^{-i\omega T/\hbar}
    \frac{\rho_2^{\ne}(E+\omega/2,E-\omega/2)}{\bar\rho(E)^2}
    F_\O^{(1|1)}(E+\omega/2|E-\omega/2).
  \label{eq:sc-multitrace-form-factor}
\end{equation}
The spectral correlations are given at leading order by
\[
\rho_2^{\neq}(E+\omega/2,E-\omega/2)
  \simeq\bar{\rho}(E)^2-\frac{1}{\pi^2\omega^2}.
\]
Following the local approximation used in~\eqref{eq:opsff-final},
we replace $F_\O^{(1|1)}$ by its equal-energy value in the term
containing the universal spectral correlations. The
$\bar\rho(E)^2$ contribution gives the smooth Fourier transform
of $F_\O^{(1|1)}$, which decays with time. The RMT term
$-1/(\pi^2\omega^2)$  instead gives the linear ramp contribution
\begin{equation}
  \left.K_\O(T;E)\right|_{\rm linear}
  \simeq
  \frac{|T|}{\pi\hbar\bar\rho(E)^3}
  F_\O^{(1|1)}(E|E)
  =
  2e^{-2S(E)}F_\O^{(1|1)}(E|E)\,\tau,
  \qquad \tau=\frac{|T|}{T_H(E)}.
\end{equation}
Thus the $1/\omega^2$ dependence is supplied by
$\rho_2^{\neq}$, while its operator coefficient is isolated by
comparison with the Sieber--Richter term
$-2I_\O(E)\tau/T_H(E)$. This fixes the semiclassical contribution
\begin{equation}
  \left.F_\O^{(1|1)}(E|E)\right|_{\rm sc}
  \simeq
  -\frac{e^{2S(E)}I_\O(E)}{T_H(E)}
  =
  -\frac{e^{S(E)}I_\O(E)}{2\pi\hbar}.
  \label{eq:sc-multitrace-eth}
\end{equation}

To match the constant term consistently, we must also retain the
same-index multitrace contribution
$e^{-2S(E)}F_\O^{(1|1)}(E|E)$ displayed
in~\eqref{eq:opsff-weighted-decomposition}.
Its contact cancels the subtraction associated with excluding
$i=j$ from $\rho_2^{\neq}$. After the smooth contribution has
decayed, the resulting expression is
\begin{equation}
  K_\O(T;E)\simeq
  2e^{-S(E)}|f_\O(E,0)|^2
  +e^{-2S(E)}F_\O^{(1|1)}(E|E)K_{\rm GOE}(\tau),
  \qquad K_{\rm GOE}(\tau)=2\tau+O(\tau^2).
\end{equation}
Within this local parametrization, matching the constant requires
$e^{-S(E)}|f_\O(E,0)|^2=I_\O(E)/T_H(E)$.
Together with~\eqref{eq:sc-multitrace-eth}, this gives
\begin{equation}
  K_\O(T;E)\simeq
  \frac{2I_\O(E)}{T_H(E)}
  \left[1-\tau+O(\tau^2)\right].
\end{equation}
The diagonal multitrace contact is essential here: for the
semiclassical contribution extracted above, it is of the same
order as the Gaussian contact.

\section{Multi-trace ETH in JT gravity} \label{sec:JT}
JT gravity coupled to matter provides a setting in which thermal correlation functions can be computed using exact diagrammatic rules. On the disk, these rules express matter correlators in terms of integrals over intermediate energies, with vertices and crossing kernels given by Gamma functions and $6j$ symbols \cite{Mertens:2017mtv, Blommaert:2018oro, Iliesiu:2019xuh}. The same diagrammatic approach extends to wormhole geometries, where matter propagators can connect different boundaries \cite{Saad:2019pqd, Jafferis:2022wez}.

In this section, we first show that the naive operator one-point spectral form factor in JT gravity is time independent and explain how multi-trace ETH correlations arise from unitarity, in subsection \ref{subsec:JT-sff}. We then turn to the JKMS matrix model, which provides an explicit realization of these corrections, in subsection \ref{subsec:jkms}, before extending the discussion to interacting matter in subsection \ref{subsec:JT-interacting}.

\subsection{A time-dependent one-point SFF in JT gravity} \label{subsec:JT-sff}
We now apply the multi-trace ETH ansatz of subsection \ref{subsec:MT} to JT gravity. 
\begin{equation}
    K_{\O}(\beta,T) \equiv \int dE_1dE_2 e^{-\beta(E_1+E_2)}e^{-\ii T(E_1-E_2)} \bE[\rho(E_1)\rho(E_2)\O_{E_1E_1}\O_{E_2E_2}]\,,
\label{eq:JT-one-point-SFF}
\end{equation}
where in JT gravity the connected matrix-element correlation is encoded by a matter geodesic connecting the operator insertions on the two boundaries. At late times, \eqref{eq:JT-one-point-SFF} has the same universal local correlations that produce the ramp and plateau of the ordinary spectral form factor \cite{Saad:2019lba, Saad:2019pqd}. Combining these spectral correlations with the multi-trace contribution of~\eqref{eq:opsff-conditional-eth} gives the late time behavior
\begin{align}\label{eq:JT-multitrace-ramp}
    K_{\O}^{\rm ramp}(\beta,T) &\simeq \int dE\, e^{-2\beta E-2S_0} F_{\O}^{(1|1)}(E|E) \min\!\qty(\frac{|T|}{2\pi},e^{S_0}\rho(E) ) \,,
\end{align}
which predicts an operator-weighted analogue of the usual ramp-to-plateau crossover. In JT gravity, we organize the large-entropy expansion in powers of $e^{-S_0}$, keeping the energy dependence of the density of states in $\rho(E)$. The spectral statistics determine the time dependence of~\eqref{eq:JT-multitrace-ramp}, while the magnitude and sign are controlled by the diagonal multi-trace kernel $F_{\O}^{(1|1)}(E|E)$.

The naive Gaussian/full ETH contraction with one matter insertion on each boundary does not produce this term. Joining the two insertions by the usual matter geodesic gives
\begin{equation} \label{eq:JT-gaussian}
    \mathcal G_{1|1}^{\rm gaussian}(\beta_L,\beta_R) = \int ds\rho(s) e^{-\frac12(\beta_L+\beta_R)s^2} \Gamma^\Delta_{s,s} = \inlinefig[5]{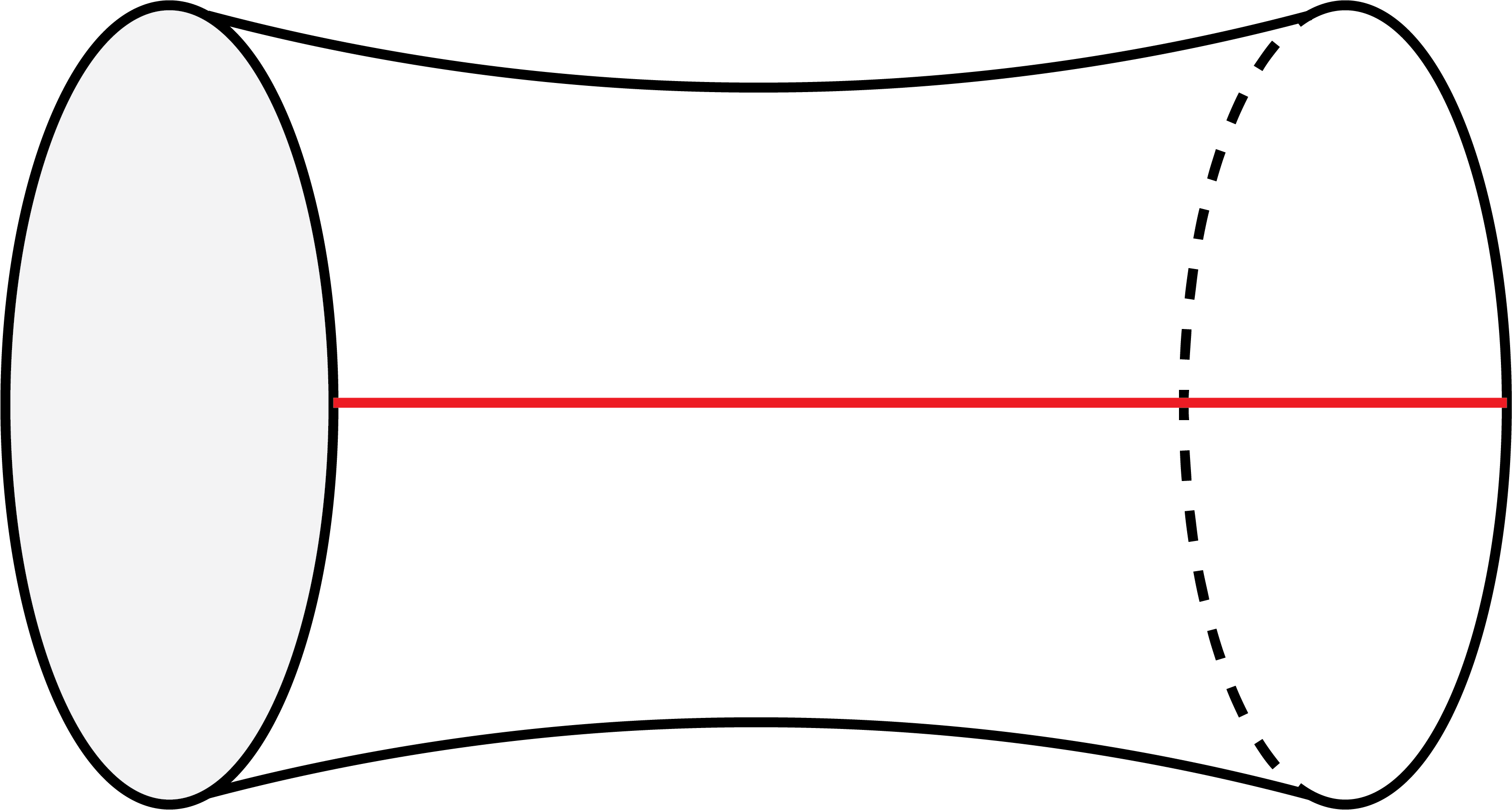} \,,
\end{equation}
where we used the JT conventions
\begin{equation} \label{eq:JT-conventions}
    E_s=\frac{s^2}{2}\,,\qquad \rho(s)=\frac{s}{4\pi^2}\sinh(2\pi s)\,,\qquad F_{\O}^{(2)}(E_{s_a},E_{s_b}) = \Gamma_{ab}^{\Delta}\equiv\frac{\Gamma\!\left(\Delta\pm\ii s_a\pm\ii s_b\right)}{2^{2\Delta+1} \Gamma(2\Delta)}\,,
\end{equation}
where the four choices of signs in the numerator are multiplied, and the density $\rho(s)$ is the reduced disk density: the full mean spectral measure is $e^{S_0}\rho(s)ds$ at leading order. In the two-boundary wormhole~\eqref{eq:JT-gaussian}, the same spectral parameter $s$ propagates on the two boundaries, which corresponds to the Gaussian ETH/single-cycle term in~\eqref{eq:mtt_eth_oo} and identifies the two energies. Then the analytic continuation $\beta_L=\beta+\ii T$ and $\beta_R=\beta-\ii T$ results in a time-independent amplitude for~\eqref{eq:JT-gaussian}. This remains true even when higher genus geometries are included, as long as the operator correlations identify the two boundary energies.

\subsubsection*{Energy-conserving Haar and the missing Weingarten channel}
A simple toy model shows that nontrivial multi-trace ETH correlations can emerge as a consequence of imposing unitary time evolution in chaotic quantum systems. This model extends the fixed-spectrum Haar orbit of subsection \ref{subsec:fixed-haar} by specifying how the operators are correlated at different times. To impose a common unitary evolution, we choose a single random Hamiltonian $H=u^\dag hu$ and use it in both factors of the spectral form factor, \cite{Roberts:2016hpo, Stanford:2021bhl, Yan:2022nod},
\begin{equation}
    U(t) = e^{-\ii(u^\dag hu)t} = u^\dag e^{-\ii ht}u \,,
\end{equation}
using the same Haar $u$ and spectrum $h$ for any time $t$. For a traceless operator, the exact Weingarten integral gives
\begin{equation}
    \int du\,\Tr(\O u^\dagger e^{\ii ht}u) \Tr(\O u^\dagger e^{-\ii ht}u) = \frac{L}{L^2-1}\Tr(\O^2)\qty(1-K_h(t))\,,
\label{eq:JT-haar-2terms}
\end{equation}
where we defined $K_h(t) \equiv \frac{1}{L^2}\left|\Tr e^{-\ii ht}\right|^2$. The first term in~\eqref{eq:JT-haar-2terms} is the Gaussian ETH correlation, and the second is the $1|1$ multi-trace correlation between diagonal matrix elements, which carries the spectral form factor $K_h(T)$.\footnote{For example, if the eigenvalues of $h$ are averaged with GUE statistics, the connected  contribution to the average of $K_h(t)$ would be given by
\begin{equation}
    \overline{K_h(t)} = \frac{1}{L^2} \int_{-2}^2dE \min\left\{\frac{|t|}{2\pi},\, \rho_h(E)\right\} \,,\qquad \rho_h(E) = \frac{L}{2\pi}\sqrt{4-E^2}\,,
\end{equation}
which at sufficiently late times simplifies to $\overline{K_h(t)}\to\frac1L$.} Here the Haar averaging determines the matrix-element fluctuations, while the common Hamiltonian fixes the energy levels whose relative phases interfere. Their double sum gives $K_h(T)$, whose minus sign in \eqref{eq:JT-haar-2terms} produces a negative ramp in the one-point SFF.

The energy-conserving Haar model therefore gives a simple reason to expect the JT answer to contain more than the usual cyclic ETH channel. A subleading correlation between different diagonal matrix elements is invisible in the naive matter-geodesic computation, but after summing over the two independent energies it contributes at the same entropy order and is precisely what allows the universal linear ramp and plateau to appear.

\subsection{Multi-trace ETH from JKMS matrix model} \label{subsec:jkms}
The construction of Jafferis, Kolchmeyer, Mukhametzhanov, and Sonner (JKMS) \cite{Jafferis:2022wez, Jafferis:2022uhu} provides a concrete gravitational realization of the matrix-model framework of subsection \ref{subsec:matrix-model}. JT gravity coupled to a free massive scalar is described by a regulated two-matrix integral for the Hamiltonian $H$ and the operator $\O$,
\begin{equation}
    Z_{\rm JKMS} = \int d H\,d\O\,e^{-L\Tr V_H(H)-V[\O;H]}\,,
\label{eq:JKMS-two-matrix-integral}
\end{equation}
where the potential $V[\O;H]$ contains the energy-dependent cyclic interactions discussed in subsection \ref{subsec:matrix-model}, here with their couplings chosen recursively so that the full planar correlators reproduce the JT disk amplitudes.\footnote{This differs only in bookkeeping from subsection \ref{subsec:matrix-model}, where the two-point kernel is fixed by $V_2$ and induced lower-point corrections are removed order by order.} 

We use the conventions of \eqref{eq:JT-conventions} throughout, and we denote the $6j$ kernel normalized with the measure $\rho(s)ds$ by braces with a subscript $\rho$.\footnote{The reduced density in (2.9) of~\cite{Jafferis:2022wez} is $\rho_{\rm JKMS}(s)=4\rho(s)$. We therefore use $\{\cdots\}_{\rho}=4\{\cdots\}_{\rm JKMS}$, where the latter is the Wilson-function kernel defined in their appendix A. This preserves their recoupling orthogonality relation $\int ds\,\rho(s)\,\{\Delta_1\ s_1\ k;\Delta_2\ s_2\ s\}_{\rho}\{\Delta_1\ s_1\ k';\Delta_2\ s_2\ s\}_{\rho}=\delta(k-k')/\rho(k)$. The Hamiltonian normalization is translated separately by $\beta_{\rm JKMS}=\beta/2$ at fixed $s$; the normalization of the external operator is fixed by $\Gamma^\Delta$ in \eqref{eq:JT-conventions}.} Then the leading connected disk matrix-element amplitudes are given by
\begin{equation}
    \overline{\O_{ab} \O_{ba}}\big|_{\rm disk}=e^{-S_0}\Gamma_{ab}^{\Delta} = \inlinefig[5]{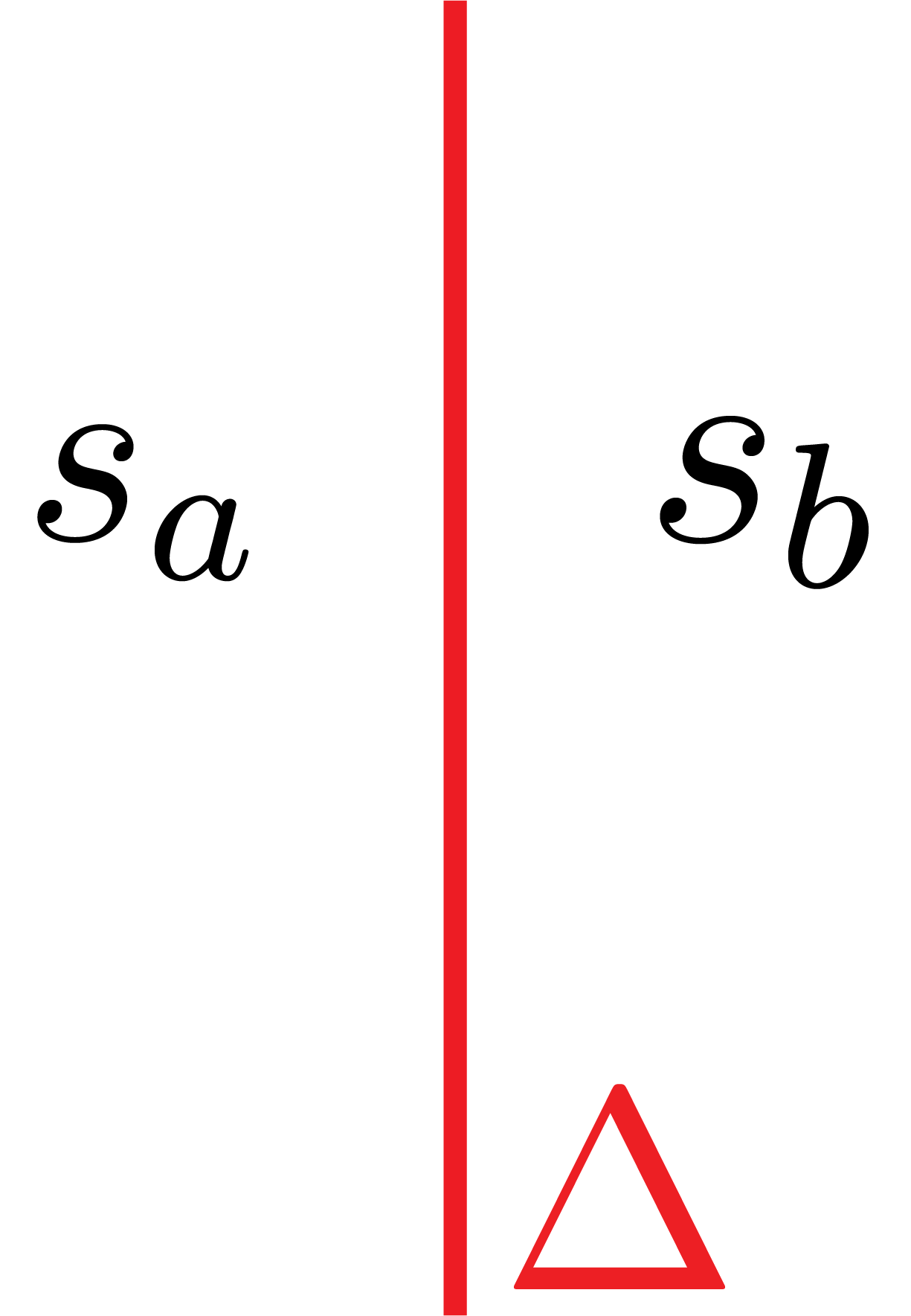}\,,
\label{eq:JKMS-2pt}
\end{equation}
\begin{equation}
    \overline{\O_{ab}\O_{bc} \O_{cd}\O_{da}}\big|_{{\rm disk}}=e^{-3S_0}\left(\Gamma_{ab}^{\Delta}\Gamma_{bc}^{\Delta}\Gamma_{cd}^{\Delta}\Gamma_{da}^{\Delta}\right)^{1/2} \left.\left\{\begin{matrix}
        \Delta & s_a & s_b \\
        \Delta & s_c & s_d
    \end{matrix} \right\}\!\!\right._\rho = \inlinefig[5]{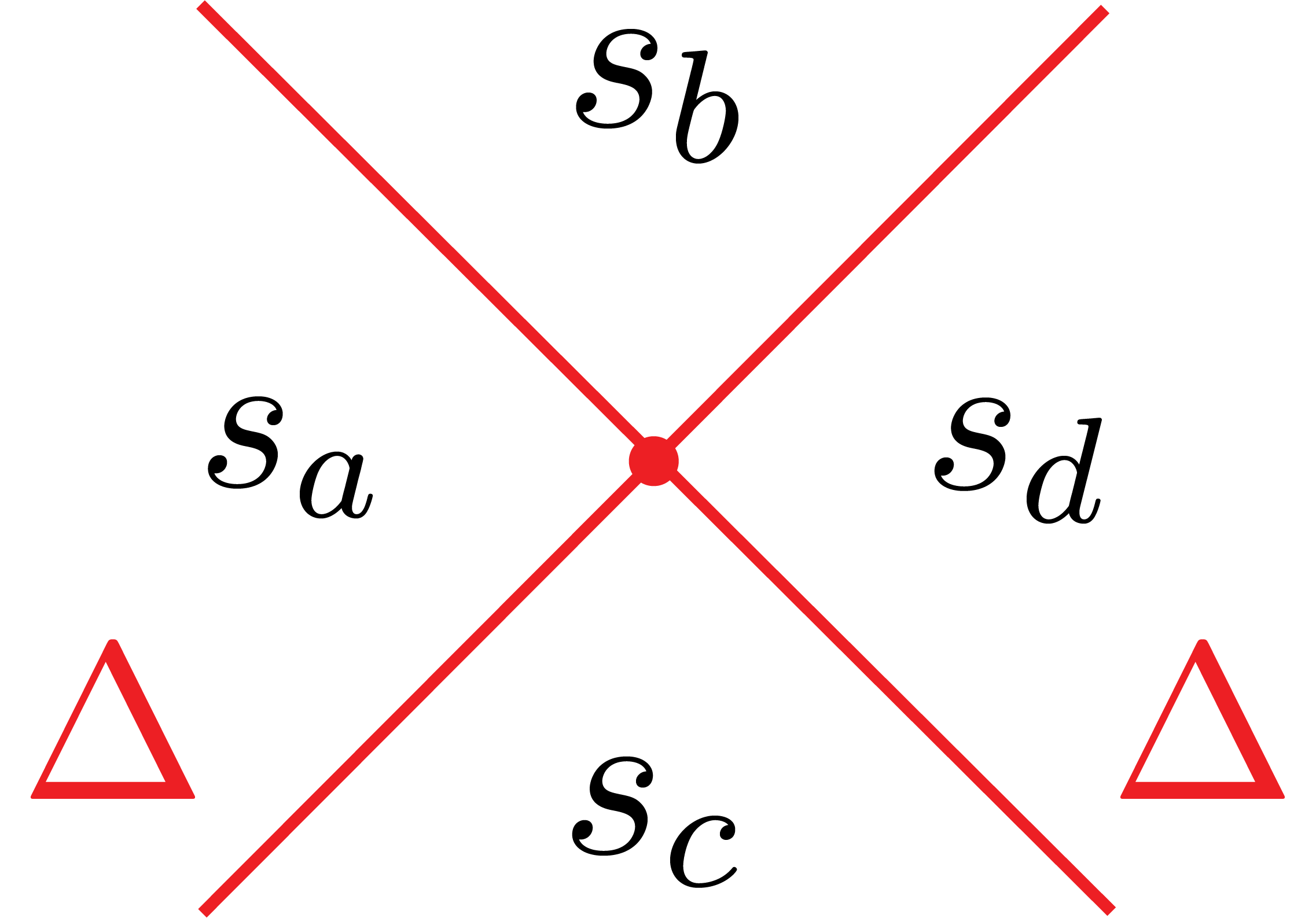}\,.
\label{eq:JKMS-4pt}
\end{equation}
For mutually distinct indices, these are the JT realizations of the physical single-cycle ETH functions $F_{\O}^{(2)}$ and $F_{\O}^{(4)}$, with
\begin{equation}
    |f_\O(E_a,E_b)|^2 = \Gamma_{ab}^\Delta \,.
\end{equation}
The four-point disk moment also has contributions from lower-order contact terms; \eqref{eq:JKMS-4pt} isolates its single-cycle contribution for mutually distinct indices.

\subsubsection*{One-point function squared}
The double trumpet with one insertion on each asymptotic boundary probes the simplest multi-cycle correlation. Since $\O$ is odd under the scalar reflection symmetry, its conditional one-point function vanishes, and we focus on the leading connected cylinder contribution. In the conventions above, the JKMS channel decomposition is \cite{Jafferis:2022wez}
\begin{equation}
\begin{aligned}
    \mathcal G_{1|1}^{\rm cyl}(\beta_L,\beta_R) &\equiv \overline{\Tr\!\left(e^{-\beta_LH}\O\right) \Tr\!\left(e^{-\beta_RH}\O\right)}\Big|_{\rm cyl} \,,\\
    &= \inlinefig[5]{imgs/jt/OO_trumpet.png} + \inlinefig[5]{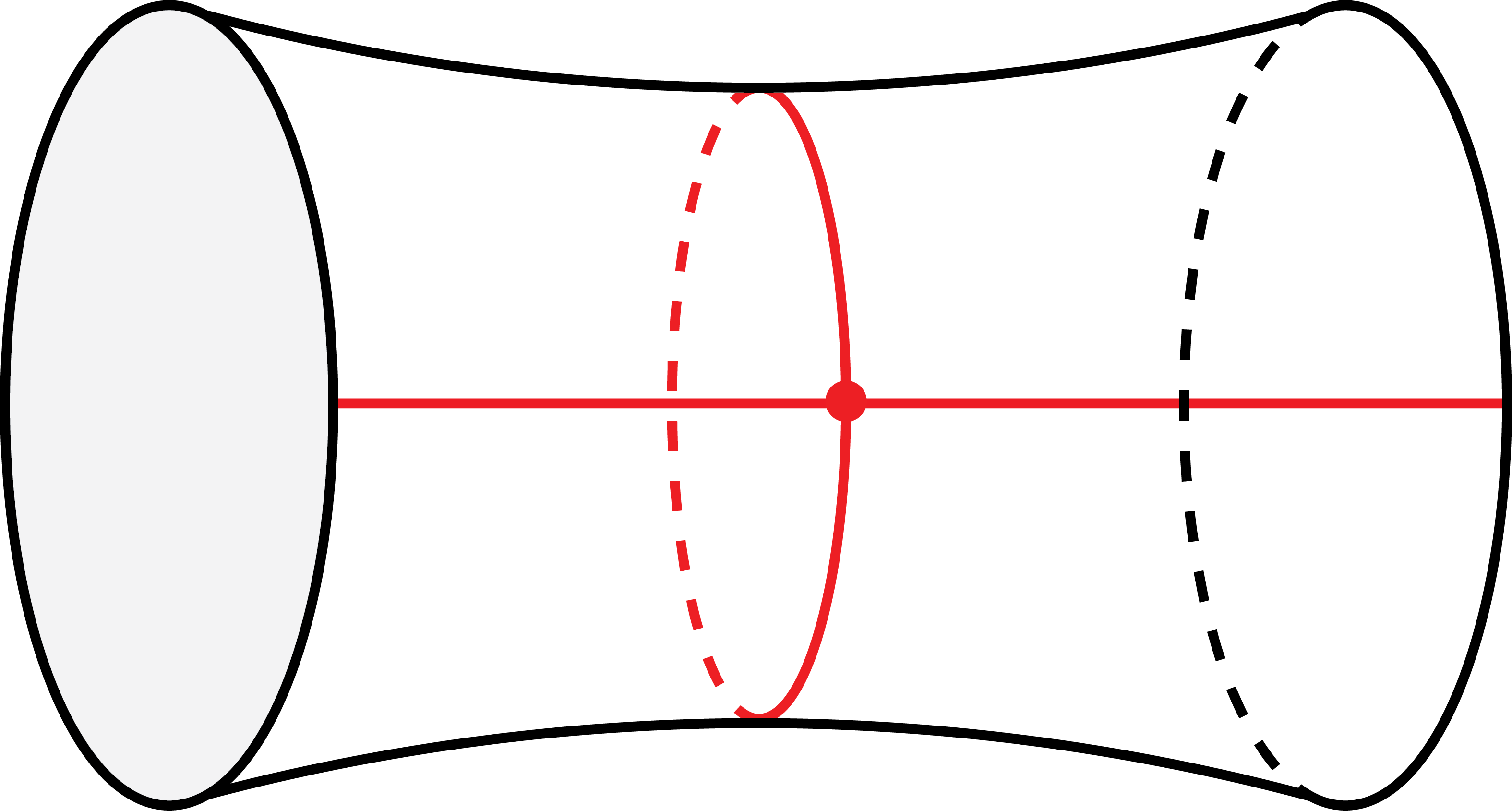} + \inlinefig[5]{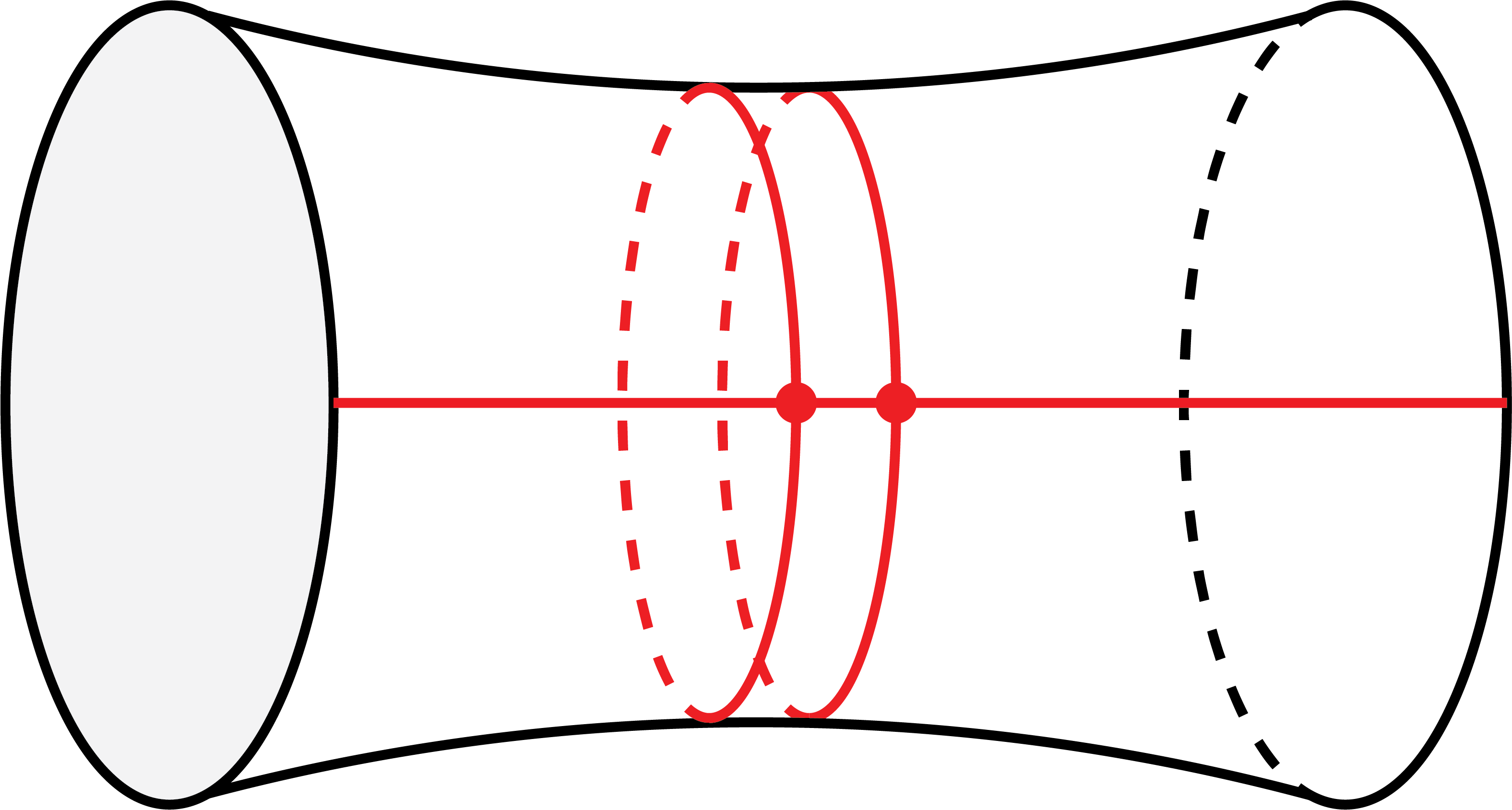}\cdots \,,\\
    &= \int_0^\infty ds_L ds_R\,\rho(s_L)\rho(s_R)e^{-\frac12\beta_Ls_L^2-\frac12\beta_Rs_R^2}\left(\Gamma_{LL}^{\Delta}\Gamma_{RR}^{\Delta}\right)^{1/2} \\
    &\qquad \times \left.\left[\frac{\delta(s_L-s_R)}{\rho(s_L)}+\left\{\begin{matrix} \Delta & s_L & s_R \\ \Delta & s_R & s_L \end{matrix} \right\}\!\!\right._\rho + \sum_{m=0}^{\infty} \left.\left\{\begin{matrix}
        2\Delta+2m & s_L & s_R \\
        \Delta & s_R & s_L
    \end{matrix} \right\}\!\!\right._\rho + \cdots\right] \,.
\end{aligned}
\label{eq:JKMS-1pt2}
\end{equation}
The first nonidentity term corresponds to the scalar of dimension $\Delta$, followed by the double-trace primaries of dimensions $2\Delta+2m$, their descendants, and higher composites. Here “double trace” refers to a composite matter state propagating around the throat, not to an additional asymptotic boundary.\footnote{The scalar determinant diverges as the closed geodesic shrinks, so the cylinder expansion in \eqref{eq:JKMS-1pt2} requires a regulator. We therefore view \eqref{eq:JKMS-1pt2} as the regulated cylinder expansion; comparing with the nonperturbative matrix integral additionally requires the JKMS double-scaling prescription.}

The correlations in the operator statistics of~\eqref{eq:JKMS-1pt2} are given by
\begin{equation}
    \overline{\O_{s_Ls_L} \O_{s_Rs_R}} = e^{-2S_0}(\Gamma_{LL}^\Delta\Gamma_{RR}^\Delta)^{1/2} \frac{\delta(s_R-s_L)}{\rho(s_R)} + e^{-2S_0}F_{\O,\rm JKMS}^{(1|1)}(E_{s_L}|E_{s_R}) \,,
\label{eq:MT-JT}
\end{equation}
where the additional $e^{-S_0}$ compared with \eqref{eq:opsff-conditional-eth} comes from converting the discrete Kronecker contact into a Dirac delta with respect to the full density $e^{S_0}\rho$, and where
\begin{equation}\label{eq:F11-jkms}
\begin{aligned}
    F_{\O,\,\rm JKMS}^{(1|1)}(E_{s_L}|E_{s_R}) &= \left(\Gamma_{LL}^{\Delta}\Gamma_{RR}^{\Delta}\right)^{1/2}\left[\left.\left\{\begin{matrix} \Delta & s_L & s_R \\ \Delta & s_R & s_L \end{matrix} \right\}\!\!\right._\rho + \sum_{m=0}^{\infty} \left.\left\{\begin{matrix}
        2\Delta+2m & s_L & s_R \\
        \Delta & s_R & s_L
    \end{matrix} \right\}\!\!\right._\rho+\cdots\right] \,.
\end{aligned}
\end{equation}
The identity term of \eqref{eq:MT-JT} has support at $E_{s_L}=E_{s_R}$ and is precisely the cyclic contribution of Gaussian ETH \eqref{eq:gaussian-eth} that results in the JT amplitude \eqref{eq:JT-gaussian}, while the second term of \eqref{eq:MT-JT} includes nonidentity matter channels with smooth support at $E_{s_L}\neq E_{s_R}$, providing the $1|1$ multi-trace correlation.

The one-particle channel makes the relation to subsection \ref{subsec:matrix-model} explicit. It is obtained by sewing the dressed connected disk four-point function and amputating the propagator along the sewn pair \cite{Jafferis:2022wez},
\begin{equation}\label{eq:F11_JKMS_one-particle}
    F_{\O,\rm JKMS}^{(1|1)}(E_{s_L}|E_{s_R})\Big|_{\text{one-particle}} = \frac{F_{\O}^{(4)}(E_{s_L},E_{s_L},E_{s_R},E_{s_R})}{F_{\O}^{(2)}(E_{s_L},E_{s_R})} = \left(\Gamma_{LL}^{\Delta}\Gamma_{RR}^{\Delta}\right)^{1/2} \left.\left\{\begin{matrix} \Delta & s_L & s_R \\ \Delta & s_R & s_L \end{matrix} \right\}\!\!\right._\rho\,.
\end{equation}
This is the dressed counterpart of the quartic boundary-splitting contraction in subsection \ref{subsec:matrix-model}. The repeated energy arguments describe the smooth single-cycle disk amplitude, not additional Kronecker delta constraints. Higher matter channels arise analogously from sewing higher disk correlators, with the appropriate amputations and lower-channel subtractions. They contribute to the same two external one-cycles and therefore build the remaining terms in $F_{\O,\mathrm{JKMS}}^{(1|1)}$. The JKMS matrix model thus explicitly shows how single-cycle disk amplitudes can generate a nonzero $F_{\O}^{(1|1)}$, correlating diagonal matrix elements at distinct energies.

\subsection{Interacting bulk matter} \label{subsec:JT-interacting}
Interacting bulk matter fields provide a generic source of multi-trace ETH correlations. We first discuss an arbitrary interacting bulk theory specified by its matter spectrum and three-point OPE coefficients $C_{ijk}$, following Penington-Witten (PW) \cite{Penington:2023dql,Penington:2024sum}. We obtain the disk four-point function and by cutting and gluing we obtain the one-point wormhole in terms of the $C_{ijk}$ and universal JT blocks. We then specialize to weakly interacting matter and as an example we explicitly compute the case of $\lambda\phi^4$ at leading order in $\lambda$. At this order two diagrams contribute, a self-energy diagram and a figure-eight vacuum diagram, producing:
\begin{equation}\label{eq:F11-interactingJT-preview}
\begin{split}
 F_{\O}^{(1|1)}\Big|_{\lambda\phi^4}
 ={}&F_{\O,\rm JKMS}^{(1|1)}
 +\delta_\lambda F_{\O}^{(1|1)}\Big|_{\rm SE}
 +\delta_\lambda F_{\O}^{(1|1)}\Big|_{\rm vac}
 +O(\lambda^2),
\end{split}
\end{equation}
where the explicit forms of the self-energy and figure-eight vacuum multi-trace functions are given by \eqref{eq:phi4-F11-SE} and \eqref{eq:phi4-F11-vac-all}, respectively.
Details on blocks and special functions can be found in  appendix \ref{app:JT-interacting-blocks}.

\subsubsection{General bulk interacting QFT}
Let us start from the three-point function of generic interacting matter derived in \cite[section 3.2 and appendix A]{Penington:2024sum}. Let $\Phi^i$, $\Phi^j$ and $\Phi^a$ be Hermitian dressed operators associated with unit-normalized matter primaries in the bulk matter theory of dimensions $\Delta_i$, $\Delta_j$ and $\Delta_a$. Let $\Pi_{E}$ denote the density of the spectral projection-valued measure with respect to $ds$, so that $\int_0^\infty ds\,\Pi_{E}=\mathbf 1$ and $E=s^2/2$. The disk amplitude carries $e^{S_0}$, as required by its Euler characteristic $\chi=1$ and we have
\begin{equation}
 \begin{aligned}
 \operatorname{Tr}\!\left[
 \Phi^i\Pi_{E_1}\Phi^j\Pi_{E_2}\Phi^a\Pi_{E_3}
 \right]_{\rm disk}
 &=e^{S_0}\rho(s_1)\rho(s_2)\rho(s_3)\,
 \mathcal T_3^{ija}(s_1,s_2,s_3),\\
 \mathcal T_3^{ija}(s_1,s_2,s_3)&=C_{ija}F_{ija}(s_1,s_2,s_3).
 \end{aligned}
 \label{eq:PW3-paper}
\end{equation} 
The coefficient $C_{ija}$ is the ordinary three-point coefficient of the matter theory on
fixed AdS$_2$. $F_{ija}$ denotes the gravitational dressing of the matter three-point vertex and defines a universal block which depends only 
on the operator dimensions $(\Delta_i,\Delta_j,\Delta_a)$ and on the three Schwarzian
energies. Its explicit integral and ${}_4F_3$ expressions, with the normalization fixed by \eqref{eq:JT-conventions}, are given in appendix \ref{app:JT-interacting-blocks}.
The same definition gives the standard two-point kernel,
\begin{equation}
 \begin{aligned}
 \operatorname{Tr}\!\left[
 \Phi^a\Pi_E\Phi^a\Pi_{E'}\right]_{\rm disk}
 &=e^{S_0}\rho(s)\rho(t)\,\mathcal T_2^a(s,t),\\
 \mathcal T_2^a(s,t)&=\Gamma_{st}^{\Delta_a},\qquad E'=t^2/2.
 \end{aligned}
 \label{eq:PW2-paper}
\end{equation}
The normalization is fixed by \eqref{eq:JT-conventions}, after converting PW's energy convention $E=s^2$ and density of states to ours.

\subsubsection*{Disk four-point function.}
The microcanonical disk four-point function can be decomposed using the OPE defined by \eqref{eq:PW3-paper} into a sum over intermediate states: 
\begin{equation}
 \operatorname{Tr}\!\left[
 \Phi^i\Pi_{E_1}
 \Phi^j\Pi_{E_2}
 \Phi^k\Pi_{E_3}
 \Phi^l\Pi_{E_4}
 \right]_{\rm disk}
 =e^{S_0}\prod_{r=1}^4\rho(s_r)\,
 \mathcal T_4^{ijkl}(s_1,s_2,s_3,s_4).
 \label{eq:PW4-def-paper}
\end{equation}
Resolving the product $\Phi^i\Pi_{E_1}\Phi^j$ into a complete set of
intermediate matter primaries $a$, using the spectral OPE in (3.59) of \cite{Penington:2024sum}, and sewing to the second pair gives
\begin{equation}
 \mathcal T_{4,\mathrm{ex}}^{ijkl}(s_1,s_2,s_3,s_4)
 =
 \sum_{a\ne\mathbf 1}
 \frac{
 C_{ija}C_{akl}\,
 F_{ija}(s_1,s_2,s_4)\,
 F_{akl}(s_2,s_3,s_4)}
 {\mathcal T_2^a(s_2,s_4)} .
 \label{eq:PW4-paper}
\end{equation}
The denominator is the propagator of the intermediate primary $a$ and is
therefore the usual amputation factor.  Equation~\eqref{eq:PW4-paper} is
the gravitationally dressed analogue of the ordinary conformal-block
decomposition of a Witten diagram
\cite{Hijano:2015zsa}.  The sum is over the physical matter primaries
appearing in the $\Phi^i\times\Phi^j$ OPE.  The identity representation is
not contained in \eqref{eq:PW4-paper} and must be added separately.

It is useful to isolate the contribution of one intermediate representation:
\begin{equation}
 \mathcal B_a^{\rm disk}(s_1,s_2,s_3,s_4)
 \equiv
 \frac{
 F_{ija}(s_1,s_2,s_4)\,
 F_{akl}(s_2,s_3,s_4)}
 {\mathcal T_2^a(s_2,s_4)} .
 \label{eq:PW-disk-block-paper}
\end{equation}
We will refer to $\mathcal B_a^{\rm disk}$ as the \emph{JT-dressed disk
block}.  It is not an additional piece of matter data: once the dimension $\Delta_a$ is specified, $\mathcal B_a^{\rm disk}$ is fixed by the universal PW functions. All matter-theory dependence in \eqref{eq:PW4-paper} therefore comes from the spectrum $\{\Delta_a\}$ and the OPE coefficients $C_{ija}$.

The generalized free field, studied by JKMS, provides a simple explicit example of the above prescription. The disk four-point function in energy space is:
\begin{equation}
 \mathcal T_4^{OOOO}\big|_{\mathrm{GFF}}
 =
 \left(
 \Gamma_{12}^{\Delta}
 \Gamma_{23}^{\Delta}
 \Gamma_{34}^{\Delta}
 \Gamma_{41}^{\Delta}
 \right)^{1/2}
 \left[
 \frac{\delta(s_1-s_3)}{\rho(s_1)}
 +
 \frac{\delta(s_2-s_4)}{\rho(s_2)}
 +
 \begin{Bmatrix}
 \Delta&s_1&s_2\\
 \Delta&s_3&s_4
 \end{Bmatrix}_{\rho}
 \right].
 \label{eq:JKMS4-paper}
\end{equation}
The first two terms are the two uncrossed Wick contractions and the $6j$ symbol is the crossed contraction \cite{Jafferis:2022wez,Mertens:2017mtv}. 

The 4-point function can be expanded using the $\cO\times \cO$ OPE in terms of normalized Wilson wavefunctions
$P_n^{\Delta,\Delta}(s;s_1,s_3)$ \cite{Jafferis:2022wez}.
For fixed $s_1,s_3$, these are degree-$n$ Wilson polynomials in $s^2$
multiplied by the gamma-function square root of the Wilson weight. We normalize them by
$\int_0^\infty ds\,\rho(s)P_n(s)P_m(s)=\delta_{nm}$.\footnote{In our convention, the wavefunctions are equal to double the JKMS wavefunctions since
$\rho_{\rm JKMS}=4\rho$.} Each $n$ labels an irreducible
$SL(2,\mathbb R)$ representation of dimension
\begin{equation}
 h_n=2\Delta+n 
 \label{eq:Wilson-hn-paper}
\end{equation}
corresponding to double trace operators $[\cO\cO]_n$. 
The completeness and crossing identities take the form
\begin{align}
 \frac{\delta(s_2-s_4)}{\rho(s_2)}
 &=
 \sum_{n=0}^{\infty}
 P_n^{\Delta,\Delta}(s_2;s_1,s_3)
 P_n^{\Delta,\Delta}(s_4;s_1,s_3),
 \label{eq:Wilson-delta-paper}
 \\
 \begin{Bmatrix}
 \Delta&s_1&s_2\\
 \Delta&s_3&s_4
 \end{Bmatrix}_{\rho}
 &=
 \sum_{n=0}^{\infty}(-1)^n
 P_n^{\Delta,\Delta}(s_2;s_1,s_3)
 P_n^{\Delta,\Delta}(s_4;s_1,s_3).
 \label{eq:Wilson-6j-paper}
\end{align}
 Adding the two equations projects onto the even
representations,
\begin{equation}
 \frac{\delta(s_2-s_4)}{\rho(s_2)}
 +
 \begin{Bmatrix}
 \Delta&s_1&s_2\\
 \Delta&s_3&s_4
 \end{Bmatrix}_{\rho}
 =
 2\sum_{m=0}^{\infty}
 P_{2m}^{\Delta,\Delta}(s_2;s_1,s_3)
 P_{2m}^{\Delta,\Delta}(s_4;s_1,s_3).
 \label{eq:Wilson-even-paper}
\end{equation}
Therefore the $\cO\times \cO$ OPE contains 
\begin{equation}
 \O\times\O
 =
 {\bf 1}
 +
 \sum_{m=0}^{\infty}
 C_{\cO\cO[\cO\cO]_m}\,[\cO\cO]_m+\cdots,
 \qquad
 h_m=2\Delta+2m ,
 \label{eq:GFF-OPE-paper}
\end{equation}

The OPE coefficient normalization is fixed by matching each Wilson wavefunction to a conformal block as in \cite[App.~C.1]{Jafferis:2022wez}. After performing the energy integrals and removing the position-space prefactor, the semiclassical, zero-temperature limit of $\left(
 \Gamma_{12}^{\Delta}
 \Gamma_{23}^{\Delta}
 \Gamma_{34}^{\Delta}
 \Gamma_{41}^{\Delta}
 \right)^{1/2}P_n(s_2)P_n(s_4)$ is
$\frac{(2\Delta)_n^2}{n!(4\Delta+n-1)_n}k_{2\Delta+n}(z)$, where $k_h(z)=z^h{}_2F_1(h,h;2h;z)\sim z^h$ is the conformal block. Thus the gravitationally dressed disk block defined above reduces for the GFF to
\[
\mathcal B_{h_m}^{\rm disk}(s_2,s_3,s_4,s_1)
=\frac{(2m)!(4\Delta+2m-1)_{2m}}{(2\Delta)_{2m}^2}
\left(
 \Gamma_{12}^{\Delta}
 \Gamma_{23}^{\Delta}
 \Gamma_{34}^{\Delta}
 \Gamma_{41}^{\Delta}
 \right)^{1/2}P_{2m}(s_2)P_{2m}(s_4),
\qquad h_m=2\Delta+2m.
\]
Comparing the Wilson expansion with the OPE expansion $\sum_m C_{OO[OO]_m}^{\,2}\mathcal B_{h_m}^{\rm disk}$ therefore gives the OPE coefficients
\[
C_{OO[OO]_m}^{\,2}
=2\,\frac{(2\Delta)_{2m}^{\,2}}
{(2m)!\,(4\Delta+2m-1)_{2m}},
\]
which are indeed the GFF double-trace OPE coefficients.
The general dressed block $F_{ijk}$  truncates to gamma-function
prefactors times a Wilson polynomial when the dimensions satisfy
$\Delta_a=\Delta_i+\Delta_j+n$, as is the case for the GFF.  The  Wilson expansion is thus the concrete GFF realization of the general energy-space OPE of  PW.

\subsubsection*{One-point wormhole with matter loop.}
The same disk data can be glued into a cylinder, producing a wormhole amplitude for a product of one-point functions. Using the standard JT gravity cutting and gluing rules for genus zero, the one-point wormhole with a matter loop can be obtained by identifying two legs of the connected disk four-point function and amputating the  propagator. Thus, with $\O=\Phi^O$, we obtain a general expression for the multi-trace ETH term:
\begin{equation}
 F_{\O}^{(1|1)}(E_L|E_R)
 =\frac{[\mathcal T_4^{OOOO}(s_L,s_R,s_R,s_L)]_c}
 {\Gamma_{LR}^{\Delta}}.
 \label{eq:disk-to-cylinder-paper}
\end{equation}
Here $c$ selects the fully connected disk amplitude, as in \eqref{eq:JKMS-4pt}.

Inserting the PW decomposition gives the explicit  result
\begin{equation}
 F_{\O}^{(1|1)}(E_L|E_R)
 =\left[\sum_{a\neq\mathbf 1} C_{OOa}C_{aOO}\,
 \mathcal B_a^{\rm cyl}(s_L,s_R)\right]_c,
 \label{eq:PW-wormhole-paper}
\end{equation}
where the cylinder block is
\begin{equation}
 \mathcal B_a^{\rm cyl}(s_L,s_R)
 \equiv\frac{F_{OOa}(s_L,s_R,s_L)F_{aOO}(s_R,s_R,s_L)}
 {\Gamma_{RL}^{\Delta_a}\Gamma_{LR}^{\Delta}}.
 \label{eq:PW-cylinder-block-paper}
\end{equation}
 Both the propagator of the exchanged primary and that of the sewn external line have been amputated, whereas the external boundary leg factors are already contained in $F_{ijk}$.  For the GFF, this reduces to the previously studied expression 
\begin{equation}\label{eq:PW-free-F11-paper}
 F_{\O}^{(1|1)}(E_L|E_R)\Big|_{GFF}
 =(\Gamma_{LL}^{\Delta}\Gamma_{RR}^{\Delta})^{1/2}
 \begin{Bmatrix}\Delta&s_L&s_R\\\Delta&s_R&s_L\end{Bmatrix}_{\rho},
\end{equation}
in agreement with \eqref{eq:MT-JT} and the one-particle channel of subsection \ref{subsec:jkms}.
Equation \eqref{eq:PW-wormhole-paper} gives the contribution from sewing a single matter line across the cylinder, rather than the full cylinder amplitude. Other contributions include vacuum diagrams and channels involving multiple matter lines around the wormhole, which require additional disk data (higher-point functions) which is in principle fixed by the $C_{ijk}$.  The calculation of the $\phi^4$ vacuum diagram below illustrates one such contribution.

\subsubsection{Weakly interacting matter.}
We now perturb away from the generalized free point, holding the physical dimension $\Delta$ and the two-point normalization of the external $\O$ fixed.   For the
double-trace tower, with $C_m=C_{\cO\cO[\cO\cO]_m}$, write
\begin{equation}
 h_m(\lambda)
 =
 h_m+\lambda\gamma_m^{(1)}+O(\lambda^2),
 \qquad
 C_m^2(\lambda)
 =
 a_m^{(0)}+\lambda a_m^{(1)}+O(\lambda^2),
 \qquad
 h_m=2\Delta+2m ,
 \label{eq:weak-data-paper}
\end{equation}
where
\begin{equation}
 a_m^{(0)}
 =
 2\,\frac{(2\Delta)_{2m}^{\,2}}
 {(2m)!\,(4\Delta+2m-1)_{2m}} .
 \label{eq:a0-paper}
\end{equation}
The quantities $\gamma_m^{(1)}$ are anomalous dimensions and
$a_m^{(1)}$ are the first-order corrections to the squared OPE
coefficients.  We will proceed analogously to the standard Witten diagram way of computing the anomalous dimensions and OPE coefficients. For example, an AdS contact diagram
has the conformal-block decomposition
\begin{equation}
 \frac{1}{\lambda}\,
 \delta_\lambda\mathcal G_4(z)
 =
 \sum_{m=0}^{\infty}
 \left[
 a_m^{(1)}\,k_{h_m}(z)
 +
 a_m^{(0)}\gamma_m^{(1)}
 \left.\partial_h k_h(z)\right|_{h=h_m}
 \right],
 \qquad
 k_h(z)=z^h{}_2F_1(h,h;2h;z),
 \label{eq:phi4-block-expansion-paper}
\end{equation}
where $z$ is the cross-ratio and $\mathcal G_4(z)$ is the ordinary  matter four-point function, with its external position-space prefactor removed \cite{Hijano:2015zsa,Mazac:2018ycv}. 
We use an analogous expansion for the AdS$_2$ Witten diagrams after including the Schwarzian dressing.  Let
$\mathcal B_h^{\rm disk}$ denote the disk block
in \eqref{eq:PW-disk-block-paper}, viewed as a function of the internal dimension $h$.  Then the disk four-point function in energy space at linear order in perturbation theory around the GFF is given by:
\begin{equation}
 \frac{1}{\lambda}\,
 \delta_\lambda\mathcal T_4
 =
 \sum_m
 \left[
 a_m^{(1)}\,\mathcal B_{h_m}^{\rm disk}
 +
 a_m^{(0)}\gamma_m^{(1)}
 \left.
 \partial_h\mathcal B_h^{\rm disk}
 \right|_{h=h_m}
 \right].
 \label{eq:phi4-PW-expansion-paper}
\end{equation}
At the free values $h_m=2\Delta+2m$ the block $B_{h_m}^{\rm disk}$ is equal to the product of Wilson wavefunctions described previously.   The anomalous-dimension term is then the derivative of the general dimension $h$ block with respect to
the intermediate dimension $h$; it is not in general given by a Wilson polynomial.

By cutting and gluing we can again derive the one-point wormhole amplitude and the form of the multi-trace ETH term 
\begin{equation}\label{eq:weak-F11-paper}
\begin{split}
 \delta_\lambda F_{\O}^{(1|1)}(E_L|E_R)\Big|_{\text{weak}}
 =\lambda\left[\sum_{m=0}^\infty\left(
 a_m^{(1)}\mathcal B_{h_m}^{\rm cyl}
 +a_m^{(0)}\gamma_m^{(1)}
 \left.\partial_h\mathcal B_h^{\rm cyl}\right|_{h=h_m}
 \right)\right]_c+O(\lambda^2).
\end{split}
\end{equation}
The cylinder block is evaluated at $(s_L,s_R)$, and the derivative acts on both trivalent functions and on the intermediate two-point denominator. If primaries are degenerate, the first-order mixing matrix must be diagonalized before applying this expansion, as is standard in the calculation of holographic correlators. 

Since we are interested in wormholes with loops around the throat, it is more natural  to use characters.  A primary of dimension $h$, together with
all of its global descendants, contributes
\begin{equation}
 \chi_h(b)
 =
 \operatorname{Tr}_{\mathcal H_h}e^{-bL_0}
 =
 \frac{e^{-hb}}{1-e^{-b}} .
 \label{eq:character-paper}
\end{equation}
 If $h_a(\lambda)=h_a^{(0)}+\lambda\gamma_a^{(1)}$, its first variation is
\begin{equation}\label{eq:generic-character-deformation-paper}
 \delta_\lambda Z_{\rm vac}(b)
 =\lambda\sum_a\gamma_a^{(1)}
 \left.\partial_h\chi_h(b)\right|_{h=h_a^{(0)}}+O(\lambda^2).
\end{equation}
The sum includes each copy of a representation, whose multiplicity remains fixed under small deformations. The derivative term gives the shift in the Boltzmann factor of $\chi_h(b)$. 
In \cite{Kraus:2020nga} the authors used K\"all\'en-Lehmann identities to expand quartic contact Witten diagrams into character derivatives and extract double-trace anomalous dimensions \cite{Kraus:2020nga}. We will now use similar methods in AdS$_2$,  computing the self-energy and figure-eight Witten diagrams on the cylinder and extracting the anomalous dimensions and OPE coefficients. 

\subsubsection{Wormhole Witten diagrams in $\lambda \phi^4$ theory}
To demonstrate this explicitly we consider the addition of a $\lambda \phi^4$ term to the scalar field theory coupled to JT gravity. We compute the leading in $\lambda$ Witten diagrams on the hyperbolic cylinder and dress them to the Schwarzian boundaries. This results in one-point wormhole amplitudes dual to multi-trace ETH correlations of boundary operators. 

As usual in JT gravity, the calculation proceeds in two steps \cite{Yang:2018gdb}. First we keep the hyperbolic cylinder of throat length $b$ fixed and compute the Witten diagram on this background.  Then we dress it to the Schwarzian boundaries using the same double trumpet gluing kernel as in ~\cite{Jafferis:2022wez}.
\subsubsection*{Witten diagrams on the fixed hyperbolic cylinder.}
Witten diagrams and their conformal-block decompositions are standard tools in AdS/CFT \cite{Hijano:2015zsa}. In AdS$_2$, analytic bootstrap methods provide a complementary treatment of contact and exchange diagrams \cite{Mazac:2018ycv}, while thermal vacuum diagrams determine anomalous dimensions through characters \cite{Kraus:2020nga}. Here the noncontractible images of the hyperbolic cylinder generate the matter corrections relevant to the one-point wormhole.
Let us start from the action:
\begin{equation}
 I[\phi]
 =\frac12\int_{X_b} d^2x\sqrt g\,
 \left[(\nabla\phi)^2+m^2\phi^2\right]
 +\frac{\lambda}{4!}\int_{X_b} d^2x\sqrt g\,\phi^4
 +I_{\rm ct},
 \qquad
 m^2=\Delta(\Delta-1),
 \label{eq:phi4-action-main}
\end{equation}
where $X_b$ is the Euclidean hyperbolic cylinder. We set the AdS radius to one and take $\Delta>1/2$ with standard scalar boundary conditions. The mass and vacuum counterterms are specified explicitly in appendix \ref{app:JT-interacting-blocks}. 
We first consider the normalized bulk matter two-point function, obtained by dividing by the vacuum partition function:
\es{}{
G^{(\lambda)}_{b,\Delta}(x,x')&= \frac{1}{Z_{\rm vac}^{(\lambda)}(b)}\int \mathcal D\phi\,\phi(x)\phi(x')e^{-I[\phi]},\\
Z_{\rm vac}^{(\lambda)}(b)&=\int\mathcal D\phi\,e^{-I[\phi]}.
}
The hyperbolic cylinder can be obtained by a quotient of the upper half-plane and has metric:
\begin{equation}
 X_b=\mathbb H_2/\langle\sigma_b\rangle,
 \qquad
 ds^2=dr^2+\cosh^2r\,dt^2,
 \qquad
 t\sim t+b,
 \label{eq:cyl-main}
\end{equation}
where $\sigma_b$ is the isometry $t\sim t+b$ that produces a hyperbolic cylinder by quotient. 
For a scalar of dimension $\Delta$, the Euclidean Green function on $\mathbb H_2$ is  \cite{Camporesi:1990wm}, \cite[Eq.~\href{https://dlmf.nist.gov/14.3.E10}{(14.3.10)}]{NIST:DLMF}
\begin{equation}
 G_{\mathbb H_2,\Delta}(d)
 =\frac{1}{2\pi}Q_{\Delta-1}(\cosh d),
 \label{eq:H2-green-main}
\end{equation}
where $d$ is geodesic distance on $\mathbb H_2$ and $Q_\nu$ is the Legendre function of the second kind.  The propagator on the quotient $X_b=\mathbb H_2/\langle\sigma_b\rangle$ is obtained by the method of images,
\begin{equation}
 G_{b,\Delta}(x,x')
 =\sum_{w\in\mathbb Z}G_{\mathbb H_2,\Delta}(d(x,\sigma_b^w x')).
 \label{eq:quotient-main}
\end{equation}
The hyperbolic distance in $(r,t)$ coordinates for the $w$-th image is
\begin{equation}
 \cosh d_w(r)
 =1+2\cosh^2r\,\sinh^2\frac{wb}{2},
 \label{eq:dw-main}
\end{equation}
so $d_w(r)\geq |w|b$, with the shortest geodesics being the ones at the throat $r=0$. Each image is a loop that winds around the cylinder. 
Below $G_b$ denotes the free quotient propagator $G_{b,\Delta}$, with the dimension suppressed, and $\delta_\lambda$ denotes its first-order change, including the coupling. The self-energy correction to the propagator can be computed by expanding the normalized correlator to first order in $\lambda$ and using Wick contractions, which gives:
\begin{equation}\label{eq:propagator}
 \delta_\lambda G_b(x,x')
 =-\int_{X_b}d^2y\sqrt g\,
 G_b(x,y)
 \left[\delta m^2+\frac{\lambda}{2}G_b^{\rm ren}(y,y)\right]
 G_b(y,x').
\end{equation}
The factor $1/2$ follows from the $12/4!$ contractions for the tadpole diagram. We have introduced a local counterterm $\delta m^2$  to renormalize the tadpole loop divergence as follows. The propagator at coincident points is obtained by regularization: 
\es{}{
G_b^{\rm ren}(y,y)=\lim_{y'\rightarrow y}\qty(G_b(y,y') +\frac{1}{2\pi}\log(\mu d(y,y')))
}
which introduces a scale $\mu$. We impose the physical condition that the dimension of the field remains $\Delta$, namely that the renormalized propagator decays as $e^{-\Delta d}$ at large distances, which fixes the counterterm. To implement this condition, we first separate the propagator into the identity  and the nontrivial windings:
\begin{equation}
 G_b^{\rm ren}(y,y)
 =G_{\mathbb H_2}^{\rm ren}(0)
 +\sum_{w\neq0}G_{\mathbb H_2,\Delta}(y,\sigma_b^w y),
\end{equation}
where the coincident limit concerns only the identity. 
Then we impose the physical condition which sets the counterterm:
\begin{equation}
 \delta m^2=-\frac{\lambda}{2}G_{\mathbb H_2}^{\rm ren}(0)
 \label{eq:mass-ct-main}
\end{equation}
and cancels the $w=0$ term in the self-energy contribution. The result is the renormalized tadpole:
\begin{equation}
 T_{\Delta,b}(r)
 =\sum_{w\neq0}G_{\mathbb H_2,\Delta}(y,\sigma_b^w y)=\frac1\pi\sum_{w=1}^{\infty}
 Q_{\Delta-1}\!\left(
 1+2\cosh^2r\,\sinh^2\frac{wb}{2}
 \right)
 \label{eq:T-exact-main}
\end{equation}
which only sums over nonzero windings around the throat. Thus the renormalized self-energy in \eqref{eq:propagator} is $(\lambda/2)T_{\Delta,b}$: the subtraction removes the contractible AdS tadpole, not the finite winding loops. The remaining vacuum counterterm is chosen to remove the AdS vacuum contribution. 

Next we attach the bulk-to-boundary propagators, obtained by the limit $r\rightarrow \pm \infty$ of the bulk-to-bulk propagator $G_{b,\Delta}$, which we denote $K_{L,b}$ and $K_{R,b}$ for left and right boundaries: 
\es{}{
K_{L,b}(v;y)=\mathcal{N}_{\Delta}\lim_{r\rightarrow -\infty} e^{-\Delta r}G_{b,\Delta}((r,v),y), \qquad K_{R,b}(u;y)=\mathcal{N}_{\Delta}\lim_{r\rightarrow \infty} e^{\Delta r}G_{b,\Delta}((r,u),y)
}
where $\mathcal N_\Delta$ is fixed by the operator normalization of \eqref{eq:JT-conventions}; its value and the explicit image sums for $K_{L,b}$ and $K_{R,b}$ are given in appendix \ref{app:JT-interacting-blocks}. 
The one-loop correction between the two asymptotic boundaries is the winding tadpole shown in figure \ref{fig:jt-quartic-winding-self-energy}:
\begin{figure}[t]
 \centering
 \includegraphics[width=.3\linewidth]{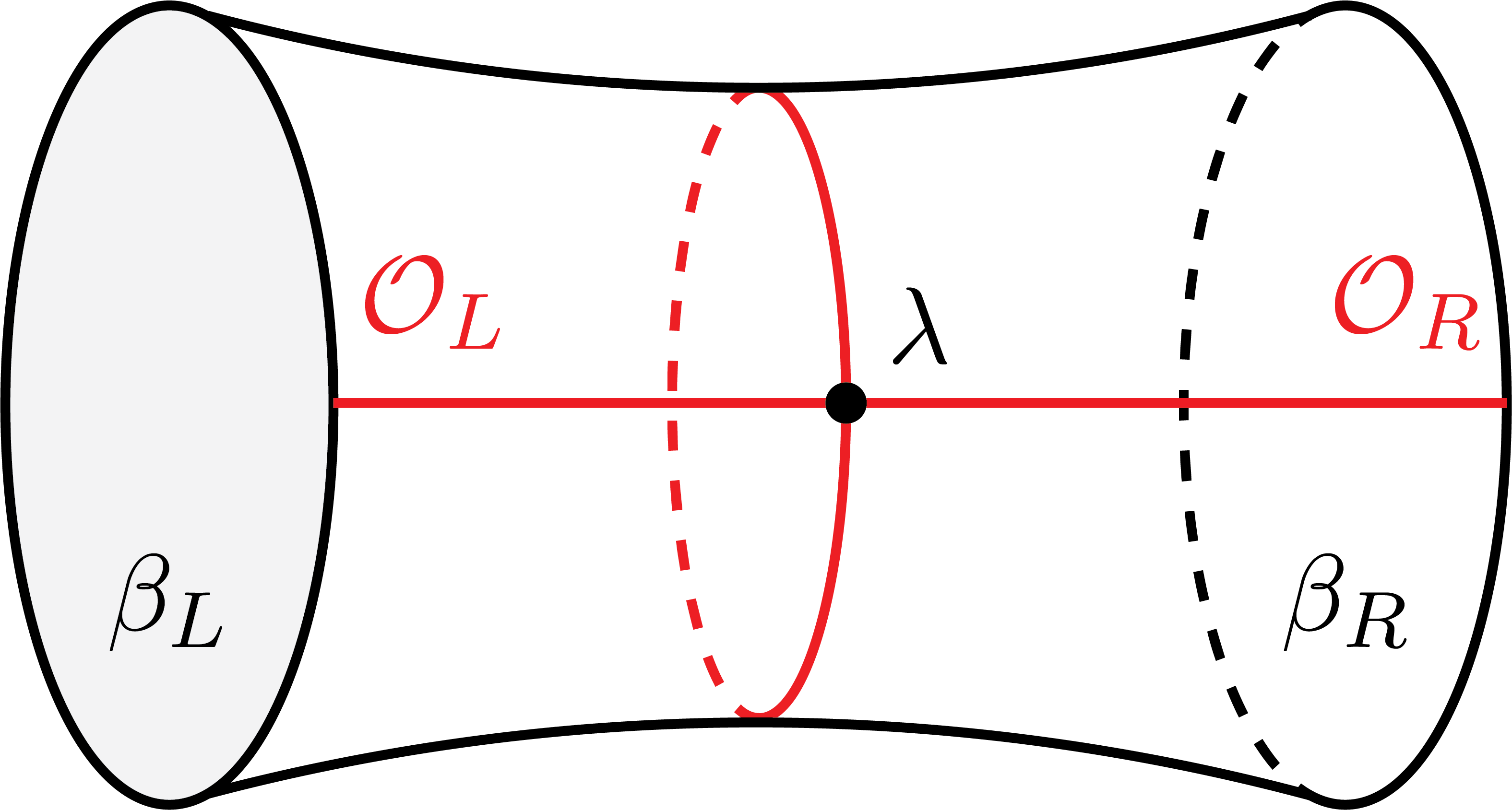}
 \caption{Self-energy diagram on the JT double trumpet. Two boundary propagators meet a winding matter loop at the quartic vertex (black dot). The figure shows a single winding, while \eqref{eq:exact-SE-main} sums over all nonzero windings and integrates the vertex over the cylinder.}
 \label{fig:jt-quartic-winding-self-energy}
\end{figure}
\begin{equation}
 \delta_\lambda G_{LR,b}(u,v)
 =-\frac{\lambda}{2}
 \int_{X_b}d^2y\sqrt g\,
 K_{L,b}(v;y)\,T_{\Delta,b}(y)\,K_{R,b}(u;y).
 \label{eq:exact-SE-main}
\end{equation}
This expression is valid at finite $\Delta>1/2$ and can be dressed to the Schwarzian boundaries to obtain the exact one-loop cylinder diagram. However instead of proceeding in this way, we first take the semiclassical $\Delta\gg1$ limit in which the expression will simplify and admit a clear physical interpretation and then dress it to the Schwarzian boundaries. 

Defining $M\equiv\Delta-\frac12$, the large-degree asymptotics of the Legendre function give  \cite[Eqs.~\href{https://dlmf.nist.gov/14.15.E14}{(14.15.14)} and~\href{https://dlmf.nist.gov/10.40.E2}{(10.40.2)}]{NIST:DLMF}
\begin{equation}
 G_{\mathbb H_2,\Delta}(d)
 \approx \frac{e^{-Md}}{\sqrt{8\pi M\sinh d}}
 \left[1+O(M^{-1})\right].
 \label{eq:heavy-G-main}
\end{equation}
This is the worldline propagator for a particle of mass $M$ moving on a geodesic of length $d$ in hyperbolic space, including the one-loop Van Vleck determinant. The leading contribution at large $M$ comes from the minimum length geodesics, localized at the throat ($r=0$). Expanding the geodesic length around $r=0$ we obtain: 
\begin{equation}
 d_w(r)
 =|w|b+\tanh\frac{|w|b}{2}\,r^2+O(r^4).
 \label{eq:dw-expand-main}
\end{equation}
Henceforth we take $w=1,2,\ldots$ and group the two orientations
$\pm w$ together. Their combined contribution is
\begin{equation}
 T_{\Delta,b}^{(w)}(r)
\approx
 \frac{e^{-Mwb}}{\sqrt{2\pi M\sinh (wb)}}
 \exp\!\left[-M\tanh\frac{wb}{2}\,r^2\right].
 \label{eq:T-gauss-main}
\end{equation}
We see the tadpole reduces to a Gaussian wavepacket of worldlines localized around the throat with a characteristic scale
\begin{equation}
 \delta r_w\sim\left[M\tanh\frac{wb}{2}\right]^{-1/2}.
\end{equation}
The expansion around $r=0$ is then under control for $M\tanh(b/2)\gg1$, when the fluctuations are small. 

In the large $\Delta\gg1$ limit we can evaluate the integral over $y$ in \eqref{eq:propagator} via saddle point. The bulk propagators give $e^{-M(d(x_L,y)+d(y,x_R))}$ so the saddle is given by $y$ lying on the geodesic $\gamma_{LR}$ that joins the two endpoints with length $d(x_L,x_R)$. We separate the integral over the geodesic $\gamma_{LR}$ and a Gaussian integral over the fluctuations transverse to the geodesic. Evaluating the transverse integral,  the result is:
\begin{equation}
 \frac{\delta_\lambda G^{}_{LR,b}}{G^{(0)}_{LR,b}}
 \simeq
 -\frac{\lambda}{4M}
 \int_{\gamma_{LR}}ds\,T_{\Delta,b}(s).
 \label{eq:eikonal-main}
\end{equation}
The same coefficient follows directly from the WKB momentum
$p(y)=\sqrt{M^2+(\lambda/2)T_{\Delta,b}(y)}
\simeq M+\lambda T_{\Delta,b}(y)/(4M)$: integrating its first variation along $\gamma_{LR}$ gives \eqref{eq:eikonal-main}. Here $M$ is the leading large-$\Delta$ WKB mass; the exact Klein-Gordon mass remains $m^2=\Delta(\Delta-1)$.
For aligned boundary angular positions, $\gamma_{LR}$ crosses the throat orthogonally, so $ds=dr$. A generic left-right geodesic is not orthogonal. In the heavy particle twist integral the aligned configuration is the leading saddle; the following multiplicative result uses this saddle as well as the radial localization condition $M\tanh(b/2)\gg1$. Using \eqref{eq:T-gauss-main},
\begin{align}
 \int_{-\infty}^{\infty}dr\,T_{\Delta,b}^{(w)}(r)
 &\simeq
 \frac{e^{-Mwb}}{\sqrt{2\pi M\sinh (wb)}}
 \sqrt{\frac{\pi}{M\tanh(wb/2)}}
 \\
 &=\frac{e^{-Mwb}}{2M\sinh(wb/2)}
 =\frac1M\frac{e^{-\Delta wb}}{1-e^{-wb}}.
 \label{eq:r-int-main}
\end{align}
We arrive at the final  result in the large $\Delta\gg1 $ approximation by summing over windings
\begin{equation}
 \frac{\delta_\lambda G_{LR,b}}{G^{(0)}_{LR,b}}
\approx
 -\frac{\lambda}{4(\Delta-\tfrac12)^2}
 \sum_{w=1}^{\infty}
 \frac{e^{-\Delta wb}}{1-e^{-wb}}.
 \label{eq:all-winding-main}
\end{equation}
In this limit, the dominant contribution is simply that of  a winding geodesic around the throat $b$  of a primary $\Delta$ and its $SL(2,\mathbb R)$ descendants. Note that the sum over windings differs  from the free scalar one-loop determinant of \cite{Jafferis:2022wez}: 
\begin{equation}
 \log Z_{\rm scalar}(b)
 =\sum_{w=1}^{\infty}\frac1w
 \frac{e^{-\Delta wb}}{1-e^{-wb}},
 \label{eq:det-main}
\end{equation}
where there is an additional $\frac{1}{w}$ factor since the loop is unmarked. The quartic vertex instead selects a point on the winding loop,  removing the $\frac{1}{w}$ factor. This corresponds to taking a derivative with respect to $\Delta$:
\es{eq:Gscalarloop}{
\frac{\delta_\lambda G_{LR,b}}{G^{(0)}_{LR,b}}\approx -\frac{\lambda}{4(\Delta-\tfrac12)^2} \qty(-\frac{1}{b} \partial_{\Delta} \log Z_{\rm scalar}(b)).
}

The approximation is not uniform as $b\to0$. In particular, the fixed-$b$ result cannot be inserted through an unrestricted small-throat region while claiming a controlled $1/\Delta$ expansion. Below we choose $b_{\min}>0$ with $M\tanh(b_{\min}/2)\gg1$ and restrict the heavy particle character formula to $b\ge b_{\min}$; the exact finite-$\Delta$ starting point remains \eqref{eq:exact-SE-main}. Other regulators are discussed in \cite{Jafferis:2022wez}.

\subsubsection*{Dressing to the Schwarzian boundaries.}
The result we obtained for the fixed geometry is a multiplicative insertion in the throat-length $b$.  The kernel that dresses a throat insertion to boundaries with energies $s_L,s_R$ is \cite{Jafferis:2022wez}
\begin{equation}
 \mathcal W_\Delta(s_L,s_R;b)
 =\left(\sinh\frac b4\right)^{2\Delta}
 \left(\cosh\frac b4\right)^{2(1-\Delta)}
 \Phi_{s_L}^{(\Delta)}(b)\Phi_{s_R}^{(\Delta)}(b).
 \label{eq:W-main}
\end{equation}
Here $\Phi_s^{(\Delta)}$ is the normalized Jacobi wavefunction of JKMS, appendix E, rescaled for our measure; its explicit formula is \eqref{eq:Jacobi-wavefunction-paper}. 
This rescaling gives $\int_0^\infty db\,\mathcal W_\Delta(s_L,s_R;b)=\delta(s_L-s_R)/\rho(s_L)$, and hence reproduces the Gaussian cylinder when no matter character is inserted.
For a character insertion, the unrestricted transform is the $SL(2,\mathbb R)$ $6j$ symbol,
\begin{equation}
 \int_0^\infty db\,\mathcal W_\Delta(s_L,s_R;b)
 \frac{e^{-hb}}{1-e^{-b}}
 =
 \begin{Bmatrix}
 h&s_L&s_R\\
 \Delta&s_R&s_L
 \end{Bmatrix}_{\rho}.
 \label{eq:JKMS-transform-main}
\end{equation}

Including the external factors $(\Gamma_{LL}^{\Delta}\Gamma_{RR}^{\Delta})^{1/2}$ of subsection \ref{subsec:jkms}, the $w=1$ self-energy correction is
\begin{align}
 \delta_\lambda\mathcal G_{1|1}^{\rm cyl}(\beta_L,\beta_R)\big|_{{\rm SE},\,w=1}\approx
 -\frac{\lambda}{4(\Delta-\tfrac12)^2}
 \int_0^\infty &ds_L\,ds_R\,
 \rho(s_L)\rho(s_R)
 e^{-\beta_LE_L-\beta_RE_R}
 \nonumber\\
 &\times
(\Gamma_{LL}^{\Delta}\Gamma_{RR}^{\Delta})^{1/2}
 \left[\begin{Bmatrix}
 \Delta&s_L&s_R\\
 \Delta&s_R&s_L
 \end{Bmatrix}_{\rho}\right]_{b\ge b_{\min}}.
 \label{eq:final-amplitude-main}
\end{align}
Only the $w=1$ term is proportional to the contribution from the free massive scalar one loop determinant. The full result, including all the windings $w$, can be obtained from \eqref{eq:Gscalarloop} by decomposing into $SL(2,\mathbb{R})$ representations. The two-point function that is dressed to the Schwarzian is the unnormalized one so we must multiply by the vacuum diagrams $Z_{\rm vac}(b)$: 
\es{}{
\mathcal{G}_{LR}^{(\lambda)}(b)=Z^{(\lambda)}_{\rm vac}(b)G^{(\lambda)}_{LR}(b)
}
The vacuum diagrams also receive an $\mathcal O(\lambda)$ correction from the figure-eight diagram in figure \ref{fig:jt-quartic-winding-figure-eight}:
\begin{figure}[t]
 \centering
 \includegraphics[width=.30\linewidth]{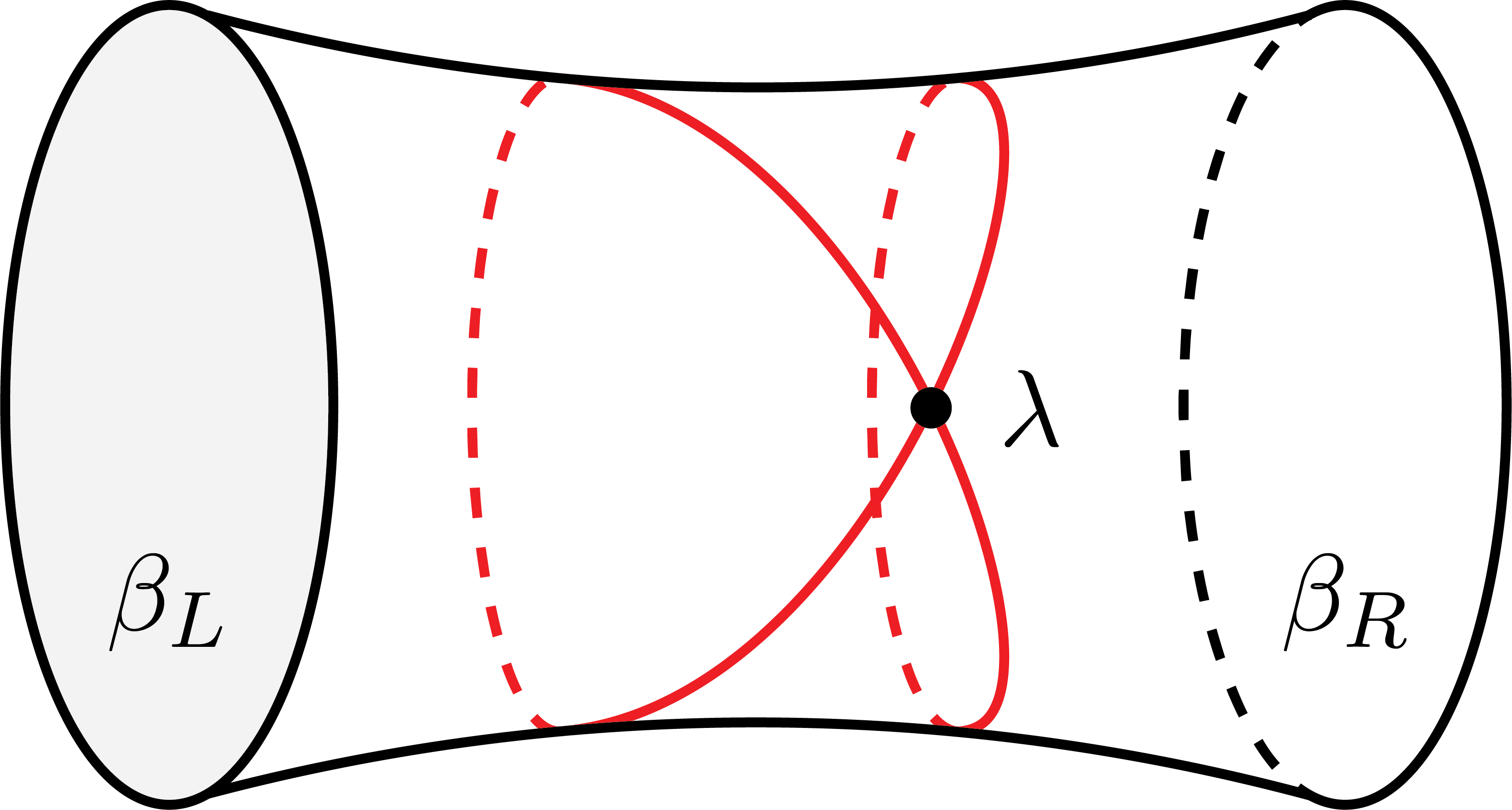}
 \caption{Figure-eight vacuum diagram on the JT double trumpet. Two matter loops wind around the cylinder and meet at a quartic vertex. The loops are drawn slightly apart for clarity, with dashed segments behind the cylinder. This diagram multiplies the free external propagator in the unnormalized one-point wormhole amplitude \eqref{eq:1pt-amplitude}. }
 \label{fig:jt-quartic-winding-figure-eight}
\end{figure} 
\es{}{
Z^{(\lambda)}_{\rm vac}(b)= Z_{\rm scalar}(b)+\delta_{\lambda}Z_{\rm vac}(b) +\mathcal{O}(\lambda^2)
}
so we have two terms at leading order in $\lambda$:
\es{}{\label{eq:1pt-amplitude}
\mathcal{G}^{(\lambda)}(b)\approx Z_{\rm scalar}(b) G^{(0)}_{LR}(b) + Z_{\rm scalar}(b) \delta_\lambda G_{LR}(b)+ \delta_\lambda Z_{\rm vac}(b)G^{(0)}_{LR}(b)
}
The first one is the free term considered in \cite{Jafferis:2022wez}.
We dress both $\mathcal O(\lambda)$ terms to the Schwarzian boundaries in turn by expressing them in $SL(2,\mathbb R)$ representations. 
\subsubsection*{Self-energy diagram}
For the self-energy contribution, it is useful to first write it in terms of the free scalar determinant. 
Introduce a fugacity $y$ for the number of particles in the free scalar determinant: 
\begin{equation}
 Z_{\rm scalar}(y,q)
=\sum_{N=0}^{\infty}y^NZ_N(q)=\prod_{n=0}^{\infty}
 \frac{1}{1-yq^{\Delta+n}},
 \qquad q=e^{-b}.
 \label{eq:fugacity-Z}
\end{equation}
The sum over windings in \eqref{eq:all-winding-main} can then be written as
\es{}{
\sum_{w=1}^{\infty} \frac{e^{-\Delta w b}}{1-e^{-wb}}= y\partial_y \log Z_{\rm scalar}(y,q)|_{y=1}= \frac{1}{Z_{\rm scalar}(b)} \sum_{N=1}^{\infty}NZ_N(q) 
}
which allows us to write the self-energy contribution as: 
\es{}{
Z_{\rm scalar}(b)\delta_\lambda G_{LR}(b)\approx -\frac{\lambda G^{(0)}_{LR}(b)}{4(\Delta-\tfrac12)^2} \sum_{N=1}^{\infty}NZ_N(q).
}
We see that the tadpole loop diagram corresponds to the insertion of the number operator $N$ in the $N$ particle Fock space representation. Decomposing $Z_N$ into $SL(2,\mathbb{R})$ characters 
\es{}{
Z_N(q)=\frac{q^{N\Delta}}{\prod_{j=1}^N(1-q^j)}
 =\sum_{\sigma\in\mathcal P_N}\chi_{h_{\sigma}}(q),\qquad
 \chi_{h_{\sigma}}(q)=\frac{q^{h_{\sigma}}}{1-q},\qquad
 h_{\sigma}=N\Delta+\sum_{j=2}^N j n_j
}
where $\mathcal P_N$ consists of tuples $\sigma=(n_2,\ldots,n_N)$ with $n_j\in\mathbb Z_{\geq0}$, each labeling one $N$-particle primary.
Dressing to the Schwarzian boundaries with the gluing kernel
\begin{equation}
 \int_0^\infty\!d b\,\mathcal W_\Delta(s_L,s_R;b)\,
 \chi_h(b)
 = \begin{Bmatrix}
 h&s_L&s_R\\
 \Delta&s_R&s_L
 \end{Bmatrix}_{\rho},
 \label{eq:JKMStransform}
\end{equation}
we obtain the self-energy contribution in the region $b\ge b_{\min}$, where the saddle is controlled: 
\es{}{
\delta_\lambda\mathcal G_{1|1}^{\rm cyl}(\beta_L,\beta_R)\big|_{\rm SE}\approx -\frac{\lambda}{4(\Delta-\tfrac12)^2}&\int_0^\infty\!d s_Ld s_R\,
 \rho(s_L)\rho(s_R)
 e^{-\beta_LE_L-\beta_RE_R}
 \\
 &\times
(\Gamma_{LL}^{\Delta}\Gamma_{RR}^{\Delta})^{1/2}\,
 \sum_{N=1}^{\infty}N
 \sum_{\sigma\in\mathcal P_N}
\left[\begin{Bmatrix}
 h_{\sigma}&s_L&s_R\\
 \Delta&s_R&s_L
 \end{Bmatrix}_{\rho}\right]_{b\ge b_{\min}}.
}
At fixed throat restriction, the free character sum acquires the particle-number weight $N$ and the displayed overall factor. The multi-trace kernel can therefore be read off directly:
\begin{equation}\label{eq:phi4-F11-SE}
\begin{split}
 \delta_\lambda F_{\O}^{(1|1)}(E_L|E_R)\Big|_{\rm SE}
 \simeq{}&-\frac{\lambda}{4(\Delta-\tfrac12)^2}
 (\Gamma_{LL}^{\Delta}\Gamma_{RR}^{\Delta})^{1/2}\times\sum_{N=1}^\infty N\sum_{\sigma\in\mathcal P_N}
 \left[\begin{Bmatrix}h_\sigma&s_L&s_R\\\Delta&s_R&s_L\end{Bmatrix}_{\rho}\right]_{b\ge b_{\min}}.
\end{split}
\end{equation}
The restriction $b\ge b_{\min}$ means the character transform in \eqref{eq:app-restricted-sixj}, with $M\tanh(b_{\min}/2)\gg1$. Formally removing this restriction makes the $N=1$ term $-\lambda/[4(\Delta-\tfrac12)^2]$ times \eqref{eq:PW-free-F11-paper}, but that removal is not a controlled step in the heavy particle approximation. The weighting by $N$ itself follows from the fixed-$b$ saddle.  The free scalar result is valid for arbitrary $\Delta$ and should be compared to the exact in $\Delta$ expression \eqref{eq:exact-SE-main}.

\subsubsection*{Vacuum figure-eight diagram}
Now we consider the vacuum figure eight correction, denoted $\delta_\lambda W_8(b)=\delta_\lambda\log Z_{\rm vac}(b)$, which in the same mass and vacuum renormalization scheme is 
\es{}{
\delta_\lambda Z_{\rm vac}(b)=Z_{\rm scalar}(b) \delta_{\lambda}W_8(b)= Z_{\rm scalar}(b)\qty(-\frac{\lambda}{8}\int_{X_b}d^2y\sqrt{g}T_{\Delta,b}(y)^2)
}
since there are two loops winding around the cylinder and ${4\choose 2}/4!=1/8$. The lightest contribution contains two winding scalar lines, so it begins at order $e^{-2\Delta b}$ in the heavy particle expansion, whereas the self-energy begins at $e^{-\Delta b}$. The interaction preserves $SL(2,\mathbb R)$ symmetry, so the vacuum trace can again be decomposed into characters, but the matter dimensions $h_\sigma$ receive order $\lambda$ corrections:
\es{}{
h_\sigma^{(\lambda)}=h_\sigma^{(0)}+\gamma_\sigma,
}
where $\gamma_\sigma=O(\lambda)$ denote the dimension shifts of the representation copies, including the coupling, and the free vacuum energy is held fixed. Their contribution can equivalently be written in terms of derivatives of characters at the classical dimension $h^{(0)}_{\sigma}$:
\es{}{
\delta_\lambda Z_{\rm vac}(b)=\sum_{N=2}^{\infty}\sum_{\sigma\in\mathcal P_N}
 \gamma_\sigma\left.\partial_h\chi_h(b)\right|_{h=h_\sigma^{(0)}}+O(\lambda^2).
}
We first isolate the two-particle contribution from $|w_1|=|w_2|=1$ in $T_{\Delta,b}^2$. The four sign combinations give the same contribution. Here $G_h$ denotes the covering-space propagator $G_{\mathbb H_2,h}$, rather than the quotient image sum:
\begin{equation}
 \delta_\lambda W_8^{(2)}(b)
 =-\frac{\lambda}{2}
 \int_{X_b}\!d^2y\sqrt g\,
 G_\Delta(y,\sigma_b y)^2 .
 \label{eq:fig8-two-particle-start}
\end{equation}
This is analogous to the two-particle thermal-AdS vacuum diagram analyzed in ~\cite{Kraus:2020nga}. Following their method, we decompose into free propagators
\begin{equation}
 G_\Delta(x,y)^2
 =\sum_{n=0}^{\infty}a_n^{\rm KL}\,
 G_{h_n}(x,y),
 \qquad h_n=2\Delta+2n,
 \label{eq:KL-two-particle}
\end{equation}
with AdS$_2$ K\"all\'en-Lehmann coefficients $a_n^{\rm KL}$ given explicitly in \eqref{eq:an-KL}. These are bulk propagator-product coefficients, not the boundary OPE coefficients $a_m^{(0)}$ of \eqref{eq:a0-paper}. For equal windings $|w_1|=|w_2|$ the same expansion applies directly. 
In AdS$_2$ the characters obey the following integral identity  \cite{Kraus:2020nga}:
\begin{equation}
 \partial_h\chi_h(b)
 =-(2h-1)
 \int_{X_b}\!d^2y\sqrt g\,
 G_h(y,\sigma_b y).
 \label{eq:character-propagator}
\end{equation}
Using \eqref{eq:KL-two-particle} and \eqref{eq:character-propagator} in \eqref{eq:fig8-two-particle-start} gives
\begin{equation}
 \delta_\lambda W_8^{(2)}(b)
 =\sum_{n=0}^{\infty}
 \gamma_n^{(2)}
\left.\partial_h\chi_h(b)\right|_{h=2\Delta+2n},
 \label{eq:fig8-character}
\end{equation}
where
\begin{equation}
 \gamma_n^{(2)}
 =\frac{\lambda\,a_n^{\rm KL}}
 {4\left(2\Delta+2n-\frac12\right)} .
 \label{eq:gamma-two-particle}
\end{equation}
These are the first-order anomalous dimensions of the double-trace primaries with dimensions $2\Delta+2n$. The figure-eight diagram thus gives the corresponding two-particle correction to the cylinder partition function. Dressing $\delta_\lambda W_8^{(2)}(b)$ with JT gravity then gives\footnote{Multiplying the connected contribution by $Z_{\rm scalar}(b)$ includes free spectator particles, but the complete higher-particle correction also requires the winding contributions omitted here.}
\begin{align}
 \delta_\lambda\mathcal G_{1|1}^{\rm cyl}(\beta_L,\beta_R)\big|_{\rm vac,\,N=2}
 \approx{}&\int_0^\infty\!d s_Ld s_R\,
 \rho(s_L)\rho(s_R)
 e^{-\beta_LE_L-\beta_RE_R}
 (\Gamma_{LL}^{\Delta}\Gamma_{RR}^{\Delta})^{1/2}
 \nonumber\\
 &\qquad\times\sum_{n=0}^{\infty}\gamma_n^{(2)}
 \left.\partial_h
 \begin{Bmatrix}
 h&s_L&s_R\\
 \Delta&s_R&s_L
 \end{Bmatrix}_{\rho}
 \right|_{h=2\Delta+2n}.
 \label{eq:fig8-energy-amplitude}
\end{align}
Similarly to ~\eqref{eq:JKMS-1pt2}, we can extract the two-particle vacuum contribution to the multi-trace ETH kernel which is
\begin{equation}\label{eq:phi4-F11-vac}
\begin{split}
 \delta_\lambda F_{\O}^{(1|1)}(E_L|E_R)\Big|_{{\rm vac},\,N=2}
 ={}&(\Gamma_{LL}^{\Delta}\Gamma_{RR}^{\Delta})^{1/2}
 \sum_{n=0}^\infty\gamma_n^{(2)}\times\left.\partial_h
 \begin{Bmatrix}h&s_L&s_R\\\Delta&s_R&s_L\end{Bmatrix}_{\rho}
 \right|_{h=2\Delta+2n}.
\end{split}
\end{equation}
Here $\gamma_n^{(2)}$ includes $\lambda$, unlike $\gamma_m^{(1)}$ in \eqref{eq:weak-data-paper}; for the same two-particle primary, $\gamma_n^{(2)}=\lambda\gamma_n^{(1)}+O(\lambda^2)$.  For all winding pairs, including $|w_1|\neq |w_2|$, the  explicit regulated result is
\begin{equation}\label{eq:phi4-F11-vac-all}
\begin{split}
 \delta_\lambda F_{\O}^{(1|1)}\Big|_{\rm vac}
 ={}&-\frac{\lambda}{8}(\Gamma_{LL}^{\Delta}\Gamma_{RR}^{\Delta})^{1/2}
 \int_{b_{\min}}^\infty db\,\mathcal W_\Delta(s_L,s_R;b)Z_{\rm scalar}(b)\\
 &\qquad\times\int_{X_b}d^2y\sqrt g\,
 \left[\sum_{w\neq0}G_{\mathbb H_2,\Delta}(y,\sigma_b^wy)\right]^2.
\end{split}
\end{equation}
The two-particle term has a finite $b_{\min}\to0$ limit for $\Delta>1/2$ and gives \eqref{eq:phi4-F11-vac}; this limit may  diverge for the full sum. This should be compared with the self-energy correction, which is proportional to the $6j$ kernels themselves, instead of their derivatives. The figure-eight instead shifts the dimensions of double-trace primaries and therefore produces derivatives with respect to the internal representation label.

\subsubsection {Relation of the $\lambda\phi^4$ diagrams to the generic weak interaction expansion.}
The two order-$\lambda$ corrections arise from different cylinder diagrams but both contribute to the same spectral decomposition.  At fixed throat length $b$ the
unnormalized matter two-point function is
\begin{equation}
 \mathcal G_{LR}^{(\lambda)}(b)
 =
 Z_{\rm vac}^{(\lambda)}(b)\,
 G_{LR}^{(\lambda)}(b),
\end{equation}
and therefore
\begin{equation}
 \delta_\lambda\mathcal G_{LR}(b)
 =
 Z_{\rm scalar}(b)\,\delta_\lambda G_{LR}(b)
 +
 \delta_\lambda Z_{\rm vac}(b)\,G_{LR}^{(0)}(b).
 \label{eq:phi4-split-paper}
\end{equation}
The first term is the self-energy correction to the through-going
propagator.  The second is the vacuum figure-eight multiplying the free
external line.

The self-energy corrects the through-going propagator. Our renormalization condition fixes the physical external dimension $\Delta$ and removes the contractible AdS tadpole. In the heavy particle saddle, \eqref{eq:all-winding-main} factorizes its remaining winding correction into a function of $b$, and the fugacity identity makes that function an insertion of the number operator:
\begin{equation}
 Z_{\rm scalar}(b)\delta_\lambda G_{LR}(b)
 \simeq-\frac{\lambda G_{LR}^{(0)}(b)}{4(\Delta-\frac12)^2}
 \sum_{N,\sigma}N\chi_{h_\sigma}(b).
 \label{eq:phi4-selfenergy-character-paper}
\end{equation}
This approximation changes the free theory weights, giving \eqref{eq:phi4-F11-SE}.

The figure-eight instead changes the matter vacuum trace. Its two-particle character derivative in \eqref{eq:fig8-character} is the usual first variation under the double-trace anomalous dimension, as in \cite{Kraus:2020nga} and JT dressing subsequently gives \eqref{eq:phi4-F11-vac}. 

The same interaction also has a disk interpretation. Cutting open the tadpole loop gives the AdS$_2$ four-point contact Witten diagram. In its $O\times O$ conformal-block expansion \eqref{eq:phi4-block-expansion-paper}, the coefficients $a_m^{(1)}$ correct the OPE weights, while $\gamma_m^{(1)}$ shift the double-trace dimensions. JT dressing replaces the conformal blocks by the analytic PW blocks in \eqref{eq:phi4-PW-expansion-paper}. At $\lambda=0$, these reduce to the JKMS Wilson-polynomial basis, while the dimension shifts introduce derivatives with respect to $h$.

Combining the two computed diagrams gives the full multi-trace ETH:
\begin{equation}\label{eq:F11-interactingJT}
\begin{split}
 F_{\O}^{(1|1)}\Big|_{\lambda\phi^4}
 ={}&F_{\O,\rm JKMS}^{(1|1)}
 +\delta_\lambda F_{\O}^{(1|1)}\Big|_{\rm SE}
 +\delta_\lambda F_{\O}^{(1|1)}\Big|_{\rm vac}
 +O(\lambda^2).
\end{split}
\end{equation}
 The vacuum term is \eqref{eq:phi4-F11-vac-all}, with the finite two-particle limit in \eqref{eq:phi4-F11-vac} and the  self-energy diagram, explicitly evaluated in the heavy particle approximation is in \eqref{eq:phi4-F11-SE}. 

\section{Multi-trace ETH in 3d gravity and holographic CFTs}\label{sec:3d}
This section presents the multi-trace ETH as a natural ansatz in 3d gravity and holographic CFTs.  We begin by describing how the full ETH is realized in these theories. Our primary tools are Virasoro TQFT (VTQFT) \cite{Collier:2023fwi, Collier:2024mgv} and the Virasoro analytic bootstrap \cite{Collier:2018exn, Collier:2019weq}. The former computes the contributions of saddles to the path integral of pure AdS$_3$ gravity and the latter allows us to describe certain universal sectors of holographic conformal field theories. Specifically, VTQFT and Virasoro mean field theory (VMFT) are used to derive multi-trace correlations in the statistics of light probe operators. 

The main text considers two limits of physical interest. At fixed central charge $c$, we study the statistics of states with large chiral dimensions $h$ and $\bar{h}$.  We also study semiclassical limits where the dimensions of operators are held fixed in units of the central charge as $c\to\infty$. These limits are specialized to the holographic study of macroscopic BTZ geometries. As a special case and whenever possible, a near-extremal limit is employed to connect back to the simple diagrammatic rules of JT gravity laid out in the previous section. Our main goal is to verify that the entropic scaling and index structure of multi-trace ETH suffices for computing correlation functions in all these limits.

In particular, for a probe operator $\cO_i$, we consider two sources of a non-trivial $F_i^{(1|1)}(E_j|E_k)$ as defined in \eqref{eq:MT}. Taking a cue from 3d gravity, we attribute these contributions to either non-trivial topology (subsection \ref{subsec:topology}) or to dynamical matter (subsection \ref{subsec:light_matter}). Topologically distinct contributions to the gravitational path integral are framed in the language of knots and links. We compute relevant limits of knot fragmentations corresponding to heavy operators, operators with large spin, and -- in the case of theories with large central charge -- states admitting a semiclassical description. The second class of non-Gaussianities come from dynamical light matter, which sources corrections to the statistics computed in pure gravity. This class includes the 3-dimensional uplift of the matter wormholes discussed in the previous section.  

For completeness, appendices \ref{app:vmft} and \ref{app:ne} contain discussions on various technical aspects of our approach. Appendix \ref{app:defects} briefly discusses why an ETH ansatz is inappropriate for describing operator statistics in the conical deficit regime.

\subsection{Primer: full ETH and the near-extremal limit}\label{subsec:full_eth_3d}
Here, following \cite{Ghosh:2019rcj}, we demonstrate the utility of the full ETH \cite{Foini:2018sdb} to computing single-boundary thermal correlation functions in 3d gravity and holographic CFTs (c.f. \cite{Belin:2020hea, Belin:2021ryy, deBoer:2024mqg, Collier:2024mgv, Espindola:2026xuc} for similar studies in 3d gravity and \cite{Kraus:2016nwo, Lashkari:2016vgj, He:2017vyf, Faulkner:2017hll, Hikida:2018khg, Maloney:2018yrz, Dymarsky:2019etq, Besken:2019bsu, Yan:2023rjh} for studies of thermalization in 2d CFTs). The partition function of 3d gravity is a functional on the set of boundaries $\Sigma$ with fixed moduli $\tau$
\begin{equation}\label{eq:z_grav}
    Z_{\rm grav}(\Sigma,\tau) = \sum_{\rm topologies}\int_{(\Sigma,\tau)} \frac{Dg}{\text{Vol}(\text{Diff}(M,\partial M))}\ e^{-S[g]} \,,
\end{equation}
computed over 3-manifolds $M$ with $\partial M=\Sigma$. The action $S$ is Einstein-Hilbert with necessary boundary and regulator terms.\footnote{
    The regulator is defined in appendix \ref{app:vmft_C0}.
}
In this subsection we consider path integrals of a single asymptotic boundary torus specified by a modulus $\tau$. Discussion of multiple boundaries and the sum over topologies will be postponed to later subsections.

Holographic CFTs have been shown to share universal properties \cite{Heemskerk:2009pn, El-Showk:2011yvt}. These theories can be parametrized in Liouville variables where
\begin{equation}
    c = 1 + 6Q^2\,,\quad Q=(b+b^{-1})\,,\quad h = \alpha(Q-\alpha) \,,\quad{\rm and}\quad \alpha=Q/2+iP \,.
\end{equation}
Our goal is to demonstrate that the full ETH is a valid ansatz in several physically relevant regimes of pure 3d gravity. One of these regimes is already familiar from the preceding section and will serve as a useful check of our computations. This is the near-extremal sector obtained in \cite{Maxfield:2020ale,Ghosh:2019rcj} by taking
\begin{equation}\label{eq:near_extremal_limit}
    \beta_L \sim c \quad{\rm and}\quad \beta_R \sim c^{-1} \quad{\rm as}\quad c\to\infty \,,
\end{equation}
with $\beta_L=2\pi \Im(\tau)$ and $\beta_R=-2\pi\Im(\bar{\tau})$.
This limit projects the right-moving chiral copy of the theory, labeled by the barred dimensions $\bar{h}$, onto the vacuum in the modularly transformed channel.\footnote{
    In section \ref{sec:JT}, $\Delta$ was reserved for the scaling dimension of $\mathfrak{sl}(2)$ blocks rather than $h+\bar{h}$. We will continue to use this definition in situations where the distinction is (hopefully) obvious like \eqref{fn:Delta_scaling}.
} AdS$_2$ universality implies that dynamics in the non-trivial chiral copy can be described by a Schwarzian particle in the limit of small left-moving temperature. A more precise definition of the near-extremal limit is provided in appendix \ref{app:ne}. Extensions to the regime of large chiral dimension and macroscopic BTZ black holes are also discussed here and in appendix \ref{app:vmft_C0}. 

\paragraph{$\Sigma_{1,2}$: The Torus 2-point function in a holographic CFT and its relation to VTQFT} 
Consider a holographic CFT whose torus two-point function is denoted as: 
\es{}{
\overline{\Tr\left(e^{2\pi i \tau\left(L_0-\frac{c}{24}\right)-2\pi i\bar{\tau}\left(\bar{L}_0-\frac{c}{24}\right)}\cO_i(0,0)\cO_i(z,\bar{z})\right)}
}
where, like in the rest of this work, the overline is intended in the sense of the ETH. Here we review the near-extremal reductions of this correlation function to the Schwarzian two-point function of JT gravity. Along the way, we arrive at a universal contribution to the two-point function that extends the validity of the ETH far outside of the near-extremal regime.

The dimensional uplifts of disk correlators in JT gravity should be compared to torus channel decompositions which take advantage of the spectral gap in holographic CFTs. The analogous geometry is clear in this case: asymptotic boundary lengths in 2d should correspond to moduli in a necklace channel decomposition along the time cycle of the torus. VTQFT can compute these amplitudes starting from the assumption that the identity flows in a necklace decomposition of the meridian spatial cycle:
\begin{equation}\label{eq:two_pt_extremal}
    \overline{\Tr\left(e^{2\pi i \tau\left(L_0-\frac{c}{24}\right)-2\pi i\bar{\tau}\left(\bar{L}_0-\frac{c}{24}\right)}\cO_i(0,0)\cO_i(z,\bar{z})\right)} \approx \abs{\braket{\tau,z}{Z_{\rm Vir}\left(\inlinefig{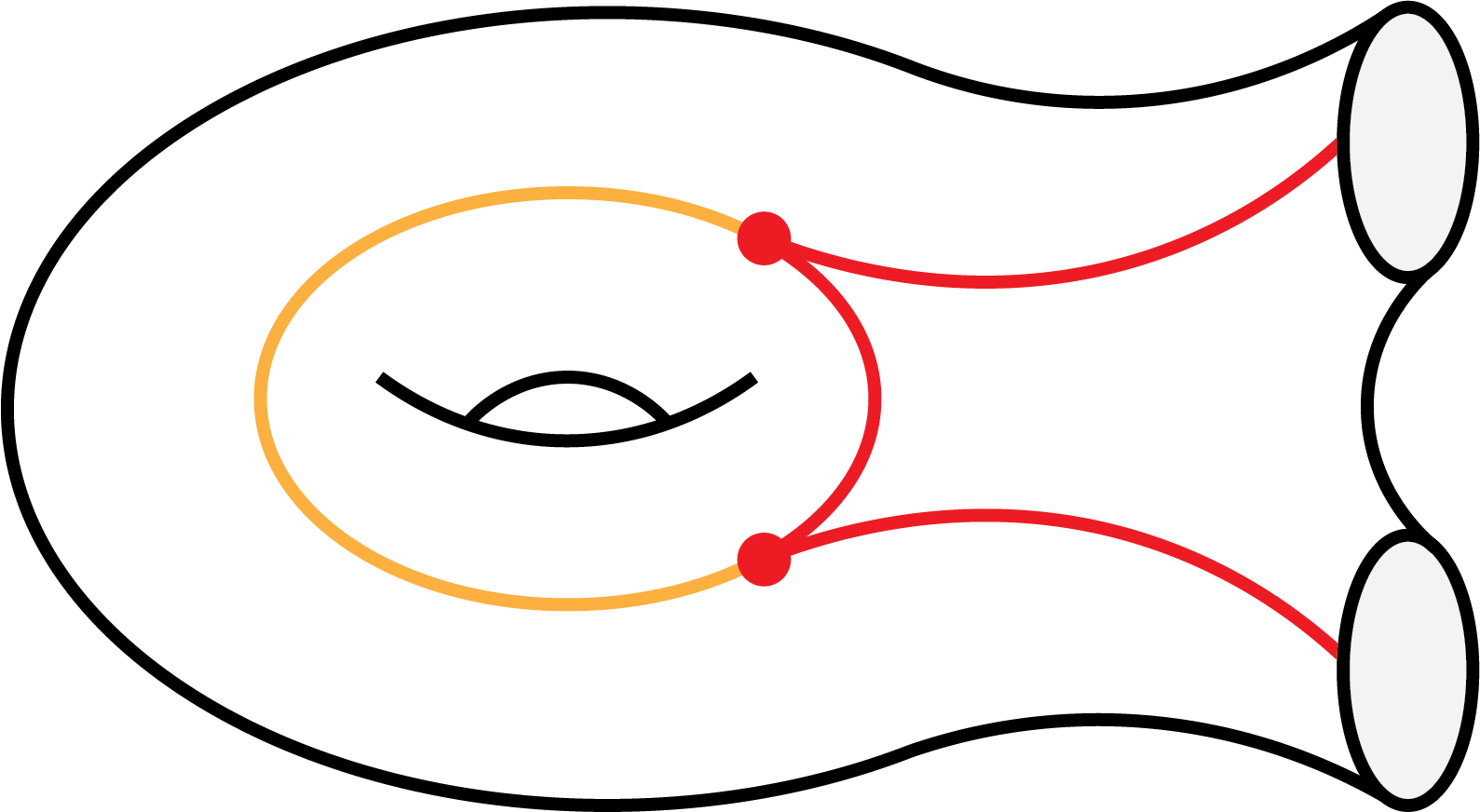}\right)}}^2 
\end{equation}
where $\ket{Z_{\rm Vir}(\cdot)}$ refers to the state prepared by Virasoro TQFT on the pictured manifold and $\approx$ denotes that the moduli $\tau$ and $z$ are chosen to ensure vacuum dominance in the CFT.\footnote{
    Vacuum dominance is guaranteed for the 2-point function given sufficiently small complex separation $z$ between operator insertions \cite{Ghosh:2019rcj}.
} For convenience, we sometimes omit ``$Z_{\rm Vir}(\cdot)$'' when labelling a state. The red lines in the channel decomposition correspond to propagation of $\cO_i$ while the orange lines denote vacuum propagation. In the near-extremal limit, only one chiral copy remains dynamical and correlation functions on this copy can be matched to disk correlators
\es{}{
    \braket{\tau,z}{\inlinefig{imgs/s6/OO_1.png}} \underset{\rm near-extremal}{\Longrightarrow} \inlinefig[6]{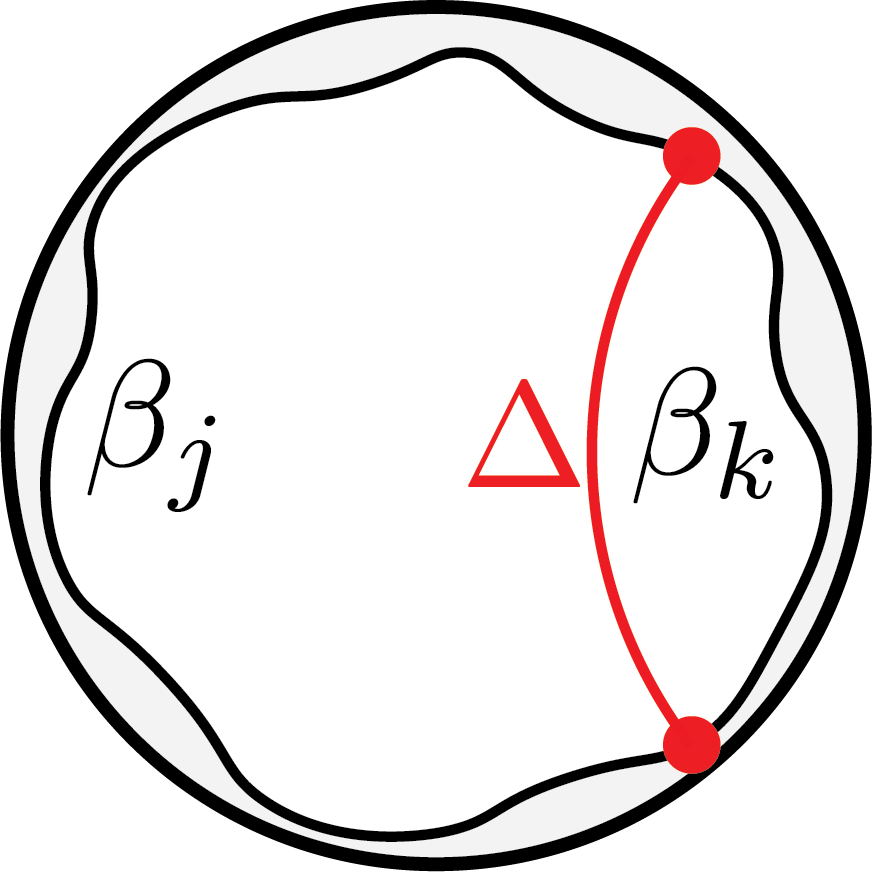}\,.
}
We will defer defining $\beta_j$, $\beta_k$ and $\Delta$ for the moment. For now, note that we are not using VTQFT as the holographic dual of a particular CFT. Rather it provides an efficient way to capture universal contributions from vacuum block dominance in a class of CFTs. 

Conformal blocks of different channels span the same Hilbert space. Our goal is to write the state on the right-hand-side of \eqref{eq:two_pt_extremal} as a superposition of heavy states that admit a near-extremal limit. The mass and spin of the black hole are determined by the holonomy around the time cycle so, for direct comparison to the near-extremal limit, operators should be inserted in a longitudinal necklace channel. We use crossing kernels \cite{Eberhardt:2023mrq} to change channels in the conformal block decomposition  
\begin{align}
    \ket{\inlinefig{imgs/s6/OO_1.png}} &= \int dP\ \fker{P_i}{P}{\mathbbm 1}{\mathbbm 1}{P_i}{P_i}\ket{\inlinefig{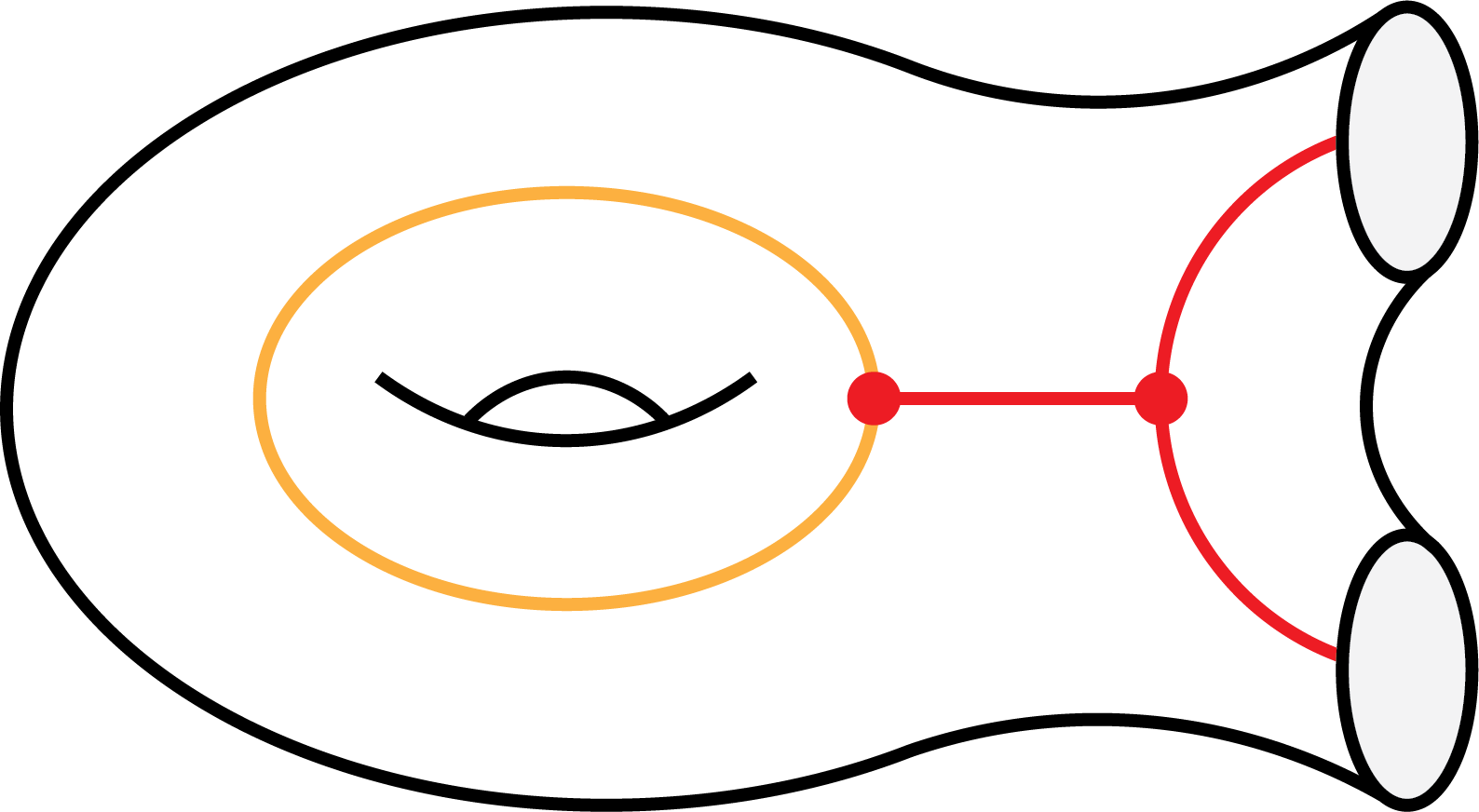}} \nn\\
    &= \int dPdP_j\ 
    \sker{\mathbbm 1}{P_j}{P}\fker{P_i}{P}{\mathbbm 1}{\mathbbm 1}{P_i}{P_i}\ket{\inlinefig{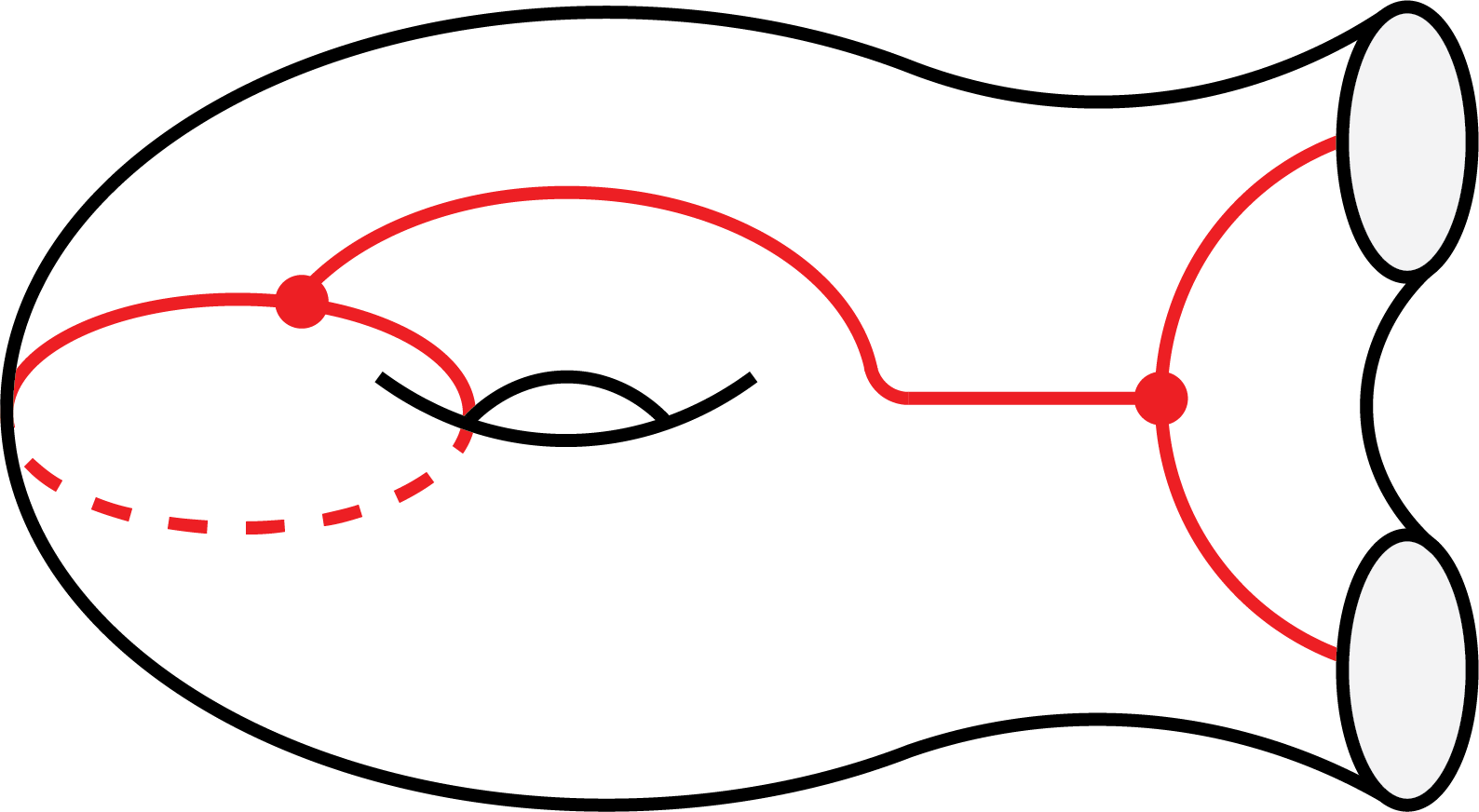}} \nn\\
    &= \int dPdP_jdP_k\ 
    \fker{P}{P_k}{P_j}{P_i}{P_j}{P_i}\sker{\mathbbm 1}{P_j}{P}\fker{P_i}{P}{\mathbbm 1}{\mathbbm 1}{P_i}{P_i}\ket{\inlinefig{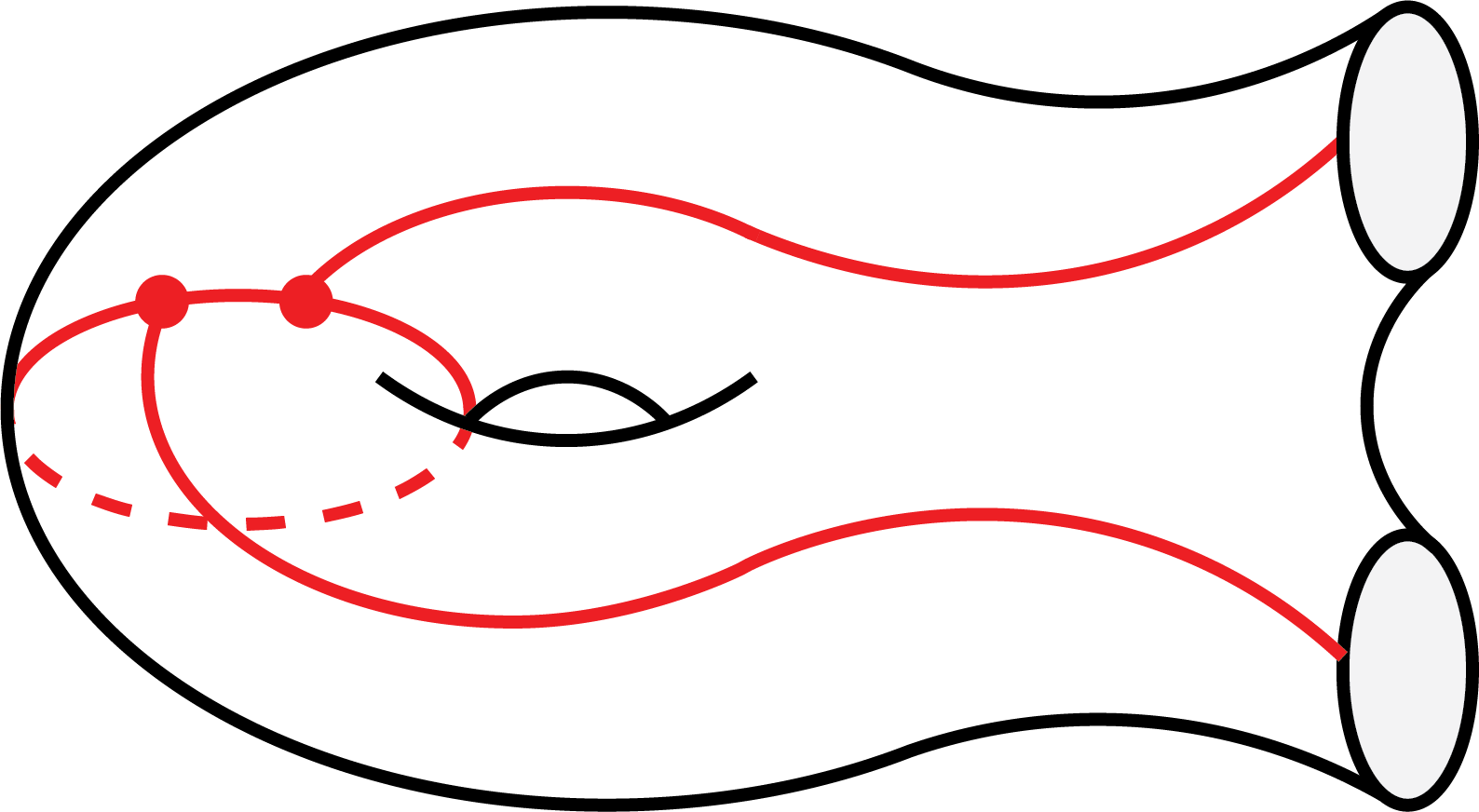}} \nn\\
    &= \int dP_jdP_k\  \fker{\mathbbm 1}{P_k}{P_j}{P_i}{P_j}{P_i}\sker{\mathbbm 1}{P_j}{\mathbbm 1}\ket{\inlinefig{imgs/s6/OO_4.png}} \,.
\end{align}
In the first line we used the following identity of the fusion kernel
\begin{equation}\label{eq:fker_id1}
    \fker{P_i}{P}{\mathbbm 1}{\mathbbm 1}{P_i}{P_i} = \delta(P-\mathbbm{1})\,,
\end{equation}
which follows, up to normalization, from the fusion rules of the identity.\footnote{
    The normalization is determined by considering norms of conformal blocks on the thrice-punctured sphere \cite{Eberhardt:2023mrq}. 
} Here $\mathbbm{1}$ refers to the vacuum momentum $iQ/2$. 

\begin{figure}
    \centering
    \includegraphics[width=0.25\textwidth]{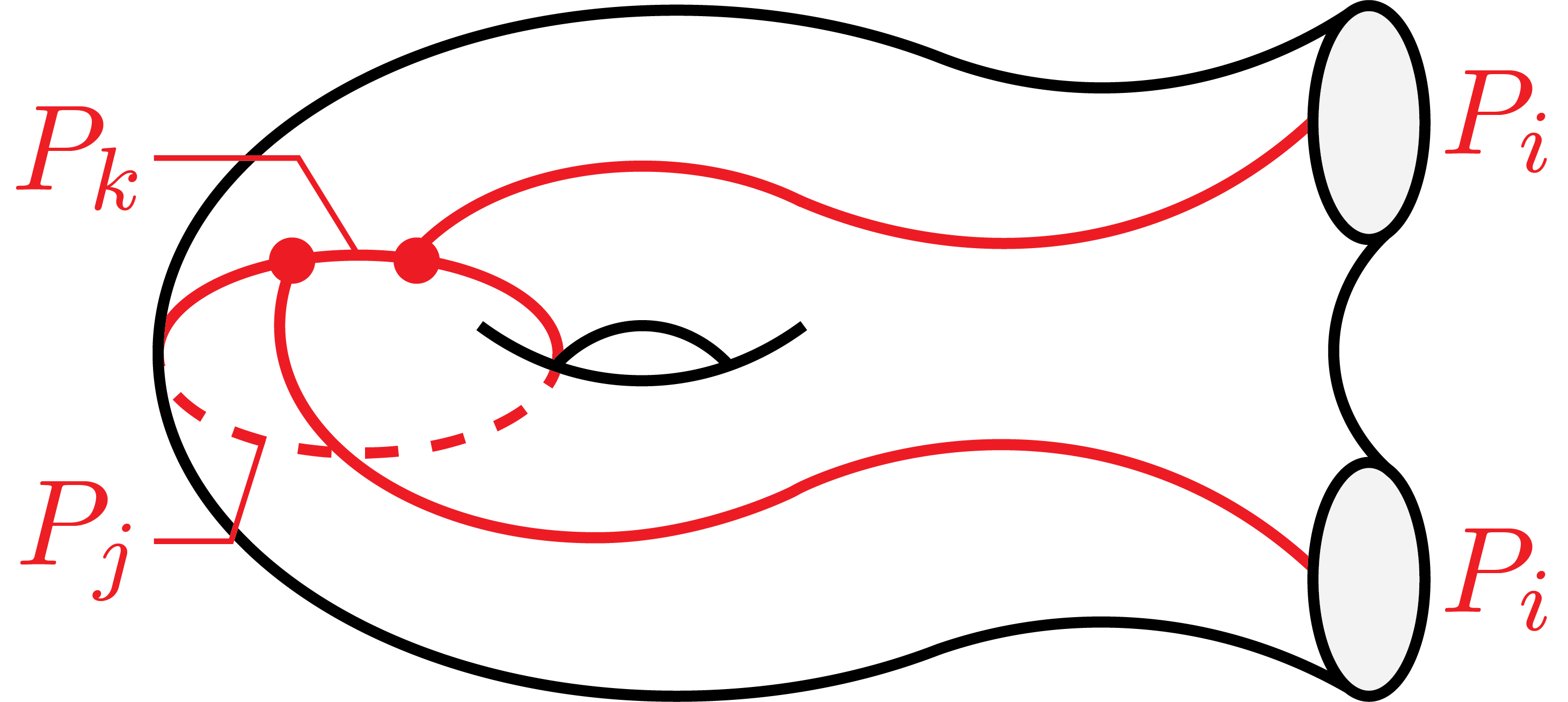}
    \caption{Vacuum dominance in the meridian necklace channel leads to the index structure pictured above in the longitudinal necklace channel.}\label{fig:O2_4_labeled}
\end{figure}

In terms of the universal structure constant and state density defined in appendix \ref{app:vmft}, the thermal two point function corresponds to the following state in VTQFT
\begin{equation}
    \ket{\inlinefig{imgs/s6/OO_1.png}} = \int \rho_0(P_j)dP_j\rho_0(P_k)dP_k\ C_0(P_i,P_j,P_k)\ket{\inlinefig{imgs/s6/OO_4.png}}\,,
\end{equation}
where momenta in the right-most state are labeled as in figure \ref{fig:O2_4_labeled}. This state also follows from the following standard ETH ansatz on matrix elements of the probe
\begin{equation}\label{eq:standard_eth_3d}
    \mel{P_j}{\cO_i}{P_k} = e^{-S(P_{jk})/2}f_i(P_{jk},\omega_{jk})R_{jk} \quad{\rm with}\quad \begin{cases}
        e^{S(P_{jk})}=\rho_0(P_{jk}) \,, \\
        |f_i(P_{jk})|^2=e^{S(P_{jk})}C_0(P_i,P_j,P_k) \,.
    \end{cases}
\end{equation}
We are most interested in the case when the individual momenta $P_j,P_k$ are either heavy or scaled appropriately with a large central charge. We verify in appendix \ref{app:vmft_rho0} that, in these limits, the universal structure constant $C_0$ is exponentially suppressed in the entropy $S(P_{jk})$ of the average energy $P_{jk}=(P_j+P_k)/2$ while the smooth function $f_i$ scales polynomially. Thus, these requirements define microcanonical regimes where the standard ETH is sufficient to compute this correlator. Explicit formulas for $f_i$ in these limits are given in appendix \ref{app:vmft_C0}.

All that is required to compute the correlation function from this ETH ansatz is a choice of moduli. One option is to specify the shape $\tau$ of the torus and the separation $z$ of the operator insertions. However, for the longitudinal necklace channel, it is more useful (c.f. appendix \ref{app:ne_blocks}) to decompose the torus into two cylinders with moduli $t_j$ and $t_k$ such that
\begin{equation}
    \tau = t_j+t_k \quad{\rm and}\quad t_j=z\,.
\end{equation}
Torus conformal blocks are recovered by inserting a complete set of states between the boundaries of the cylinders
\begin{align}
    \braket{t_j,t_k}{\inlinefig{imgs/s6/OO_4.png}} &= \cF(P_i;P_j,P_k|t_j,t_k) \nn\\
    &= \frac{1}{C_0(P_i,P_j,P_k)}\Tr_{\cV_j,\cV_k}(e^{2\pi it_j (L_0-c/24)} \cO_i e^{2\pi it_k (L_0-c/24)} \cO_i) \,,
\end{align}
where the traces are taken over the Verma modules $\cV_j$ and $\cV_k$. 

The chiral correlation function is then
\begin{equation}
    \braket{t_j,t_k}{\inlinefig{imgs/s6/OO_1.png}} = \int \rho_0(P_j)dP_j\rho_0(P_k)dP_k\ C_0(P_i,P_j,P_k)\cF(P_i;P_j,P_k|t_j,t_k)\,.
\end{equation}
We now confirm that, in the near-extremal limit, this function recovers the ETH statistics of JT gravity described in section \ref{sec:JT}. Appendix \ref{app:ne_blocks} shows that it suffices to focus on the contributions of primaries in this regime giving 
\begin{equation}
    \braket{t_j,t_k}{\inlinefig{imgs/s6/OO_4.png}} \approx e^{-\beta_j s_j^2-\beta_k s_k^2} \,,
\end{equation}
where the $\beta$'s and $P$'s are related to the $t$'s and $s$'s such that $\beta,s=O(1)$ while
\begin{equation}
    \beta = 2\pi b^2\Im(t) \quad{\rm and}\quad s = b^{-1}P \,.
\end{equation}
Here ``$\approx$'' denotes the leading term in a small $b$ expansion. The conventions established in appendix \ref{app:ne_ops} give\footnote{\label{fn:Delta_scaling}
    Here, like in section \ref{sec:JT}, $\Delta$ is the scaling dimension of an $\mathfrak{sl}(2)$ block, not a Virasoro block. 
}
\begin{equation}
    \braket{\tau,z}{\inlinefig{imgs/s6/OO_1.png}} \underset{\rm near-extremal}{=} \int \rho(s_j)ds_j\rho(s_k)ds_k\  \Gamma^\Delta_{jk}e^{-\beta_js_j^2-\beta_ks_k^2}\,,
\end{equation}
which matches the expected result from the diagrammatic rules of JT.

\paragraph{$\Sigma_{1,2n}$: Torus $2n$-point function and higher order moments} We now move to computing the higher order cyclic moments which appear in the $2n$-point function
\es{}{
    \overline{\Tr\left(e^{2\pi i \tau\left(L_0-\frac{c}{24}\right)-2\pi i\bar{\tau}\left(\bar{L}_0-\frac{c}{24}\right)}\cO_i(0,0)\cO_i(z_2,\bar{z}_2)\cdots\cO_i(z_{2n},\bar{z}_{2n})\right)}\,.
}
As before, our goal will be to compute the VTQFT state associated to operator insertions in a necklace channel decomposition of the time cycle. Note that we have neglected to compute correlation functions with an odd number of operator insertions. There is a physical reason for this omission. Though we can always use VTQFT to compute a kernel between the meridian and longitudinal necklace channels, there is in general no guarantee that the vacuum is dominant in the meridian channel for some choice of moduli. We verify in appendix \ref{app:ne_blocks} that this is the case for an even number of operator insertions. We also argue that vacuum dominance for an odd number of insertions requires additional non-universal assumptions on the OPE coefficients of the theory. This fact has previously been noted in the AdS$_3$/CFT$_2$ literature (e.g. \cite{Kraus:2016nwo, Chang:2016ftb, Chandra:2022bqq, deBoer:2024mqg}).

\begin{figure}
    \centering
    \begin{subfigure}{0.2\linewidth}
        \center
        \includegraphics[width=0.6\linewidth]{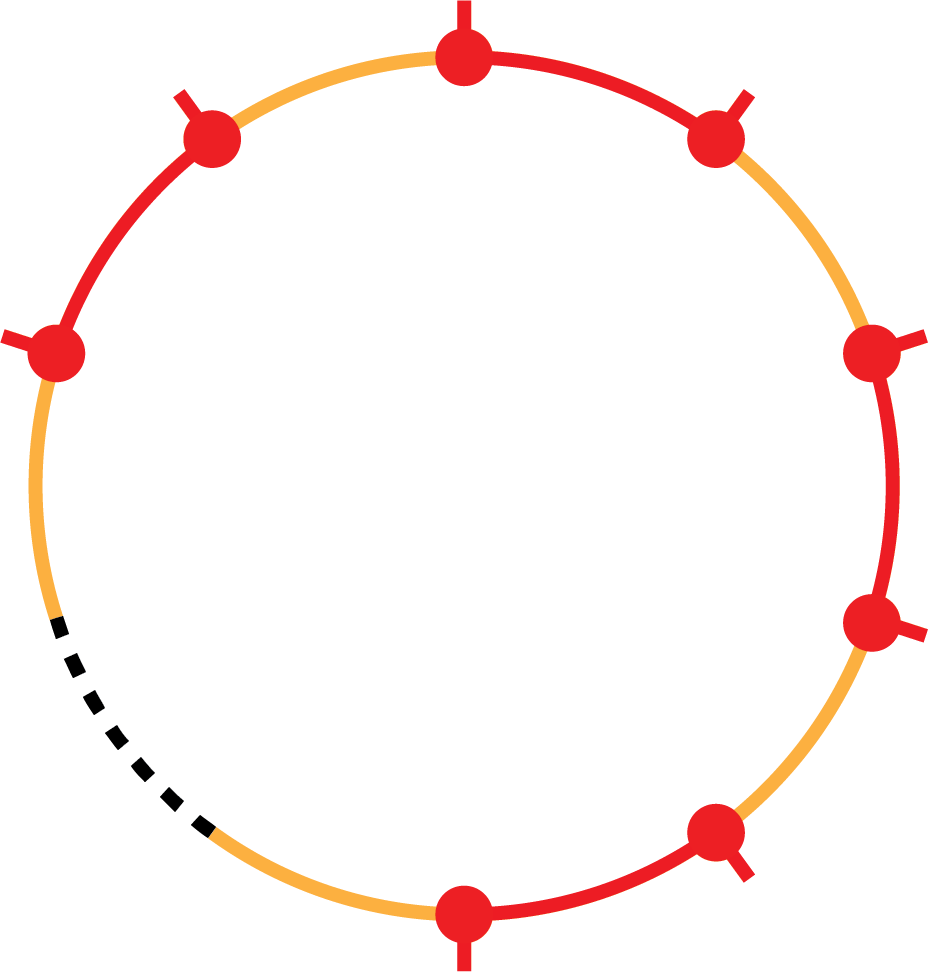}
        \caption{$\ket{\cF^{\cC_1}}$}
    \end{subfigure}
    \begin{subfigure}{0.2\linewidth}
        \center
        \includegraphics[width=0.6\linewidth]{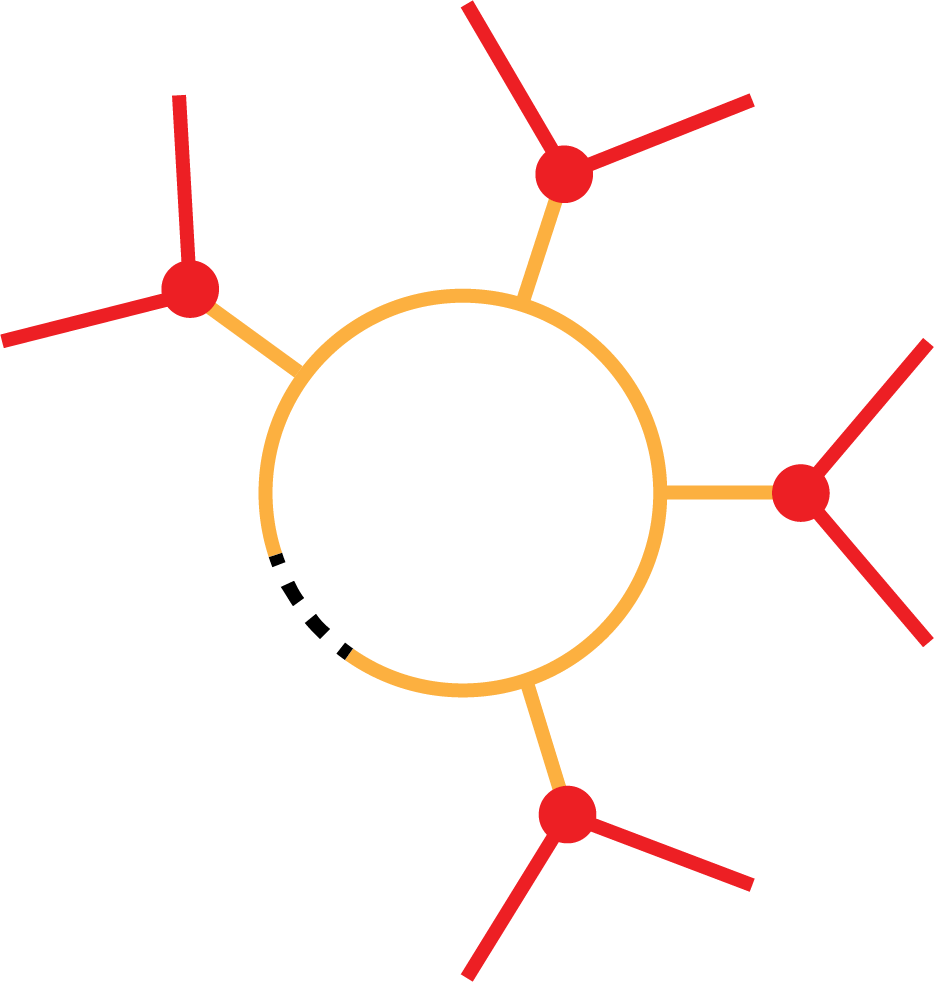}
        \caption{$\ket{\cF^{\cC_2}}$}
    \end{subfigure}
    \begin{subfigure}{0.2\linewidth}
        \center
        \includegraphics[width=0.6\linewidth]{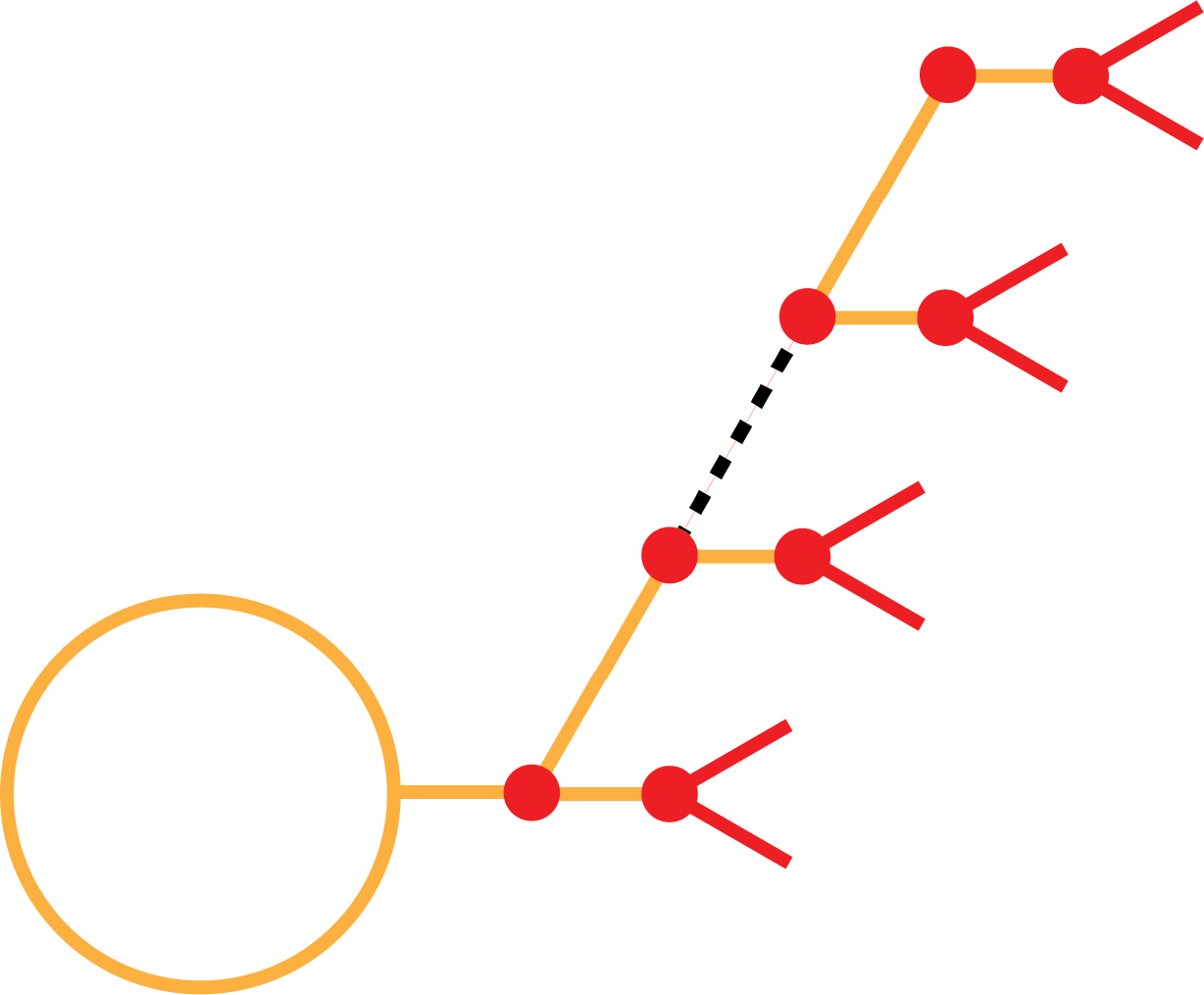}
        \caption{$\ket{\cF^{\cC_3}}$}
    \end{subfigure}
    \begin{subfigure}{0.2\linewidth}
        \center
        \includegraphics[width=0.6\linewidth]{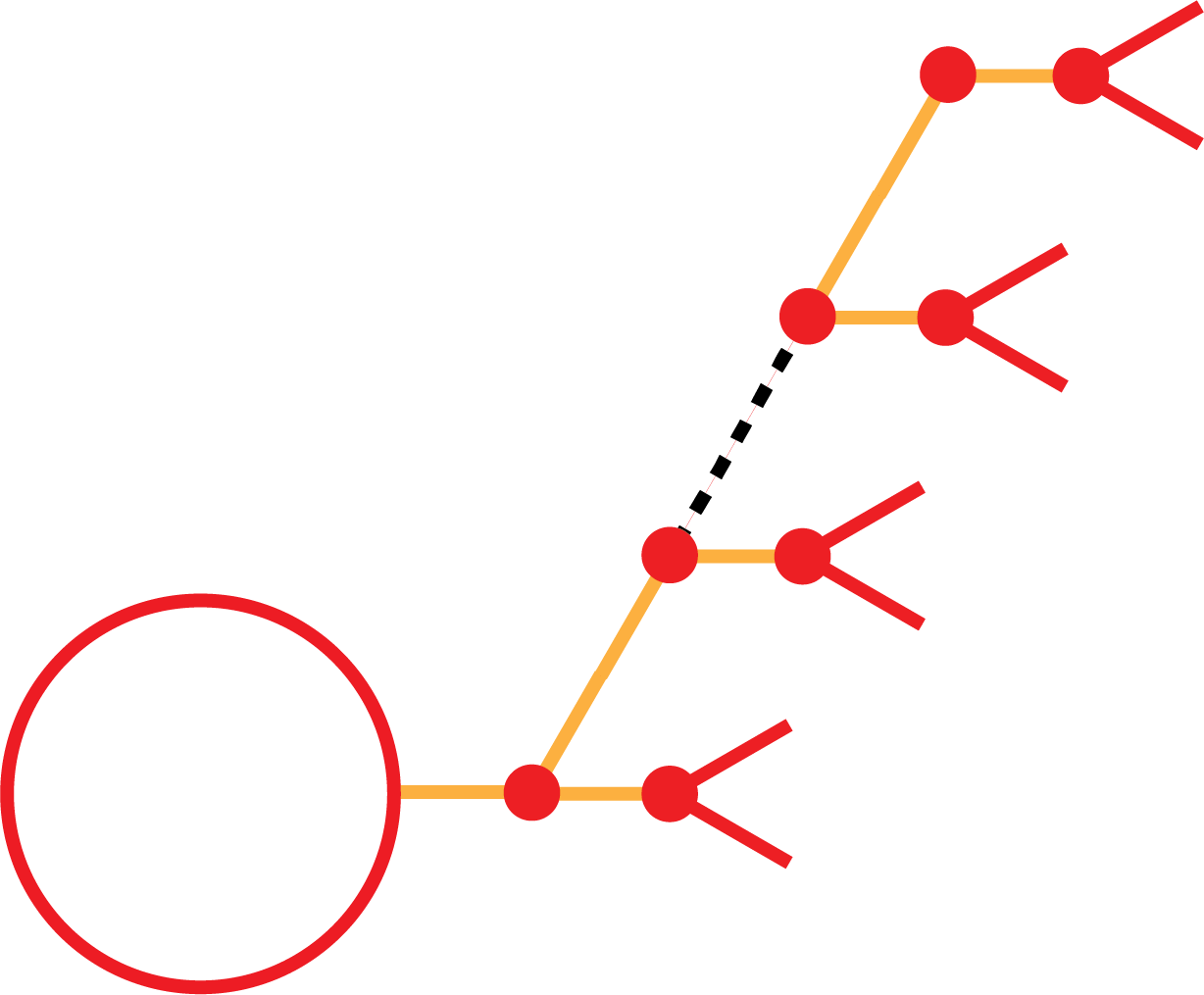}
        \caption{$\ket{\cF^{\cC_4}}$}
    \end{subfigure}

    \begin{subfigure}{0.2\linewidth}
        \center
        \includegraphics[width=0.6\linewidth]{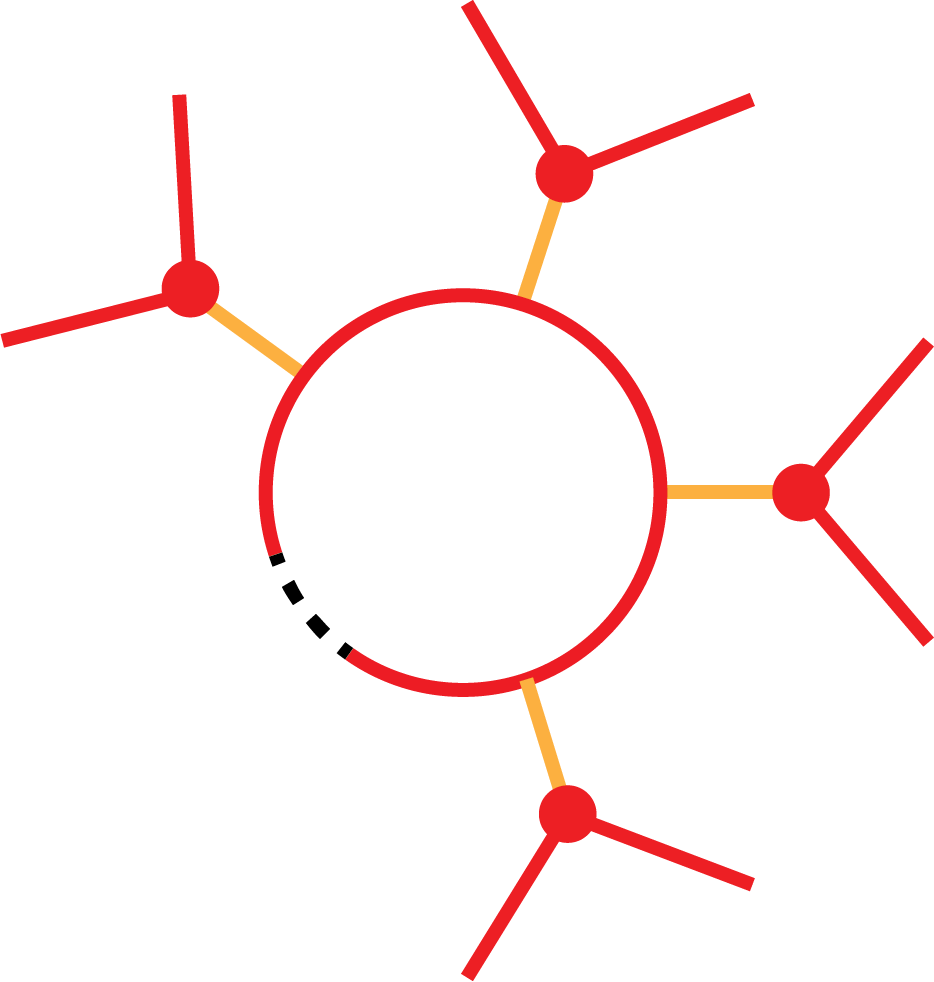}
        \caption{$\ket{\cF^{\cC_5}}$}
    \end{subfigure}
    \begin{subfigure}{0.2\linewidth}
        \center
        \includegraphics[width=0.6\linewidth]{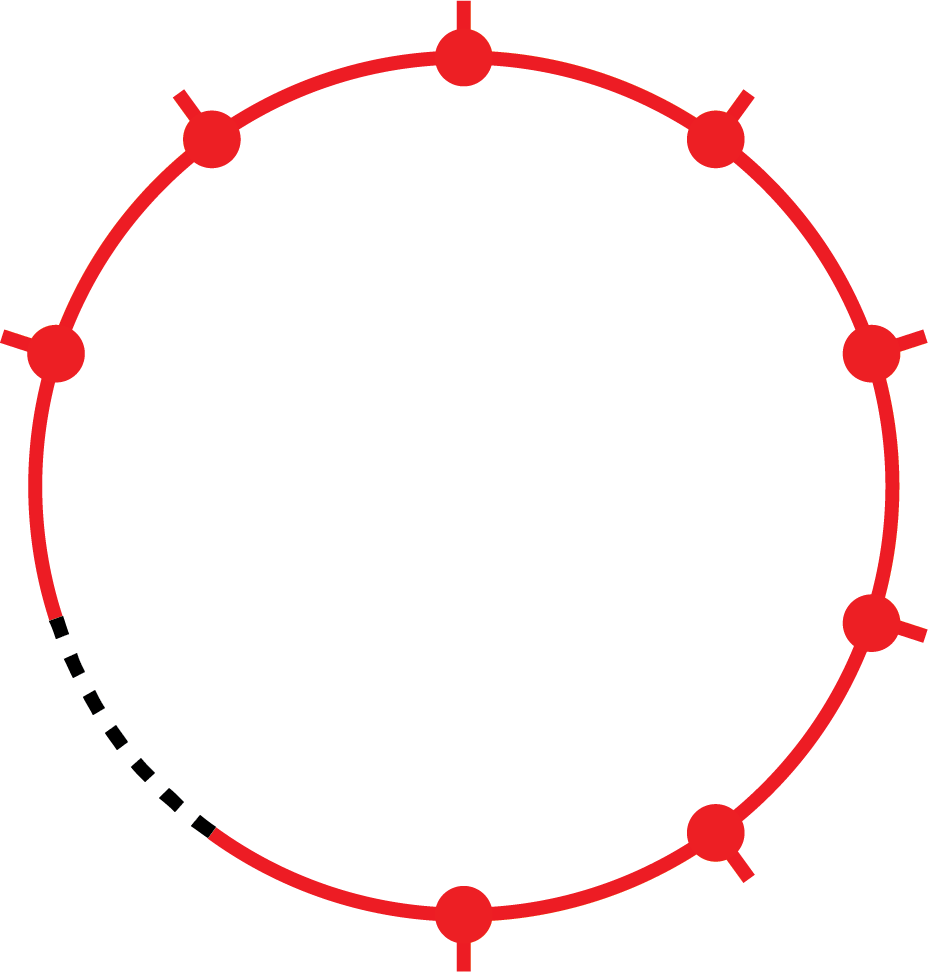}
        \caption{$\ket{\cF^{\cC_6}}$}
    \end{subfigure}

    \caption{Above we visually describe the various channel decomposition used in \eqref{eq:OO^n_decomps}. First, repeated use is made of \eqref{eq:fker_id1} to express $\cC_1$ as the set of $n$ degree 2 binary trees attached to the meridian cycle in $\cC_2$. This identity is a specific case of the more general \eqref{eq:fker_id} and the second line uses a trivial version of this equality to collect the separate trees into a single degree $(n+2)$ binary tree attached to the meridian cycle at a single point of $\cC_3$. We then $\mathbb{S}$-transform from the spatial cycle to the time cycle to arrive at $\cC_4$ and perform all fusion moves in reverse to arrive at $\cC_5$. \eqref{eq:fker_id} clarifies that there is no penalty for recovering the $n$ degree 2 binary trees and this is accomplished in line 4 of \eqref{eq:OO^n_decomps}. Finally, we arrive at the longitudinal necklace channel $\cC_6$ by $n$ actions of the $\mathbb{F}$-kernel. }
    \label{fig:2n_channels}
\end{figure}

For now, start with vacuum dominance in the meridian necklace decomposition of the torus with 2$n$ punctures
\begin{equation}
    \overline{\Tr\left(e^{2\pi i \tau\left(L_0-\frac{c}{24}\right)-2\pi i\bar{\tau}\left(\bar{L}_0-\frac{c}{24}\right)}\cO_i(0,0)\cO_i(z_2,\bar{z}_2)\cdots\cO_i(z_{2n},\bar{z}_{2n})\right)} \approx \abs{\braket{\tau,\vec{z}}{\inlinefig[6]{imgs/s6/O2n_1.png}}}^2\,.
\end{equation}
A particular torus is here specified by a modulus $\tau$ and a list of complex numbers $\vec{z}=\{z_i\}$ giving the locations of $(2n-1)$ of the insertions. We refer to the channel decomposition on the right-hand side as $\cC_1$. The choice to place the ``first'' insertion at the origin is purely cosmetic. We will verify that the holomorphic component of the overlap reduces to a disk correlator under the near-extremal limit
\begin{equation}
    \braket{\tau,\vec{z}}{\inlinefig[6]{imgs/s6/O2n_1.png}} \underset{\rm near-extremal}{\Longrightarrow} \inlinefig[6]{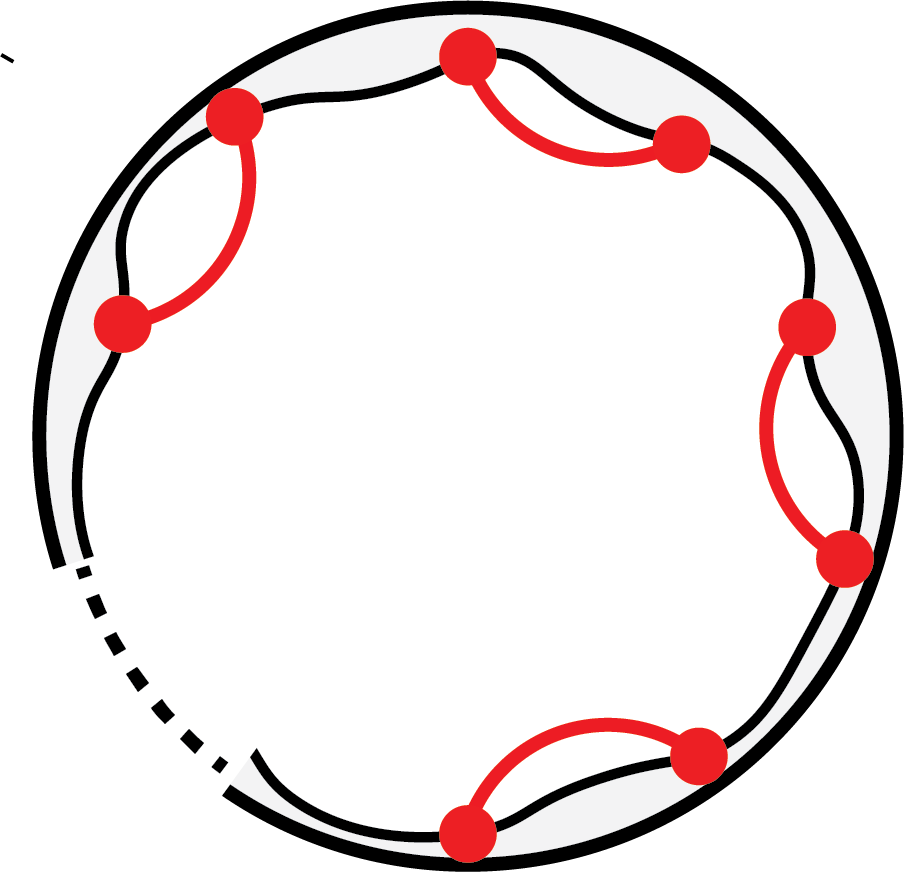} ~.
\end{equation}
As detailed in appendix \ref{app:ne_blocks}, the CFT fusion rules imply that dominance of the vacuum along one orange arc imposes vacuum propagation along every other arc. 

\begin{figure}
    \centering
    \includegraphics[width=0.25\textwidth]{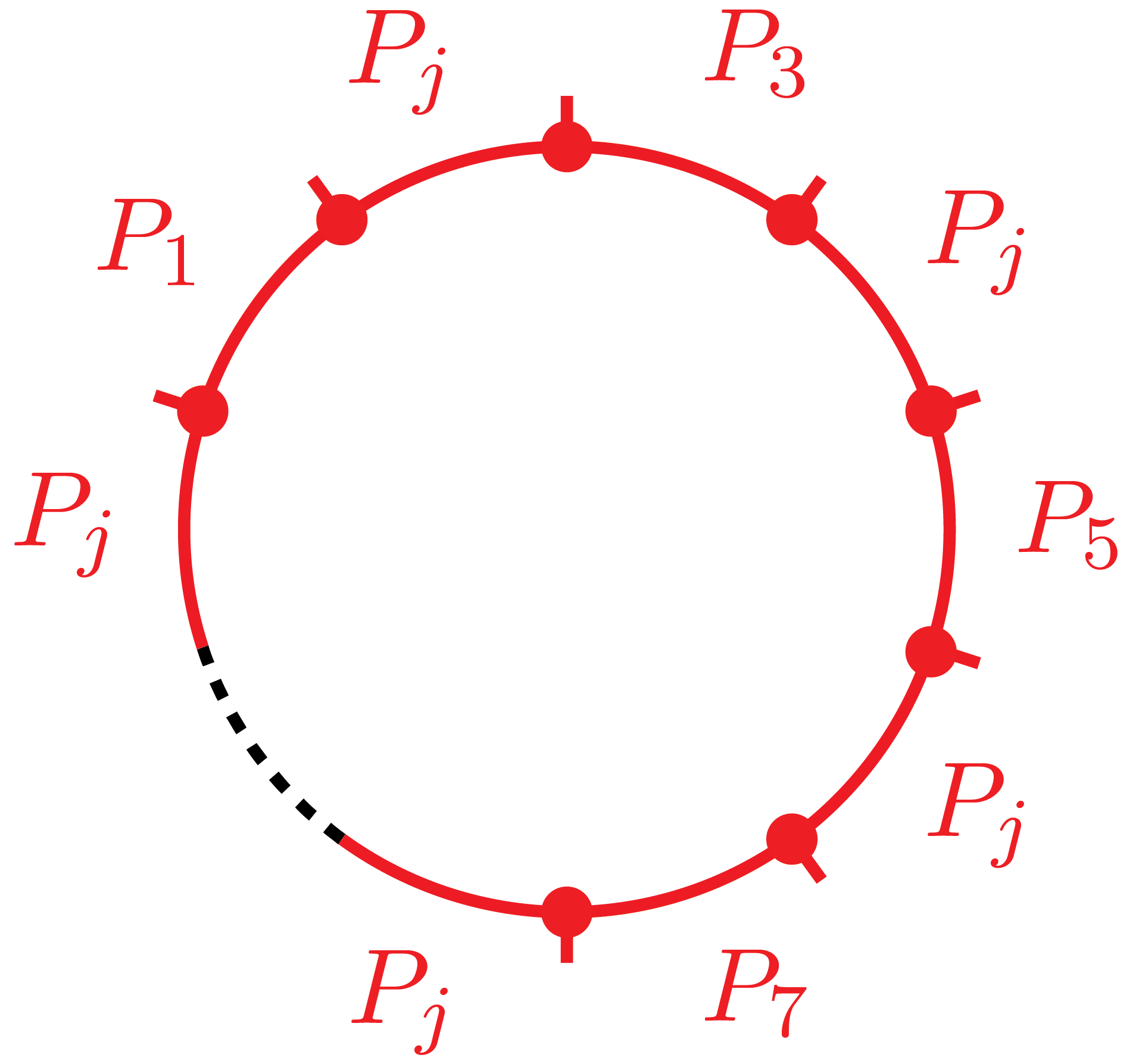}
    \caption{Vacuum dominance in the meridian necklace channel leads to the index structure pictured above in the longitudinal necklace channel. The red dots label insertions of $\cO_i$ and carry a weight $P_i$ which is not pictured.}\label{fig:O2n_6_labeled}
\end{figure}

The change-of-basis kernel to the longitudinal necklace channel is computed as follows
\begin{align}\label{eq:OO^n_decomps}
    \ket{\cF^{\cC_1}(\mathbbm 1)} &= \ket{\cF^{\cC_2}(\mathbbm 1)} \nn\\
    &= \ket{\cF^{\cC_3}(\mathbbm 1)} \nn\\
    &= \int dP_j\ \sker{\mathbbm 1}{P_j}{\mathbbm 1}\ket{\cF^{\cC_4}(P_j)} \nn\\
    &= \int dP_j\ \sker{\mathbbm 1}{P_j}{\mathbbm 1}\ket{\cF^{\cC_5}(P_j)} \nn\\
    &= \int dP_j\ \sker{\mathbbm 1}{P_j}{\mathbbm 1}\prod_{k=1}^{n}\int dP_{2k-1}\ \fker{\mathbbm 1}{P_{2k-1}}{P_j}{P_i}{P_j}{P_i}\ket{\cF^{\cC_6}(\{P_{a}\})} \,.
\end{align}
Here we have made repeated use of \cite{Eberhardt:2023mrq} 
\begin{equation}\label{eq:fker_id}
    \fker{P_{13}}{P_{12}}{P_1}{P_2}{P_3}{\mathbbm 1} = \delta(P_{12}-P_3) \,,
\end{equation}
and chosen to label all but one momenta using the first $n$ odd positive integers. The various conformal block channels are diagrammed in figure \ref{fig:2n_channels}. Rewriting the state above gives
\begin{equation}\label{eq:O2n_vacuum_dominance}
    \ket{\inlinefig[6]{imgs/s6/O2n_1.png}} = \int \rho_0(P_j)dP_j\prod_{k=1}^n\rho_0(P_{2k-1})dP_{2k-1}\ C_0(P_i,P_j,P_{2k-1}) \ket{\inlinefig[6]{imgs/s6/O2n_6.png}}\,,
\end{equation}
where momenta in the right-most state are labeled as in figure \ref{fig:O2n_6_labeled}. We would like to confirm that this formula takes the form of an ETH ansatz in the limits of interest. 

First, let's move to the near-extremal limit by specifying the torus moduli. We choose a decomposition of the torus into $2n$ cylinders with moduli $\{t_a\}$ such that
\begin{equation}
    \tau = \sum_{a=1}^{2n} t_a \quad{\rm and}\quad t_a=z_a-z_{a-1} \,,
\end{equation}
where the second equation holds for all $a>2$. Setting $\beta_a=2\pi b^2\Im(t_a)$ gives
\begin{align}
    &\braket{\tau,z_2,\cdots,z_{2n}}{\inlinefig[6]{imgs/s6/O2n_1.png}}  \nn\\
    &\qquad\qquad\underset{\rm near-extremal}{=} \int\rho(s_j)ds_{j}\prod_{k=1}^n\rho(s_{2k-1})ds_{2k-1}\ \Gamma^\Delta_{j,2k-1}e^{-\beta_{2k-1}s_{2k-1}^2}e^{-\beta_{2k-1}s_j^2} \nn\\
    &\qquad\qquad\underset{\rm near-extremal}{=} \int\rho(s_j)ds_{j}\ e^{-(\beta-\sum_k^n\beta_{2k-1})s_j^2}\prod_{k=1}^n\rho(s_{2k-1})ds_{2k-1}\ \Gamma^\Delta_{j,2k-1}e^{-\beta_{2k-1}s_k^2} \nn\\
    &\qquad\qquad\underset{\rm near-extremal}{=} \inlinefig[6]{imgs/s6/O2n_disk.png} \,.
\end{align}
This specific disk correlator can be computed from the hyperbolic polygon with $2n$ geodesic edges as defined in \cite{Boruch:2023trc}. All edges are attached to sections of the hyperbolic boundary with respective length $\beta_{a}$. The gluing is performed by integrating the geodesic lengths and for even $a$ this identifies the energy of the boundary with the common $s_j$ of the polygon. However, for odd $a=2k-1$, the integration measure is modified by the weight $\Delta$ of the probe $\cO_i$ and so the energies on either side of these geodesics can be different. Hence, generically $s_j\neq s_k$.

Moving beyond the near-extremal limit, \eqref{eq:O2n_vacuum_dominance} extends the structure identified in \eqref{eq:standard_eth_3d} to the correlator of $2n$ operator insertions
\begin{align}\label{eq:O2n_3d_stats}
    \overline{\bra{P_1}\cO_i\ketbra{P_2}\cO_i\ketbra{P_3}\cdots\ketbra{P_{2n}}\cO_i\ket{P_1}} = \left(\prod_{j=2}^n\frac{\delta(P_{2j}-P_2)}{\rho_0(P_{2j})}\right)\times\left(\prod_{k=1}^n C_0(P_i,P_{2k},P_{2k-1})\right)\,.
\end{align}
Note that the product of delta-functions in parenthesis does not contribute for the special case $n=1$ discussed in the previous section. This leaves just the single factor of $C_0$ identified in \eqref{eq:standard_eth_3d}. More generally, set $\bar{P}=(\sum P_a)/2n$ to the average momentum and let the differences $\omega_{a}=\bar{P}-P_a$ be small in the relevant parameter (either $\bar{P}$ or $b^{-1}$ as discussed in appendix \ref{app:vmft_rho0}). Then this equation takes the form of the full ETH ansatz \eqref{eq:full-eth} since the spectral statistics of $\cO_i$ obey
\begin{gather}\label{eq:full_eth_3d}
    \overline{\bra{P_1}\cO_i\ketbra{P_2}\cO_i\ketbra{P_3}\cdots\ketbra{P_{2n}}\cO_i\ket{P_1}} = e^{-(2n-1)S(\bar{P})}F_i^{(2n)}(P_1,\cdots,P_{2n}) \nn\\
    {\rm with}\quad \begin{cases}
        e^{S(\bar{P})} = \rho_0(\bar{P}) \,,\\
        F_i^{(2n)} = e^{nS(\bar{P})}\times\prod_{j=2}^n\delta(P_{2j}-P_2)\times\prod_{k=1}^{n}C_0(P_i,P_2,P_{2k-1}) \,.
    \end{cases}
\end{gather}
Above we have grouped the $(n-1)$ factors of $\rho_0$ appearing in \eqref{eq:O2n_3d_stats} into the overall exponential scaling in front of the function $F_i^{(2n)}$. Each factor of $C_0$ also scales exponentially given either large central charge or appropriately scaled kinematics as discussed in appendix \ref{app:vmft_C0}. A function that scales polynomially with the momenta is achieved by the $e^{nS(\bar{P})}$ in the definition of $F_i^{(2n)}$. This is compensated by setting the exponential factor to $e^{-(2n-1)S(\bar{P})}$ as expected from \eqref{eq:full-eth}. We conclude that this family of single-boundary time-ordered correlators obey the entropy scaling and index structure of the full ETH.

\subsection{Multi-trace ETH from topology change}\label{subsec:topology}
So far we have demonstrated that a full ETH ansatz is sufficient for computing universal properties of time-ordered $2n$-point thermal correlators in AdS$_3$/CFT$_2$. Going beyond these observables will require the multi-trace ETH. In this section we discuss multi-trace ETH correlations resulting from non-trivial topology. We will use knot theory as an organizing principle for understanding these statistics in both CFT$_2$ and AdS$_3$.\footnote{
    TQFT techniques have long been used to compute knot invariants and hyperbolic volumes (e.g. \cite{Kashaev:1996kc, murakami2002kashaevsconjecturechernsimonsinvariants, Gukov:2003na, Dimofte:2009yn, Witten:2010cx, murakami2010introductionvolumeconjecture,  Dijkgraaf:2010ur, EllegaardAndersen:2011vps}). Connections have also been made between knot theory and AdS$_3$/CFT$_2$ duality \cite{Maloney:2016kee, Benjamin:2021wzr}. For example, \cite{Collier:2023fwi} introduced the refined volume conjecture for hyperbolic knots, \cite{Collier:2024mgv} tested it on a well-known example, and recently \cite{Chandra:2026awz} extended the conjecture to spherical knot complements. We follow the approach of \cite{Chandra:2025fef} which associated the amplitudes of knot fragmentations to operator statistics. We are particularly thankful to J. Chandra for many helpful discussions on these new and thought-provoking developments.
} In particular, we find that links describe both operator matrix element statistics in a particular theory \textit{and} contributions to the sum over topologies in the gravitational path integral. While links are fundamentally 3-dimensional objects, their relevance to correlation functions in CFT$_2$ makes no appeal to holography: it is possible and useful to apply the language of 3-manifolds to the study of correlation functions in 2d CFTs. 

\begin{figure}
    \centering
    \begin{subfigure}{0.4\linewidth}
        \centering
        \includegraphics[width=0.8\linewidth]{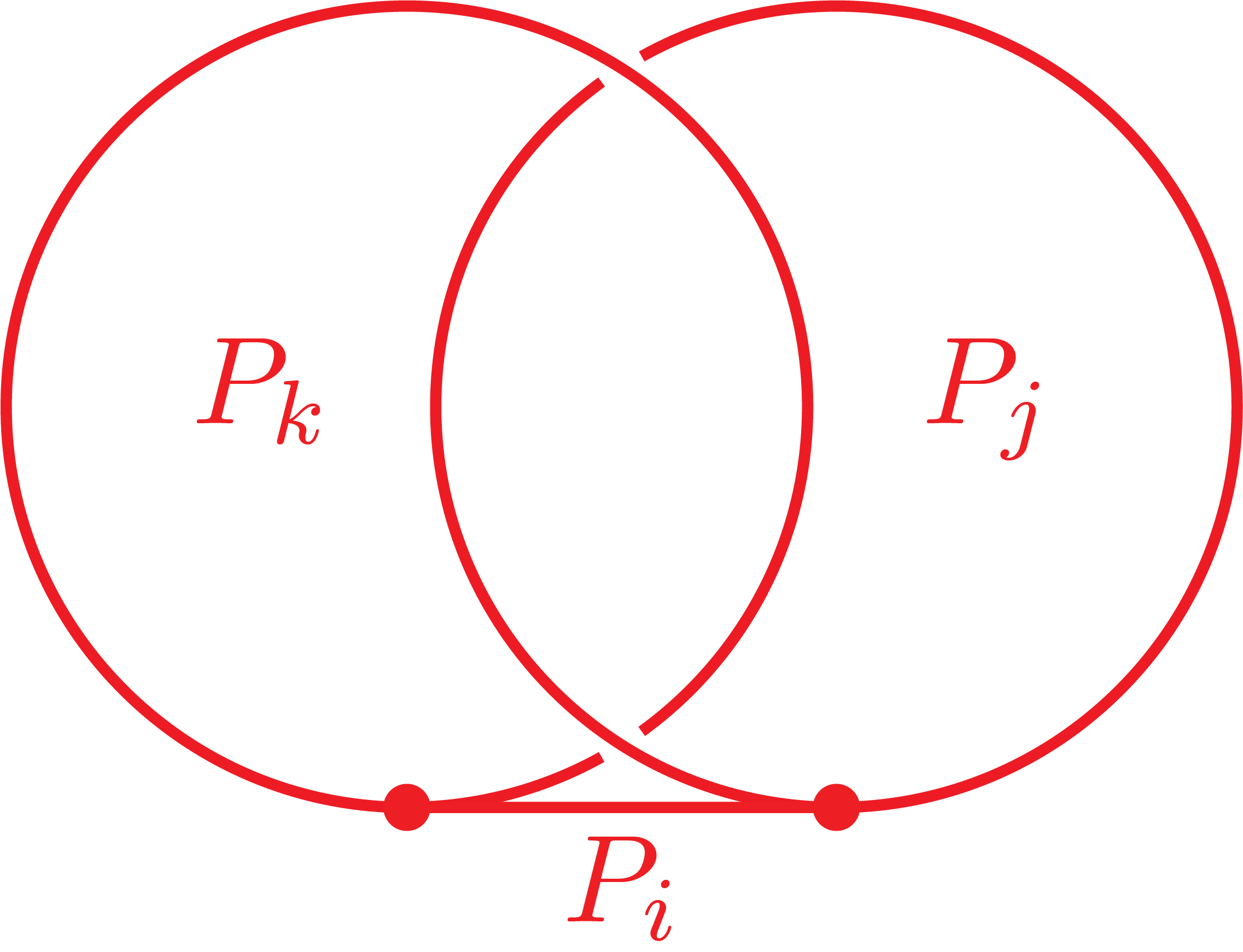}
        \caption{A fragmentation of the Hopf link}
    \end{subfigure}

    \begin{subfigure}{0.3\linewidth}
        \centering
        \includegraphics[width=0.8\linewidth]{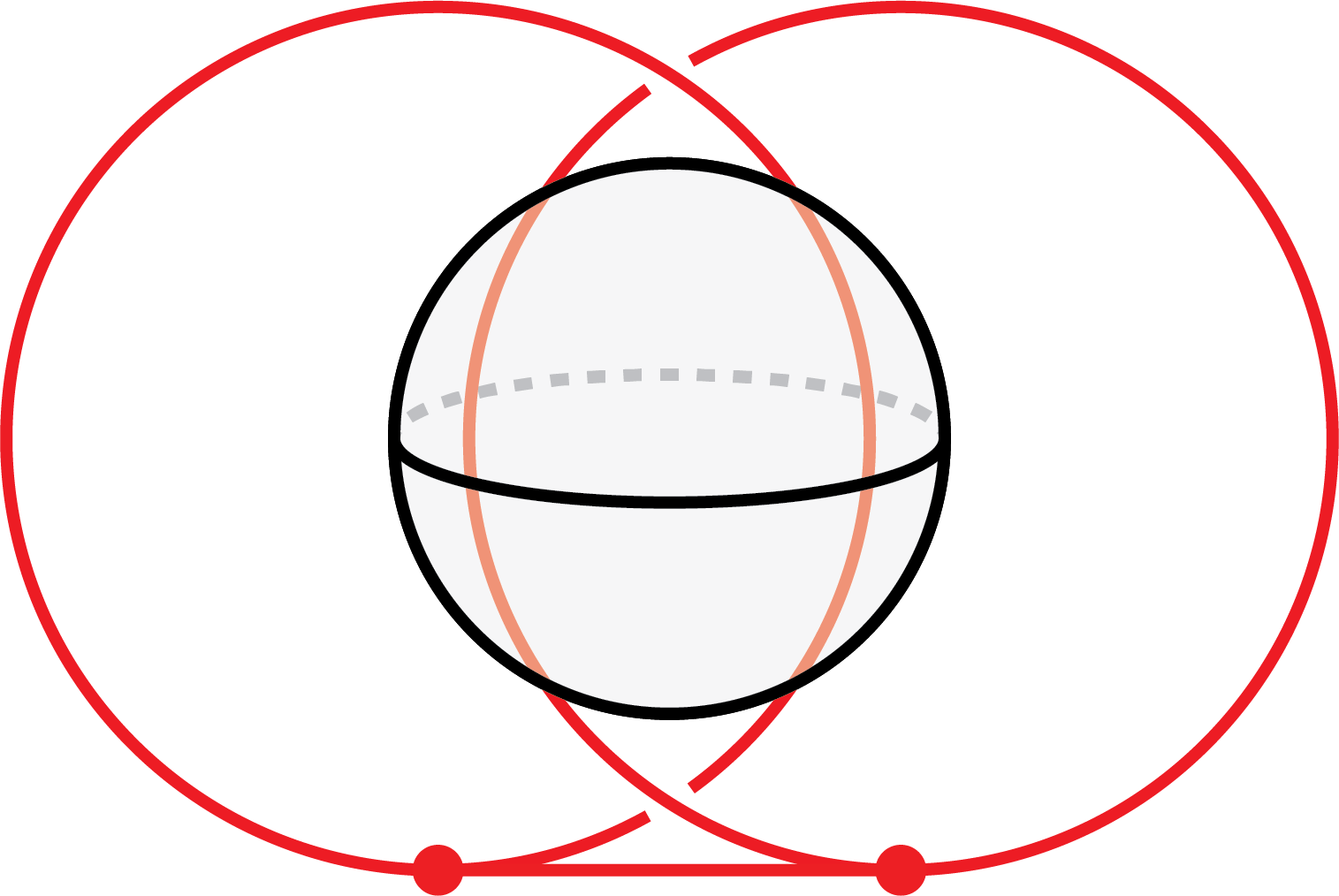}
        \caption{$\Sigma_{0,4}$}
    \end{subfigure}
    \begin{subfigure}{0.3\linewidth}
        \centering
        \includegraphics[width=0.8\linewidth]{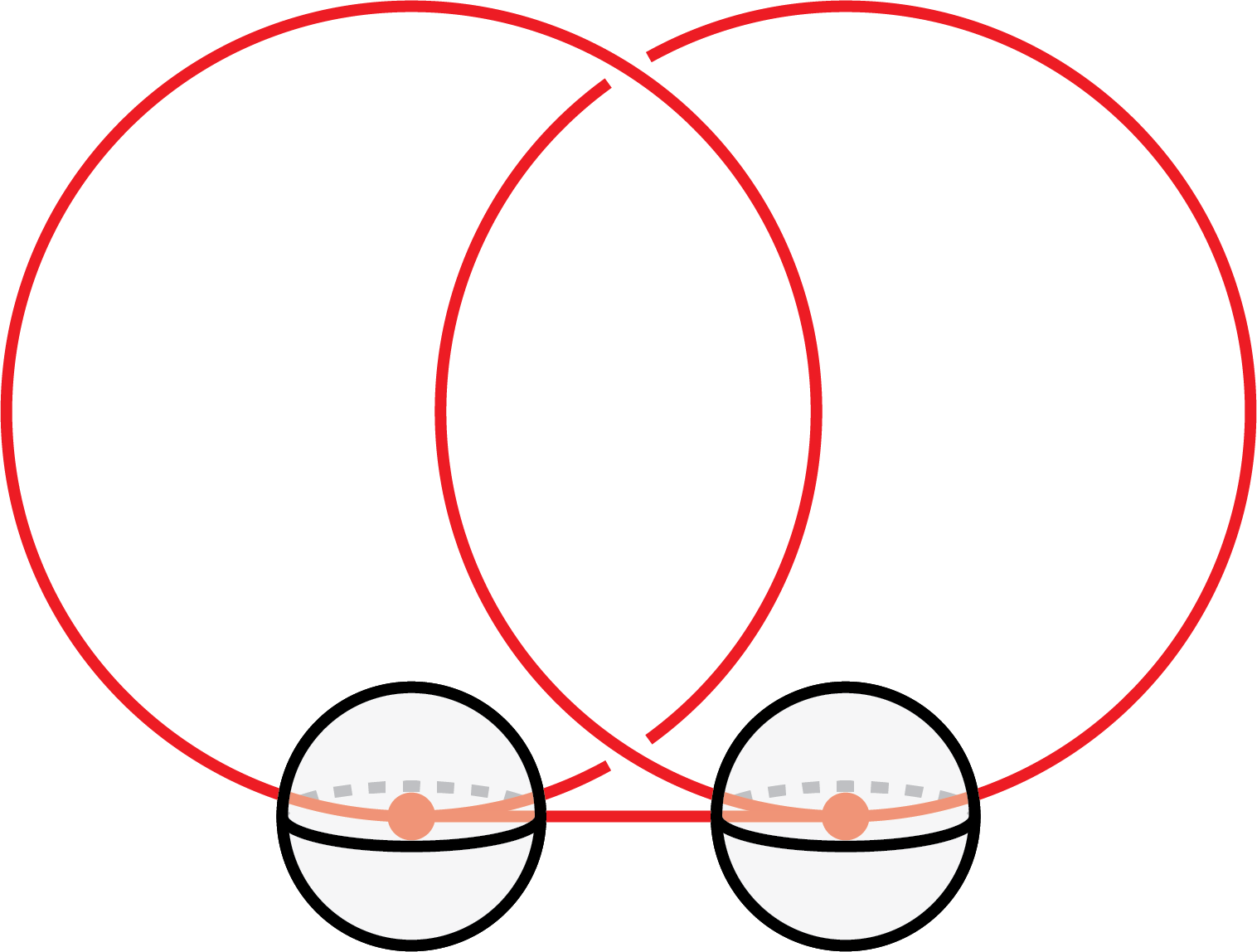}
        \caption{$\Sigma_{0,3}\sqcup\Sigma_{0,3}$}
    \end{subfigure}
    \begin{subfigure}{0.3\linewidth}
        \centering
        \includegraphics[width=0.8\linewidth]{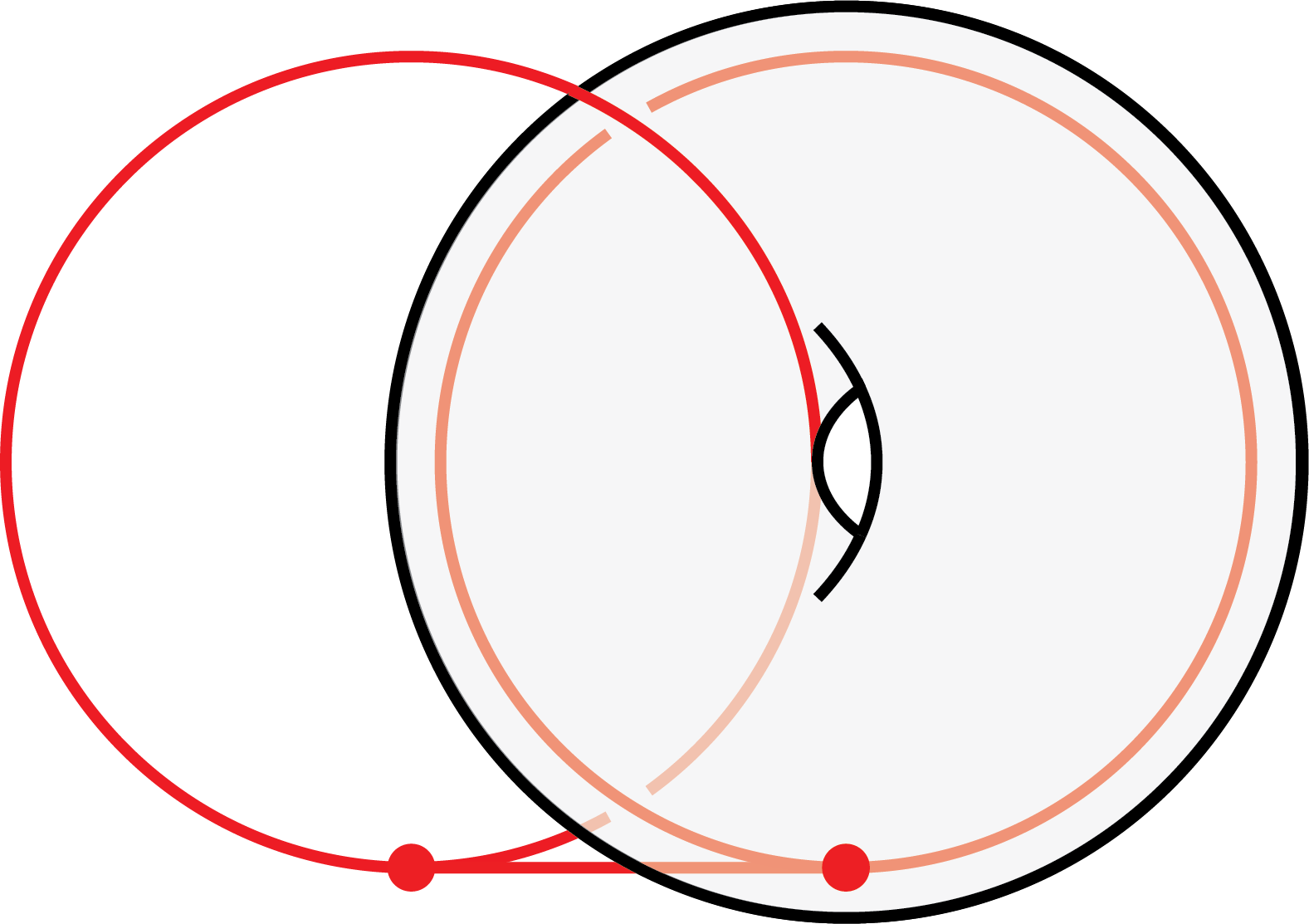}
        \caption{$\Sigma_{1,1}$}
    \end{subfigure}

    \caption{A non-trivial multi-trace ETH ansatz is realized in 3d gravity and holographic CFTs by considering Heegaard splittings of the fragmented Hopf link. The statistics of operators in individual CFTs are derived by splitting the link along $\Sigma_{0,4}$. Statistics of probe operators in a black hole background are derived by splitting along $\Sigma_{1,1}$. In both cases, operator statistics also follow from a wormhole geometry connecting two thrice-punctured spherical boundaries.}
    \label{fig:heegaard_splittings}
\end{figure}

An $n$-link is an embedding of $n$ disjoint circles into the 3-sphere $S^3$. The complement of a link $L$ in its embedding space defines a 3-manifold $M_L$ and a VTQFT amplitude $Z_{\rm Vir}(M_L)$. Each cycle $j$ of the link $L$ may additionally be assigned a weight $P_j$ such that the boundary of $M_L$ is marked. The simplest link is the trivial unlink
\begin{equation}
    [\infty] = \inlinefig{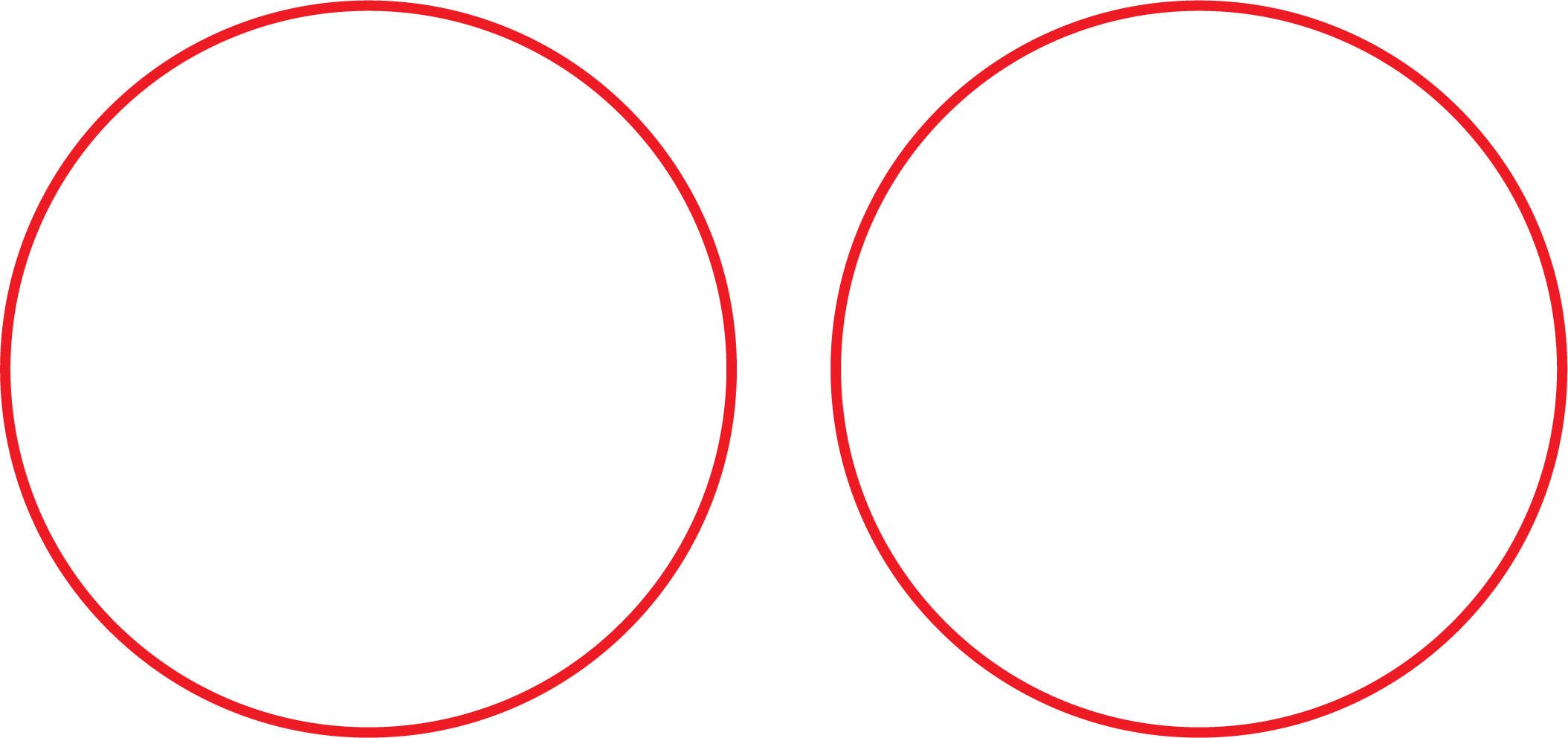}
\end{equation}
to which we assign the label $L=[\infty]$ (c.f. footnote \ref{fn:t_gamma}). The utility of 2-links follows from their various Heegaard splittings, which are divisions of the 3-manifold $M_L$ along a 2-surface. In particular, we assign physical observables to the two classes of Heegaard splittings pictured for the Hopf link in figure \ref{fig:heegaard_splittings}.

\paragraph{Splitting along $\mathbf{\Sigma_{0,4}}$} The first Heegaard splitting divides the 3-manifold $M_L$ along the 4-punctured sphere $\Sigma_{0,4}$. The interior and exterior of the sphere are solid 3-balls and the link $L$ defines a tangle in each of these balls. For simplicity, assume the sphere is situated such that the exterior tangle is trivial and focus on the non-trivial interior tangle. Given a CFT$_2$, we assign an operator to each of the punctures and a correlation function 
\begin{align}\label{eq:g(x)}
    \tilde{G}(x) &= \expval{\cO_j(x_1)\cO_j(x_2)\cO_k(x_3)\cO_k(x_4)} \nn\\
    &= G(x)\left(\prod_{a<b}x_{ab}^{(\sum_c \Delta_c)/3-\Delta_a-\Delta_b}\right)\,
\end{align}
to the splitting surface $\Sigma_{0,4}$. The quantity in parenthesis neatly packages the kinematics so that $G$ can depend exclusively on the crossing ratio
\begin{equation}
    x = \frac{(x_1-x_2)(x_3-x_4)}{(x_1-x_3)(x_2-x_4)} \,,
\end{equation}
and we choose $x_2=0$, $x_3=\infty$ and $x_4=1$ so that $x=x_1$ \cite{Ribault:2014hia}. The fixed punctures are so labeled by the weights of the operator insertions. The crossing ratio is restricted to $x\in\Sigma_{0,3}$ since operators cannot lie on coincident points. The interior tangle corresponds to an identity conformal block in the decomposition
\begin{equation}\label{eq:G}
    G(x) = \sum_{\{i\}} c_{ijj}c_{ikk}\abs{\braket{x}{\cF^{\cC_L}_{0,4}(P_i;P_j,P_k)}}^2 \,.
\end{equation}
The central charge of the theory fixes the blocks, the spectrum $\{i\}$ determines the weights that appear in the sum, and the OPE coefficients $c_{ijk}$ weigh the contribution of each primary. Holomorphic and antiholomorphic sectors are differentiated by conjugation. By definition, the channel $\cC_L$ is given by the interior tangle with the additional fictitious vacuum line $P_i=\mathbbm{1}$.\footnote{
    In fact, $\cC_L$ might refer to any one of a number of channels since we can always draw arbitrary fictitious vacuum lines. We will eventually specify our choice by drawing the full trivalent diagram on the sphere.
}

We say that the vacuum is dominant in the channel $\cC_L$ when there is a function $G_L$ such that
\begin{equation}
    \lim_{x\to x_L}G(x) = \lim_{x\to x_L} G_L(x) \,,
\end{equation}
with
\begin{equation}\label{eq:G_L}
     G_L(x) \equiv c_{\mathbbm{1}jj}c_{\mathbbm{1}kk}\abs{\braket{x}{\cF^{\cC_L}(\mathbbm{1};P_j,P_k)}}^2 \,,
\end{equation}
and $x_L$ in the closure of the thrice-punctured sphere. Appearance of the trivial line $P_i=\mathbbm{1}$ is associated to vacuum dominance in the channel $\cC_L$. Generally, $G_L$ will be a multivalued function of $x$ and only approximates $G$ in a neighborhood of $x_L$. Following \cite{Maloney:2007ud} and \cite{Keller:2014xba}, \cite{Maloney:2016kee} and \cite{Benjamin:2021wzr} consider that the full correlation function $G(x)$ might be recovered from a sum over images of $G_L(x)$. Crossing symmetry characterizes the invariance of the correlation function under permutations of operator insertions. It dictates
\begin{equation}
    G(\gamma\cdot x) = (\gamma\cdot G)(x) \quad{\rm or}\quad G(x)=(\gamma\cdot G)(\gamma^{-1}\cdot x) \,,
\end{equation} 
for any crossing transformation $\gamma$. By $(\gamma\cdot G)$ we mean a different correlation function related to $G$ by operator permutations. \cite{Maloney:2016kee} and \cite{Benjamin:2021wzr} conjecture that 
\begin{equation}\label{eq:poincare}
    G(x) \stackrel{?}{=} \sum_{\{\gamma\}} (\gamma\cdot G_L)(\gamma^{-1}\cdot x) \,,
\end{equation}
where the sum over transformations is taken so as to achieve a crossing invariant function. We will characterize this sum more precisely in a moment. For now, let's accept this conjecture and consider what it implies for the statistics of operators. 

The action $U[\gamma]$ of a transformation $\gamma$ on the space of $\Sigma_{0,4}$ conformal block defines a distributional kernel $\K^\gamma$. This acts as a change-of-basis matrix relating blocks in the transformed channel $\C_{\gamma L}$ to blocks in the original channel $\cC_L$
\begin{align}\label{eq:3d-Ugamma}
    U[\gamma]\ket{\cF^{\cC_L}(\mathbbm{1})} = \ket{\cF^{\cC_{\gamma L}}(\mathbbm 1)} = \int dP\ \K^\gamma_{\mathbbm{1}P} \ket{\cF^{\cC_L}(P)} \,.
\end{align}
Here $\gamma L$ is the link created when the channel $\cC_{\gamma L}$ is placed in the interior of the splitting surface. A convention for the special case $\gamma L=L$ is discussed in subsection \ref{subsec:unlink}.  Notice that, in the final expression of \eqref{eq:3d-Ugamma}, the internal momentum is not trivial ($P\neq\mathbbm{1}$) meaning that there is an additional Wilson line with endpoints lying on $L$. Rewriting \eqref{eq:poincare} as
\begin{equation}\label{eq:G_ansatz}
    G(x) = c_{\mathbbm{1}jj}c_{\mathbbm{1}kk}\left|\int \rho_0(P)dP\ \left(\sum_{\{\gamma\}}\frac{1}{\rho_0(P)}\K^\gamma_{\mathbbm{1}P}\right) \braket{x}{\cF^{\cC_L}(P)}\right|^2\,,
\end{equation}
allows us to interpret the expression in parenthesis as a proposal for OPE coefficient statistics of the operator $\cO$ with associated momentum $P$. Following \cite{Chandra:2025fef}, we say that the Wilson line $P$ fragments the link $L$. 

The support of the OPE density above depends on the choice of contour in \eqref{eq:G_ansatz} and does not clearly match the spectrum $\{i\}$ in \eqref{eq:G}. However, if \eqref{eq:poincare} is correct, then the structure constants associated to a particular operator $\cO_i$ flowing in the channel $\cC_{L'}$ are computed by the projection 
\begin{equation}\label{eq:Oi_ope_ansatz}
    \overline{\bra{P_j}\cO_i\ketbra{P_m}{P_k} \cO_i\ket{P_n}} = \sum_{\{\gamma\}}\abs{\cN_{L'}\mel{\cF^{\cC_{L'}}(P_i)}{U[\gamma]}{\cF^{\cC_L}(\mathbbm{1})}}^2 \,.
\end{equation} 
In particular, we can compute the multi-trace ETH contraction using
\begin{align} \label{eq:F11-3dgen}
    &\overline{\bra{P_j}\cO_i\ketbra{P_j}{P_k} \cO_i\ket{P_k}} = \nn\\
    &\qquad\qquad\abs{\int \rho_0(P)dP\ \left(\sum_{\{\gamma\}}\frac{C_0(P_i,P_j,P_j)C_0(P_i,P_k,P_k)}{\rho_0(P)}\K^\gamma_{\mathbbm{1}P}\right) \braket{\cF^{\cC_\infty}(P_i)}{\cF^{\cC_\infty}(P)}}^2\,,
\end{align}
where we have unit normalized the identity OPE coefficients $c_{\mathbbm{1}jj}=c_{\mathbbm{1}kk}=1$.\footnote{
    When the spectrum is discrete, blocks can always be normalized such that $c_{\mathbbm{1}jj}=1$. This choice is possible because the identity always appears in the fusion of an operator with itself. The structure constant may be distributional when the spectrum is continuous.
} Each term in the sum over $\{\gamma\}$ is proportional to the VTQFT amplitude $\abs{Z_{\rm Vir}(M_{\gamma L})}^2$ associated to the link $\gamma L$ fragmented by the Wilson line $P_i$. The proportionality constant is an artifact of the normalization of $\cC_L$ blocks specified in appendix \ref{app:ne_blocks}. Above we specialized to
\begin{equation}
    |\cN_\infty|^2 = \abs{C_0(P_i,P_j,P_j)C_0(P_i,P_k,P_k)}^2\,.
\end{equation}
OPE coefficients in a different channel $\cC_{L'}$ can be similarly computed by swapping $\cN_{L'}\bra{\cF^{\cC_{L'}}(P_i)}$ for $\cN_{\infty}\bra{\cF^{\cC_\infty}(P_i)}$ above. This justifies our earlier statement that link fragmentations compute operator statistics in individual CFTs. 

\paragraph{Splitting along $\mathbf{\Sigma_{1,1}}$} The second Heegaard splitting is performed along a toric boundary $\Sigma_{1,1}$ such that each cycle of $L$ defines a channel decomposition of $\Sigma_{1,1}$. To picture the splitting, start by drawing two (likely intersecting) channel decompositions on $\Sigma_{1,1}$. Situate the splitting surface so that one cycle of $L$ wraps once around the noncontractible meridian cycle of $\Sigma_{1,1}$. Then, slightly deform this cycle so that it lies in the exterior of the splitting surface. Finally, deform the other cycle into the splitting surface. The result is a link $L$ fragmented by a Wilson line that intersects the splitting surface at a marked point. 

Like the previous splitting, we associate to $L$ a change-of-basis matrix $U[L]$ which acts on the Hilbert space $\cH_{1,1}$ of once-punctured torus conformal blocks.\footnote{
    The fragmentation is not strictly necessary but it helps verify that $U[L]$ is well-defined on the space of normalizable states.  
} Labelling the interior cycle with a weight $P_j$ and the exterior cycle with a weight $P_k$, the matrix elements of $U[L]$ are denoted
\begin{equation}\label{eq:U_elems}
    U[L]_{jk} = \braket{\cF^{\cC_{\infty}}_{1,1}(P_i;P_j)}{\cF^{\cC_{L}}_{1,1}(P_i;P_k)}  \,,
\end{equation}
with $P_i$ labeling the weight of the puncture. Note that the bra and ket are presented in different channels. The channel $\cC_L$ is the trivalent diagram formed by projecting the interior cycle onto the splitting surface. Likewise, $\cC_\infty$ is the trivial channel where the exterior cycle propagates. This again is the VTQFT amplitude $Z_{\rm Vir}(M_L)$. As before, the matrix elements of $U[L]$ may be written in a single channel at the cost of some kernel. 

The upshot is that the link $L$ also describes a modular transformation of the boundary manifold $\Sigma_{1,1}$.\footnote{
    It is conceivable that different modular transformations might lead to the same link once the splitting surface is removed.
} In a theory of gravity, some subset of these transformations contribute to the path integral in \eqref{eq:z_grav}. Here we depart from Virasoro TQFT. The essential modification is a new inner product on the space of conformal blocks\footnote{
    The initiated reader may be tempted to think of the link $\gamma$ as a wormhole connecting disconnected asymptotic boundaries of spacetime decorated by trivalent diagrams. While technically correct, one should be careful to note that the topology of $M_\gamma$ can differ widely from conventional wormhole solutions. For example, $M_\infty$ is the connected sum of genus 1 handlebodies which is a different topology from those of \cite{Maldacena:2004rf}.
}
\begin{equation}\label{eq:ip_poincare}
    \overline{\braket{\cF(P_j)}{\cF(P_k)}} \equiv \sum_{\{\gamma\}}\mel{\cF(P_j)}{U[\gamma]}{\cF(P_k)}\,, 
\end{equation}
where the sum is taken over the set of links $\{\gamma\}$ which appear in the sum over topologies.\footnote{
    It is perhaps premature to refer to \eqref{eq:ip_poincare} as an inner product. In particular, as far as we are aware, positive-definiteness has not been confirmed.
} Like \eqref{eq:Oi_ope_ansatz}, each term in this sum computes the VTQFT amplitude of a link. Gravitational amplitudes are computed using the modified inner product
\begin{equation}
    Z_{\rm grav}(\tau|\Sigma_{1,1}) = \overline{\left|\braket{\tau}{Z_{\rm Vir}(\Sigma_{1,1})}\right|^2} \equiv \sum_{\{\gamma\}} \left|\mel{\tau}{U[\gamma]}{Z_{\rm Vir}(\Sigma_{1,1})}\right|^2 \,,
\end{equation}
where contributions from the holomorphic and antiholomorphic sectors are here differentiated by conjugation. 

The set $\{\gamma\}$ must, at a minimum, contain the relative mapping class group\footnote{\label{fn:topologies}
    It is not clear whether all modular transformations of the boundary should contribute to the sum over topologies in the gravitational path integral. Consider for example the single-boundary non-handlebody geometries of \cite{Belin:2026pko}. These manifolds are on-shell for genus $g>2$ but not strictly required for modular invariance or by the sum over ${\rm MCG}(M_g)$. At a minimum, orientation-preserving diffeomorphisms are gauged in general relativity. This forces us to identify individual saddles with their orbits under boundary-preserving diffeomorphisms which is accomplished here through a modular sum. 
} 
\begin{equation}
    {\rm MCG}(M) = \frac{{\rm Diff}(M,\partial M)}{{\rm Diff}_0(M,\partial M)} \,,
\end{equation}
which describes the set of boundary preserving diffeomorphisms that are not continuously connected to the identity. Supposing the presence of some $L\in\{\gamma\}$ will allow us to deduce statistical properties of black hole microstates with weights $P_j$ and $P_k$. Crucially, while all links are associated to a unitary $U[L]$, not all $L$ need lie in $\{\gamma\}$. Section \ref{subsec:hopf} describes non-Gaussianities coming from $\gamma\in{\rm MCG}(M_{1,1})$ and we refer to discussion of the Whitehead link in \cite{Chandra:2025fef} for an example of a more exotic contribution.

\paragraph{Summing Over Images} We have identified two seemingly distinct criteria for the importance of a particular link $L$ to AdS$_3$/CFT$_2$ duality: appearance in the sum over modular images needed to complete the ``seed'' function $G_L$ and appearance in the sum over topologies required by gravity. These requirements are related and we now develop the remaining technology needed to express their relation precisely.  

The mapping class group of the punctured torus ${\rm MCG}(\Sigma_{1,1})$ describes redundancies in the choice of torus modulus $\tau\in\H$. It is the matrix group
\begin{equation}\label{eq:psl2z}
    PSL(2,\Z) = \left\{\gamma=\begin{bmatrix}
        a & b \\
        c & d
    \end{bmatrix}\ \mid\ a,b,c,d\in\Z,\ |\gamma|=1 \right\}/(I\sim -I) \,.
\end{equation}
Equivalently, the set of mapping class group actions is the dictionary of words that can be formed from two letters:
\begin{equation}
    \S = U\begin{bmatrix}
        0 & -1 \\
        1 & 0
    \end{bmatrix} \quad{\rm and}\quad \T= U\begin{bmatrix}
        1 & 1 \\
        0 & 1
    \end{bmatrix}\,.
\end{equation}
The torus modulus transforms as
\begin{equation}
    \S\ket{\tau}=\ket{-1/\tau} \quad{\rm and}\quad \T\ket{\tau}=\ket{\tau+1} \,.
\end{equation}
The Hilbert space of moduli is also spanned by the set of torus blocks and the mapping class group acts nicely on these states
\begin{align}
    \S\ket{\inlinefig{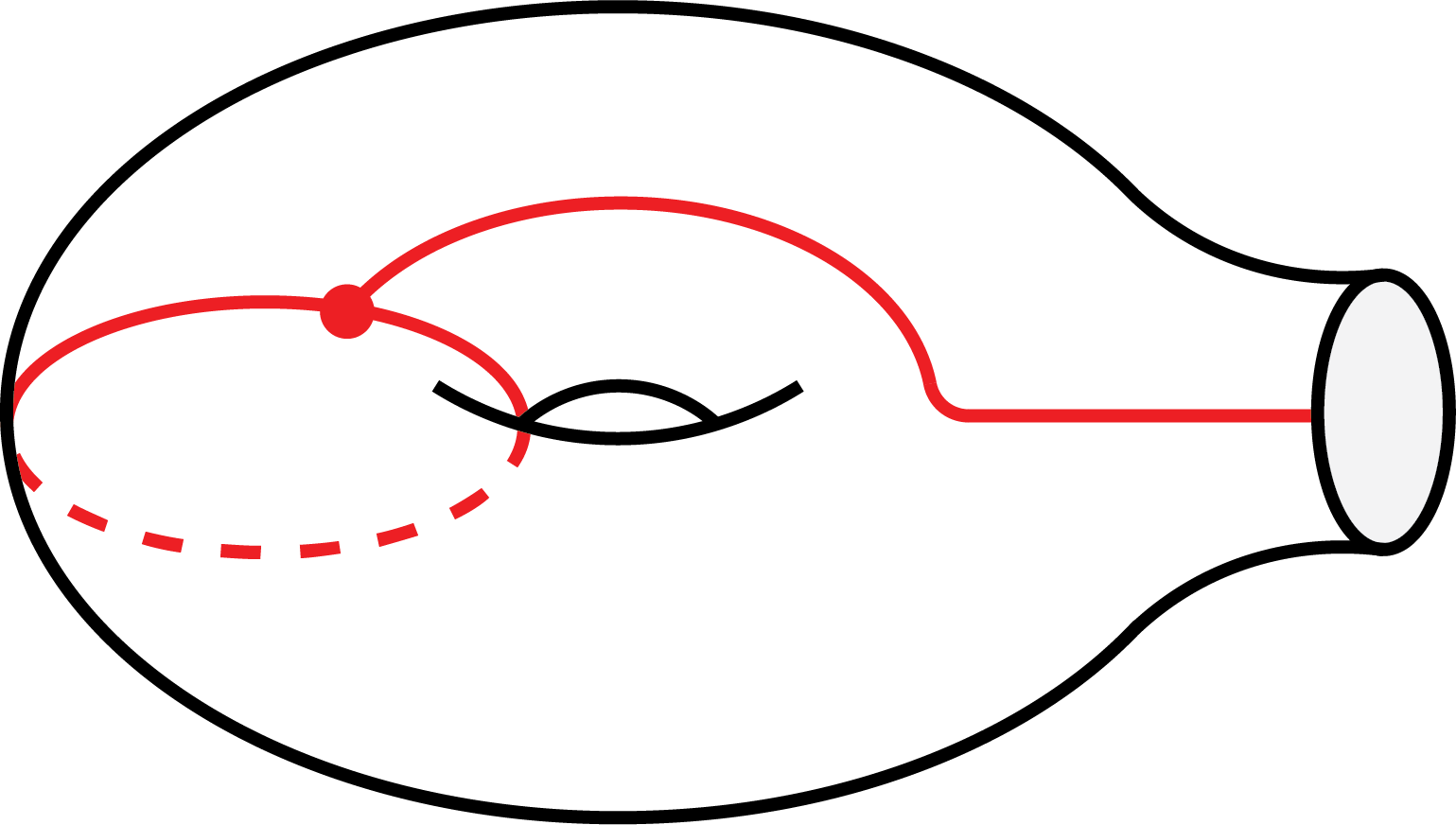}} &= \ket{\inlinefig{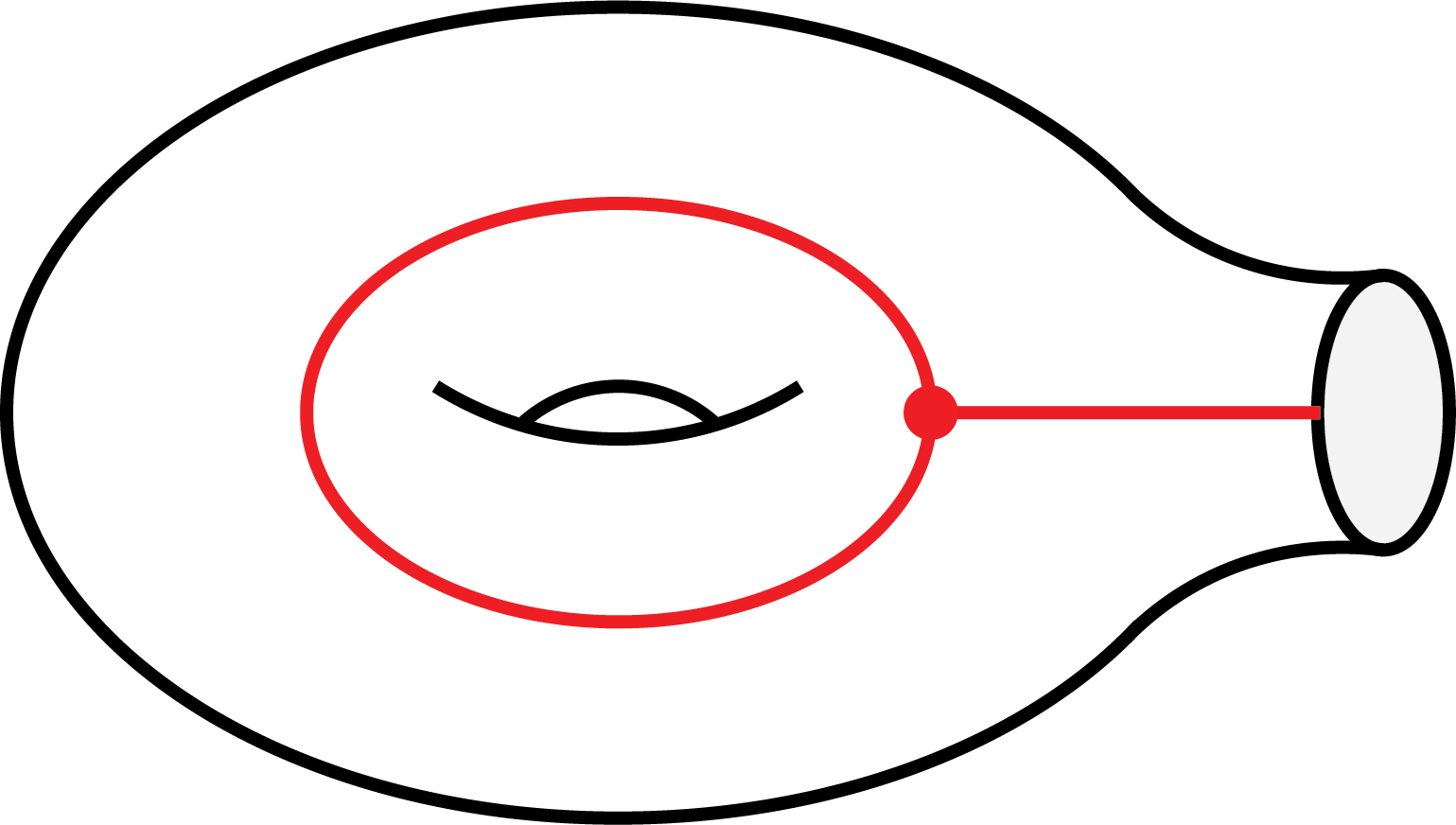}} = \int dP_k\ \sker{P_j}{P_k}{P_i}\ket{\inlinefig{imgs/s6/SO.png}}\,,\nn\\
    \T\ket{\inlinefig{imgs/s6/O.png}} &= \ket{\inlinefig{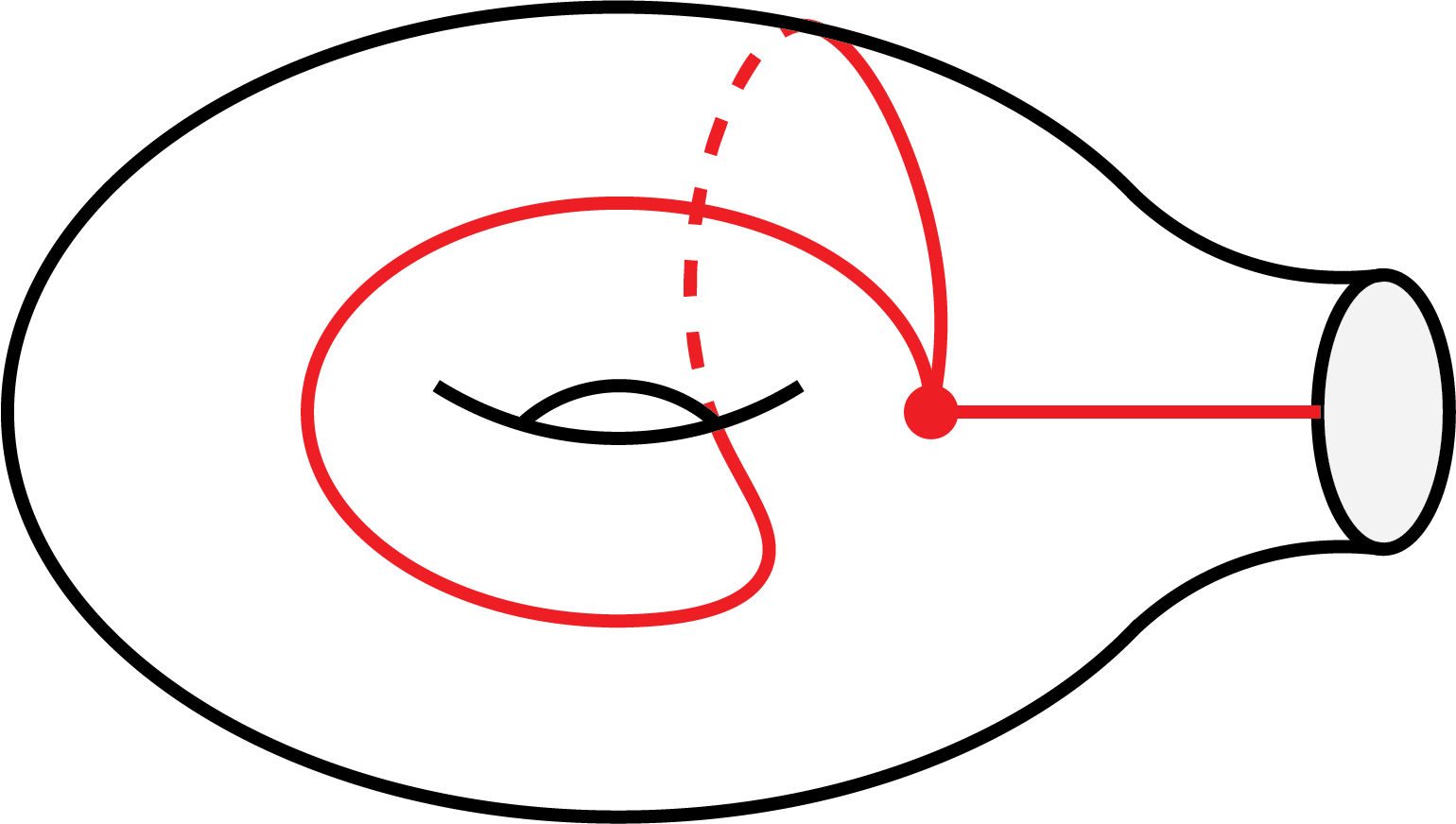}} = \bker{P_i}{P_j}{P_j}\ket{\inlinefig{imgs/s6/O.png}}\,.
\end{align}
We will repeatedly abuse the notation by referring to the action of links on various classes of states using the same map $U$.\footnote{
    The upside of this notation is that it makes modular invariance clear:
    \begin{equation}
        Z(\tau) = \int\rho_0(P)dP\ \braket{\tau}{\cF_{1,0}(P)} = \int\rho_0(P)dP\ \mel{\tau}{\S^\dag\S}{\cF_{1,0}(P)} = Z(-1/\tau) \,.
    \end{equation}
} 

Recall that the crossing ratio $x$ lives on the thrice-punctured sphere $\Sigma_{0,3}$. By the uniformization theorem, this hyperbolic manifold can be realized as a quotient of the upper-half-plane. In particular, 
\begin{equation}
    \Sigma_{0,3} = \H / \Gamma(2)\,,
\end{equation}
where $\Gamma(2)$ is the index-6 subgroup of $PSL(2,\Z)$ corresponding to odd $a,d$ and even $b,c$ in \eqref{eq:psl2z}.\footnote{
    The subgroup $H\subset PSL(2,\Z)$ giving the crossing ratio moduli space must have been of index 6 since we need $PSL(2,\Z)/H=S_3$, the 6-dimensional symmetry group of 3 indistinguishable fixed points corresponding to $x_2$, $x_3$, and $x_4$. 
} Equivalently, it is the free group formed by the words
\begin{equation}
    \Gamma(2) = \langle\T^2,\ \S\T^2\S\rangle \,.
\end{equation}
The crossing ratio is related to the torus modulus by the function
\begin{equation*}
    x(\tau) = \left(\frac{\theta_2(\tau)}{\theta_3(\tau)}\right)^4 \,,
\end{equation*}
which is invariant under the action of any $\gamma\in\Gamma(2)$ on $\tau$ but otherwise transforms non-trivially
\begin{equation}
    \S\ket{x}=\ket{1-x} \quad{\rm and}\quad \T\ket{x}=\ket{\frac{x}{x-1}} \,.
\end{equation}
The action of these letters on sphere 4-point blocks is
\begin{align}
    \S\ket{\inlinefig[4.5]{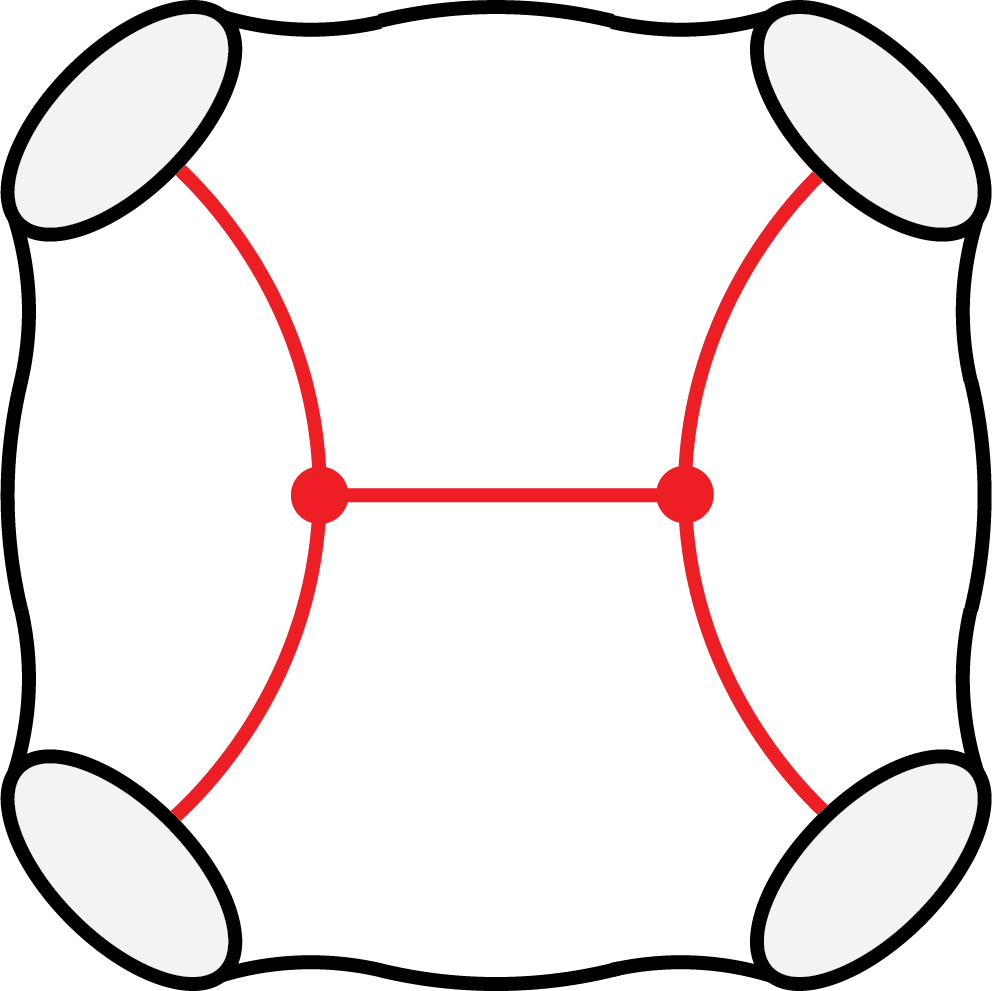}} &= \ket{\inlinefig[4.5]{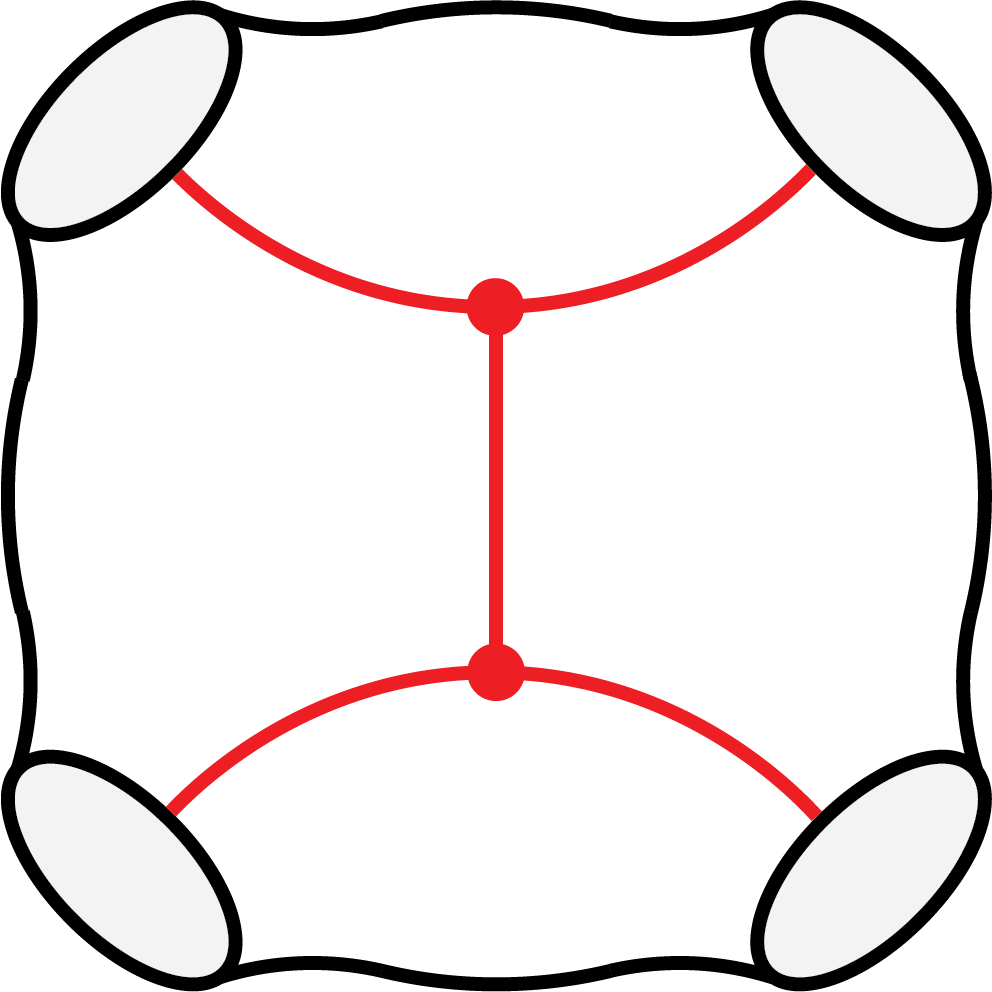}} = \int dP_i'\ \fker{P_i}{P_i'}{P_j}{P_j}{P_k}{P_k}\ket{\inlinefig[4.5]{imgs/s6/G.png}}\,,\nn\\
    \T\ket{\inlinefig[4.5]{imgs/s6/G.png}} &= \ket{\inlinefig[4.5]{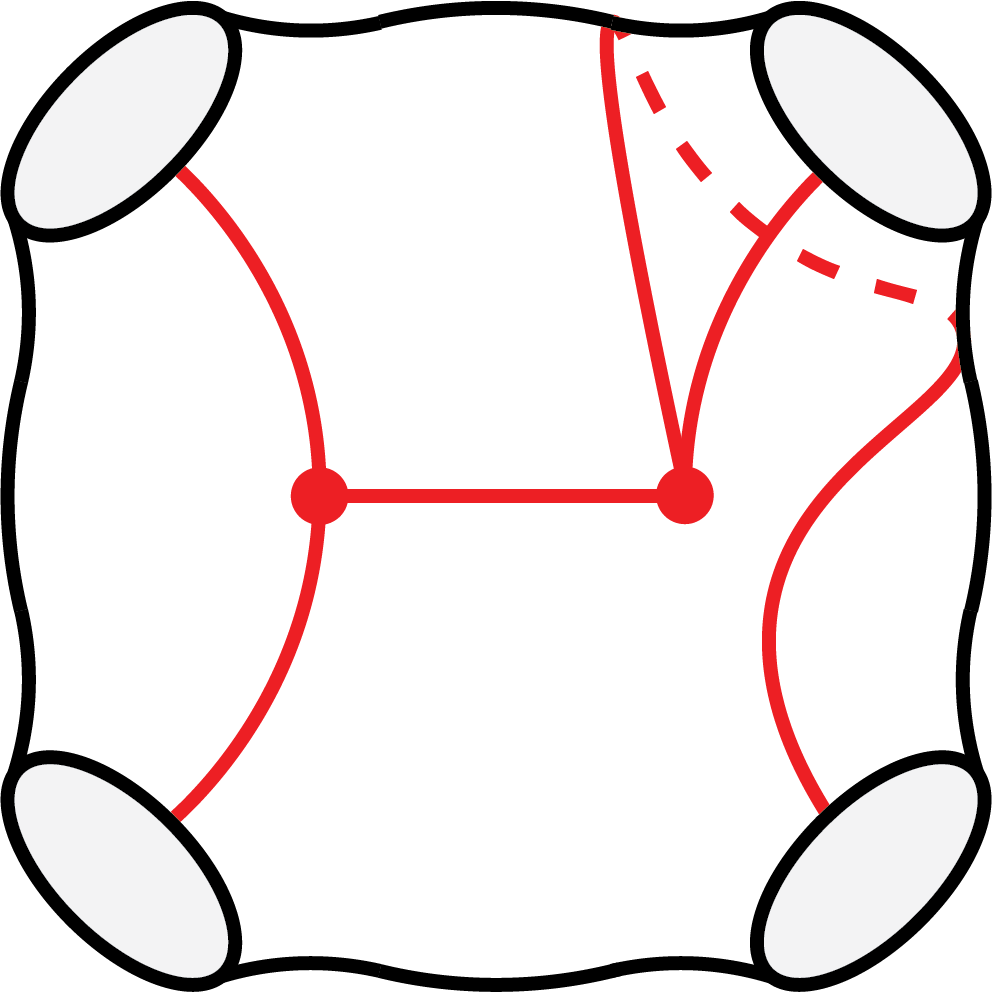}} = \bker{P_i}{P_k}{P_k}\ket{\inlinefig[4.5]{imgs/s6/G.png}} \,.
\end{align}
Simply put, $\S$ acts by rotating the block and $\T$ twists boundary insertions about each other. The former operation permutes distinguishable operator insertions meaning that $\S\cdot G\neq G$. 

A single-valued ansatz for $G(x)$ is achieved in \eqref{eq:G_ansatz} by summing non-trivial images of the seed function $G_L$. For convenience, assume that all external operators are scalars. When these insertions are distinguishable, the sum can be performed over (some subset of) $\Gamma(2)$. In our case, operator insertions are only pair-wise distinguishable which forces the correlation function to be single-valued on a restricted domain. Specifically, $G(x)$ must be well-defined for $x\in\H/\Gamma_0(2)$ where
\begin{equation}
    \Gamma_0(2) = \langle \T,\ \S\T^2\S \ |\ (\T\cdot\S\T^2\S)^2=1 \rangle \,.
\end{equation}
This is the index-3 subgroup of $PSL(2,\Z)$ corresponding to odd $a,d$ and even $c$.\footnote{
    Permutations of three distinguishable points on the 2-sphere are the 6 elements of $PSL(2,\Z)/\Gamma(2)\cong S_3$. When 2 elements are indistinguishable, we need only concern ourselves with $|S_3/S_2|=3$ of these elements. It follows that the subgroup $H\subset PSL(2,\Z)$ defining the domain of $G$ satisfies $PSL(2,\Z)/H\cong(S_3/S_2)$ as sets. This justifies choosing $H=\Gamma_1(2)=\Gamma_0(2)$ which is an index 3 subgroup but not normal. 
} Now recall that the vacuum is dominant in the channel $\cC_\infty$ as the crossing ratio approaches $x_\gamma=x(\tau_\gamma=\infty)=0$. The inner product is invariant under $\T$ since
\begin{equation}
    \lim_{x\to 0}(\T\cdot G_\infty)(\T^{-1}\cdot x) = \abs{\bker{\mathbbm 1}{P_k}{P_k}\braket{0}{\inlinefig[4.5]{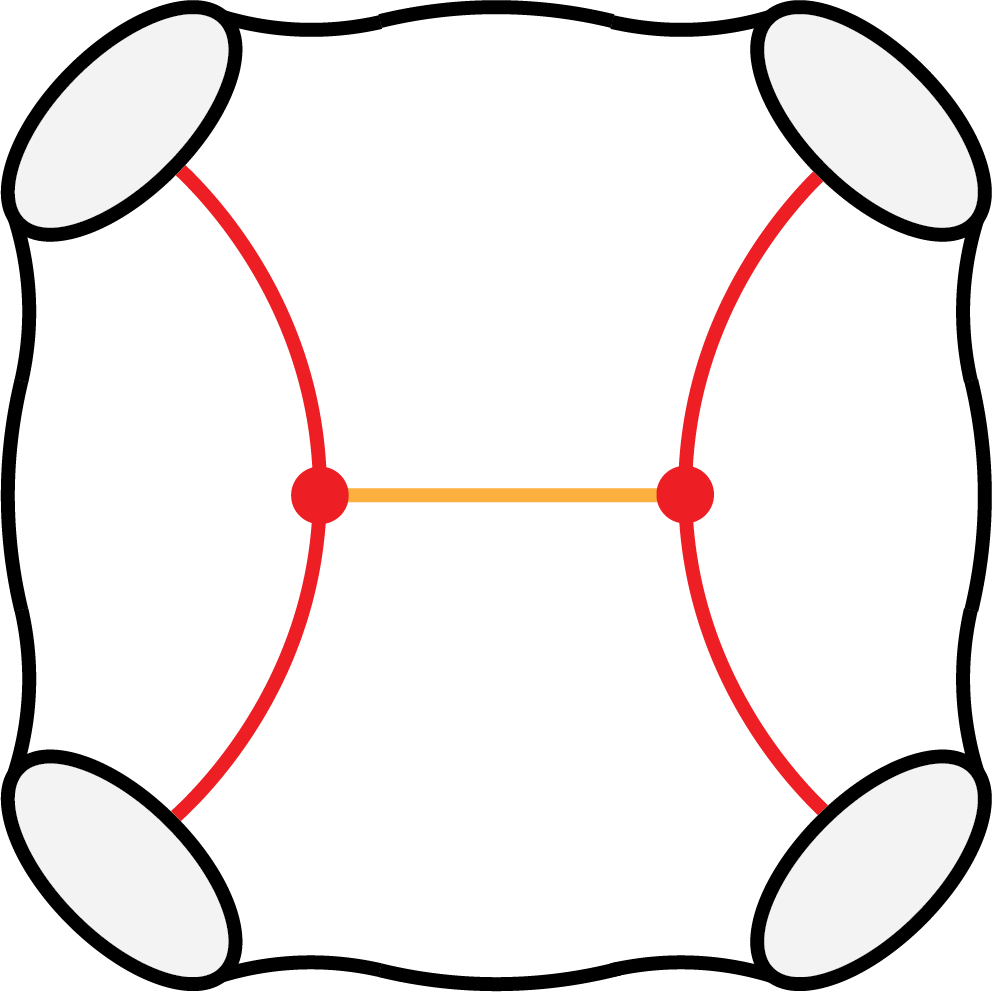}}}^2 = G_\infty(0)\,,
\end{equation}
where the first equality used $\T^{-1}\ket{0}=\ket{0}$ and the second used the assumption that external operators are scalar. To each $\gamma\in\Gamma_0(2)/\langle\T\rangle$ we associate the rational number $\tau_\gamma=a/c$ where $a$ and $c$ are coprime.\footnote{\label{fn:t_gamma}
    For us, the identification with the rational numbers is mostly a novelty. It is useful insofar as it allows us to easily specify links. For example, this justifies our choice of the label $[\tau_\gamma=\infty]$ for the trivial link which has $c=0$.
} The remaining matrix elements lie in the family of solutions to $|\gamma|=ad-bc=1$ and so our sum runs over the cosets $\gamma=[a/c]$ with odd $a$ and even $c$.

Note however that $G_\infty$ is not invariant under $\S\T^2\S$. This leads to the non-trivial realization of the multi-trace ETH discussed in section \ref{subsec:hopf}.

\subsubsection{Unlink}\label{subsec:unlink}
The unlink characterizes both the degeneration limit of CFT correlation functions and, after fragmentation, the conventional VTQFT inner product on the space of conformal blocks 
\begin{equation}
    \mel{\cF_{1,1}(P_j)}{U\left[\infty\right]}{\cF_{1,1}(P_k)} = \braket{\cF_{1,1}(P_j)}{\cF_{1,1}(P_k)}\ = \frac{\delta(P_j-P_k)}{\rho_0(P_j)C_0(P_i,P_j,P_k)} \,.
\end{equation}
This inner product was used in section \ref{subsec:full_eth_3d} to match correlation functions in a BTZ background to a full ETH ansatz. On the CFT$_2$ side, the unlink describes the identity contribution 
\begin{equation}
    G_\infty(x) = c_{\mathbbm{1}jj}c_{\mathbbm{1}kk}\abs{\braket{x}{\cF^{\cC_\infty}(\mathbbm{1};P_j,P_k)}}^2 \,,
\end{equation}
which is dominant in the degeneration limit. We avoid a non-sensical divergence by defining $\mathbb{K}^{\infty}_{\mathbbm{1}P}=\epsilon\delta(P-\mathbbm{1})$ with $\epsilon$ a small constant such that
\begin{equation}
    \overline{\bra{P_j}{\cO_{\mathbbm 1}}\ketbra{P_j}{P_k}\cO_{\mathbbm 1}\ket{P_k}} = 1 \,.
\end{equation}
Alternatively, we recognize that the coefficient in front of the $\cC_{\infty\cdot L}$ term in \eqref{eq:poincare} is just $c_{\mathbbm{1}jj}c_{\mathbbm{1}kk}$ and that identity OPE coefficients are unit normalizable.

A more interesting relation is revealed in the $\S\cdot[\infty]=[0]$ channel. In a neighborhood of $x(\tau=0)=1$, the holomorphic component of $G$ is well-approximated by
\begin{align}
    \S\ket{\inlinefig[4.5]{imgs/s6/G_inf.png}} = \ket{\inlinefig[4.5]{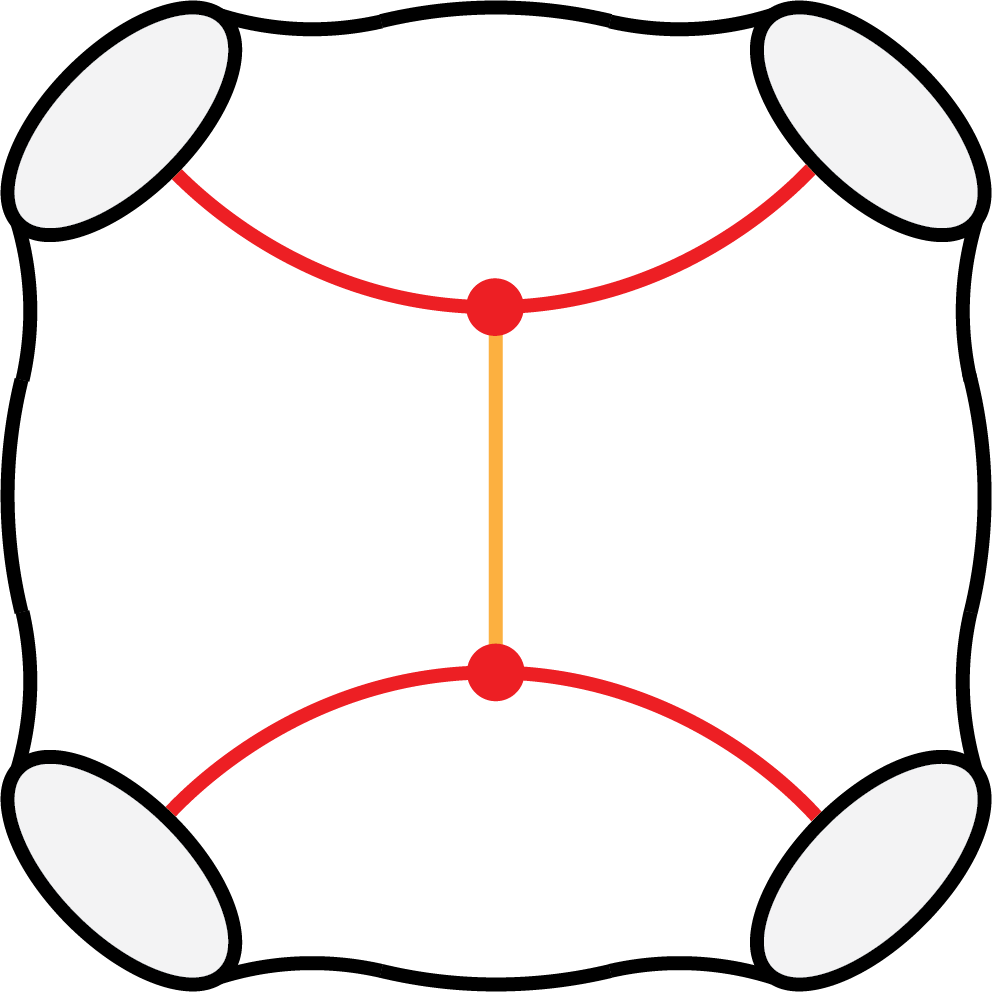}} = \int \rho_0(P)dP\ C_0(P,P_j,P_k)\ket{\inlinefig[4.5]{imgs/s6/G.png}} \,,
\end{align}
which corresponds to the kernel
\begin{equation}
    \K^{0}_{\mathbbm{1}P}=\rho_0(P)C_0(P,P_j,P_k) \,,
\end{equation}
and the total OPE density
\begin{equation}
    \overline{\bra{P_j}{\cO}\ketbra{P_k}\cO\ket{P_j}} = C_0(P,P_j,P_k)C_0(\bar{P},\bar{P}_j,\bar{P}_k) \,.
\end{equation}
Alternatively, we computed \eqref{eq:Oi_ope_ansatz} using the state 
\begin{equation}
    \cN_0\bra{\cF^{\cC_0}} = \cN_0\bra{\inlinefig[4.5]{imgs/s6/G.png}}\,.
\end{equation}
This implies an extended range of validity for the standard ETH ansatz \eqref{eq:standard_eth_3d} beyond the limits discussed in appendix \ref{app:vmft}. 

\subsubsection{Hopf Link}\label{subsec:hopf}
Here we study the simplest topological correction to the Gaussian OPE statistics of the unlink. The action of $\T$ on $\ket{\cF^{\cC_\infty}_{0,4}}$ is trivial when $P=\bar{P}$ and $P_j=P_k$ so the first novel topological correction to operator statistics is given by the Hopf link
\begin{equation}
    \left(\S\T^2\S\right)\cdot[\infty] = [-1/2] = \inlinefig{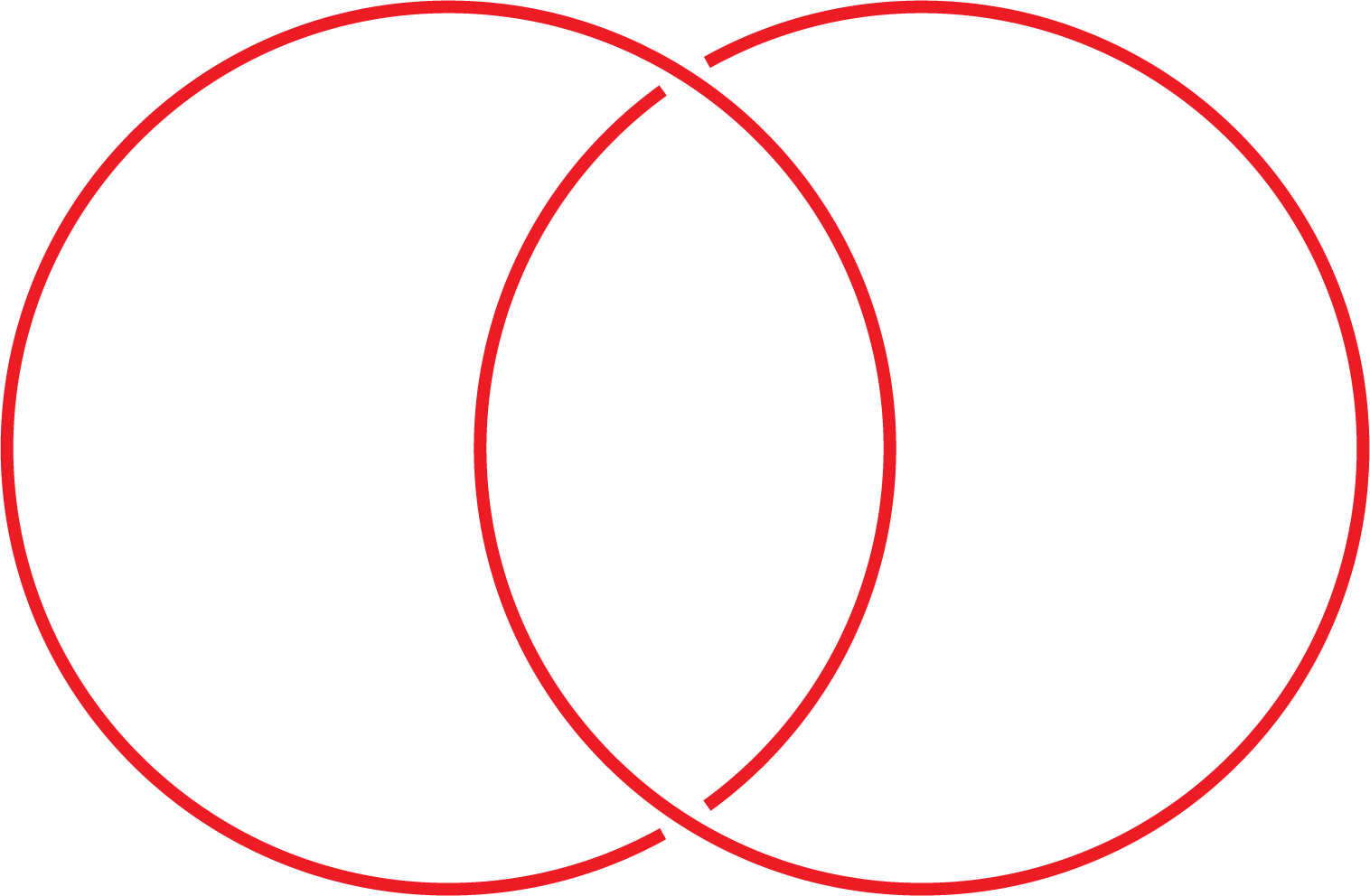} \,,
\end{equation} 
first studied in VTQFT in \cite{Collier:2024mgv}. On the gravity side, the Hopf link gives a correction of the form in \eqref{eq:U_elems} to the inner product \eqref{eq:ip_poincare} on the space of conformal blocks
\begin{align}\label{eq:torus_hopf}
    \mel{\cF^{\cC_\infty}_{1,1}(P_i;P_k)}{U[-1/2]}{\cF^{\cC_\infty}_{1,1}(P_i;P_j)} &= \braket{\cF^{\cC_\infty}_{1,1}(P_i;P_k)}{\cF^{\cC_{-1/2}}_{1,1}(P_i;P_j)} \nn\\
    &= \bra{\cF_{1,1}^{\cC_\infty}(P_i;P_k)}\int dP_j'\ \sker{P_j}{P_j'}{P_i} \ket{\cF_{1,1}^{\cC_\infty}(P_i;P_j')} \nn\\
    &= \int dP_j'\ \sker{P_j}{P_j'}{P_i}\frac{\delta(P_k-P_j')}{\rho_0(P_k)C_0(P_i,P_k,P_k)} \nn\\
    &= \frac{\sker{P_j}{P_k}{P_i}}{\rho_0(P_k)C_0(P_i,P_k,P_k)} \,.
\end{align}
Note the need to fragment the link by the Wilson line $P_i$ to ensure that the inner product is taken over normalizable blocks. 

On the CFT$_2$, the Hopf link is associated to the action of a non-trivial kernel on $[0]$-channel blocks
\begin{align}
    \S\T^2\S\ket{\inlinefig[4.5]{imgs/s6/G_inf.png}} = \ket{\inlinefig[4.5]{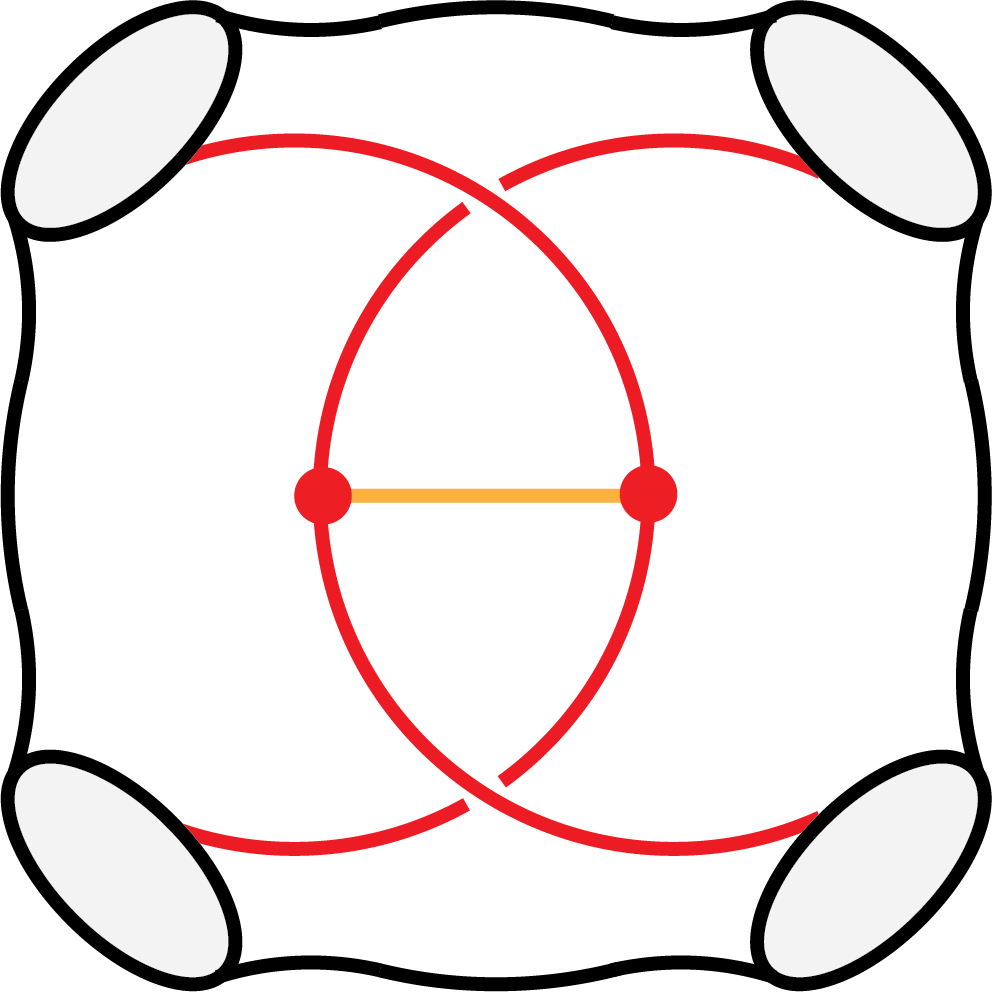}} = \int dP\ (\bker{P}{j}{k})^2\fker{\mathbbm 1}{P}{j}{k}{j}{k}\ket{\inlinefig[4.5]{imgs/s6/G_inf.png}}\,.
\end{align}
This gives a correction to OPE statistics in the [0]-channel
\begin{equation}
    \overline{\bra{P_j}\cO\ketbra{P_j}{P_k}\cO\ket{P_k}} \supset \abs{e^{2\pi i(P^2-P_j^2-P_k^2-Q^2/4)}C_0(P,P_j,P_k)}^2 \,,
\end{equation}
where here ``$\supset$'' highlights that this is just one summand in a series.

In fact, vacuum propagation in the $\cC_{\infty}$ channel of $\Sigma_{0,4}$ is a special case of a more general class of so-called vertical rational tangles generated by the Hopf link
\begin{equation}
    (\S\T^{2n}\S)\cdot[\infty] = [-1/2n] = \inlinefig{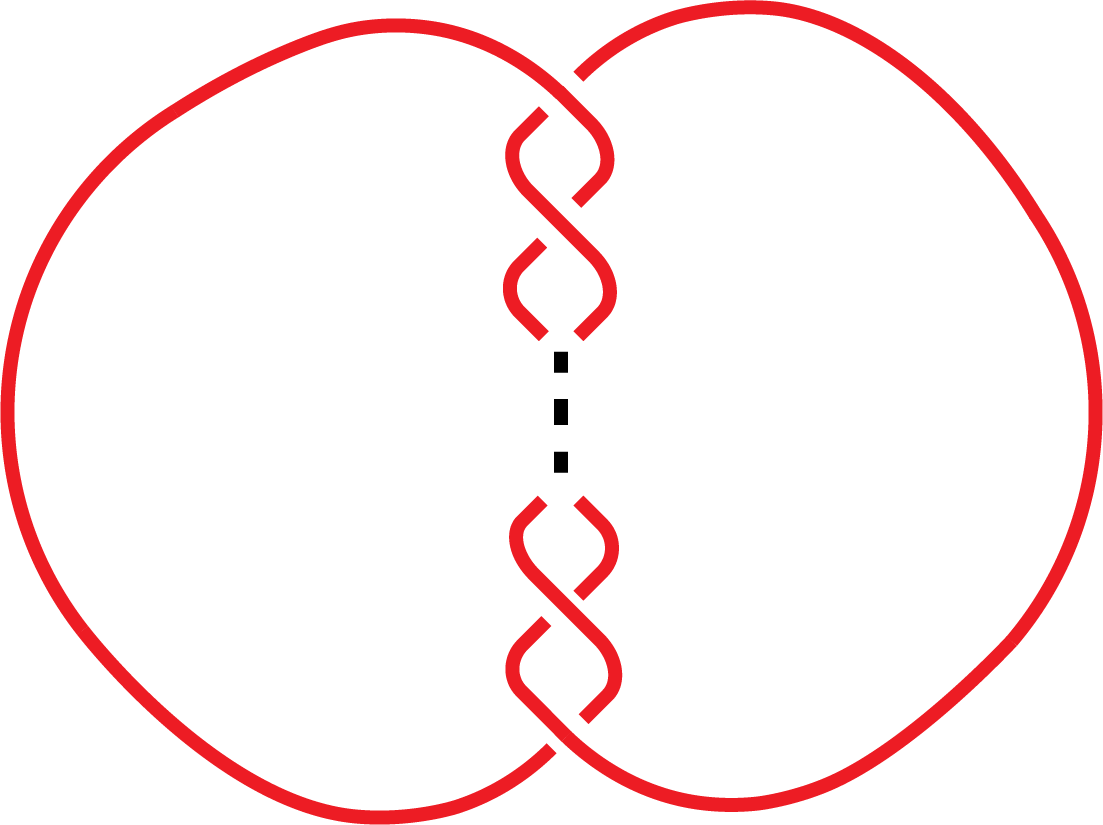} \,,
\end{equation}
whose action on $[0]$-channel blocks amounts to multiplication by the braiding phase $(\bker{P}{j}{k})^{2n}$ for any $n\in\Z$. In particular, the action of the unlink corresponds to the special choice $n=0$ while $n<0$ corresponds to braidings of opposite chirality. Summing over all such links in both sectors gives
\begin{align}
    \sum_{n\in\Z} \S\T^{2n}\S\ket{\inlinefig{imgs/s6/G_inf.png}}\ket{\inlinefig{imgs/s6/G_inf.png}} &= \sum_{n\in\Z}\ket{\inlinefig{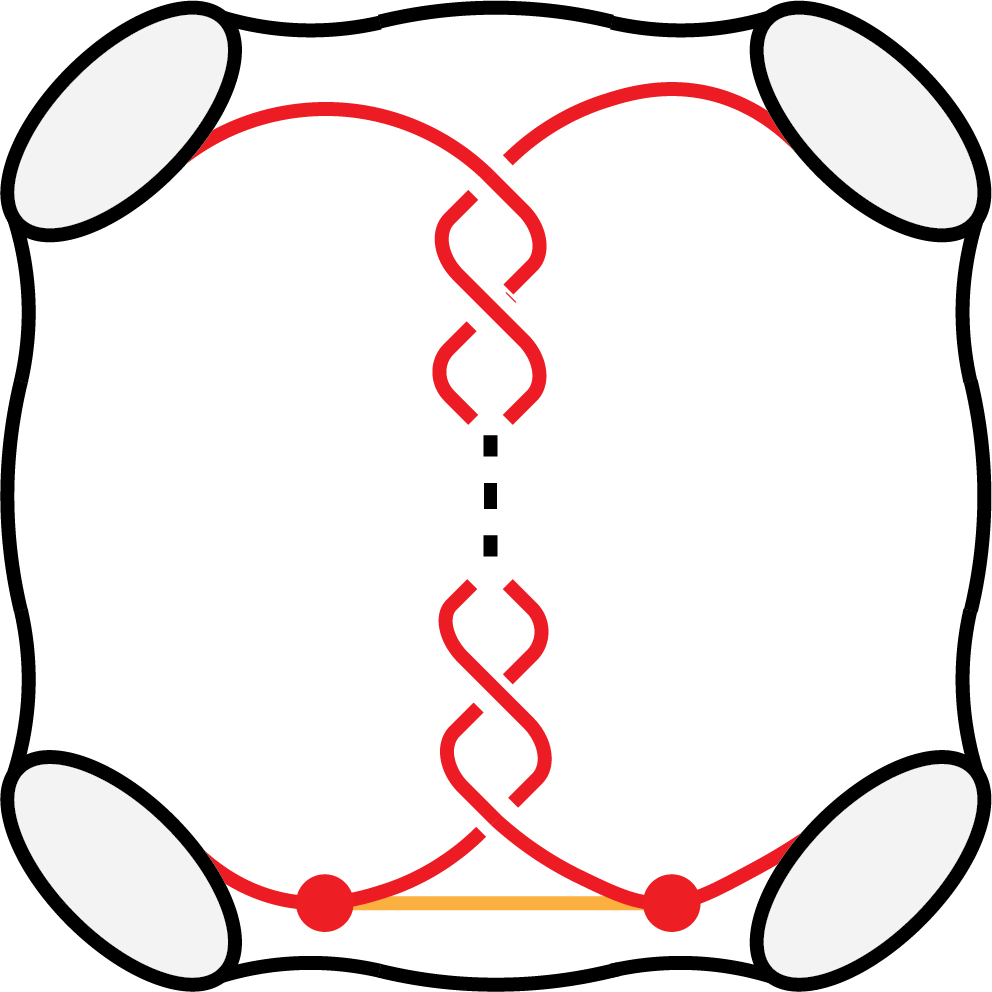}}\ket{\inlinefig{imgs/s6/ST2nSG_inf.png}} \nn\\
    &= \int\rho_0(P)dP\int\rho_0(\bar{P})d\bar{P}\ \left|\sum_{n\in\Z}(\bker{P}{j}{k})^{2n} C_0(P,P_j,P_k)\ket{\inlinefig{imgs/s6/G.png}}\right|^2 \nn\\
    &= \sum_{n\in\Z}\int\frac{ds}{4s}\ \left|\rho_0(P_n)C_0(P_n,P_j,P_k)\ket{\inlinefig{imgs/s6/G.png}}\right|^2\,,
\end{align}
where
\begin{equation}
    P_n(s) = \frac{s+\ell_n/s}{2} \quad{\rm and}\quad \bar{P}_n(s) = \frac{s-\ell_n/s}{2}.
\end{equation}
The sum over vertical tangles imposes a quantization condition 
\begin{equation}
    \ell \in n+\ell_j+\ell_k \,,
\end{equation}
on the spin of operators that can flow through the intermediate channel.

The statistics of intermediate operators in the $\cC_{\infty}$ channel are subsequently given by the projection \eqref{eq:Oi_ope_ansatz}. Letting $\cO_i$ flow in this channel gives 
\begin{align}\label{eq:sphere_hopf}
    \overline{\bra{P_j}\cO_i\ketbra{P_j}{P_k}\cO_i\ket{P_k}} &\supset \sum_{n\in\Z}\abs{\cN_\infty\mel{\inlinefig[4.5]{imgs/s6/G.png}}{U[1/2n]}{\inlinefig[4.5]{imgs/s6/G_inf.png}}}^2 \nn\\
    &= \sum_{n\in\Z}\abs{\cN_\infty\int dP\ \fker{i}{P}{j}{k}{j}{k}\frac{(\bker{P}{j}{k})^{2n}}{C_0(P,P_j,P_k)}}^2 \nn\\
    &= \sum_{n\in\Z}\int\frac{ds}{4s}\abs{\rho_0(P_n)\ \cN_\infty\frac{\fker{P_i}{P_n}{P_k}{P_j}{P_k}{P_j}}{\fker{\mathbbm 1}{P_n}{P_k}{P_j}{P_k}{P_j}}}^2.
\end{align}
where in the last line we exchanged the sum over braiding phases for a sum over Regge trajectories of fixed spin as described above.

The $n=-1$ term in the sum over braiding phases is proportional to the Hopf link \eqref{eq:torus_hopf}. Focusing on just this contraction gives
\begin{equation}\label{eq:hopf_contraction}
    \overline{\bra{P_j}\cO_i\ketbra{P_j}{P_k}\cO_i\ket{P_k}} \supset \abs{\frac{\sker{P_j}{P_k}{P_i}C_0(P_i,P_j,P_j)}{\rho_0(P_k)}}^2 \,,
\end{equation}
by \eqref{eq:torus_hopf} and the uniqueness of VTQFT amplitudes computed via hyperbolic splitting surfaces. Let us temporarily pause to consider that the Hopf contraction is not obviously symmetric under the exchange $j\leftrightarrow k$. Equation 2.49 of \cite{Collier:2023fwi} demonstrates that this is only a cosmetic ambiguity. The main ingredient is the following property of the modular S-kernel
\begin{equation}
    \int dP\ \sker{P_1}{P}{P_0}\overline{\sker{P}{P_2}{P_0}} = \delta(P_1-P_2) \,.
\end{equation}
We can use this to rewrite the normalization of torus conformal blocks
\begin{align}
    \frac{\delta(P_j-P_k)}{\rho_0(P_j)C_0(P_j,P_j,P_i)} &= \braket{\cF_{1,1}(P_i,P_j)}{\cF_{1,1}(P_i,P_k)}\nn\\
    &= \int dP_j'dP_k'\ \sker{P_j}{P_j'}{P_i}\overline{\sker{P_k}{P_k'}{P_i}}\braket{\cF_{1,1}(P_i,P_j')}{\cF_{1,1}(P_i,P_k')}\nn\\
    &= \int dP\ \frac{\sker{P_j}{P}{P_i}\overline{\sker{P_k}{P}{P_i}}}{\rho_0(P)C_0(P,P,P_i)}\,.
\end{align}
Relabeling the momenta and integrating both sides gives the desired result
\begin{align}
    \int dP'\ \sker{P_j}{P'}{P_i} \frac{\delta(P_k-P')}{\rho_0(P_k)C_0(P_k,P_k,P_i)} &= \frac{\sker{P_j}{P_k}{P_i}}{\rho_0(P_k)C_0(P_k,P_k,P_i)} \nn\\
    &= \int dP'\ \sker{P_j}{P'}{P_i}\int dP\ \frac{\sker{P_k}{P}{P_i}\overline{\sker{P'}{P}{P_i}}}{\rho_0(P)C_0(P,P,P_i)} \nn\\
     &= \int dP\ \frac{\sker{P_k}{P}{P_i}}{\rho_0(P)C_0(P,P,P_i)}\int dP'\ \sker{P_j}{P'}{P_i}\overline{\sker{P'}{P}{P_i}} \,.
\end{align}
We can rewrite the Hopf link contribution to OPE statistics as
\begin{align}\label{eq:hhl_hopf}
    \overline{\bra{P_j}\cO_i\ketbra{P_j}{P_k}\cO_i\ket{P_k}} &\supset \abs{C_0(P_i,P_j,P_j)C_0(P_i,P_k,P_k)\frac{\sker{P_j}{P_k}{P_i}}{\rho_0(P_k)C_0(P_i,P_k,P_k)}}^2 \nn\\
    &= \abs{\sqrt{C_0(P_i,P_j,P_j)C_0(P_i,P_k,P_k)}\times\left(\sqrt{\frac{C_0(P_i,P_j,P_j)}{C_0(P_i,P_k,P_k)}}\times\frac{\sker{P_j}{P_k}{P_i}}{\rho_0(P_k)}\right)}^2 \,.
\end{align}
The first line manifests the $j\leftrightarrow k$ symmetry of this non-Gaussianity and the second will be useful for determining its scaling behavior relative to the Gaussian Hopf-link contraction. The symmetric combination in parenthesis is familiar as the Racah-Wigner normalization of the modular S-kernel. 

Ultimately, we are interested in whether the Hopf contraction can be accommodated by a multi-trace ETH ansatz. We proceed to studying this question in two limits of physical interest.

\paragraph{Ultraheavy limit} Whether the Hopf contraction \eqref{eq:hopf_contraction} is subleading depends on the input momenta. Here we consider the situation where $P_i$ is light and both $P_j$ and $P_k$ are heavy. As discussed in appendix \ref{app:vmft}, both of our universal formulas scale exponentially in this limit
\begin{equation}
    \lim_{P\to\infty}\rho_0(P)=e^{2\pi Q P} \quad{\rm and}\quad \lim_{P\to\infty}\log C_0(P_i,P,P)=-2\pi Q P+2h_i\log P \,.
\end{equation}
It follows that the ratio
\begin{equation}\label{eq:C0_ratio}
    \sqrt{\frac{C_0(P_i,P_j,P_j)}{C_0(P_i,P_k,P_k)}} \approx \left(\frac{P_j}{P_k}\right)^{h_i}\times e^{-\pi Q\omega_{jk}} \,,
\end{equation}
when the difference $\omega_{jk} = P_j-P_k$ is kept $O(1)$ in units of the average $P_{jk}=(P_j+P_k)/2$. Above and in the remainder of our discussion of the ultraheavy limit we use ``$\approx$'' to denote the leading answer in the limit of large $P_{jk}$.

Next, we look at the scaling behavior of the torus one-point kernel. The exact kernel is given by a contour integral \cite{Eberhardt:2023mrq}\footnote{
    There are many identical ways to write this kernel. We choose this particularly symmetric form because it is well-equipped to treating $P_j$ and $P_k$ on equal footing asymptotically.
}
\begin{equation}\label{eq:S_contour}
    \sker{P_j}{P_k}{P_i} = Q_b(P_i,P_j,P_k)\times e^{2\pi i(P_j^2+P_k^2)}\int_{\cC}\frac{dx}{2i}\ e^{2\pi i x^2} \frac{S_b(Q/4+i P_i/2\pm iP_j\pm iP_k\pm x)}{S_b(\pm2x)}\,,
\end{equation}
where
\begin{equation}
    Q_b = e^{i\pi Q\alpha_i^*/2}\rho_0(P_k)S_b(Q/2-iP_i)\frac{\Gamma_b(Q\pm 2iP_j)\Gamma_b(Q/2-iP_i\pm 2iP_k)}{\Gamma_b(Q\pm 2iP_k)\Gamma_b(Q/2+iP_i\pm 2iP_j)}\,,
\end{equation}
$\alpha_i^*=Q-\alpha_i$, and the special functions $\Gamma_b$ and $S_b$ are as defined in appendix \ref{app:vmft}. Our asymptotics allow us to choose a purely imaginary contour $\cC$ for the $x$-integral. Plugging in \eqref{eq:gammab_largex}, the pre-factor becomes
\begin{equation}
    \log \frac{e^{-i\pi Q\alpha_i^*/2}Q_b(P_i,P_j,P_k)}{\rho_0(P_k)} \approx 2\pi\alpha_i^*(P_j+P_k)+h_i\log\frac{P_k}{P_j}+\log S_b(\alpha_i^*)\,.
\end{equation}
The $j\leftrightarrow k$ exchange symmetry is already evident. The asymmetric factor in parenthesis of \eqref{eq:C0_ratio} is canceled in the heavy limit by its inverse in $Q_b$. Similarly, the asymmetric difference in the phase factor of \eqref{eq:C0_ratio} becomes a symmetric sum due to the factor of $\rho_0$ in $Q_b$.

To confirm that the Hopf contraction satisfies the expected multi-trace ETH scaling behavior in this limit, we need to also check that the term proportional to the average momentum $P=(P_j+P_k)/2$ above is cancelled by a similar term in the integral. This would leave only a polynomial dependence in $j$ and $k$ that could be absorbed into the definition of $F_i^{(1|1)}$. It will be helpful to scale all relevant quantities with $P$ such that
\begin{equation}
    P_j = P+\omega/2,\quad P_k=P-\omega/2,\quad x=i\sigma P\,.
\end{equation}
By assumption, $\omega$ does not scale with $P$. The integration contour now runs over $\sigma\in\R$. The resulting integrand is continuous but not differentiable because of several log branch-cuts. We find
\begin{align}
    &\log \frac{S_b(Q/4+i P_i/2\pm i(P+\omega/2)\pm i(P-\omega/2)\pm i\sigma P)}{S_b(\pm i\sigma P)}  \nn\\
    &\hspace{16em} \approx\begin{cases}
        4\pi i\sigma P_iP &\quad{\rm when}\ \sigma\geq 2\,,\\
        2\pi (\sigma\alpha_i-2\alpha_i^*)P &\quad{\rm when}\ 0\leq\sigma<2\,,\\
        -2\pi (\sigma\alpha_i+2\alpha_i^*)P &\quad{\rm when}\ -2\leq\sigma<0\,,\\
        -4\pi i\sigma P_iP &\quad{\rm when}\ \sigma<-2\,.\\
    \end{cases}
\end{align}
Evidently the integrand is symmetric about $\sigma=0$. Plugging this back in gives
\begin{equation}
    \int_{i\R}\frac{dx}{2i}\ e^{2\pi i x^2} \frac{S_b(Q/4+i P_i/2\pm iP_j\pm iP_k\pm x)}{S_b(\pm2x)} \approx \frac{e^{-i\pi/4}}{2\sqrt{2}}e^{-2\pi (\alpha_i^*P+i\alpha_i^2/4)}\left(1+\erf\left[\sqrt{\frac{\pi}{2}}e^{-i\pi/4}\alpha_i\right]\right)\,.
\end{equation}
The desired factor of $e^{-2\pi\alpha_i^*P}$ cancels a similar factor which entered in the heavy limit of $Q_b$ leaving only $P_i$ dependent exponential terms. In total we have
\begin{equation}
    \sker{P_j}{P_k}{P_i} \approx \frac{e^{i\pi(Q\alpha_i^*-\alpha_i^2)/2}}{2\sqrt{2}}\rho_0(P_k)S_b(\alpha_i^*)\left(\frac{P_k}{P_j}\right)^{h_i}\left(1+\erf\left[\sqrt{\frac{\pi}{2}}e^{-i\pi/4}\alpha_i\right]\right)\,,
\end{equation}
The exponential factor is a pure phase for a light probe. Note that, unlike 3.20 of \cite{Collier:2019weq}, here we have taken both $P_j$ and $P_k$ heavy. 

Overall the relative suppression with respect to the Gaussian piece is: 
\begin{equation}
    \left|\sqrt{\frac{C_0(P_i,P_j,P_j)}{C_0(P_i,P_k,P_k)}}\times\frac{\sker{P_j}{P_k}{P_i}}{\rho_0(P_k)}\right| = O(e^{-S(P)})\,.
\end{equation}
Subleading factors in the momenta give one-loop corrections to this contraction and can be reassembled to compute the contribution to $F_i^{(1|1)}$. For our purposes, it suffices to confirm that the non-Gaussianity is suppressed relative to the leading Gaussian piece. In terms of the microcanonical entropy, this contribution gives an $O(e^{-S(P)})$ relative correction to the matrix element statistics of $\cO_i$, as predicted by the multi-trace ETH ansatz of \eqref{eq:MT}.

\paragraph{Macroscopic BTZ} Next we consider what the Hopf contraction implies for the statistics of light probe operators in a semiclassical limit. We should take care when evaluating equation \eqref{eq:hopf_contraction} in this regime since the Cardy density of states specifically applies to above-threshold operators. Its analytic continuation is highly oscillatory when the operator weights are taken outside of the normalizable Liouville spectrum. Instead, we use \eqref{eq:sphere_hopf} which has two advantages: (1) it makes manifest the $j\leftrightarrow k$ symmetry of this contraction just like equation \eqref{eq:hhl_hopf} and (2) this form is amenable to analytic continuation below the black hole threshold. The analytic continuation of $\mathbb{F}$ was studied extensively in \cite{Collier:2018exn}. 

So motivated, let us proceed to study the integral form of the Hopf contraction. Using \eqref{eq:C0} and \eqref{eq:fker_contour}, the integrand becomes
\begin{equation}
    \frac{\fker{P_i}{P_n}{P_k}{P_j}{P_k}{P_j}}{\fker{\mathbbm 1}{P_n}{P_k}{P_j}{P_k}{P_j}} = \frac{1}{C_0(P_i,P_j,P_k)}\times\int_{\cC}\frac{ds}{i}\prod_{a\in\{1,2,3,4\}}\frac{S_b(s+U_a)}{S_b(s+V_a)}
\end{equation}
where
\begin{equation}
    \begin{array}{c c c}
        U_1 = U_4 = 0\,, &\quad& V_1 = Q/2+i(P_k-P_j-P)\,, \\
        U_2 = -2iP_j\,, &\quad& V_2 = Q/2+i(P_k-P_j+P)\,, \\
        U_3 = 2iP_k\,, &\quad& V_3=V_4^*=Q/2+iP_i\,.
    \end{array}
\end{equation}
The statement $V_3=V_4^*$ is only strictly true when $P_i\in\R$. The reflection symmetry between $V_3$ and $V_4$ around the imaginary $P_i$-axis becomes a split along the real axis centered around $Q/2$ when the operator $\cO_i$ corresponds to a below-threshold state. The integrand has poles wherever
\begin{equation}
    s+U_k,Q-s-V_k\in -mb-nb^{-1}\,,
\end{equation}
for $m,n\in\Z_{\geq 0}$. The contour $\cC$ is anchored at $\pm i\infty$ and passes to the right of the $U_a$ poles and to the left of the $V_a$ poles. 

In the macroscopic BTZ regime, the external momenta scale with the central charge while their difference is kept small. Take
\begin{equation}
    p_{jk} = b(P_j+P_k), \quad \omega_{jk}=b^{-1}(P_j-P_k),\quad p=bP_n,\quad{\rm and}\quad \Delta=b^{-1}\alpha_i\,,
\end{equation}
with $p_{jk},p,\omega_{jk},\Delta_i\sim O(1)$. In these limits,
\begin{equation}
    \frac{1}{S_b(s+V_1)S_b(s+V_2)} \approx \left(2\cosh(\pi p)\right)^{2(\omega_{jk}-s/b)} \,,
\end{equation}
and
\begin{equation}
    \prod_{a}S_b(s+U_a) \approx \frac{e^{-2\pi p_{jk}/b^2-2s/b}}{(2\pi s)^{1+2s/b}}(2\sinh(2\pi p_{jk}))^{-2(\omega_{jk}-s/b-1/2)} \,.
\end{equation}
The integrand decays exponentially with $s$ whenever $p\gg 2p_{jk}$. It follows that the dominant contribution to the integral over $p$ comes from poles minimizing $\Re s$. The left-most poles come from $S_b(s+V_3)$ and occur when $s=\alpha_i+mb$ for $m\in\Z_{\geq 0}$. Higher order terms are not exponentially suppressed in the relevant large parameter so it would seem that summing the discrete series
\begin{equation}
    \int_{\cC}\frac{ds}{i}\prod_{a\in\{1,2,3,4\}}\frac{S_b(s+U_a)}{S_b(s+V_a)} \approx \sum_{m\in\Z_{\geq 0}}\Res_{s=\alpha_i+mb} \frac{S_b(s+U_1)S_b(s+U_2)S_b(s+U_3)S_b(s+U_4)}{S_b(s+V_1)S_b(s+V_2)S_b(s+V_3)S_b(s+V_4)}\,,
\end{equation}
is inevitable. Luckily, each summand also carries a factor of $b^{2m}$. As $b\to 0$, only the $m=0$ term remains
\begin{equation}
    \int_{\cC}\frac{ds}{i}\prod_{a\in\{1,2,3,4\}}\frac{S_b(s+U_a)}{S_b(s+V_a)} \approx e^{-2\pi p_{jk}/b^2}\left(\frac{\pi b^2}{e}\right)^{2\Delta-3/2}\left(\frac{\cosh(\pi p)}{\sinh(2\pi p_{jk})}\right)^{2\omega_{jk}-2\Delta}\sech(\pi p_{jk})\csch(\pi p_{jk})\,,
\end{equation}
since higher order terms are polynomially suppressed.

We now have all the necessary ingredients to confirm that the sum over vertical rational tangles can be accounted for by the multi-trace ETH in a semiclassical limit. Combining the prefactor to the integral with the block normalization factor $\cN_\infty$ gives
\begin{align}
    \frac{C_0(P_i,P_j,P_j)C_0(P_i,P_k,P_k)}{C_0(P_i,P_j,P_k)} &\approx  e^{-S(p_{jk})}\times\left|\frac{f_i(p_j)f_i(p_k)}{f_i(p_{jk},\omega_{jk})}\right|^2 \nn\\
    &= e^{-2\pi p_{jk}/b^2}\times(2\pi b)^{-5/4}\left[\frac{(2p_{jk})^{2\Delta}}{2\sinh(2\pi p_{jk})}\right]\frac{\Gamma(\Delta)^2}{\Gamma(\Delta+i\omega)\Gamma(\Delta-i\omega)}\,,
\end{align}
where above we plugged in the exact formula for the smooth function $f_i$ provided in \eqref{eq:btz_fi}. The important thing is that this function scales polynomially with the average momentum $p_{jk}$. Taking into account the exponential scaling of the contour integral above confirms that these contributions are subdominant to the unlink by an overall factor of $O(e^{-S(p_{jk})})$, as anticipated by \eqref{eq:MT}.

\subsection{Multi-trace ETH from dynamical matter}\label{subsec:light_matter}

The next class of non-Gaussianities we consider come from the inclusion of dynamical matter. The topological non-Gaussianities discussed in the previous section were fundamentally different from the gravitating matter loops which gave rise to a multi-trace ETH in JT gravity. Here we consider the dimensional uplifts of the multi-trace ETH contributions encountered in section \ref{sec:JT}. In particular, we identify a 3d wormhole geometry containing a bulk matter Wilson loop that reproduces the 1-particle bulk matter loop contribution in JT with matter upon dimensional reduction. 

The inclusion of dynamical matter is a departure from the realm of pure gravity. One motivation for this excursion is that pure AdS$_3$ gravity is known to suffer from pathologies that can be cured by the inclusion of non-perturbative corrections to the spectrum \cite{Maloney:2007ud, Keller:2014xba}. These corrections have been referred to as long strings \cite{Kim:2014bga}, strongly-coupled strings \cite{DiUbaldo:2023hkc}, spinning strings \cite{Li:2026fxo, Maxfield:2022rry}, or simply conical defects \cite{Benjamin:2020mfz} (also see \cite{Alday:2019vdr} and \cite{Maxfield:2020ale}). Here we demonstrate that the inclusion of these states also leads to a non-trivial realization of the multi-trace ETH.

\begin{figure}
    \centering
    \includegraphics[width=0.8\textwidth]{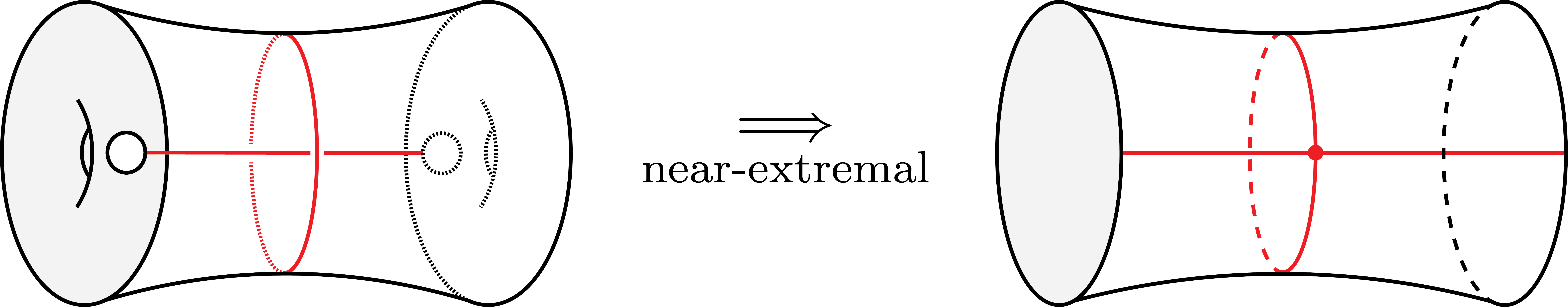}
    \caption{The marked 3-manifold $M_\chi$ has two torus boundaries that are connected at the their punctures by a light operator $\cO_i$. A bulk matter loop, homologous to both boundaries and sourced by an operator $\cO_\chi$ dual to a dynamical bulk matter field, winds once around the line sourced by $\cO_i$. In the near-extremal limit, this geometry becomes the one-particle contribution to the squared JT one-point function computed in \cite{Jafferis:2022uhu}. Intersecting Wilson lines in the 2d geometry become non-trivially linked in 3d.}
    \label{fig:matter_wh_reduction}
\end{figure}

\paragraph{One Matter Loop} First, let us qualitatively discuss the 3d geometry giving rise to the desired wormhole. To match the computation in \eqref{eq:F11_JKMS_one-particle}, the matter loop must wrap a cycle that does not degenerate in the near-extremal limit. This ensures the existence of a bulk matter loop in the dimensional reduction. Additionally, the matter loop is homologous to both boundaries in 2d. This, in conjunction with the discussion in section \ref{subsec:full_eth_3d}, helps identify a preferred channel for this computation. The near-extremal limit is performed in a channel where momenta wrap the longitudinal cycles of the boundary tori. The bulk matter loop should be homologous to these momenta. As in section \ref{subsec:full_eth_3d}, in the near-extremal regime, the theory is only dynamical in the left-moving sector where saddles are computed by VTQFT.

The resulting marked 3-manifold is referred to as $M_\chi$.\footnote{
    We are very grateful to B. Post for useful discussions on this geometry in VTQFT. 
} A similar 3-manifold was studied in section 5 of \cite{deBoer:2024mqg}. Whereas that calculation involved thrice-punctured spherical boundaries, we use once-punctured toric boundaries. This makes the connection to chaotic dynamics in black hole backgrounds more immediate. Throughout, it will be crucial to remember that we are using torus conformal blocks; not all cycles are contractible. A schematic visual is provided in figure~\ref{fig:matter_wh_reduction}. 

Now let us compute the VTQFT amplitude associated to $M_\chi$. Start by moving to a channel where the internal matter Wilson loop wraps the non-contractible meridian cycle of the boundary tori
\begin{equation}\label{eq:spool_candidate}
    Z_{\rm Vir}(M_\chi) = \int dP\ \sker*{P_\chi}{P}{\mathbbm 1} Z_{\rm Vir}\left(\inlinefig[4.5]{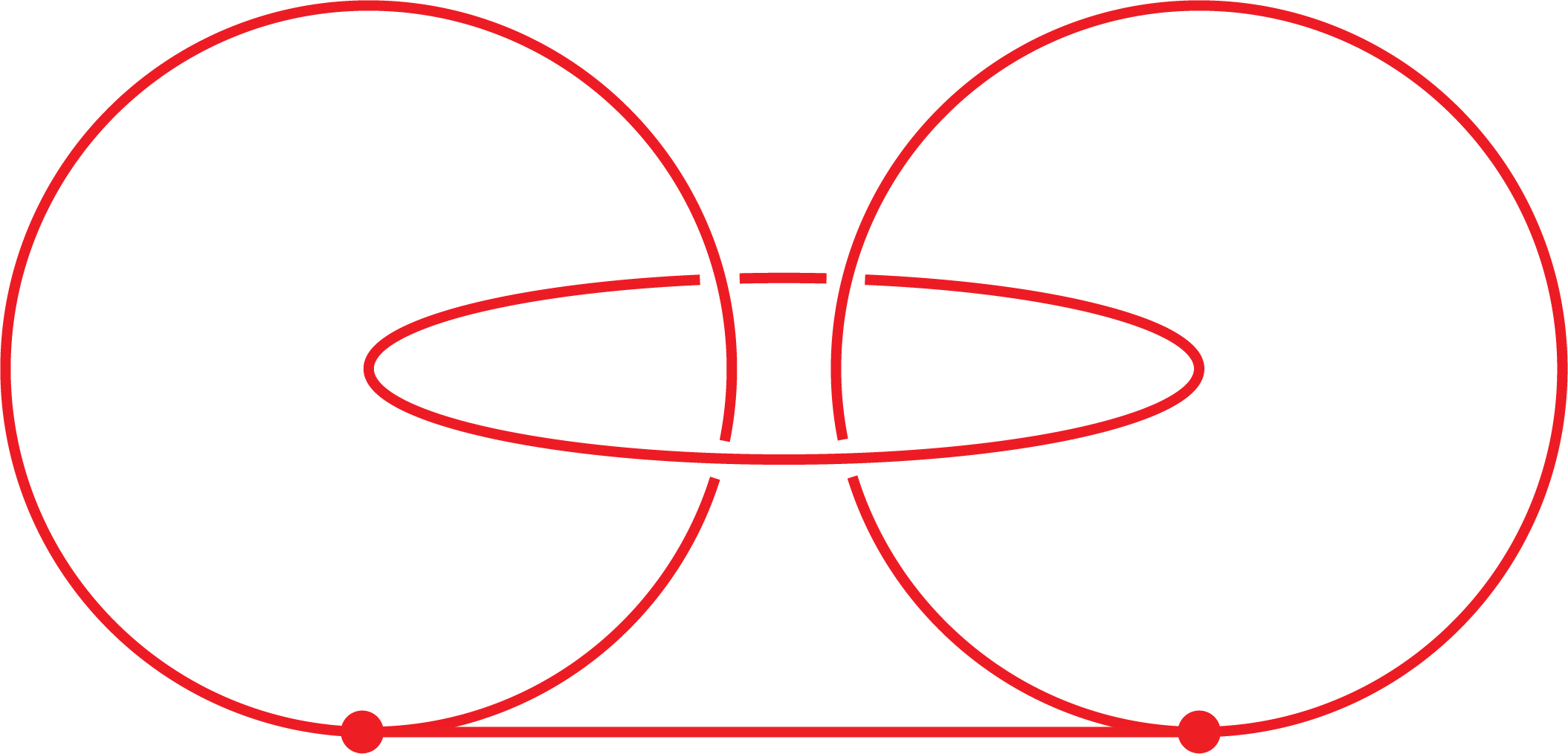}\right)\,.
\end{equation} 
The new spectator momentum label $P$ wraps a cycle which degenerates in the near-extremal limit. The result is a link between $P$ and the asymptotic boundary energies $P_j$ and $P_k$. This link is fragmented by $P_i$, the chiral momentum label of the light operator $\cO_i$. An identity on the thrice-punctured sphere allows us to remove the $P_j$ and $P_k$ cycles\footnote{
    Taking $P_i\to\mathbbm{1}$ gives the Verlinde-loop operator. The twice-punctured sphere is not a hyperbolic manifold so care should be taken when using its conformal blocks in a Virasoro TQFT computation since the latter only accurately computes on-shell quantities. In this case, the formula follows from a degeneration limit of the thrice-punctured sphere and comports well with results in rational TQFTs. 
} 
\begin{equation}
    \inlinefig[4.5]{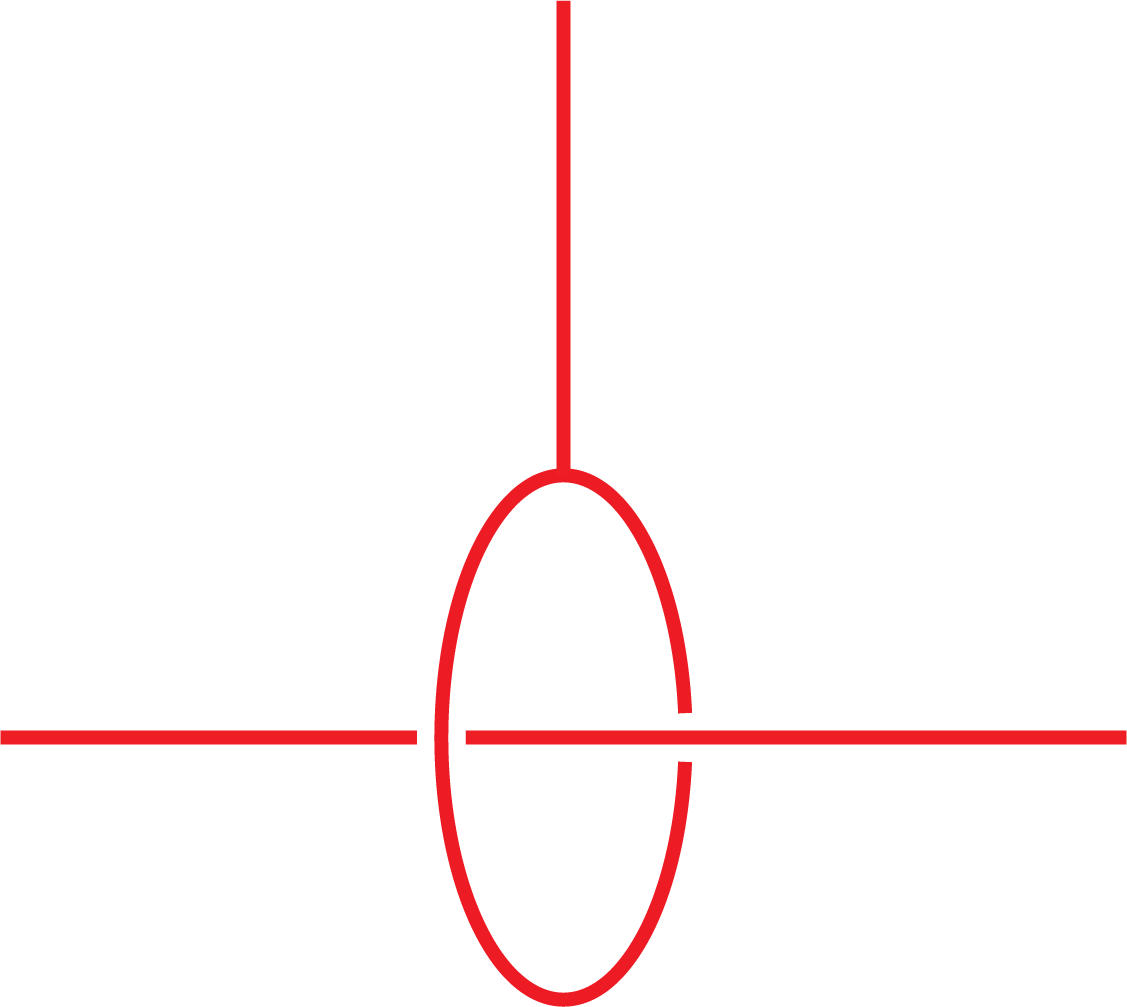} = \frac{\sker{P_j}{P}{P_i}}{\sker{\mathbbm 1}{P}}\inlinefig[3]{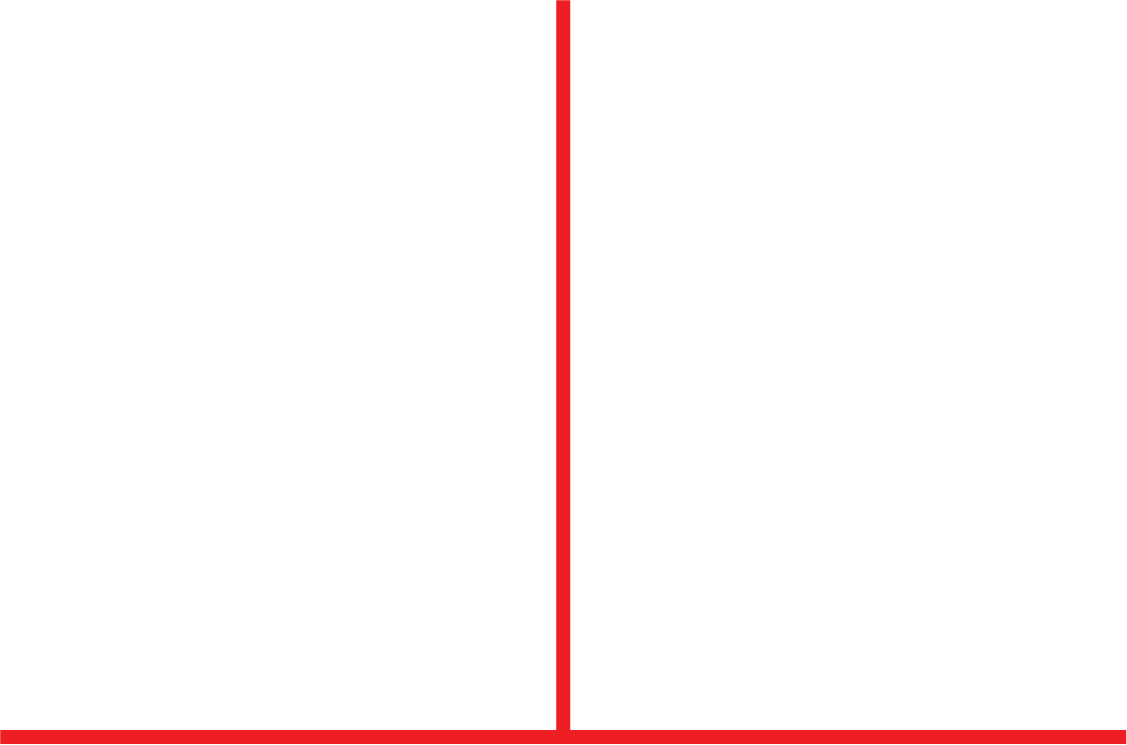} \quad{\rm and}\quad \inlinefig[4.5]{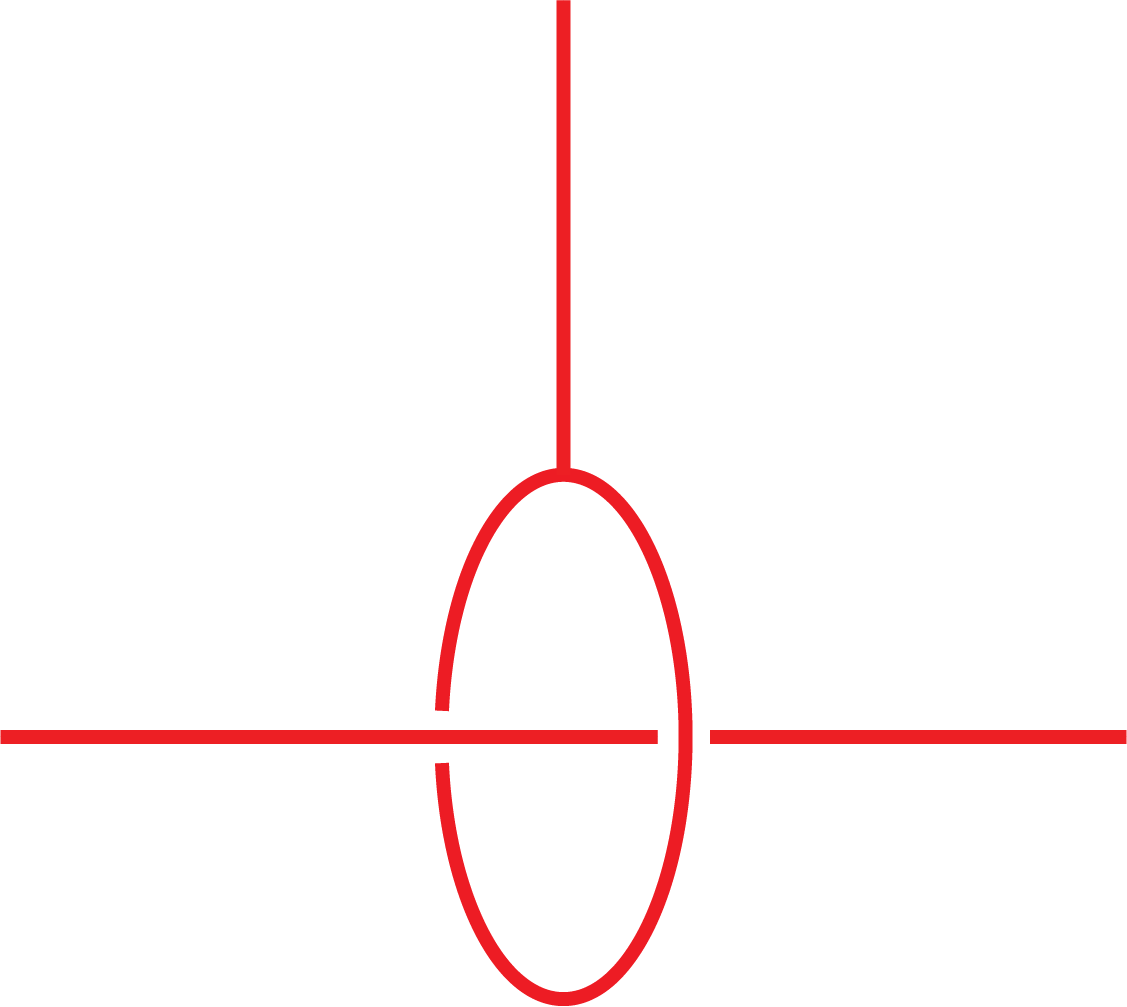} = \frac{\sker*{P_k}{P}{P_i}}{\sker{\mathbbm 1}{P}}\inlinefig[3]{imgs/s6/3pt.png}\,.
\end{equation}
The modular kernel is real except for a braiding phase of $e^{\pi ih_i}$. Without the relative complex conjugation, we study an inequivalent link
\begin{equation}
    \inlinefig[4.5]{imgs/s6/SM_chi_left.png} \neq \inlinefig[4.5]{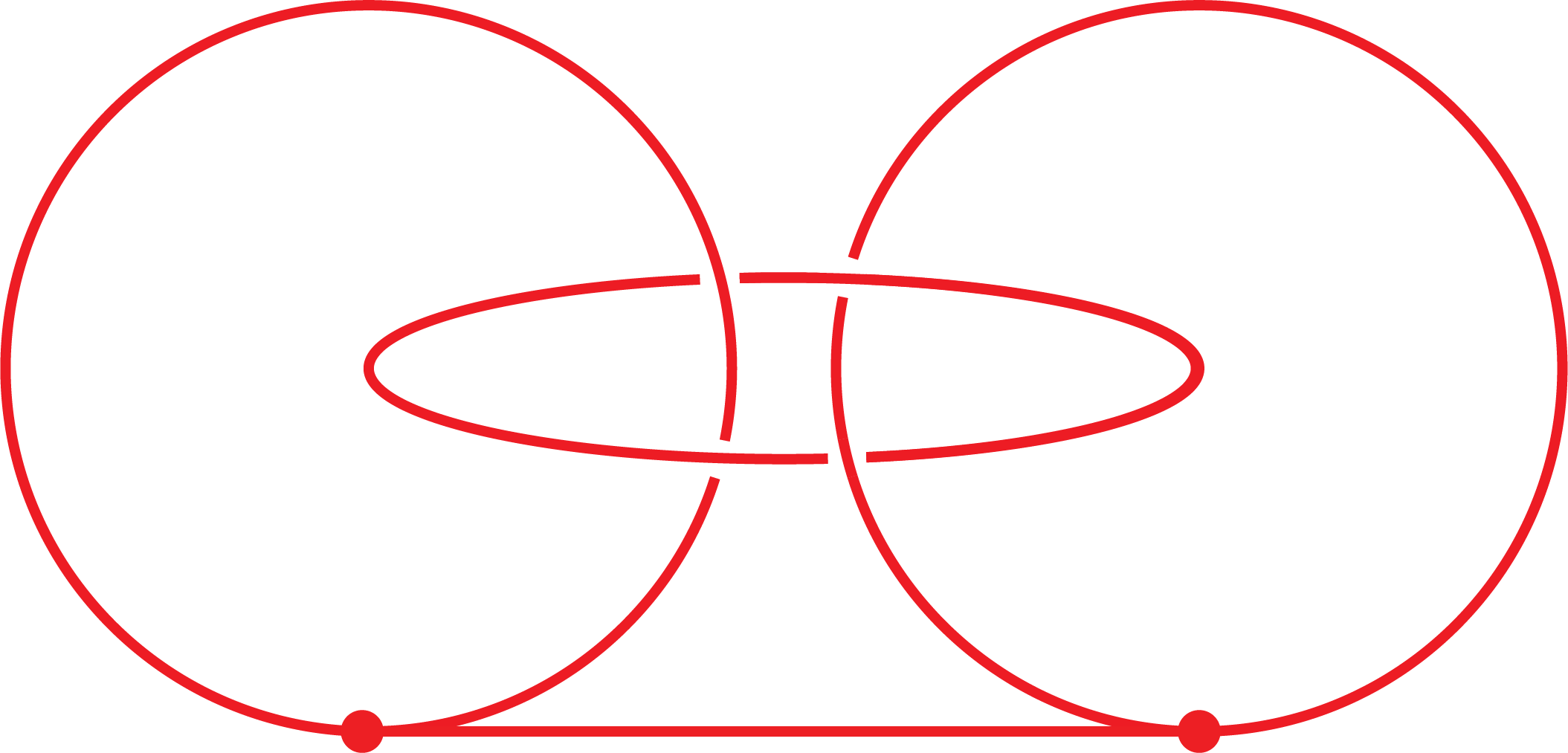}\,,
\end{equation}
where the $P$ matter loop winds a non-contractible figure 8 around the asymptotic boundaries.

What remains after removing the internal loops is a sunset diagram embedded in the two-boundary torus wormhole. The usual normalization of the thrice-punctured sphere gives
\begin{equation}
    Z_{\rm Vir}\left(\inlinefig[4.5]{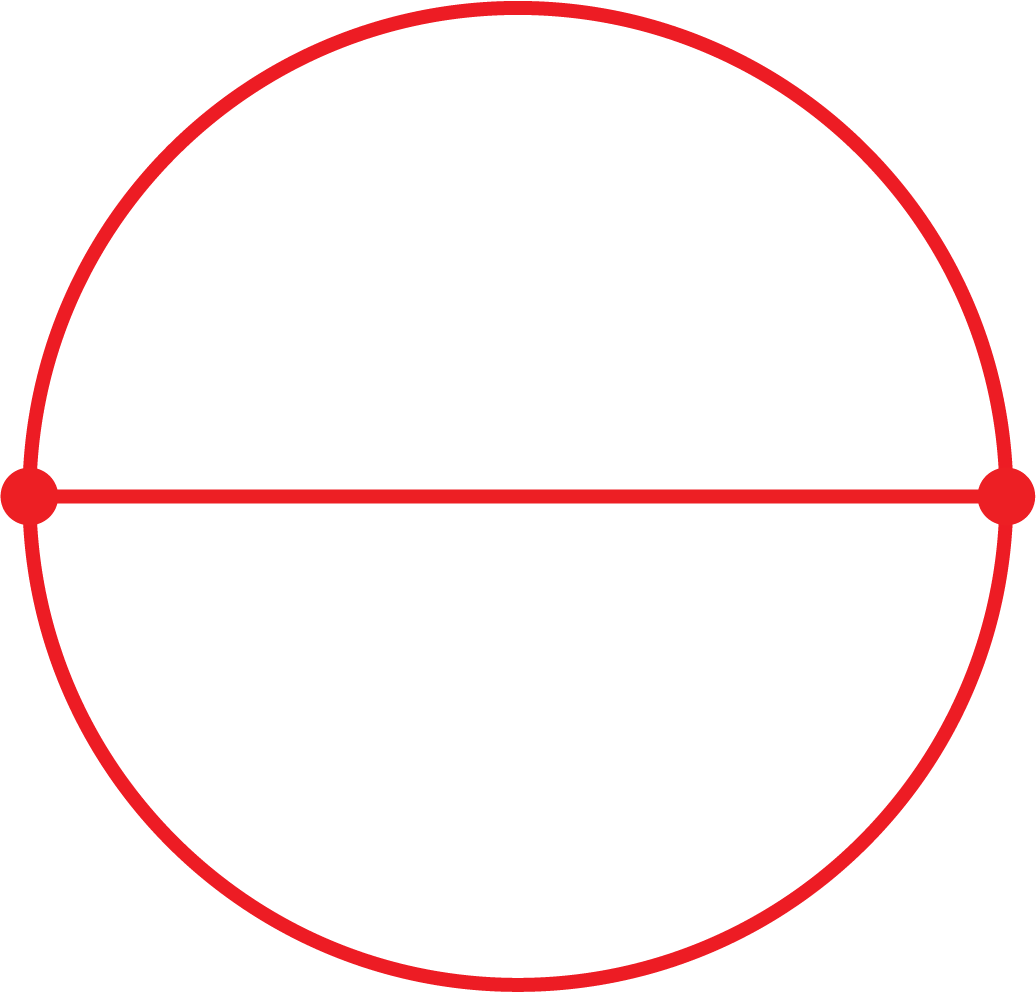}\right) = \frac{1}{C_0(P_i,P,P)}\braket{\cF_{1,0}(\mathbbm{1})}\,.
\end{equation}
The inner product is formally infinite but well-defined as a degeneration limit of a norm on once-punctured torus blocks \cite{Collier:2023fwi}. Removing this infinite pre-factor gives
\begin{align}
    Z_{\rm Vir}(M_\chi) &= \int dP\ \frac{\sker{P_j}{P}{P_i}\sker*{P_k}{P}{P_i}\sker*{P_\chi}{P}}{\sker{\mathbbm 1}{P}\sker{\mathbbm 1}{P}} \nn\\
    &= \frac{1}{\rho_0(P_j)C_0(P_i,P_j,P_j)}\int dP\ \frac{\sker{P}{P_j}{P_i}\sker*{P_k}{P}{P_i}\sker*{P_\chi}{P}}{\sker{\mathbbm 1}{P}}\,,
\end{align}
where in the last line we used the symmetry of the Hopf contraction in \eqref{eq:hopf_contraction} and the fact that $\sker{\mathbbm 1}{P}=\rho_0(P)$. 

Evidently, the presence of a bulk matter Wilson loop induces a non-Gaussianity in the statistics of heavy operators. This particular contribution is familiar as the fusion density identified in \cite{Post:2024itb}.\footnote{
    It appears that our formula is the complex conjugate of the fusion density presented in \cite{Post:2024itb}. It is, but the density is also real in this parameter regime. 
} Specifically, 
\begin{align}\label{eq:zvir_Mchi}
    Z_{\rm Vir}(M_\chi) &= \frac{\cN_{i}[P_\chi,P_j,P_k]}{\rho_{1,1}(P_i;P_j)} = \frac{1}{\sqrt{C_0(P_i,P_j,P_j)C_0(P_i,P_k,P_k)}}
    \begin{Bmatrix}
        P_i & P_j & P_j \\
        P_\chi & P_k & P_k
    \end{Bmatrix}_b\,.
\end{align}
The torus one-point density $\rho_{1,1}(P_i;P_j)=\rho_0(P_j)C_0(P_i,P_j,P_j)$ as defined in \cite{Collier:2023fwi}. The Virasoro fusion kernel $\cN_{i}[P_\chi,P_j,P_k]$ was introduced in \cite{Post:2024itb} as a non-rational extension of the fusion coefficient $N^{\chi}_{jk}$ familiar from rational CFTs and the Verlinde formula
\begin{equation}
    N^{\chi}_{jk} = \sum_{l}\frac{\sker{j}{l}\sker{k}{l}\sker*{h}{\chi}}{\sker{\mathbbm 1}{l}}\,.
\end{equation}
The Virasoro fusion kernel shares many desirable properties with fusion coefficients including reality, positivity, and a sensible degeneration limit. Most importantly, the fusion kernel is also diagonalized by the modular kernel once the probe momentum $P_i$ is introduced to regulate the otherwise distributional density. 

To check this result, we compute the near-extremal limit of $M_\chi$. The denominator of \eqref{eq:zvir_Mchi} cancels nicely with the disconnected density of states, giving a simple expression for the state prepared by the two-boundary wormhole
\begin{align}\label{eq:zvir_Mchi_wh}
    &\ket{Z_{\rm Vir}(M_\chi)} = \int \rho_0(P_j)dP_j\rho_0(P_k)dP_k \nn\\
    &\hspace{2cm} \sqrt{C_0(P_i,P_j,P_j)C_0(P_i,P_k,P_k)}\begin{Bmatrix}
        P_i & P_j & P_j \\
        P_\chi & P_k & P_k
    \end{Bmatrix}_b\ket{\cF^\cC_{1,1}(P_i;P_j)}\ket{\cF^\cC_{1,1}(P_i;P_k)}\,.
\end{align}
The near-extremal limit of once-punctured torus conformal blocks, 
\begin{equation}
    \braket{\tau}{\cF^\cC_{1,1}(P_i;P_k)} \approx e^{-\beta s_k^2} \,,
\end{equation}
is discussed in appendix \ref{app:ne_blocks}, where $\beta$ and $s_k$ are defined relative to $\tau$ and $P$ as in appendix \ref{app:ne_ops}. All in all, we find
\begin{equation}
    \braket{\tau_j,\tau_k}{Z_{\rm Vir}(M_\chi)} \underset{\rm near-extremal}{=} \int \rho(s_j)ds_j\rho(s_k)ds_k\ e^{-\beta_j s_j^2}e^{-\beta_k s_k^2}\times\sqrt{\Gamma^{i}_{jj}\Gamma^{i}_{kk}}\times\begin{Bmatrix}
        \Delta_i & s_j & s_j \\
        \Delta_\chi & s_k & s_k
    \end{Bmatrix}
\end{equation} 
which matches the one-particle contribution to the JKMS channel decomposition in \eqref{eq:JKMS-1pt2}. In particular, we have here recovered the multi-trace ETH term in \eqref{eq:F11_JKMS_one-particle}.

More broadly, the fusion kernel appears as a contribution to a modified inner product on the space of conformal blocks. Here it is helpful to recall the usual resolution of the identity in Virasoro TQFT which is computed by the wormhole geometry \cite{Collier:2023fwi}
\begin{align}\label{eq:vtqft_id_res}
    Z_{\rm Vir}(\Sigma_{g,n}\times I) &= \int \rho_{g,n}^\cC(\vec{P})\vec{P}\int\rho_{g,n}^\cC(\vec{P'})d\vec{P'}\ \braket{\cF_{g,n}^\cC(\vec{P})}{\cF_{g,n}^\cC(\vec{P'})}\ket{\cF_{g,n}^\cC(\vec{P})}\otimes\ket{\cF_{g,n}^\cC(\vec{P'})}\nn\\
    &= \int \rho_{g,n}(\vec{P})d\vec{P}\ \ket{\cF_{g,n}^\cC(\vec{P})}\otimes\ket{\cF_{g,n}^\cC(\vec{P})}\,,
\end{align}
defined in terms of the channel density of states
\begin{equation}
    \rho_{g,n}^\cC(\vec{P}) = \prod_{\rm internal} \rho_0(P_a) \prod_{\rm pants}C_0(P_i,P_j,P_k)\,.
\end{equation}
The modification to the inner product 
\begin{equation}
    \overline{\braket{\cF^\cC_{1,1}(P_i,P_j)}{\cF^\cC_{1,1}(P_i,P_k)}} = \frac{\delta(P_j-P_k)}{\rho_{1,1}(P_i;P_k)} + \frac{1}{\sqrt{C_0(P_i,P_j,P_j)C_0(P_i,P_k,P_k)}}
    \begin{Bmatrix}
        P_i & P_j & P_j \\
        P_\chi & P_k & P_k
    \end{Bmatrix}_b + \cdots\,
\end{equation}
augments \eqref{eq:vtqft_id_res} by addition of the wormhole amplitude \eqref{eq:zvir_Mchi_wh}. Recall that multi-trace contributions derived from topology could be similarly attributed to a modified inner product \eqref{eq:ip_poincare}. It follows that this wormhole also isolates a modification to matrix element statistics 
\begin{equation}\label{eq:mtt_eth_cft2}
    \overline{\bra{P_j}\cO_i\ketbra{P_m}{P_k}\cO_i\ket{P_n}} = e^{-S(P_{jk})}|f_i(P_{jk},\omega_{jk})|^2\delta_{jn}\delta_{mk} + e^{-S(P_j)-S(P_k)}F_i^{(1|1)}(P_j|P_k)\delta_{jm}\delta_{kn} \,,
\end{equation}
where $S$ and $f_i$ are as described in section \ref{subsec:full_eth_3d} and
\begin{equation} \label{eq:F11-3dinteracting}
    F_i^{(1|1)}(P_j|P_k) = \sqrt{C_0(P_i,P_j,P_j)C_0(P_i,P_k,P_k)}\times
    \begin{Bmatrix}
        P_i & P_j & P_j \\
        P_\chi & P_k & P_k
    \end{Bmatrix}_b\,.
\end{equation}
This places the amplitude of $M_\chi$ in the form of the multi-trace ETH ansatz in \eqref{eq:two-point-MT}. Such a correction only appears when dynamical matter, here exemplified through bulk Wilson loops, is turned on.

\section{Discussion} 
\paragraph{Extending to multiple matter loops in 3d gravity.}
An arbitrary number of Wilson loops are allowed to proliferate through spacetime in a truly dynamical theory of gravity coupled to matter. One example of this is the JKMS channel decomposition in \eqref{eq:JKMS-1pt2} which contains double-trace operators and higher composites. In 3d gravity, these complicated interactions have traditionally been treated via a heat kernel analysis \cite{Giombi:2008vd}. Unfortunately, the directness of the heat kernel method also obscures the topological nature of 3d gravity. The Wilson spool has recently emerged as a technology that captures the contributions of dynamical matter by taking advantage of the topological approach \cite{Castro:2023bvo, Bourne:2024ded}. The spool is defined in the Chern-Simons formulation where the coupling of free matter to the gauge fields $A$ and $\bar{A}$ is determined by a pair of chiral representations $P$ and $\bar{P}$. These representations are in turn determined by the mass and spin of the matter fields. Specializing to the BTZ geometry, an individual spool $\mathbb{W}_{P,\bar{P}}(A,\bar{A})$ accounts for the contribution of a single Wilson line winding an arbitrary number of times around the non-contractible spatial cycle. The matter partition function $Z_{\rm matter}$ is the sum over arbitrary insertions of indistinguishable spools
\begin{equation}
    Z_{\rm matter} = \exp\left(\frac{1}{4} \mathbb{W}_{P,\bar{P}}(A,\bar{A})\right) \,.
\end{equation}
The individual spool bears a striking resemblance to the JKMS sum over composite operators in \eqref{eq:MT-JT} and to the VTQFT amplitude on the right-hand side of \eqref{eq:spool_candidate}. It would be interesting to see if $\mathbb{W}$ can be computed in VTQFT.  

A promising development towards this end is the recent work \cite{Bourne:2026jbx} which computes the spool from an exact quantization of Teichmuller space. We hope that fully bridging these disparate approaches will bring us closer to the goal of deriving exact statistics for matter coupled to 3d gravity.

\paragraph{Symmetries and multi-trace ETH.}
It would be interesting to extend our analysis to systems with additional symmetries. Fermion parity, for example, leads to a distinction between correlations of matrix elements in different parity sectors, which could be probed by comparing ordinary and $(-1)^F$-inserted one-point spectral form factors. Time-reversal symmetry introduces additional index contractions, familiar from real and quaternionic second-order free probability \cite{Redelmeier_2012,redelmeier2015}, suggesting a corresponding extension of multi-trace ETH to orthogonal and symplectic ensembles. In gravity, fermion parity, time-reversal, and corresponding anomalies can modify spectral statistics even without propagating fermionic matter \cite{Stanford:2019vob,Boruch:2026hbr}. This raises the question of whether spin structures and anomalies impose additional constraints on multi-trace cumulants \cite{Onoda:2025yue}, and how phases from invertible topological quantum field theories affect the corresponding wormhole amplitudes.

\paragraph{Higher genus in JT gravity with matter.}
The multi-trace $F_{\O}^{(1|1)}$ correlations in JT allow the one-point SFF to depend on time, unlike the Gaussian ETH term, but do not by themselves produce the universal late-time ramp. This further requires correlations between the energy levels in the JT spectrum, as discussed in subsection \ref{subsec:rmt-eth}. It would be interesting to study these effects in higher-genus JT gravity with matter, starting from a cylinder containing a matter geodesic that connects the two boundaries with an additional handle, such as 
\begin{equation}
    K_{\O}^{\rm ramp}(\beta,T) \quad\overset{?}{=}\quad \inlinefig[5]{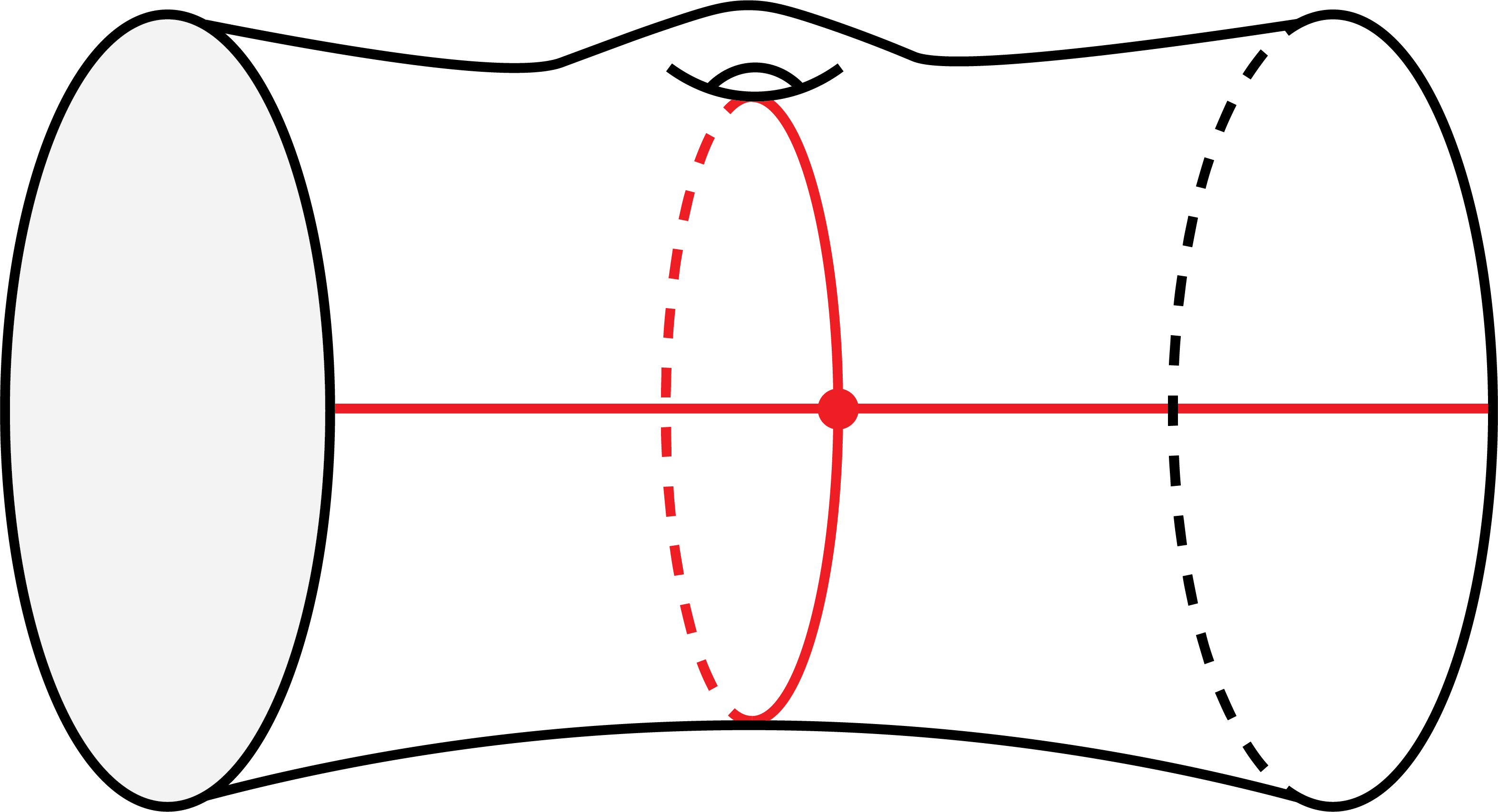} + \inlinefig[5]{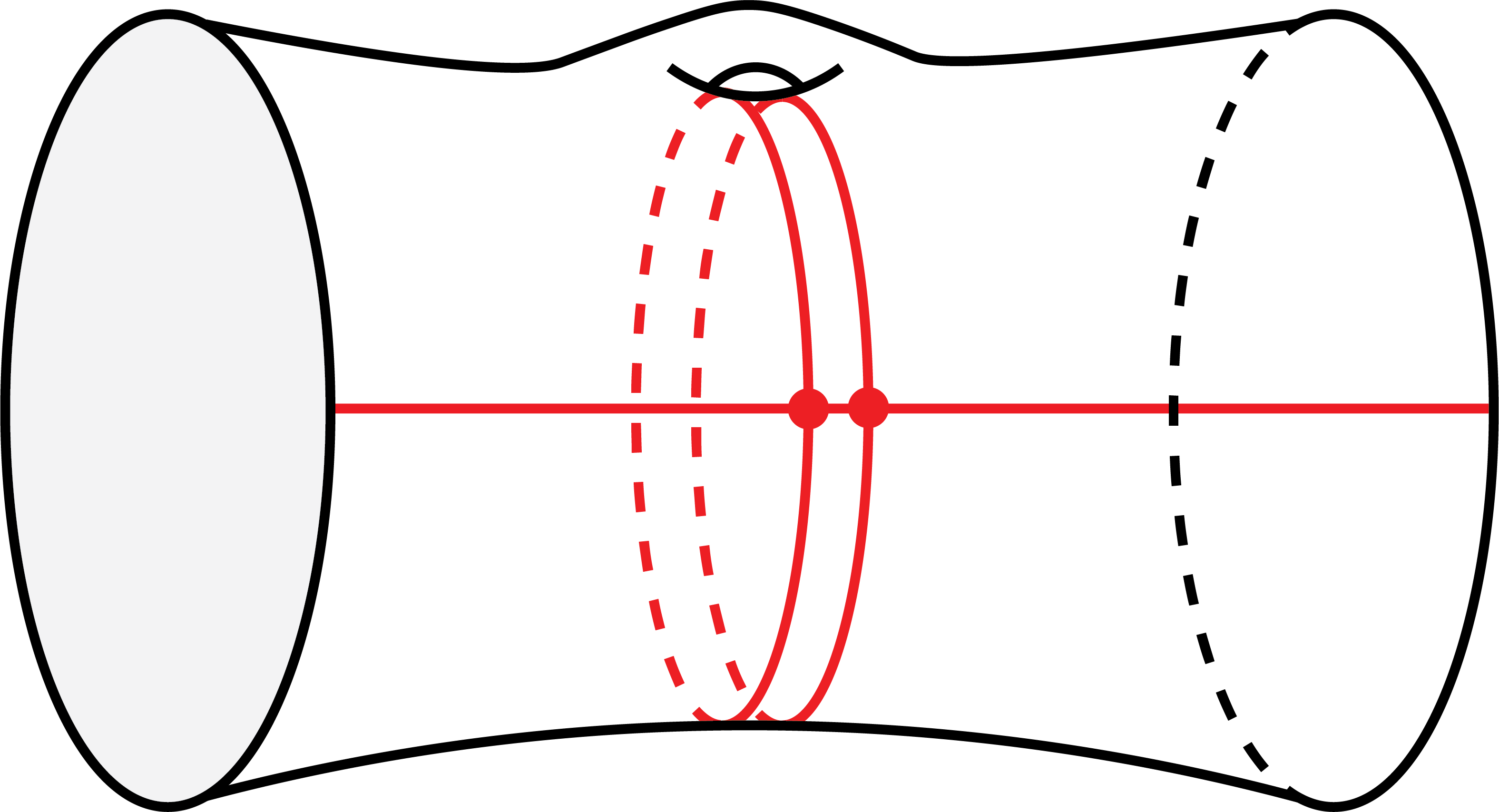} + \cdots\,.
\end{equation}
The spectral correlations must be included before integrating over the two energies, rather than multiplying the resulting cylinder amplitude by a separate spectral form factor.

\paragraph{Operator statistics below the black hole threshold.}
It is a striking fact that states below the black hole threshold can exhibit ensemble averaging \cite{Chandra:2022bqq,Schlenker:2022dyo}. In 3d gravity, this was demonstrated for conical defects whose scaling dimension lies between the multitwist threshold and the black hole threshold \cite{Chandra:2022bqq,Chang:2016ftb}. The mass of such states scales with an $O(1)$ fraction of the central charge in a semiclassical limit. A natural follow-up question is whether probe operators thermalize in a microcanonical window of these states. Unfortunately, one quickly runs into an obstacle when trying to study probe limits of these geometries since, below the multitwist threshold, the semiclassical saddles become complex \cite{Chang:2016ftb}. This reflects non-analyticities in a naive contour prescription for the vacuum fusion kernel. In appendix \ref{app:defects}, we consider the analytic properties of the fusion kernel in a regime of a light probe coupled to a background of heavy defects. We find that such dynamics do not admit even a standard ETH ansatz. For this claim we present two pieces of evidence: (1) the microcanonical degeneracy $e^S$ becomes $O(1)$ in this limit and (2) the smooth function $f_i$ has normal modes rather than the quasinormal modes typical of thermal systems.

\section*{Acknowledgements}
We thank Micha Berkooz, Jan Boruch,  Jeevan Chandra, Anatoly Dymarsky, Laura Foini, Luca Iliesiu, Ohad Mamroud, Silvia Pappalardi, Geoff Penington, Anatoly Polkovnikov, Boris Post, Klaus Richter, Martin Sasieta, Steve Shenker, Julian Sonner, Joaquin Turiaci, Juan Diego Urbina, and Cynthia Yan for valuable discussions and feedback. 
The authors thank the Leinweber Institutes for Theoretical Physics at UC Berkeley and Stanford University for their support.
AA is supported by the NSF Graduate
Research Fellowship Program under grant no. DGE 2146752.
GDU is supported
by the Japan Science and Technology Agency (JST) as
part of Adopting Sustainable Partnerships for Innovative
Research Ecosystem (ASPIRE), Grant No. JPMJAP2318. GDU is partly supported by
the Department of Energy, Office of Science, Office of
High Energy Physics through the award DE-SC0025522,
and by the Department of Energy through QuantISED
award DE-SC0019380. GDU thanks the Mainz Institute for Theoretical Physics (MITP) of the DFG Cluster of Excellence PRISMA+ (Project ID 390831469) for its hospitality during the workshop “Quantum Chaos and Holography”, where this work was first presented. 

\paragraph{Statement on AI Use:} We acknowledge assistance from OpenAI's GPT-6 and Anthropic's Claude Sonnet 5. These models were particularly helpful in manuscript revisions, checking mathematical derivations, and design of figure 5. The original ideas and main computations were carried out by the authors who take full responsibility for the validity of their work. 

\appendix

\section{Explicit JT blocks and cylinder formulas}\label{app:JT-interacting-blocks}
This appendix collects the technical formulas used in subsection \ref{subsec:JT-interacting}. The AdS radius is one. We keep the conventions of \eqref{eq:JT-conventions}:
\begin{equation}\label{eq:app-JT-conventions}
 E=\frac{s^2}{2},\qquad E_a=\frac{s_a^2}{2},\qquad
 \rho(s)=\frac{s\sinh(2\pi s)}{4\pi^2},\qquad
 \Gamma_{ab}^{h}=\frac{\Gamma(h\pm\ii s_a\pm\ii s_b)}{2^{2h+1}\Gamma(2h)}.
\end{equation}
Each independent sign is multiplied. The density is with respect to $ds$: the full disk measure is $e^{S_0}\rho(s)ds$, or $e^{S_0}\rho(s)dE/s$ in energy variables. In particular, the projector density used in the text is $\Pi_E=s\delta(E-H)$, with $s=\sqrt{2E}$; it satisfies $\int_0^\infty ds\,\Pi_E=\mathbf1$. An unstripped disk trace is $e^{S_0}\prod_r\rho(s_r)$ times its order-one kernel $\mathcal T_n$. The cylinder has $\chi=0$ and no overall $e^{S_0}$ factor.

\subsection*{The universal trivalent function and sewn blocks}
For positive matter dimensions, an explicit representation of the density-stripped PW trivalent function in our normalization is
\begin{equation}\label{eq:app-PW-F-integral}
\begin{split}
 F_{ijk}(s_1,s_2,s_3)
 ={}&\frac{\Gamma(\Delta_i\pm\ii s_1\pm\ii s_3)
 \Gamma(\Delta_j\pm\ii s_1\pm\ii s_2)}
 {2^{\Delta_i+\Delta_j+\Delta_k+1/2}
 \Gamma(\Delta_i+\Delta_j-\Delta_k)\Gamma(2\Delta_i)\Gamma(2\Delta_j)}\\
 &\times\int_0^\infty dx\,x^{\Delta_i+\Delta_j-\Delta_k-1}\\
 &\times{}_2F_1(\Delta_i+\ii s_1+\ii s_3,\Delta_i+\ii s_1-\ii s_3;2\Delta_i;-x)\\
 &\times{}_2F_1(\Delta_j-\ii s_1+\ii s_2,\Delta_j-\ii s_1-\ii s_2;2\Delta_j;-x).
\end{split}
\end{equation}
This integral converges when $\Delta_i+\Delta_j>\Delta_k$. Otherwise one first cyclically permutes $(i,j,k)$ and $(s_1,s_2,s_3)$ so that the inequality holds. The resulting function is cyclically invariant and gives its analytic continuation to the other orderings. The identity is excluded and is treated separately by the two-point contact term.

Equation~\eqref{eq:app-PW-F-integral} follows from the Bessel overlaps in appendix~A of \cite{Penington:2024sum}, with the densities and operator factors stripped before sewing. For example, the overlap underlying the calculation is
\begin{equation}\label{eq:app-Bessel-overlap}
\begin{split}
 &\int_{-\infty}^{\infty}d\chi\,
 K_{2\ii s}(2\sqrt{1+x}\,e^\chi)e^{2h\chi}K_{2\ii t}(2e^\chi)\\
 &\qquad=\frac{\Gamma(h\pm\ii s\pm\ii t)}{8\Gamma(2h)}
 (1+x)^{\ii s}
 {}_2F_1(h+\ii s+\ii t,h+\ii s-\ii t;2h;-x).
\end{split}
\end{equation}
The normalized Bessel states have measure $32\rho(s)$ in the $\chi$ coordinate. Relative to a unit-normalized matter primary in this overlap, the dressed operator is rescaled by $2^{-h-3/2}$; the two-point kernel is then exactly $\Gamma^h$ in \eqref{eq:app-JT-conventions}. Applying the same factors to the three-point overlap gives the power of two in \eqref{eq:app-PW-F-integral}. This fixes the normalization of $F_{ijk}$, rather than introducing an independent vertex constant.

For completeness, the closed expression is a sum of three ${}_4F_3(1)$ functions. Define, only in the following formula,
\begin{equation}\label{eq:app-PW-parameters}
\begin{gathered}
 p_1=\Delta_i+\ii s_1+\ii s_3,\quad
 p_2=\Delta_i+\ii s_1-\ii s_3,\quad
 p_3=\Delta_i+\Delta_j-\Delta_k,\\
 q_1=\Delta_k-\Delta_i-\ii s_1+\ii s_2,\quad
 q_2=\Delta_k-\Delta_i-\ii s_1-\ii s_2,\\
 c_1=2\Delta_i,\qquad c_2=\Delta_j+\Delta_k-\Delta_i.
\end{gathered}
\end{equation}
Then
\begin{equation}\label{eq:app-PW-F-hypergeometric}
\begin{split}
 F_{ijk}={}&\frac{\Gamma(\Delta_i-\ii s_1\pm\ii s_3)
 \Gamma(\Delta_j+\ii s_1\pm\ii s_2)}
 {2^{\Delta_i+\Delta_j+\Delta_k+1/2}\Gamma(p_3)}\\
 &\times\sum_{r=1}^3
 \frac{\Gamma(p_r)\prod_{\ell=1}^2\Gamma(q_\ell+p_r)
 \prod_{t\ne r}\Gamma(p_t-p_r)}
 {\Gamma(c_1-p_r)\Gamma(c_2+p_r)}\\
 &\times{}_4F_3\!\left(\begin{matrix}
 p_r,\ q_1+p_r,\ q_2+p_r,\ 1-c_1+p_r\\
 (1-p_t+p_r)_{t\ne r},\ c_2+p_r
 \end{matrix};1\right).
\end{split}
\end{equation}
Here $(1-p_t+p_r)_{t\ne r}$ lists the two denominator parameters; it is not a Pochhammer symbol. Coincident pole families are evaluated by a limit of the complete sum, not term by term. The real integral~\eqref{eq:app-PW-F-integral} is usually preferable at such points. To obtain \eqref{eq:app-PW-F-hypergeometric}, apply the Mellin transform
$\int_0^\infty dx\,x^{z-1}{}_2F_1(a,b;c;-x)=\Gamma(z)\Gamma(a-z)\Gamma(b-z)\Gamma(c)/[\Gamma(a)\Gamma(b)\Gamma(c-z)]$
to both factors. The remaining Barnes integrand is
\begin{equation}\label{eq:app-PW-Barnes}
 \frac{1}{2\pi\ii}\int_{c-\ii\infty}^{c+\ii\infty}dz\,
 \frac{\Gamma(z)\prod_{r=1}^3\Gamma(p_r-z)
 \prod_{\ell=1}^2\Gamma(q_\ell+z)}
 {\Gamma(c_1-z)\Gamma(c_2+z)},
 \quad
 \max(0,\Delta_i-\Delta_k)<c<\min(\Delta_i,p_3).
\end{equation}
Closing to the right gives the three terms in \eqref{eq:app-PW-F-hypergeometric}; each ${}_4F_3$ has parametric excess one and converges at unity for generic parameters.

For fixed external primaries $i,j,k,l$, the disk and cylinder blocks are therefore explicitly
\begin{align}
 \mathcal B_a^{\rm disk}(s_1,s_2,s_3,s_4)
 &=\frac{F_{ija}(s_1,s_2,s_4)F_{akl}(s_2,s_3,s_4)}{\Gamma_{24}^{\Delta_a}},
 \label{eq:app-PW-disk-block}\\
 \mathcal B_a^{\rm cyl}(s_L,s_R)
 &=\frac{F_{OOa}(s_L,s_R,s_L)F_{aOO}(s_R,s_R,s_L)}
 {\Gamma_{RL}^{\Delta_a}\Gamma_{LR}^{\Delta}}.
 \label{eq:app-PW-cylinder-block}
\end{align}
The connected subtraction is applied to the full OPE sum before repeated energies are imposed, as specified in the main text. In particular it is not applied separately to each matter block. When differentiating these blocks at fixed external dimensions, differentiate both trivalent factors and their internal propagator. The latter obeys
\begin{equation}\label{eq:app-Gamma-log-derivative}
 \partial_h\log\Gamma_{ab}^h
 =\sum_{\epsilon,\eta=\pm1}\psi(h+\epsilon\ii s_a+\eta\ii s_b)
 -2\log2-2\psi(2h),\qquad
 \psi(z)=\partial_z\log\Gamma(z).
\end{equation}

\subsection*{Wilson wavefunctions and the Racah kernel}
The explicit normalization of the Wilson wavefunctions is useful for checking the preceding constants. For $a_1,a_2,a_3,a_4$ with positive real parts and conjugate pairs, put $A=\sum_{j=1}^4a_j$ and
\begin{align}
 W_n(x;a_1,a_2,a_3,a_4)
 &=(a_1+a_2)_n(a_1+a_3)_n(a_1+a_4)_n\nonumber\\
 &\quad\times{}_4F_3\!\left(\begin{matrix}-n,n+A-1,a_1+\ii x,a_1-\ii x\\
 a_1+a_2,a_1+a_3,a_1+a_4\end{matrix};1\right),
 \label{eq:app-Wilson-polynomial}\\
 r_n&=\frac{n!(n+A-1)_n}{\Gamma(2n+A)}
 \prod_{j<k}\Gamma(a_j+a_k+n),
 \label{eq:app-Wilson-norm}\\
 P_n^{\Delta,\Delta}(s;s_1,s_3)
 &=\frac{2W_n(s;\Delta+\ii s_1,\Delta-\ii s_1,\Delta+\ii s_3,\Delta-\ii s_3)}{\sqrt{r_n}}
 \left[\prod_{j=1}^4\Gamma(a_j\pm\ii s)\right]^{1/2}.
 \label{eq:app-Wilson-wavefunction}
\end{align}
The last line uses $a_j=\Delta\pm\ii s_1,\Delta\pm\ii s_3$, and $(a)_n=\Gamma(a+n)/\Gamma(a)$ throughout. These formulas give $\int_0^\infty ds\,\rho(s)P_n(s)P_m(s)=\delta_{nm}$; the factor two converts the JKMS measure $4\rho$ to ours \cite{Jafferis:2022wez}. At $h_m=2\Delta+2m$, \eqref{eq:app-PW-F-integral} gives the check
\begin{equation}\label{eq:app-free-vertex-check}
 C_{OO[OO]_m}F_{OO[OO]_m}(s_1,s_2,s_3)
 =\big(2\Gamma_{12}^{\Delta}\Gamma_{13}^{\Delta}\Gamma_{23}^{h_m}\big)^{1/2}
 P_{2m}^{\Delta,\Delta}(s_1;s_2,s_3),
\end{equation}
with the phase of $[OO]_m$ chosen consistently with its Wilson wavefunction. 

The $\rho$-normalized $6j$ symbol is four times the JKMS symbol. Its explicit Wilson-function form is
\begin{equation}\label{eq:app-sixj-explicit}
\begin{split}
 \begin{Bmatrix}\Delta_1&k_1&x\\\Delta_2&k_2&\nu\end{Bmatrix}_{\rho}
 ={}&4\Big[\Gamma(\Delta_1\pm\ii k_1\pm\ii x)
 \Gamma(\Delta_1\pm\ii k_2\pm\ii\nu)\\
 &\qquad\times\Gamma(\Delta_2\pm\ii k_1\pm\ii\nu)
 \Gamma(\Delta_2\pm\ii k_2\pm\ii x)\Big]^{1/2}\\
 &\times W_\nu(x;\Delta_1+\ii k_1,\Delta_2+\ii k_2,
 \Delta_2-\ii k_2,\Delta_1-\ii k_1).
\end{split}
\end{equation}
To evaluate the Wilson function, set $g=(a+b+c+d)/2$, $\widetilde a=g-d$ and $\widetilde d=g-a$. Then \cite{Jafferis:2022wez}
\begin{equation}\label{eq:app-Wilson-function}
\begin{split}
 W_\nu(x;a,b,c,d)
 ={}&\frac{\Gamma(d-a)}{\Gamma(a+b)\Gamma(a+c)
 \Gamma(d\pm\ii x)\Gamma(\widetilde d\pm\ii\nu)}\\
 &\times{}_4F_3\!\left(\begin{matrix}
 a+\ii x,a-\ii x,\widetilde a+\ii\nu,\widetilde a-\ii\nu\\
 a+b,a+c,1+a-d
 \end{matrix};1\right)+(a\leftrightarrow d).
\end{split}
\end{equation}
The interchange also exchanges $\widetilde a$ and $\widetilde d$; singular individual terms are again combined before taking limits.

\subsection*{The Jacobi transform and the small-throat restriction}
The wavefunction used in \eqref{eq:W-main} is
\begin{equation}\label{eq:Jacobi-wavefunction-paper}
\begin{split}
 \Phi_s^{(\Delta)}(b)
 &=\frac{2^{2\Delta+1}\Gamma(\Delta)}{\Gamma(2\Delta)}
 \big[\Gamma(\Delta+2\ii s)\Gamma(\Delta-2\ii s)\big]^{1/2}\\
 &\quad\times{}_2F_1\!\left(\tfrac12+2\ii s,\tfrac12-2\ii s;
 \Delta+\tfrac12;-\sinh^2\tfrac b4\right).
\end{split}
\end{equation}
It is twice the JKMS wavefunction, so the kernel $\mathcal W_\Delta$ is four times theirs. The unrestricted identities are
\begin{align}
 \int_0^\infty db\,\mathcal W_\Delta(s_L,s_R;b)
 &=\frac{\delta(s_L-s_R)}{\rho(s_L)},
 \label{eq:app-Jacobi-orthogonality}\\
 \int_0^\infty db\,\mathcal W_\Delta(s_L,s_R;b)\chi_h(b)
 &=\begin{Bmatrix}h&s_L&s_R\\\Delta&s_R&s_L\end{Bmatrix}_{\rho}.
 \label{eq:app-Jacobi-character}
\end{align}
The first is distributional. The second converges for positive $h$ and $\Delta$ and is the JKMS character transform, appendix~E of \cite{Jafferis:2022wez}.

The self-energy saddle additionally requires $M\tanh(b/2)\gg1$, with $M=\Delta-1/2$. A controlled use of that approximation therefore keeps $b\ge b_{\min}>0$ and chooses $M\tanh(b_{\min}/2)\gg1$. The restriction on a $6j$ transform in the main text means explicitly
\begin{equation}\label{eq:app-restricted-sixj}
\begin{split}
 \left[\begin{Bmatrix}h&s_L&s_R\\\Delta&s_R&s_L\end{Bmatrix}_{\rho}\right]_{b\ge b_{\min}}
 &\equiv\int_{b_{\min}}^\infty db\,\mathcal W_\Delta(s_L,s_R;b)\chi_h(b)\\
 &=\begin{Bmatrix}h&s_L&s_R\\\Delta&s_R&s_L\end{Bmatrix}_{\rho}
 -\int_0^{b_{\min}}db\,\mathcal W_\Delta(s_L,s_R;b)\chi_h(b).
\end{split}
\end{equation}
This is an integration restriction, not a new matter block. Its $b_{\min}\to0$ limit is the ordinary $6j$ symbol, but the self-energy approximation is not uniform in that limit. For the full scalar determinant, loop-dependent terms also require a short-throat regulator. A concrete fixed-topology prescription is to integrate those terms from $b_{\min}$, leaving the no-loop Gaussian contact unrestricted. Different topology-changing completions need not give the same full cylinder amplitude. Character derivatives at fixed $b_{\min}$ commute with this restricted transform because the cutoff is independent of $h$.

\subsection*{Bulk propagators, counterterms, and the figure-eight}
For the canonically normalized bulk field and standard boundary conditions, take $\Delta>1/2$ in the $\phi^4$ calculation. The propagator in \eqref{eq:H2-green-main} can also be evaluated as
\begin{equation}\label{eq:app-bulk-hypergeometric}
\begin{split}
 G_{\mathbb H_2,h}(d)
 ={}&\frac{\Gamma(h)}{2\sqrt\pi\Gamma(h+1/2)}e^{-hd}
 {}_2F_1\!\left(\tfrac12,h;h+\tfrac12;e^{-2d}\right),\qquad d>0.
\end{split}
\end{equation}
In particular its large-distance coefficient is
$A_\Delta=\Gamma(\Delta)/[2\sqrt\pi\Gamma(\Delta+1/2)]$.
The boundary normalization in the radial limits of the main text is
\begin{equation}\label{eq:app-boundary-normalization}
 \mathcal N_\Delta=\frac{2^{-2\Delta-3/2}}{\sqrt{A_\Delta}}.
\end{equation}
Indeed a unit-normalized matter boundary primary has the factor $2^{-\Delta}/\sqrt{A_\Delta}$ in these radial coordinates, and the dressed-operator factor fixed above contributes $2^{-\Delta-3/2}$. The image sums are therefore explicitly
\begin{align}
 K_{R,b}(u;r,t)&=\mathcal N_\Delta A_\Delta
 \sum_{n\in\mathbb Z}[\cosh r\cosh(t-u+nb)-\sinh r]^{-\Delta},
 \label{eq:app-K-right}\\
 K_{L,b}(v;r,t)&=\mathcal N_\Delta A_\Delta
 \sum_{n\in\mathbb Z}[\cosh r\cosh(t-v+nb)+\sinh r]^{-\Delta}.
 \label{eq:app-K-left}
\end{align}
The free boundary-to-boundary correlator is
\begin{equation}\label{eq:app-free-LR}
 G_{LR,b}^{(0)}(u,v)
 =2^{-2\Delta-3}\sum_{n\in\mathbb Z}
 \left[2\cosh\frac{u-v+nb}{2}\right]^{-2\Delta}.
\end{equation}
The tadpole correction divided by this free correlator is independent of the external normalization.

Point splitting gives
\begin{equation}\label{eq:app-renormalized-AdS-tadpole}
 G_{\mathbb H_2}^{\rm ren}(0)
 =\frac{\log(2\mu)-\gamma_{\rm E}-\psi(\Delta)}{2\pi},
\end{equation}
where $\gamma_{\rm E}$ is Euler's constant and $\mu$ is the scale in the main-text subtraction. Writing $G_0=G_{\mathbb H_2}^{\rm ren}(0)$ locally in the following equation, the order-$\lambda$ counterterms are
\begin{equation}\label{eq:app-explicit-counterterms}
 I_{\rm ct}=\int_{X_b}d^2y\sqrt g\left[\frac{\delta m^2}{2}\phi^2+\delta\Lambda\right],
 \qquad \delta m^2=-\frac\lambda2G_0,
 \quad \delta\Lambda=\frac\lambda8G_0^2.
\end{equation}
These implement normal ordering relative to the AdS vacuum, fix the external dimension, and remove its vacuum-energy correction. For $G_b^{\rm ren}(y,y)=G_0+T_{\Delta,b}(y)$, Wick contraction gives
\begin{equation}\label{eq:app-vacuum-subtraction}
 -\frac\lambda8(G_0+T)^2-\frac{\delta m^2}{2}(G_0+T)-\delta\Lambda
 =-\frac\lambda8T^2.
\end{equation}
This proves the renormalized figure-eight integrand used in the main text.

The propagator-product coefficients are
\begin{equation}\label{eq:an-KL}
 a_n^{\rm KL}=\frac{(\frac12)_n}{2\sqrt\pi\,n!}
 \frac{(2\Delta+2n)_{1/2}(2\Delta+n)_n}
 {(\Delta+n)_{1/2}^{\,2}(2\Delta+n-\frac12)_n}.
\end{equation}
They apply to $G_{\mathbb H_2,\Delta}(d)^2=\sum_n a_n^{\rm KL}G_{\mathbb H_2,2\Delta+2n}(d)$, for $d>0$, in the canonical bulk normalization \cite{Kraus:2020nga}; they are not boundary OPE coefficients. The required image integral is
\begin{equation}\label{eq:app-image-character-integral}
 \int_{X_b}d^2y\sqrt g\,G_{\mathbb H_2,h}(y,\sigma_b^w y)
 =\frac{b}{2h-1}\chi_h(|w|b),\qquad w\ne0.
\end{equation}
For $w=1$, this yields \eqref{eq:character-propagator}, and hence
$\gamma_n^{(2)}=\lambda a_n^{\rm KL}/[2(2h_n-1)]$
with $h_n=2\Delta+2n$. At $\Delta=1$ it gives $a_0^{\rm KL}=3/(2\pi)$ and $\gamma_0^{(2)}=\lambda/(4\pi)$, providing an independent normalization check.

\section{VMFT and heavy asymptotics}\label{app:vmft}
The main primitives in our study of thermalization in AdS$_3$/CFT$_2$ are a universal state density $\rho_0$ and a universal 3-point function $C_0$. Both are defined through a degeneration limit of the Ponsot-Teschner fusion kernel \cite{Ponsot:1999uf}\footnote{\label{fn:sq2}
    This only defines $\rho_0$ and $C_0$ up to an overall constant. We use the conventions of \cite{Eberhardt:2023mrq} times a factor of $\sqrt{2}$. 
}
\begin{equation}
    \fker{\mathbbm 1}{P_i}{P_j}{P_k}{P_j}{P_k} = \rho_0(P_i)C_0(P_i,P_j,P_k)\,.
\end{equation}
In gapped theories, these functions describe the universal kinematics of the leading Regge trajectory in Virasoro mean field theory. This appendix describes  limits of interest to the main text. A more complete discussion can be found in \cite{Collier:2018exn} (see also \cite{Ponsot:2000mt} for a more mathematical approach). 

A relevant tool in the study of these primitives is the meromorphic Barnes double gamma function, defined recursively by
\begin{equation}
    \Gamma_b(Q/2)=1 \quad{\rm and}\quad  \Gamma_b(x+b^{\pm 1})=\frac{\sqrt{2\pi}b^{\pm (b^{\pm 1}x-1/2)}}{\Gamma(b^{\pm 1}x)}\Gamma_b(x)\,,
\end{equation}
along with the related quantity
\begin{equation}
    S_b(x) = \frac{\Gamma_b(x)}{\Gamma_b(Q-x)} \,.
\end{equation}
These functions are useful because there are known asymptotics for $\Gamma_b$:

\paragraph{Ultraheavy Limit} If $\Re (x)>0$, then, as $|x|\to\infty$, 
\begin{equation}\label{eq:gammab_largex}
    \log\Gamma_b(x)\sim -\frac{1}{2}x^2\log x+\frac{3}{4}x^2+\frac{Q}{2}x\log x-\frac{Q}{2}x-\frac{Q^2+1}{12}\log x+\log\Gamma_0(b)\,,
\end{equation}
where $\Gamma_0(b)$ is a constant with no $x$ dependence. This corresponds to an ultraheavy chiral dimension $P\gg 1$. 

\paragraph{Macroscopic BTZ Limit} Alternatively, we can take a large central charge limit ($b\to 0$) of $\Gamma_b$ to get
\begin{equation}\label{eq:gammab_largeb}
    \log\Gamma_b(b^{-1}x+b/2) \sim \frac{1}{2b^2}\left(\frac{1}{2}-x\right)^2\log b+\frac{2x-1}{4b^2}\log(2\pi)-\frac{1}{b^2}\int_{1/2}^x dt\log\Gamma(t)\,.
\end{equation}
Equivalently,
\begin{equation}
    \log\Gamma_b(y) \sim \frac{(1+b^2-2yb)^2}{8b^2}(\log(b)+\frac{y}{2b}\log(2\pi)-\frac{1}{b^2}\psi_{-2}(yb-b^2/2))+\cdots\,,
\end{equation}
where $\psi_{(n)}(z)$ is the Polygamma function of order $n$ and $y\sim O(b^{-1})$. The ellipsis obscures constant terms that are irrelevant for our purposes.\footnote{
    The exact value is
    \begin{equation}
        \frac{1}{b^2}\left(\frac{5}{24}\log(2)+\frac{3}{2}\log(A)+\frac{1}{4}\log(\pi)\right)-\frac{1}{4}\log(2\pi)
    \end{equation}
    where $A$ is Glaisher's constant.
} 
The universal formula calls for taking $x=1/2+iPb$ or $x=1+2iP/b+b^2/2$. These are cases where the chiral dimension is macroscopic $P\sim O(b^{-1})$ in units of the central charge.

\paragraph{$\mathfrak{sl}(2)$ Limit} Finally consider the limit
\begin{equation}\label{eq:gammab_sl2}
    \lim_{b\to 0}\Gamma_b(b\Delta) = (2\pi b)^{\Delta/2-1}b^\Delta\Gamma(\Delta) \,,
\end{equation}
which is useful for the study of $\mathfrak{sl}(2)$ blocks. These blocks describe parametrically light matter in a semiclassical or near-extremal limit of an asymptotically AdS$_3$ background.

\subsection{A Universal State Density}\label{app:vmft_rho0}
The universal density of states is the Plancherel measure of the quantum group $\cU_q(\mathfrak{sl}(2,\R))$ \cite{Ponsot:1999uf}
\begin{equation}\label{eq:rho0}
    \rho_0(P) = |S_b(Q+2iP)|^2 = 4\sinh(2\pi bP)\sinh(2\pi b^{-1}P) \,,
\end{equation}
with deformation parameter $q=e^{i\pi b^2}$. The momentum $P$ labels a continuous series representation of the $q$-deformed group. Equivalently, the density is a degeneration limit of the modular kernel $\rho_0(P)=\sker{\mathbbm 1}{P}/\sqrt{2}$ (c.f. footnote \ref{fn:sq2}). 

\paragraph{Ultraheavy Limit} A central assumption of local rotational invariance is that the number of states $L_I$ needed to compute a correlation function in a small microcanonical window $I$ only depends on the average energy in the window. We demonstrate this property on the space of representations in two limits of physical interest. First, consider the asymptotics of large chiral dimension. Here the density exhibits the expected Cardy behavior
\begin{equation}
    \lim_{P\to\infty}\rho_0(P) = e^{2\pi QP} \,.
\end{equation}
Let $\bar{P}=\frac{1}{|I|}\sum_{a\in I}P_a$ be the average momentum in the window $I$. At large chiral dimension,
\begin{equation}
    \prod_I\frac{1}{\rho_0(P_a)} \approx e^{-2\pi Q\sum_I P_a} \,, 
\end{equation}
where here ``$\approx$'' denotes the leading behavior in a large $P_a$ expansion. Stated more simply, the product of densities is expressible as a function of the density of the average. For example, 
\begin{equation}
    L_I = e^{2\pi Q P_{jk}}\,,
\end{equation}
if the window contains two momenta with average $P_{jk}=(P_j+P_k)/2$. 

\paragraph{Macroscopic BTZ} A similar result holds within a microcanonical window in a semiclassical limit we refer to as the macroscopic BTZ regime. This limit is defined by taking
\begin{equation}\label{eq:macroscopic_btz}
    P_j = p_j / b \quad{\rm and}\quad P_j - P_k = b \omega_{jk} \,, 
\end{equation}
where the $\{P_a\}$ are chosen such that $p_j,p_k,\omega_{jk}\sim O(1)$ as $b\to 0$. Define the average semiclassical energy $\bar{p}=\sum_a p_a$. In this limit
\begin{equation}
    \prod_{I}\frac{1}{\rho_0(p_a/b)} \approx e^{-2\pi (\sum_I p_a)/b^2} \,,
\end{equation}
where here ``$\approx$'' denotes the leading behavior in a small $b$ expansion. It is clear that the window size is a function of the average semiclassical energy. For example,
\begin{equation}
    L_I = e^{2\pi p_{jk}/b^2}
\end{equation}
for an average rescaled average energy $p_{jk}=(p_j+p_k)/2$. Near-extremal windows are slightly more subtle and we defer further discussion to appendix \ref{app:ne_ops}.

\subsection{A Universal 3-Point Function}\label{app:vmft_C0}
The universal structure constant is \cite{Dorn:1994xn, Zamolodchikov:1995aa}
\begin{equation}\label{eq:C0}
    C_0(P_i,P_j,P_k) = \frac{\Gamma_b(2Q)}{\Gamma_b(Q)^3}\frac{\prod_{\pm}\Gamma_b(Q/2\pm iP_i\pm iP_j\pm iP_k)}{\prod_{a\in\{i,j,k\}}\Gamma_b(Q+2iP_a)\Gamma_b(Q-2iP_a)}\,.
\end{equation}
Below we consider interesting limits of this formula. In these limits, $C_0$ fixes all the necessary primitives to determine the functional dependence of a standard ETH ansatz (c.f. footnote \ref{fn:sq2}). 

\paragraph{Ultraheavy Limit} When at least two momenta are heavy, say $P_j,P_k\gg 1$, the universal structure constant gives an estimate for the matrix statistics of the operator $\cO_i$ dual to the momentum $P_i$
\begin{equation}
    \log C_0(P_i,P_j,P_k) = -2\pi Q P + 2(P_i^2+Q^2/4-\omega^2)\log(2P) + O(P^0)\,,
\end{equation}
where $P = (P_j+P_k)/2$ and $\omega=P_j-P_k$. This formula follows from large argument asymptotics of $\Gamma_b$ in \ref{eq:gammab_largex}. In terms of the standard ETH ansatz, it corresponds to the following smooth functions over finite energy windows
\begin{equation}\label{eq:heavy}
    S(P)=2\pi QP \quad{\rm and}\quad |f_i(P)|^2 \propto (2P)^{4h_i^2}\,.
\end{equation}
The leading correction to this formula describes contributions to the universal coefficient that are not enhanced by taking $P$ large
\begin{align}
    &\log C_0(P_i,P_j,P_k) + 2\pi QP - 2 (h_i-\omega^2)\log(2P) = \nn\\
    & -\frac{1}{2}\log(2) + \log\Gamma_b(2Q) - 3\log\Gamma_b(Q) + 4\log|\Gamma_b(Q/2+iP_i)| - 2\log|\Gamma_b(Q+2iP_i)| + O(P^{-1}) \,,
\end{align}
and further contributions are suppressed in large $P$ and small $\omega$. 

\paragraph{Macroscopic BTZ} The three-point function exhibits a universal divergence when we take the central charge large without also scaling the momenta 
\begin{equation}\label{eq:naive_semiclassical_C0}
    \log C_0(P_i,P_j,P_k) = \frac{1}{b^2}\left(1+\frac{1}{3}\log(\frac{2}{A^{36}})\right) -8(P_i^2+P_j^2+P_k^2)\log(2)-\frac{3}{2}\log(2\pi)+O(b^2)\,.
\end{equation}
It is difficult to see how such a formula could be placed in the form of an ETH ansatz since, even if the divergent constant were removed by introducing a regulator to the action \eqref{eq:z_grav}, kinematics still only contribute at order $O(1)$. A more interesting $b\to 0$ limit requires defining a semiclassical scaling procedure. We consider three such limits. The near-extremal regime is covered in appendix \ref{app:ne_ops} and the non-universal case of heavy conical defects is discussed in appendix \ref{app:defects}. Here we consider momenta in the macroscopic BTZ regime defined in \eqref{eq:macroscopic_btz}. The operator $\cO_i$ is treated as a light probe by taking its scaling dimension only slightly above the vacuum. 

Such a limit was considered in \cite{Collier:2019weq} to smooth accumulation kinks appearing in $C_0$ when the difference $\omega$ is not also appropriately scaled.  Restricting to an $O(b)$ difference gives
\begin{equation}
    C_0(\alpha_i=b\Delta,\ p=b(P_j+P_k)/2,\ \omega=b^{-1}(P_j-P_k)) = e^{-\tilde{S}(p)}|f_i(p,\omega)|^2 \,.
\end{equation}
Like \eqref{eq:naive_semiclassical_C0}, this limit contains a universal divergent constant at leading order which can be regulated. Removing this constant reveals an exponential scaling in the regulated entropy
\begin{equation}\label{eq:btz_s}
    S(p) = \tilde{S}(p)-S_0 = 2\pi p/b^2 \,,
\end{equation}
and the smooth function 
\begin{equation}\label{eq:btz_fi}
    |f_i(p,\omega)|^2 = (2\pi b)^{-5/4}\left[\frac{(2p)^{2\Delta}}{2\sinh(2\pi p)}\right]\frac{\Gamma(\Delta+i\omega)\Gamma(\Delta-i\omega)}{\Gamma(2\Delta)}\,,
\end{equation}
here computed to leading order in a large central charge.\footnote{ 
    The exact value of the regulator is
    \begin{equation}\label{eq:C0_regulator}
        b^2S_0 = \frac{3}{2 }\log(A)-\frac{1}{24 }\log(2/b^3)\,.
    \end{equation}
}
These formulas were computed using the semiclassical limits \eqref{eq:gammab_largeb} for the heavy sector and \eqref{eq:gammab_sl2} for the light sector. The latter reveals the expected spectrum of quasinormal modes, here appearing at imaginary $\omega$. 

\section{Near-extremal limits}\label{app:ne}
At fixed spin, a 2d CFT with large central charge and no chiral symmetries contains a Schwarzian sector \cite{Ghosh:2019rcj, Pal:2023cgk}. This is the CFT equivalent of AdS$_2$ universality. Mathematically, this follows because the quantization of certain Virasoro coadjoint orbits reduces to the dynamics of a particle on the $SL(2,\R)$ group manifold semiclassically. The latter can be described by the Schwarzian.\footnote{
    There are many references for derivations of the kinematics of the Schwarzian particle. Here are a few: \cite{Stanford:2017thb, Mertens:2017mtv, Kitaev:2017hnr, Mertens:2018fds, Kitaev:2017awl, Yang:2018gdb}.
} Thus, results in this sector are computed by a Schwarzian quantum mechanics to leading order in the central charge. In Liouville variables, this is accomplished by taking the Liouville coupling $b$ to be small. Equivalently, the central charge of the theory is taken large. This section clarifies how the semiclassical regime isolates the Schwarzian kinematics of a fixed spin sector. Specifically, we outline how the main primitives required to study the multi-trace ETH --- the density of states, the cyclic contraction, and the non-cyclic contraction --- reduce in the near-extremal limit. 

\subsection{States, Operators, and Bulk Interactions}\label{app:ne_ops}
\paragraph{States} The relevant sector is accessed by considering Liouville momenta parametrically close to extremality
\begin{equation}\label{eq:ne_state}
    P = bs \,,
\end{equation}
where $s$ is taken $O(1)$ as $b\to 0$. This labels a state with large chiral dimension ($h\sim O(1/b^2)$) in the semiclassical limit. To leading order in the spectral gap, the density of states is given by the Plancherel measure \eqref{eq:rho0}. In the Schwarzian sector this becomes
\begin{equation}\label{ne_density}
    \rho_0(bs) \approx 8\pi b^2s\sinh(2\pi s) \,,
\end{equation}
which is the $\mathfrak{sl}(2,\R)$ Plancherel measure. The approximation ``$\approx$'' denotes the leading behavior in a small $b$ expansion. This matches the Schwarzian density of states up to a $b$-dependent pre-factor. For simplicity, this factor is absorbed into the state-normalization
\begin{equation}
    \braket{s}{s'} \equiv 32\pi^3b^2\frac{\delta(s-s')}{\rho_0(bs)}\,,
\end{equation}
recovering the conventions in section \ref{sec:JT}.

\paragraph{Operators} The Schwarzian sector can be probed using light operators with complex momentum
\begin{equation}\label{eq:ne_op}
    \alpha = b\Delta \,,
\end{equation}
where $\Delta$ is taken $O(1)$ as $b\to 0$. The 3-point vertex follows by application of the $\mathfrak{sl}(2)$ limit \ref{eq:gammab_sl2} to the universal formula \eqref{eq:C0}
\begin{equation}\label{eq:ne_3pt}
    \mel{P_1=bs_1}{\O}{P_2=bs_2} \approx \frac{b^{4\Delta}}{(2\pi b)^3}\frac{\Gamma(\Delta\pm is_1\pm is_2)}{\Gamma(2\Delta)} \,,
\end{equation}
but there are some subtleties. The $b$-dependent prefactor can be absorbed into the operator normalization. More importantly, this matrix element appears to be the square of the 3-point vertex discussed in section \ref{sec:JT}. This is because $\cO$ corresponds to a bilocal operator on the disk. Bilocals were discussed in \cite{Mertens:2017mtv} where JT correlators were computed using cylinder transition amplitudes in Liouville theory with ZZ boundary conditions. That computation is matched to the torus two-point function in section \ref{subsec:full_eth_3d} by Cardy's unfolding trick. The upshot is that \eqref{eq:ne_3pt} should be identified with the square of the matrix element on the disk. 

Still, there is a missing factor of $2^{-\Delta}$ relative to the matrix elements of operators on the disk (as seen for example in (6.80) of \cite{Yang:2018gdb}). This is because of the change in coordinates $z=-2t$ between the complex modulus $z$ used in conformal blocks and the asymptotic time $t$ used to parametrize the Schwarzian.\footnote{ 
    Since operators are modular forms, they transform covariantly. For example, an operator with chiral weight $h$ transforms as
    \begin{equation}
        \cO(f) = \left(\frac{\partial f}{\partial z}\right)^{-h}\cO(z) \,,
    \end{equation}
    under reparametrizations, which accounts for the missing factor. 
} Taking all of these factors into account, we can adjust the normalization of each probe such that
\begin{equation}
    \mel{s_1}{\cO}{s_2} \equiv \Gamma^\Delta_{12}\,,
\end{equation}
in the near-extremal limit. The reader should note that, unless explicitly stated, such operators are only used in a fixed probe limit. The inclusion of parametrically light dynamical matter moves us out of the realm of pure gravity. 

Now recall that JT gravity is dual to a double-scaled matrix integral. The size of the matrix $L_J$ is fixed in a sector of spin $J$ \cite{Ghosh:2019rcj}. This size is parametrized by the zero-point shift in the entropy 
\begin{equation}
    L_J = e^{S_0(J)} \quad{\rm with}\quad S_0(J) = 2\pi\sqrt{\frac{|J|}{4G_N}} \,,
\end{equation}
and is the physical parameter which controls the topological expansion. This draws a subtle but important contrast with the universal density $\rho_0$ discussed in appendix \ref{app:vmft_rho0}. Here
\begin{equation}
    \mel{s_1}{\cO}{s_2} = \sqrt{\frac{1}{L_J}}f_\cO(s_1,s_2)R_{12} \quad{\rm with}\quad |f_\cO(s_1,s_2)|^2=\Gamma^\Delta_{12} \,,
\end{equation}
where $\cO$ refers to a single insertion on the disk (not to the bilocal discussed earlier). Unlike $\rho_0$, the Schwarzian density of states does not compute the window size $L_J$. This amounts to bookkeeping when computing correlators but is important for verifying the validity of free probability to the ETH.

\paragraph{Bulk Interactions} Given the density of states and the 3-point function, one can compute any time-ordered correlator in the theory. Computing out-of-time-order correlators requires defining a near-extremal limit of the fusion kernel. It suffices to define a limit of the Virasoro 6j-symbol since the two are related by the universal structure constant defined in \eqref{eq:C0}. Taking 
\begin{equation}
    \begin{tabular}{ c c c c }
        $\alpha_i = b\Delta_i$, & & $P_j = bs_j$\,, \\
        $\alpha_\chi = b\Delta_\chi$, & & $P_k = bs_k$\,,
    \end{tabular}
\end{equation}
the quantum 6j symbol admits a limit
\begin{equation}\label{eq:ne_otoc}
    \begin{Bmatrix}
        \Delta_i & s_j & s_j \\
        \Delta_\chi & s_k & s_k
    \end{Bmatrix} = \lim_{b\to 0}2\pi b^3\begin{Bmatrix}
        P_i & P_j & P_j \\
        P_\chi & P_k & P_k
    \end{Bmatrix}_b\,,
\end{equation}
to the $SL(2,\R)$ 6j symbol denoted on the left. See B.2 of \cite{Mertens:2017mtv} for a derivation of this limit and appendix \ref{app:JT-interacting-blocks} for a more thorough discussion of the Racah kernel. 

\subsection{Blocks}\label{app:ne_blocks}
The three-point function and 6j symbol were the main inputs needed to explain the non-Gaussianities of JT gravity coupled to dynamical matter discussed in the main text. To explain the near-extremal reduction of 3d saddles we also need to understand how to take a near-extremal limit of the conformal blocks. 

Conformal blocks nicely package the data associated to the descendants of conformal primaries \cite{Eberhardt:2023mrq}. A block 
\begin{equation}
    \cF^{\cC}_{g,n}(\{P_{\rm ext}\}; \{P_{\rm int}\}\ |\ \vec{z})
\end{equation}
is associated to a trivalent diagram $\cC$ on a Riemann surface $\Sigma_{g,n}$ of genus $g$ with $n$ boundaries. One may equivalently label the block by the dual graph of $\cC$ on $\Sigma_{g,n}$. Each edge of $C$ is labeled by a primary $P\in\{P_{\rm int}\}$ while the boundaries are labeled by some $P\in\{P_{\rm ext}\}$. The block is a function of the moduli $\vec{m}$ of $\Sigma_{g,n}$. Conformal blocks can be constructed recursively in the degeneration limit of cycles in the dual graph of $\cC$. There are only two possibilities in such a limit. Either $\Sigma_{g,n}$ is split into two disjoint manifolds $\Sigma_{g_1,n_1}\sqcup\Sigma_{g_2,n_2}$ such that $g_1+g_2=g$ and $n_1+n_2=n+2$,
or the manifold gains two extra boundaries yielding some $\Sigma_{g-1,n+2}$.
Any descendant of $P$ can flow across the cut cycle. The gluing is performed by tracing the choice of descendant across the cut over the Verma module $\cV_P$. 

Below we discuss near-extremal limits of two classes of conformal blocks relevant for the main text. We will often drop the channel label $\cC$ or the momenta labels $P$ when the implication is clear. 

\paragraph{$\mathbf{\Sigma_{1,1}}$}
Consider the once-punctured torus conformal block
\begin{equation}
    \cF_{1,1}(P_i,P_j\ |\ \tau) = \braket{\tau}{\inlinefig{imgs/s6/O.png}} \,.
\end{equation}
The torus is a complex manifold with periodic identifications $z\sim z+1\sim z+\tau$. The exponential map $w = e^{2\pi iz}$ relates this to a complex structure on the plane with identification $w \sim qw$ where $q=e^{2\pi i\tau}$ is the nome of the torus. There is a conformal anomaly associated to this transformation
\begin{equation}
    T(w) = T(w(z)) - \frac{c}{12}\{w,z\} = T(z) - \frac{c}{24} \,,
\end{equation}
which shifts all Virasoro generators and the action of chiral dilations by a constant
\begin{equation}
    L_n \mapsto L_n - \frac{c}{24}\delta_{n0} \quad{\rm and}\quad q^{L_0} \mapsto q^{L_0-c/24}\,,
\end{equation}
meaning the Hamiltonian on the plane is not a simple reparametrization of the Hamiltonian on the torus.

In the degeneration limit $q\to 0$, the torus conformal block becomes a sum over blocks of the thrice-punctured sphere. The anomaly is a small price to pay since 3-point blocks are exceedingly simple on the plane.  The torus block is recovered by tracing over the choice of descendant through the degenerate cycle
\begin{equation}
    \cF_{1,1}(P_i;P_j|q) = \sum_{n\in\Z} q^{h+n-c/24}\sum_{p_j\in\cV^n_j}\cN(P_i,p_j,p_j)\cF_{0,3}(P_i,p_j,p_j)\,.
\end{equation} 
The first sum is taken over each level of the Verma module $\cV_j=\bigoplus_{n\geq 0}\cV^n_j$. The levels form orthogonal subspaces of the module and the power of the nome associated to each summand depends only on the level
\begin{equation}
    q^{h+n-c/24} = \exp\left[2\pi i\tau\left(P^2+n-1/24\right)\right] \,.
\end{equation}
The second sum labels a basis of states of $\cV^n_j$. The function
\begin{equation}
    \cN(P_i,p_j,p_k) = \frac{\mel{p_j}{\cO_i}{p_k}}{C_0(P_i,P_j,P_k)} \,,
\end{equation}
computes the ratio of three-point functions between arbitrary descendents $p_j\in\cV^n_j,p_k\in\cV^n_k$ and their respective primaries $P_j$ and $P_k$. From hereon we unit normalize the conformal block on $\Sigma_{0,3}$ (ie: $\cF_{0,3}=1$).

Now let us consider the near-extremal limit. The central charge is taken large by sending $b\to 0$ while operators and states are scaled as in equations \ref{eq:ne_state} and \ref{eq:ne_op}. This regime is realized as a controlled approximation by setting
\begin{equation}
    2\pi i \tau b^2 = -\beta\,,
\end{equation}
with $\beta=O(1)$. Powers of the nome simplify the summands above greatly
\begin{equation}
    q^{h+n-c/24}\cF = e^{-\beta s^2}\times(1 + O({\rm poly}(1/b))e^{-\#/b^2})\,,
\end{equation}
where by convention we have stripped the Casimir factor. We have also used that the 3-point function of the descendents $\mel{p_j}{\cO_i}{p_j}$ is related to the primary 3-point function $C_0(P_i,P_j,P_j)$ by a polynomial function of the weights. This function may be determined exactly using local Ward identities but, for our purposes, it suffices to note that this function scales sub-exponentially. It follows that, up to corrections that are exponentially small in the central charge, 
\begin{equation}
    \cF_{1,1}\left(b\Delta_i ;bs_j\ \Big|\ i\frac{\beta}{2\pi b^2}\right) = e^{-\beta s_j^2} \,.
\end{equation}
Note that the dependence on the external operator $\cO_i$ appears to have dropped out. This is due to our choice of normalization for $\cF$, here captured in the definition of $\cN$. We elect to absorb all external operator dependence into the three-point functions computed in equation \ref{eq:ne_3pt}. 

\paragraph{$\mathbf{\Sigma_{1,2n}}$}
The plumbing and sewing construction of conformal blocks extends to arbitrary punctured Riemann surfaces \cite{Cho:2017oxl}. In the case of torus blocks, it can be argued that the vacuum is dominant for some choice of moduli when the number of operator insertions is even.\footnote{
    Treating the case with an odd number of punctures requires additional assumptions on the OPE coefficients of the theory. The associated correlation functions do not have a clear near-extremal limit. 
} Here we explicitly demonstrate vacuum dominance of the 4-punctured time-ordered torus conformal block and leave the general case implicit. We work in the right-moving sector of section \ref{subsec:full_eth_3d} where $\beta_R\sim O(1/c)$. See section 3.5 of \cite{Ghosh:2019rcj} for a similar discussion.

Choose a representative for each operator insertion with $0\leq\Re(z)<1$ and order the insertions according to their real parts.  Let $a,b\in\{1,2,3,4\}$, $w_a=e^{2\pi iz_a}$, and $\cC$ the longitudinal necklace channel decomposition of $\Sigma_{1,4}$. Then
\begin{align}
    &C_0(P_i,P_1,P_2)C_0(P_i,P_2,P_3)C_0(P_i,P_3,P_4)C_0(P_i,P_4,P_1)\times\cF_{1,4}^{\cC}(P_i;P_1,P_2,P_3,P_4|w_1,w_2,w_3,w_4) \nn\\
    &\quad = \sum_{n_a\in\Z}\sum_{p_a\in\cV_a^n}\bra{p_1}q_1^{L_0-c/24}\cO_i \ketbra{p_2}q_2^{L_0-c/24}\cO_i\ketbra{p_3}q_3^{L_0-c/24}\cO_i\ketbra{p_4}q_4^{L_0-c/24}\cO_i\ket{p_1} \nn\\
    &\quad = \sum_{n_a\in\Z}\left(\prod_{b}q_b^{h_b+n_b-c/24}\right)\times\sum_{p_a\in\cV_a^n}\bra{p_1}\cO_i\ketbra{p_2}\cO_i\ketbra{p_3}\cO_i\ketbra{p_4}\cO_i\ket{p_1} \,.
\end{align}
Like before, $\cV_a^n$ is the level $n$ subset of the Verma module with highest weight $h_a$. Dividing the right-hand side by the factors of $C_0$ recovers the function $\cN$ defined above. We therefore focus on the term in parenthesis since it is the only part of the final expression which scales exponentially as $b\to 0$. 

The moduli of this block specify the locations $q_a=e^{2\pi i t_a}$ of each operator insertion. Place the initial insertion at the origin so that $\sum t_a\sim 1$ with $t_2=z_2$ and $t_1=1-z_4$. In the near-extremal limit, $\tau$ approaches the cusp at $0$ and
\begin{align}
    \prod_{b}q_b^{h_b+n_b-c/24} &\approx \exp\left[2\pi i\left(\sum_a^{2n} t_aP_a^2-1/24\right)\right] 
    \,.
\end{align}
It is useful to $\mathbb{S}$-transform $\tau$ to the dual meridian necklace channel $\tilde{\cC}$ where the torus is the lattice quotient
\begin{equation}
    z\sim z-1/\tau\sim z-1 \quad{\rm or}\quad w\sim \tilde{q}w\,.
\end{equation}
with $\tilde{q}=e^{-2\pi i/\tau}$. This rescales the argument of the exponential giving
\begin{equation}
    \prod_{b}\tilde{q}_b^{h_b-n_b-c/24} \approx \exp\left[-2\pi i/\tau\left(\sum_a^{2n} t_a P_a^2-1/24\right)\right] 
\end{equation}
The exponential can be maximized by minimizing the term in parenthesis. As $\Im\tau\to 0$ such terms are heavily preferred in the sum over states defining the block.

To guarantee vacuum dominance, it is necessary to specify how the remaining moduli scale in the near-extremal limit. Recall that the vacuum corresponds to the imaginary Liouville momentum $iQ/2$. 
In the right-moving sector, we scale odd $t_a$ as
\begin{equation}
    2\pi it_a  = -b^2\beta_a\,,
\end{equation}
with $\beta_a\sim O(1)$. Letting $k$ be a positive integer, we can equivalently state that $t_{2k-1}$ is taken parametrically small in the central charge. This should be contrasted with the left-moving sector where $t_{2k-1}\sim O(c)$. The comparatively large value of the even $t_{2k}$ moduli makes it preferable to set the associated momenta as small as possible. In particular this sets $P_{2k}=iQ/2$.  
Consequently, we cannot set $P_1$ and $P_3$ to the identity since the fusion rules dictate $\mel{\mathbbm 1}{\cO_i}{\mathbbm 1} = 0$. Indeed, only the operator $\cO$ can flow in the fusion between $\cO$ and the identity so we are forced to set $P_1=P_3=P_i$ giving
\begin{align}
    \prod_{b}\tilde{q}_b^{h_b-n_b-c/24} &\approx \exp\left[-2\pi i/\tau\left( \sum_{k}^nt_{2k}P_{2k}^2 -1/24\right)\right] \nn\\
    &\underset{n.e}{=} \exp\left[-\frac{\pi^2}{b^4\beta_R}\left(\sum_{k}^n t_{2k}-b^2/6\right)\right] \,,
\end{align}
where we have rescaled $2\pi\Im(\tau)=b^2\beta_R$ so that $\beta_R\sim O(1)$. The first line has dropped terms of the form $t_{2k-1}P_{i}$ which vanish as $b\to 0$ due to the choice of small odd moduli. In the second line we took the near-extremal limit. The combination of the $b$-dependent prefactor and the constant offset in the parenthesis make this term increasingly relevant to defining the block as the central charge is taken large.

The extension to any even number of operator insertions is simple, just let $a,b\in\{1,2,\cdots,2n\}$. However, while the blocks are simple in this channel, the match to JT disk correlators is more obscure. In the main text, we work in the left-moving sector and start with the assumption of vacuum dominance in $\tilde{\cC}$ before showing that the equivalent expression in $\cC$ has a clear near-extremal interpretation.

\section{Conical deficits and non-universal data}\label{app:defects}
The universal 3-point function discussed in appendix \ref{app:vmft} has an extended regime of validity below the black-hole threshold. Specifically we consider matrix elements associated to momenta $P_j$ and $P_k$ such that there is no $m\in\Z_{\geq 0}$ such that
\begin{equation}
    \alpha = \alpha_j + \alpha_k + mb < \frac{Q}{2}\,. 
\end{equation}
These states are dual to operators whose dimension scales with the central charge 
\begin{equation}
    P = i\left(\frac{1-2\eta}{2b}+b/2\right) \quad{\rm or}\quad \alpha=\eta/b \,,
\end{equation}
with $\eta\in(1/4,1/2)$. The upper-bound corresponds to the black-hole threshold. The lower bound ensures that no multi-twist operators appear in the fusion of any two defects \cite{Collier:2018exn}. In this regime, the universal structure constant limits to
\begin{align}\label{eq:C0_defects}
    &b^2\log C_0(P_i,P_j,P_k) \approx \nonumber\\
    &\quad \left(\psi(2 \eta_i) + \psi(2-2 \eta_i) - \psi(1+\eta_i-\eta_j-\eta_k) - \psi(-\eta_i+\eta_j+\eta_k)\right)  \\
    &\quad +(\text{2 cyclic permutations}) - \psi(2-\sum \eta) - \psi(-1+\sum \eta) - \psi(2)\,,\nonumber
\end{align}
as $b\to 0$ and when the operator $\cO_i$ also has momenta in this range. The special function $\psi$ is a polygamma function of negative order
\begin{equation}
    \psi(z) = \int_0^z \log\Gamma(t)dt\,.
\end{equation}
The universal formula is not real outside of this regime. As noted in \cite{Collier:2019weq}, it picks up the imaginary part
\begin{equation}\label{eq:im_C0}
    \Im(b^2\log C_0(P_i,P_j,P_k)) \approx \pi(\sum \eta-1)\,.
\end{equation}
This limit can also be derived from a saddle-point analysis of 3d gravity with conical defects \cite{Chandra:2022bqq}.

We suspect that only above-threshold matrix elements of probe operators exhibit signatures of chaos like adherence to an ETH ansatz. As a test of this conjecture, we investigate whether the unlink statistics computed in section \ref{subsec:topology} take the form of an ETH ansatz in the conical deficit regime. A direct comparison with the limits considered in the main text requires computing matrix elements of a parametrically light probe in a below-threshold microcanonical window. Similarly to the macroscopic BTZ regime outlined in appendix \ref{app:vmft_C0}, the difference in the momenta can be fixed to $\eta_j-\eta_k=\omega b$ with $\omega=O(1)$. However, taking $\eta_i\to 0$ is more delicate given the lower bound on the $\eta$s. In fact, there is no longer a wormhole saddle when $\sum\eta<1$. Instead the leading topology is the product of spheres studied in \cite{Chang:2016ftb}. The bulk geometry contains a short-distance singularity that is surrounded by a horizon in Lorentzian signature. An unknown structure constant is associated to this singularity making this a non-universal contribution to operator product statistics. This stands in contrast to the bulk geometries studied in the main text which lacked singular intersections of below-threshold particles.

\begin{figure}
    \centering
    \begin{subfigure}{0.49\linewidth}
        \centering
        \includegraphics[width=\linewidth]{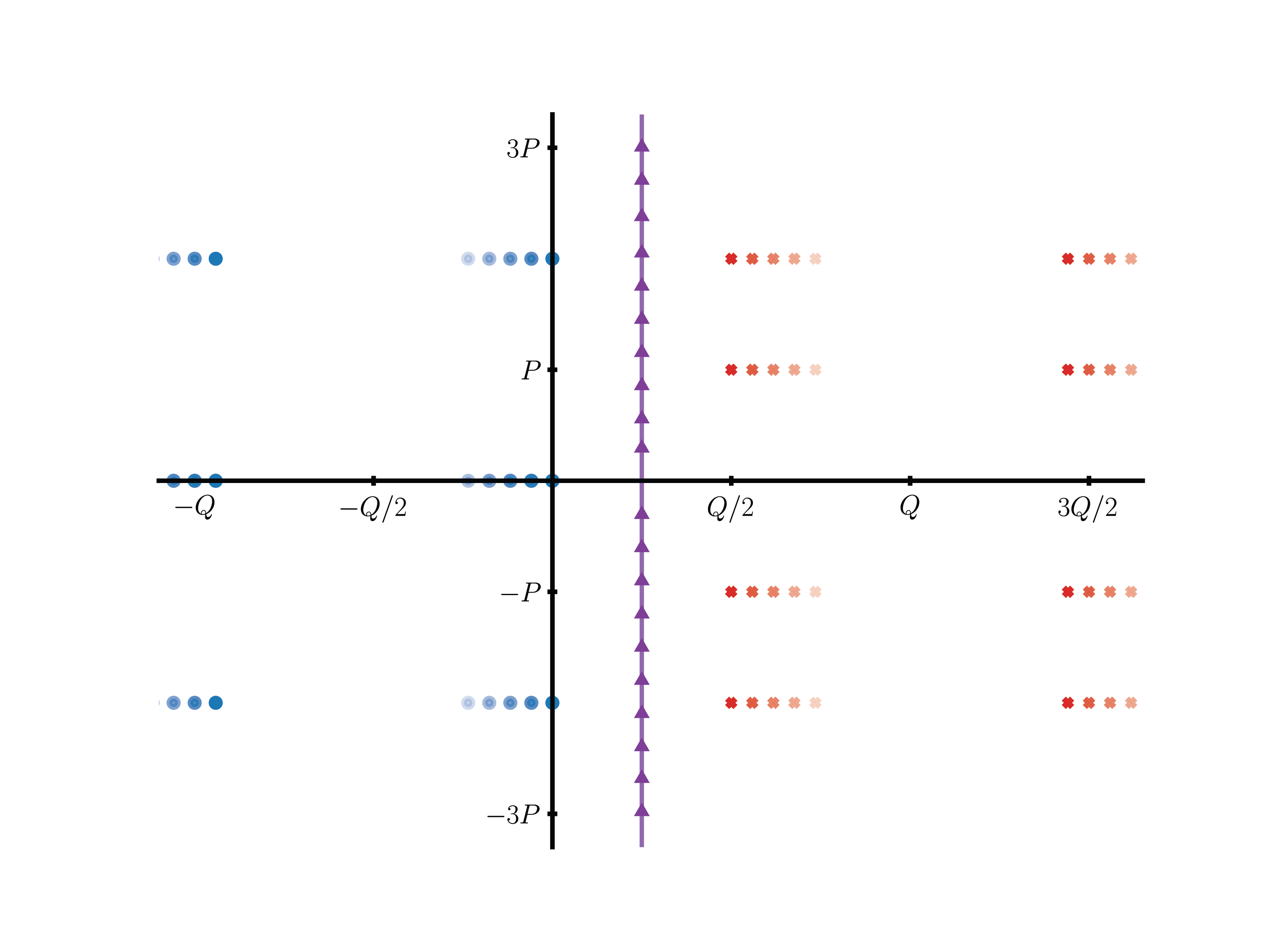}
        \caption{\textit{(1)}}
    \end{subfigure}
    \begin{subfigure}{0.49\linewidth}
        \includegraphics[width=\linewidth]{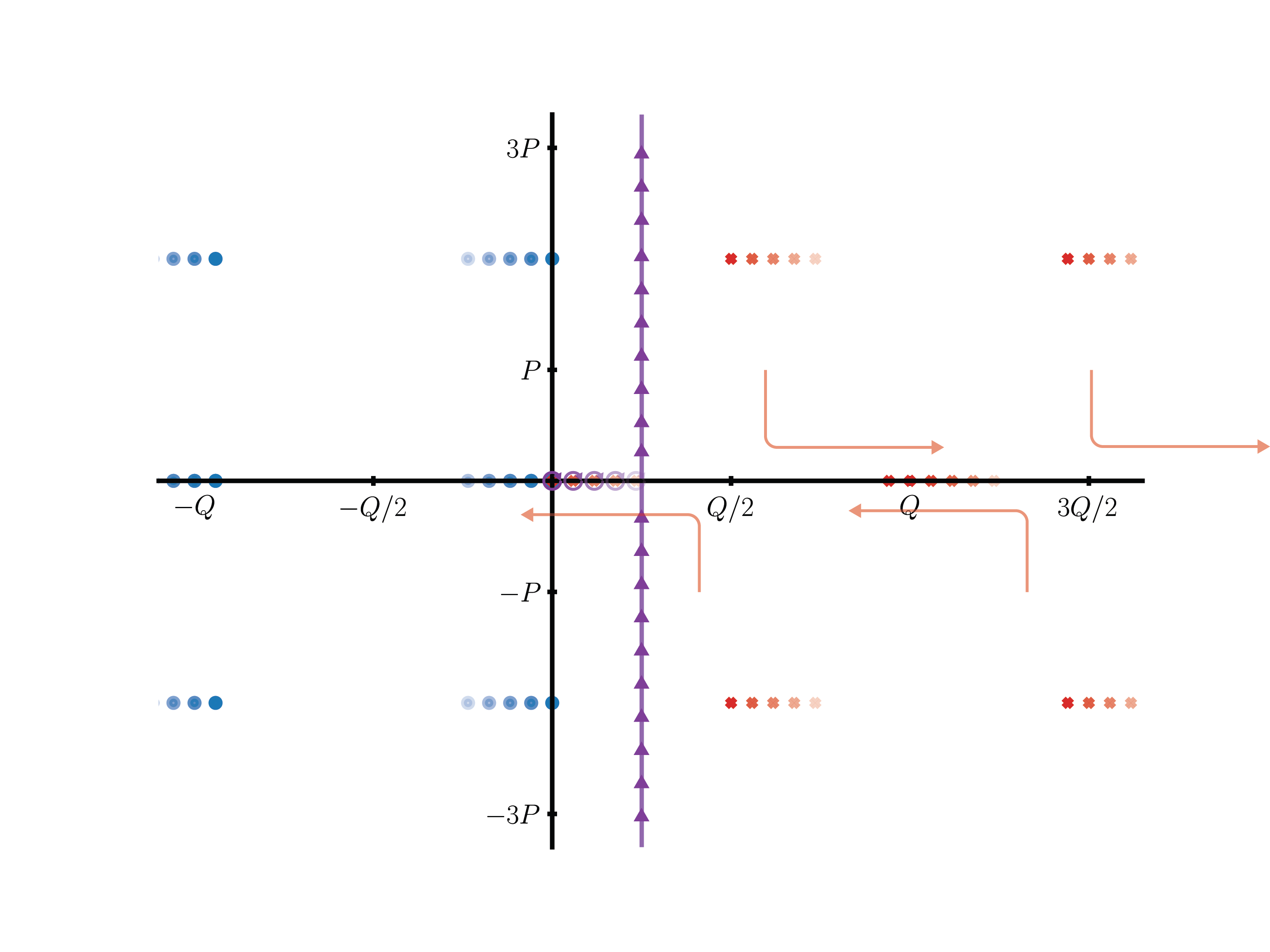}
        \caption{\textit{(2)}}
    \end{subfigure}

    \begin{subfigure}{0.49\linewidth}
        \includegraphics[width=\linewidth]{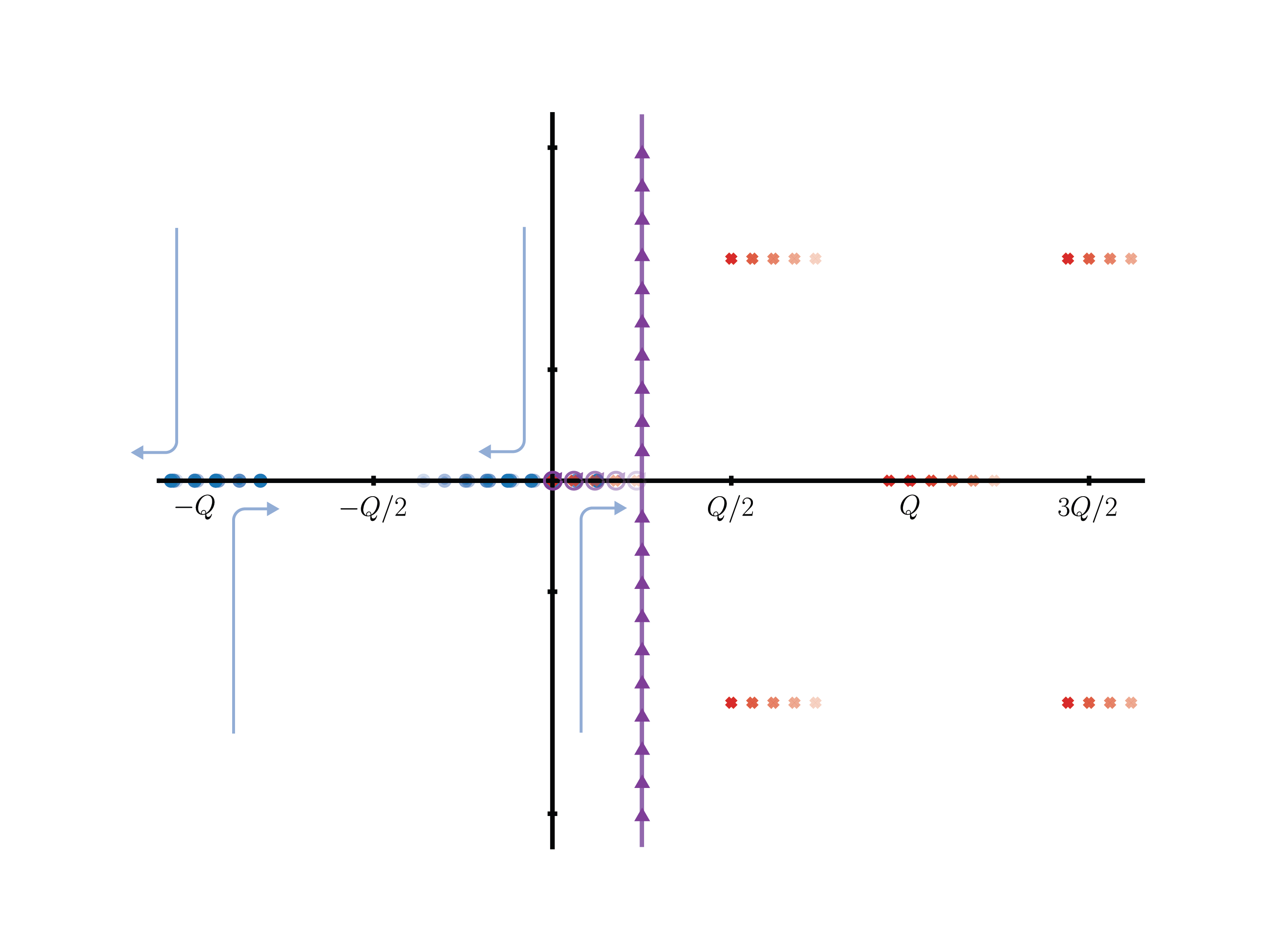}
        \caption{\textit{(3)}}
    \end{subfigure}
    \begin{subfigure}{0.49\linewidth}
        \includegraphics[width=\linewidth]{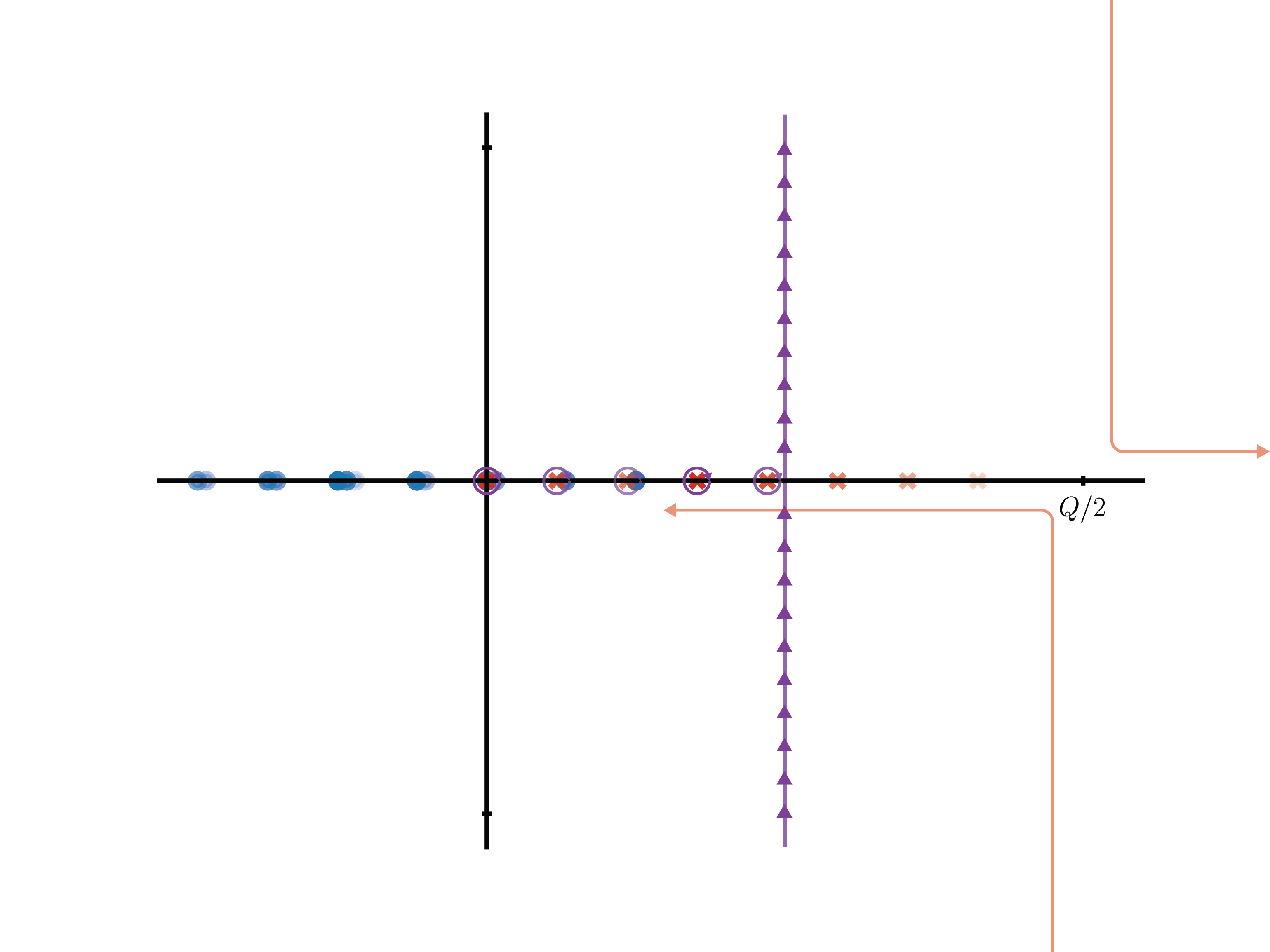}
        \caption{\textit{(4)}}
    \end{subfigure}
    \caption{Here we describe contour deformations in the complex $s$-plane of \eqref{eq:fker_contour} as momenta are tuned below the black-hole threshold. \textit{(1)} In the case of pair-wise identical external operators relevant for this work, above-threshold momenta result in 7 semi-finite lines of poles. The middle line of poles to the left of the integration contour appear with multiplicity 2. \textit{(2)} As we tune the t-channel momentum to the vacuum, the middle pair of poles on the right-hand side intersect on the real axis then separate. A triple pole at $s=0$ gives the degeneration limit of the $\mathbb{F}$-kernel. \textit{(3)} The contour is unchanged if the external operators are tuned below the black-hole threshold but above the multitwist threshold. \textit{(4)} Finally, we deform the s-channel momentum to that of a light probe. Analyticity again requires the introduction of new poles to the integration contour. However, the triple-pole at $s=0$ continues to be the dominant contribution to the kernel.}
    \label{fig:conical_contour_defs}
\end{figure}

Despite our lack of bulk understanding, we can continue to make progress using analytic properties of the fusion kernel. In integral form, the kernel is
\begin{equation}\label{eq:fker_contour}
    \fker{P}{P_i}{P_j}{P_k}{P_j}{P_k} = \cP\bar{\cP}\times\int_\cC\frac{ds}{i}\ \prod_{a\in\{1,2,3,4\}}\frac{S_b(s+U_a)}{S_b(s+V_a)}
\end{equation}
where
\begin{equation}
    \begin{array}{c c c}
        U_1 = 0\,, &\quad& V_1 = Q-\alpha_i+\alpha_k-\alpha_j\,, \\
        U_2 = Q-2\alpha_j\,, &\quad& V_2 = \alpha_i+\alpha_k-\alpha_j\,, \\
        U_3 = 2\alpha_k-Q\,, &\quad& V_3 = \alpha\,,\\
        U_4 = 0\,, &\quad& V_4 = Q-\alpha \,,
    \end{array}
\end{equation}
and $\alpha=Q/2+iP$. The coefficients $\cP$ and $\bar{\cP}$ are as in (2.11) of \cite{Collier:2018exn} and the integration contour anchors at $\pm i\infty$ while skirting to the right of $U_a$ poles and to the left of $V_a$ poles. To arrive at \eqref{eq:C0_defects}, we tune the momenta below threshold such that
\begin{equation}
    \begin{array}{c c c}
        U_1 = 0\,, &\quad& V_1 = Q+\eta_k-\eta_j-\eta_i\,, \\
        U_2 = Q-2\eta_j/b\,, &\quad& V_2 = \eta_k-\eta_j+\eta_i\,, \\
        U_3 = 2\eta_k/b-Q\,, &\quad& V_3 = Q/2+iP \,. \\
        U_4 = 0\,, &\quad& V_4 = Q/2-iP \,.
    \end{array}
\end{equation}
As all 4 momenta are tuned to the black hole threshold, the 8 semi-finite lines of poles in the complex $s$-plane associated to each factor of $S_b$ collapse onto the real axis. Further tuning into the discrete below-threshold range causes poles to cross the integration contour as pictured in figure \ref{fig:conical_contour_defs}. The residues of these poles must be picked up to maintain analyticity of the kernel. 

Universal behavior is described by assigning the vacuum to the t-channel momentum $P$. In the conical deficit regime of interest, the integral is dominated by the intersection of three poles at the origin. Specifically, the poles of $U_1$ and $U_4$ at $s=0$ intersect the pole of $V_3$ at $s=\alpha$ as $P\to iQ/2$. This same intersection gives the degeneration limit of the $\mathbb{F}$-kernel studied in appendix \ref{app:vmft}, albeit for widely different kinematics. Crucially, this contribution stays dominant as the s-channel momentum is tuned to a light probe ($\alpha_i=b\Delta$), meaning that we can continue to study the same universal primitives in this regime. Figure \ref{fig:conical_contour_defs} tracks the movement of the $V_3$ pole through these deformations. 

Setting $\alpha_i=b\Delta$, the three-point function simplifies greatly
\begin{equation}
    C_0(\alpha_i=b\Delta, \eta=b(\alpha_j+\alpha_k)/2, \omega=b^{-1}(\alpha_j-\alpha_k)) = e^{-\tilde{S}(\eta)}(f_i(\eta,\omega))^2\,.
\end{equation}
As is expected in the semiclassical limit, the function $\tilde{S}(\eta)$ contains a divergence that should be regulated. Subtracting the regulator term $S_0$ defined in \eqref{eq:C0_regulator} gives
\begin{equation}
    S(\eta) = \tilde{S}(\eta) - S_0 = -2\pi i\eta/b^2\,.
\end{equation}
Unlike the macroscopic BTZ limit discussed in appendix \ref{app:vmft_C0}, the regulated exponential term in the conical deficit regime is imaginary and so does not admit an interpretation as a microcanonical entropy. We take this as confirmation of our earlier conjecture that conical defects do not exhibit chaos in the sense of a standard ETH ansatz. The smooth function is
\begin{equation}
    (f_i(\eta,\omega))^2 = (2\pi b)^{-5/4}\left[\frac{e^{\pi i\Delta}(1-2\eta)^{2\Delta}}{\sin(2\pi\eta)}\right]\frac{\Gamma(\Delta+\omega)\Gamma(\Delta-\omega)}{\Gamma(2\Delta)}\,.
\end{equation}
Cosmetically, this expression is similar to \eqref{eq:btz_fi}. The scaling with $b$ and the spectrum of complex poles are shared (although now poles are associated to real normal $\omega$) since both follow from the global $\mathfrak{sl}(2)$ limit of the light sector. However, the quantity in brackets behaves very differently to the analogous term in \eqref{eq:btz_fi}. Generically, this term is imaginary, reflecting the non-trivial imaginary part of $C_0$ noted in \eqref{eq:im_C0}. 

\newpage
\bibliography{ref}
\bibliographystyle{utphys2}

\end{document}